\documentclass[numberedappendix,appendixfloats]{emulateapj}

\usepackage{amsmath}
\usepackage{amssymb}
\usepackage{graphicx}
\usepackage{appendix}
\usepackage{color}
\usepackage{multirow}
\usepackage{soul}

\newcommand{\lSect}[1]{{\label{sec:#1}}}
\newcommand{\lFig}[1]{{\label{fig:#1}}}
\newcommand{\lEq}[1]{{\label{eq:#1}}}
\newcommand{\lTab}[1]{{\label{tab:#1}}}
\newcommand{\FIGFF}[2]{{\ref{fig:#2}{#1}}}

\newcommand{\FIG}[2]{{Fig.~\FIGFF{#1}{#2}}}
\newcommand{\Fig}[1]{{\FIG{}{#1}}}

\newcommand{\Sectff}[1]{{\ref{sec:#1}}}
\newcommand{\Sect}[1]{{\S\Sectff{#1}}}

\newcommand{\Eqref}[1]{{\ref{eq:#1}}}

\newcommand{\eqff}[1]{{\Eqref{#1}}}

\newcommand{\eq}[1]{{equation~\eqff{#1}}}

\newcommand{\Msun}{\ensuremath{\mathrm{M}_\odot}}
\newcommand{\Lsun}{\ensuremath{\mathrm{L}_\odot}}

\newcommand{\Tab}[1]{{Table~\ref{tab:#1}}}

\newcommand{\msf}{{\ensuremath{M_\mathrm{4}}}}
\newcommand{\scc}{{\ensuremath{\mu_{4}}}} 

\def\gtaprx {\lower .1ex\hbox{\rlap{\raise .6ex\hbox{\hskip .3ex
	{\ifmmode{\scriptscriptstyle >}\else
		{$\scriptscriptstyle >$}\fi}}}
	\kern -.4ex{\ifmmode{\scriptscriptstyle \sim}\else
		{$\scriptscriptstyle\sim$}\fi}}}
\def\ltaprx {\lower .1ex\hbox{\rlap{\raise .6ex\hbox{\hskip .3ex
	{\ifmmode{\scriptscriptstyle <}\else
		{$\scriptscriptstyle <$}\fi}}}
	\kern -.4ex{\ifmmode{\scriptscriptstyle \sim}\else
		{$\scriptscriptstyle\sim$}\fi}}}

\graphicspath{{tmp_figs/}}

\begin{document}

\title{The Explosion of Helium Stars Evolved with Mass Loss}

\author{T.\ Ertl\altaffilmark{1}, 
        S.\ E.\ Woosley\altaffilmark{2},
        Tuguldur Sukhbold\altaffilmark{3,4,5},
        and H.-T. Janka \altaffilmark{1}}
\altaffiltext{1}{Max-Planck-Institut f{\"u}r Astrophysik, Postfach
  1317, 85741 Garching, Germany}
\altaffiltext{2}{Department of Astronomy and Astrophysics, University
  of California, Santa Cruz, CA 95064}
\altaffiltext{3}{Department of Astronomy, Ohio State University,
  Columbus, Ohio, 43210}
\altaffiltext{4}{Center for Cosmology and AstroParticle Physics, Ohio State University, Columbus OH 43210}
\altaffiltext{5}{NASA Hubble Fellow}

\begin{abstract}
Light curves, explosion energies, and remnant masses are calculated
for a grid of supernovae resulting from massive helium stars that have
been evolved including mass loss. These presupernova stars should
approximate the results of binary evolution for stars in interacting
systems that lose their envelopes close to the time of helium core
ignition. Initial helium star masses are in the range 2.5 to
40\,\Msun, which correspond to main sequence masses of about 13 to
90\,\Msun.  Common Type Ib and Ic supernovae result from stars whose
final masses are approximately 2.5 to 5.6\,\Msun.  For heavier stars, a
large fraction of collapses lead to black holes, though there is an
island of explodability for presupernova masses near 10\,\Msun.  
The median neutron star mass in binaries is
1.35--1.38\,\Msun \ and the median black hole mass is between 9 and 11\,\Msun. 
Even though black holes less massive than 5 \Msun\ are rare, they are predicted down to the maximum neutron star mass. There is no empty ``gap'', only a less populated mass range.
For standard assumptions
regarding the explosions and nucleosynthesis, the models predict light
curves that are fainter than the brighter common Type Ib and Ic
supernovae. Even with a very liberal, but physically plausible increase in
$^{56}$Ni production, the highest energy models are fainter, at peak, than
10$^{42.6}$\,erg\,s$^{-1}$, and very few approach that limit.
The median peak luminosity ranges from 10$^{42.0}$ to 10$^{42.3}$\,erg\,s$^{-1}$.
Possible alternatives to the standard neutrino-powered and radioactive-illuminated 
models are explored. Magnetars are a promising alternative. 
Several other unusual varieties of Type I supernovae at both high and 
low mass are explored.
\end{abstract}


\section{Introduction}
\lSect{intro}

A large fraction, perhaps half or more of all stars massive enough to
experience iron-core collapse, also interact with a close binary
companion some time during their lives \citep{San11,San12}. This
interaction alters their evolution and the sorts of supernovae and
remnants they produce \citep{Pod92,Wel99,Lan12,Dem17}. The effect on
the birth function for compact remnants is particularly important
since the most accurate stellar mass determinations come from stars in
relatively close binary systems.

A common outcome for strongly interacting binaries is for one or both
stars to lose most of their hydrogen envelopes when they first become
red supergiants. Continued mass loss by binary interaction and winds
then removes any residual hydrogen and a portion of the helium (He) and
carbon-oxygen (CO) core as well. As a result when the star dies, its
presupernova mass is smaller than the He core mass would have been had 
the star evolved in isolation. Its composition also differs and 
the braking of the rotation of the exploding core by a low density 
hydrogen envelope does not occur. Mixing and fall back are altered 
and of course, if all the hydrogen is lost, the supernova is Type I, 
not Type II.

Many studies of binary evolution have been and are being carried out,
spurred on most recently by the burgeoning field of gravitational wave
astronomy. The actual outcome from any given binary depends upon its
initial masses, orbital separation, and the intricacies of mass loss
and transfer, especially if a common envelope forms. Population
synthesis and a large number of models are required to explore 
broad range of possibilities realistically. Here, however, we continue
to adopt a gross simplification to binary evolution \citep[see
  also][]{Woo19}. It is assumed that the consequence of binary
interaction is to promptly remove the entire hydrogen envelope of the
star at helium ignition. This surely does not happen all the
time. Some Type IIb supernovae still have hydrogen despite
experiencing a binary interaction, though that hydrogen mass is
generally small. Some stars do not become red giants at the beginning
of helium burning, but later on, and the evolution of the secondary
will be different from that of the primary. Changing the metallicity
will alter the radii of the stars, the binding energies of their
envelopes, and their mass transfer history.

Nevertheless, this assumption allows us to explore some of the major
consequences of binary evolution using a simple, minimally
parametrized grid of models.  The price paid is that the connection
between a given helium star mass and its main sequence mass is
imprecise.  The mass loss during helium burning may also be distorted
since the core may actually have burned some helium before losing its
last hydrogen. On the other hand mass loss rates for helium stars are
uncertain and, for the time being, that uncertainty dominates the
outcome.

Our main purpose is to explore the different sorts of outcomes
that might occur in binaries with a large range of masses. What
distribution of black hole and neutron star remnant masses do they
produce? What are their explosion energies and how much $^{56}$Ni do
they make? What do their supernovae look like? These questions were
answered for non-rotating single stars of solar metallicity by
\citet{Suk16}. Here we determine the first order corrections to that
study for the effects of binary membership. 

An important related goal is understanding the origin of Type Ib and
Ic supernovae.  It is thought that most of these are derived
from stripped stars in binaries \citep{Lan12,Bra17}. Here the
appropriate mass range and energies are determined assuming that the
supernovae result exclusively from neutrino-heating and the light
curves are powered by radioactivity. Perhaps surprisingly, for current
observational statistics, this simple prescription can explain only about
half of the common events. Inadequate $^{56}$Ni is produced to explain
the brighter half. This leads us to conclude that some other energy
source could be frequently powering or, at least, illuminating these
brighter events. The intriguing possibility of heating common Type Ib and Ic supernovae using a newly formed magnetar is explored.

Other mass ranges, both smaller and larger than inferred for common
Type Ib and Ic events, give rise to different sorts of events that
may or may not have been detected yet. The lighter stars experience
radius expansion and have bright initial peaks from envelope
recombination or circumstellar interaction. The heavier ones produce
faint, red supernovae with broad light curve peaks that may have
escaped discovery. In the magnetar paradigm, these big stars could
also occasionally be superluminous supernovae, broad-line Type Ic
supernovae, or even gamma-ray bursters.

In \Sect{physandmods}, the progenitor stars used in the study are
discussed.  They are taken from the recent survey of \citet{Woo19}. In
\Sect{expl}, we discuss the simulation of the explosions using the
one-dimensional (1D) neutrino-transport code P-HOTB, postprocessed
using KEPLER to get details of nucleosynthesis and calculate light
curves. This is the same approach used by \citet{Suk16} to survey the
evolution of single stars, but there have been some substantial
changes to P-HOTB and the way $^{56}$Ni yields are calculated. These
revisions are discussed in \Sect{expl} and in the Appendix.
\Sect{results} gives our general results and discusses their
dependence on the choice of a model for the central
engine. \Sect{nucleo} discusses the nucleosynthesis of iron, oxygen,
and magnesium. Observations limit the amount of iron that
core-collapse supernovae can eject relative to oxygen and magnesium
and that, in turn, limits the radioactive contribution to the light
curve.  \Sect{remnants} summarizes our results for the birth functions
for neutron stars and black holes.  \Sect{litephys} and \Sect{lite}
give our results for light curves and their dependence on assumed
physics, especially the evaluation of the $^{56}$Ni yield. As
previously mentioned, it proves difficult to obtain the full
brightness for some of the Type Ib and Ic supernovae in the literature
using just the yields from the neutrino-driven explosion models and
possible alternatives are discussed. In \Sect{conclude}, we summarize
our findings and discuss their consequences.

\section{Presupernova Models}
\lSect{physandmods}

\begin{deluxetable*}{ccccccccc}
\tablecaption{Properties of progenitor models (subset)}
\tablehead{ \colhead{${M_{\rm He,i}}$}  &
            \colhead{${M_{\rm preSN}}$}  &
            \colhead{${M_{\rm ZAMS}}$}  &
            \colhead{${M_{\rm CO}}$}  &
            \colhead{${M_{\rm Fe}}$}  &
            \colhead{${\log\ L_{\rm preSN}}$}  &
            \colhead{${\log\ T_{\rm eff}}$}  &
            \colhead{${Y_{\rm s}}$}\vspace{1mm}
            \\
            \colhead{[\Msun]}  &
            \colhead{[\Msun]}  &
            \colhead{[\Msun]}  &
            \colhead{[\Msun]}  &
            \colhead{[\Msun]}  &
            \colhead{[${\rm erg\,s^{-1}}$]}  &
            \colhead{[K]}  &
            \colhead{ }
            }\\

\startdata
\multicolumn{8}{c}{Si-flash}\\
\multicolumn{8}{c}{}\\
2.5  &  2.10  &  13.5  &  1.37  &  1.29  &  4.37  &  4.73  &  0.99  \\
3.0  &  2.45  &  15.1  &  1.61  &  1.32  &  4.61  &  3.99  &  0.99  \\
3.1  &  2.52  &  15.5  &  1.66  &  1.34  &  4.63  &  4.16  &  0.99  \\
3.2  &  2.59  &  15.8  &  1.70  &  1.37  &  4.65  &  4.34  &  0.99  \\
\multicolumn{8}{c}{}\\
\multicolumn{8}{c}{inflated envelope}\\
\multicolumn{8}{c}{}\\
2.6  &  2.15  &  13.9  &  1.41  &  1.30  &  4.37  &  3.85  &  0.99  \\
2.7  &  2.22  &  14.2  &  1.46  &  1.29  &  4.51  &  3.87  &  0.99  \\
2.8  &  2.30  &  14.5  &  1.51  &  1.31  &  4.55  &  3.89  &  0.99  \\
2.9  &  2.37  &  14.8  &  1.56  &  1.34  &  4.59  &  3.92  &  0.99  \\
\multicolumn{8}{c}{}\\
\multicolumn{8}{c}{standard $\dot{M}$}\\
\multicolumn{8}{c}{}\\
3.3  &  2.67  &  16.1  &  1.75  &  1.33  &  4.66  &  4.29  &  0.99  \\
3.4  &  2.74  &  16.4  &  1.80  &  1.35  &  4.68  &  4.29  &  0.99  \\
3.5  &  2.81  &  16.7  &  1.85  &  1.36  &  4.69  &  4.55  &  0.99  \\
3.6  &  2.88  &  17.0  &  1.90  &  1.34  &  4.72  &  4.58  &  0.99  \\
3.7  &  2.95  &  17.3  &  1.96  &  1.35  &  4.74  &  4.62  &  0.99  \\
3.8  &  3.02  &  17.5  &  2.01  &  1.37  &  4.75  &  4.65  &  0.99  \\
3.9  &  3.09  &  17.8  &  2.07  &  1.37  &  4.77  &  4.67  &  0.99  \\
4.0  &  3.15  &  18.1  &  2.12  &  1.38  &  4.78  &  4.69  &  0.99  \\
4.1  &  3.22  &  18.4  &  2.17  &  1.43  &  4.80  &  4.70  &  0.99  \\
4.2  &  3.29  &  18.7  &  2.22  &  1.39  &  4.81  &  4.72  &  0.99  \\
4.3  &  3.36  &  19.0  &  2.27  &  1.39  &  4.83  &  4.73  &  0.99  \\
4.4  &  3.42  &  19.2  &  2.33  &  1.40  &  4.85  &  4.74  &  0.99  \\
4.5  &  3.49  &  19.5  &  2.38  &  1.47  &  4.85  &  4.75  &  0.99  \\
5.0  &  3.82  &  20.8  &  2.65  &  1.55  &  4.91  &  4.86  &  0.99  \\
5.5  &  4.14  &  22.1  &  2.90  &  1.57  &  4.96  &  4.92  &  0.99  \\
6.0  &  4.45  &  23.3  &  3.15  &  1.52  &  5.02  &  4.95  &  0.99  \\
6.5  &  4.75  &  24.5  &  3.39  &  1.61  &  5.06  &  4.97  &  0.99  \\
7.0  &  5.05  &  25.7  &  3.63  &  1.50  &  5.10  &  4.99  &  0.99  \\
7.5  &  5.35  &  26.8  &  3.87  &  1.48  &  5.13  &  5.00  &  0.99  \\
8.0  &  5.64  &  27.9  &  4.11  &  1.41  &  5.17  &  5.00  &  0.99  \\
9.0  &  6.20  &  29.7  &  4.59  &  1.45  &  5.21  &  5.01  &  0.99  \\
10  &  6.75  &  31.7  &  5.07  &  1.44  &  5.27  &  5.02  &  0.99  \\
11  &  7.05  &  33.7  &  5.45  &  1.63  &  5.27  &  5.24  &  0.64  \\
12  &  7.27  &  35.7  &  5.55  &  1.61  &  5.28  &  5.22  &  0.38  \\
14  &  8.04  &  39.7  &  6.11  &  1.68  &  5.36  &  5.24  &  0.19  \\
16  &  8.84  &  43.7  &  6.79  &  1.69  &  5.39  &  5.25  &  0.22  \\
18  &  9.62  &  47.7  &  7.42  &  1.41  &  5.42  &  5.25  &  0.23  \\
20  &  10.4  &  51.7  &  8.01  &  1.43  &  5.47  &  5.30  &  0.25  \\
24  &  12.1  &  59.7  &  9.40  &  1.60  &  5.58  &  5.34  &  0.22  \\
28  &  13.9  &  67.7  &  10.9  &  1.81  &  5.68  &  5.36  &  0.21  \\
32  &  15.7  &  75.7  &  12.4  &  1.95  &  5.76  &  5.37  &  0.20  \\
36  &  17.6  &  83.7  &  14.0  &  1.96  &  5.83  &  5.39  &  0.20  \\
40  &  19.6  &  91.7  &  15.7  &  2.27  &  5.89  &  5.40  &  0.19  \\
\multicolumn{8}{c}{}\\
\multicolumn{8}{c}{$1.5\times \dot{M}$}\\
\multicolumn{8}{c}{}\\
5.0  &  3.43  &  20.8  &  2.37  &  1.44  &  4.85  &  4.89  &  0.99  \\
6.0  &  3.96  &  23.3  &  2.77  &  1.61  &  4.95  &  4.98  &  0.99  \\
7.0  &  4.45  &  25.7  &  3.17  &  1.51  &  5.03  &  5.05  &  0.99  \\
8.0  &  4.92  &  27.9  &  3.56  &  1.54  &  5.09  &  5.09  &  0.98  \\
10  &  4.96  &  31.7  &  3.73  &  1.58  &  5.08  &  5.06  &  0.22  \\
12  &  5.43  &  35.7  &  4.04  &  1.44  &  5.12  &  5.07  &  0.21  \\
14  &  5.86  &  39.7  &  4.40  &  1.44  &  5.18  &  5.15  &  0.22  \\
16  &  6.34  &  43.7  &  4.79  &  1.43  &  5.20  &  5.15  &  0.21  \\
18  &  6.84  &  47.7  &  5.19  &  1.53  &  5.26  &  5.16  &  0.20  \\
20  &  7.39  &  51.7  &  5.62  &  1.48  &  5.29  &  5.17  &  0.19  \\
24  &  8.54  &  59.7  &  6.55  &  1.69  &  5.37  &  5.18  &  0.18  \\
28  &  9.77  &  67.7  &  7.50  &  1.41  &  5.43  &  5.19  &  0.17  \\
32  &  11.1  &  75.7  &  8.55  &  1.50  &  5.53  &  5.18  &  0.16  \\
36  &  12.4  &  83.7  &  9.60  &  1.64  &  5.60  &  5.18  &  0.16  \\
\enddata
\tablecomments{The quantities are evaluated at oxygen ignition. 
  $Y_{\rm s}$ is the helium mass fraction at the surface of the 
  presupernova star. A full version of the table that includes all 
  progenitors is available in the online version.}  \lTab{bigtable}
\end{deluxetable*}

\begin{figure}[h] 
\centering 
\includegraphics[width=\columnwidth]{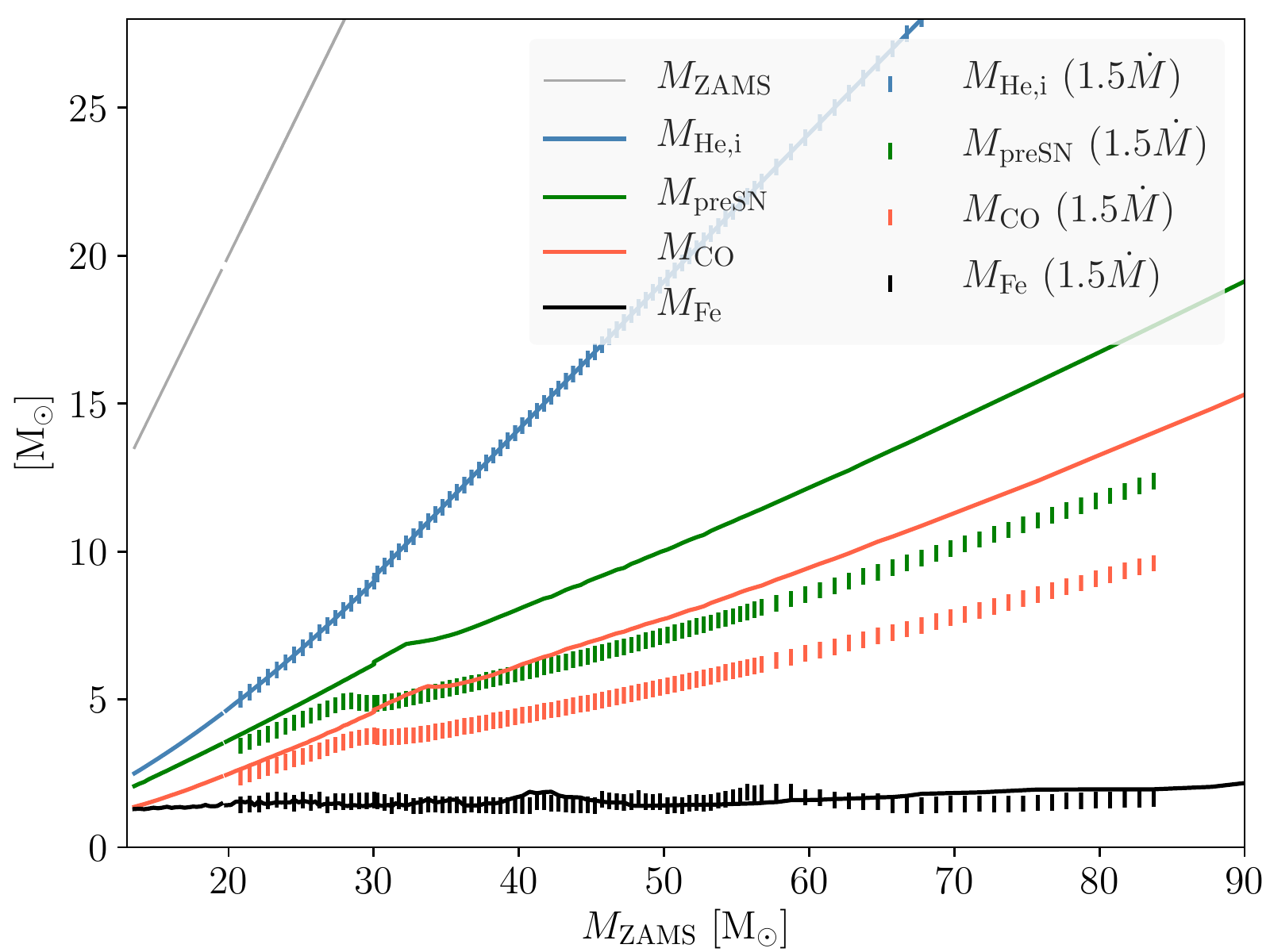}
\includegraphics[width=\columnwidth]{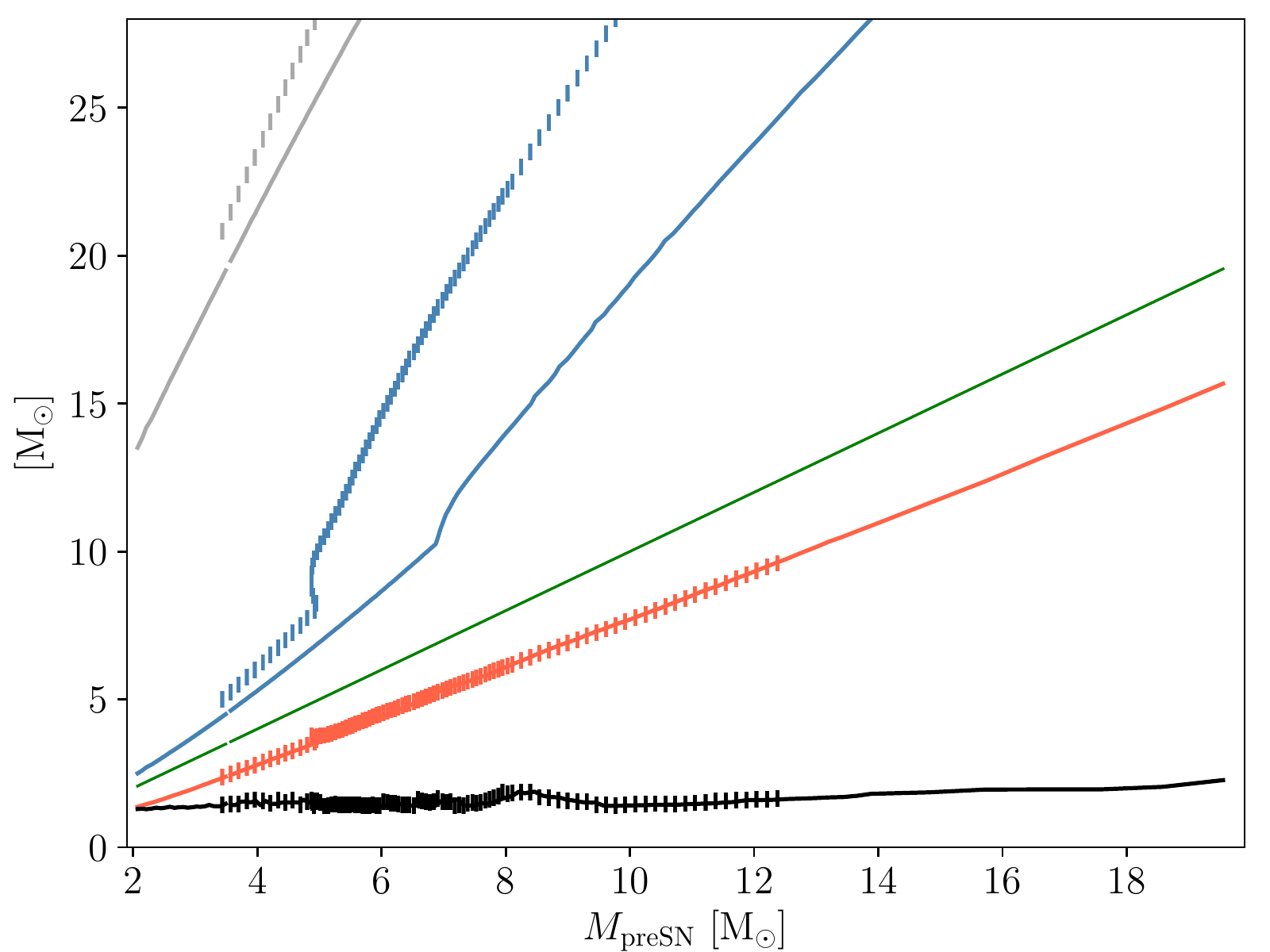}
\caption{{\em Top:} Initial helium star mass ($M_{\rm He,i}$), presupernova
  mass ($M_{\rm preSN}$), carbon-oxygen core mass ($M_{\rm CO}$), and
  iron core mass ($M_{\rm Fe}$) are given as a function of the
  estimated initial main sequence mass of the star just before its
  envelope is lost. Vertical hatching shows models that used 1.5 times
  the standard mass loss rate. {\em Bottom:} The same core masses as a function of the
  final (presupernova) mass of the star.  \lFig{cores}}
\end{figure}

All presupernova models were calculated using the KEPLER code
\citep{Wea78}, the physics of which has been extensively discussed in
the literature \citep[e.g.,][]{Woo02,Woo07,Woo15,Suk14,Suk18}.  All
stars were non-rotating and had solar metallicity ($Z$ = 0.0145, with a total mass fraction of iron-group species $X_{\rm Fe}=1.46\times 10^{-3}$). Models were
defined by their initial mass, $M_{\rm He,i}$, which was composed
mostly of helium. In the paper models that used the standard mass loss
rate will often be referred to by the name ``Hexx'' where ``xx'' is
$M_{\rm He,i}$. The lower bound to the survey, 2.5\,\Msun, was set by
the lightest star to experience iron-core collapse in the KEPLER code
\citep{Woo19}.  The evolution of still lighter stars 1.6--2.5\,\Msun,
evolved by Woosley was not studied here because none were successfully
evolved to iron core collapse.  Most will become CO or neon-oxygen
white dwarfs. Some uncertain fraction may evolve to electron-capture
supernovae and contribute to the supernova rate and remnant
distribution, but they are not included in the analysis or statistics
presented here.

\Tab{bigtable} and \Fig{cores} summarize the models that were
studied. $M_{\rm He,i}$ and $M_{\rm preSN}$ are the masses of the
initial helium star and the presupernova mass, the difference being
mass loss as a Wolf-Rayet (WR) star. The initial composition was taken
to be the ashes of hydrogen burning in a massive star of solar
metallicity. For details and formulae relating $M_{\rm He,i}$ to the
zero age main sequence (ZAMS) mass, see \citet{Woo19}. The mass loss
rate employed was that of \citet{Yoo17} which specifies different
parameters for WNE and WC/WO stars. For comparison, a smaller grid of
models was computed that used 1.5 times that mass loss rate.  Results
for those models are given at the end of the table. While the focus in
this paper is on stars calculated using the ``standard'' rate, models
with enhanced mass loss gave very similar presupernova stars when
adjusted to the same final presupernova mass. They may have the
advantage of producing a larger fraction of Type Ic supernovae, since
the CO core is more frequently uncovered.

$M_{\rm CO}$ and $M_{\rm Fe}$ give the CO- and iron-core masses for
the presupernova star. The CO-core mass is where the helium mass
fraction falls below 0.01 moving inwards in the star. The iron core
mass is where the mass fraction of iron exceeds 0.5 moving inwards in
the star. CO-core masses are larger, for a given presupernova mass,
than for single stars due to the removal of helium by WR winds during
helium burning. This and the larger carbon mass fraction following
helium burning result in the presupernova stars having more compact
structures, i.e, smaller compactness parameters than their single star
counterparts \citep{Woo19}.

Also given are the luminosity and effective temperature of the
presupernova star. These quantities are evaluated at oxygen ignition
to avoid complications introduced by the silicon flash in lower mass
(M $\le$ 3.2\,\Msun) models. $Y_{\rm s}$ is the helium mass fraction at the
surface of the presupernova star. Values less than 0.9 show the loss
of the entire helium shell and may be relevant for making a Type Ic
instead of a Ib supernova.

Explosions for helium stars smaller than 40\,\Msun \ were calculated
using both KEPLER and P-HOTB (\Sect{expl}). More massive stars would
not have exploded using the present prescription for neutrino-energy deposition following iron
core collapse. Their remnant masses are included in the statistics
assuming full collapse of the presupernova star and nucleosynthesis
ejected in the winds of these ``failures'' is also included in computing
the averages (\Sect{nucleo}). For the standard mass loss rate, helium
stars above 70\,\Msun \ produce presupernova stars over 35\,\Msun\ at
oxygen ignition \ and experience a violent pulsational pair
instability that modifies their final mass. The final collapse of the
iron core was not modeled here, but it is assumed that no outgoing
shock will be launched. The pulsations were also not studied here
\citep[though see][]{Woo19}. When computing remnant mass averages it
is assumed that no black hole over 46\,\Msun \ was formed.


\section{Explosion Modeling}
\lSect{expl}

As in \citet{Suk16}, explosions were calculated using two different
one-dimensional hydrodynamics codes, Prometheus-Hot Bubble (henceforth
P-HOTB; \citealt{Jan96,Kif03}) and KEPLER \citep{Wea78,Woo02}. There
were few changes in the KEPLER postprocessing, except for the
introduction of an alternate way of evaluating iron-group yields using
a deeper mass cut more consistent with P-HOTB. The changes in P-HOTB
were extensive and are discussed both here and in the Appendix.

\subsection{Modeling with P-HOTB}
\lSect{PHOTB}

P-HOTB is a 1D, 2D, or 3D neutrino-hydrodynamics code for simulating
the explosion of supernovae. The original version was derived from the
\textsc{Prometheus} hydrodynamics code \citep{Fry89,Fry91,Mue91}. It
was first developed and employed in 1D and 2D simulations by
\citet{Jan96}, but later upgraded to include more microphysics and a
more generalized numerical grid
\citep[e.g.,][]{Kif03,Sch06,Arc07,Ugl12,Ert16a}. Since 2010, P-HOTB
has also been applied to 3D supernova simulations by \citet{Ham10} and
--supplemented for an axis-free Yin-Yang grid-- by
\citet{Won10a,Won10b,Won13,Won15,Won17} and \citet{Ges18}.

P-HOTB is optimized to efficiently simulate neutrino-driven supernovae
continuously from the onset of stellar core collapse, through core
bounce and the onset of explosion, and onto the late stages of
the ejecta expansion. In order to cover this large range of spatial
and temporal scales for large sets of models, the treatment of the
problem and of the relevant microphysics is approximated in many
aspects, but efforts are made to preserve the essential physics of the
neutrino-driven mechanism.

The chief characteristics of P-HOTB are its use of an approximate,
gray treatment of neutrino transport in the outer layers (optical
depths below about ${\cal{O}}(1000)$) of the new-born neutron star
(NS) and the replacement of the central, high-density core of
the neutron star by an inner grid boundary that shrinks with time.  The movement
of the grid boundary mimics the contraction of the cooling and
deleptonizing compact remnant. Excising the core of the neutron star allows larger
time steps, facilitating the efficient long-time simulation of the
explosion including the central neutrino source.  It also permits a
choice of neutrino luminosities and mean spectral energies as inner
boundary conditions that enable neutrino-driven explosions to be
calculated with a chosen energy in spherical symmetry or in higher
dimensions. The replacement of the inner core of the nascent neutron star is
justified by the fact that the high-density nuclear equation of state
(EOS) remains uncertain despite considerable theoretical and
experimental progress \citep[see, e.g.,][]{Sch19}.

At the onset of a calculation, the rest-mass density $\rho$, radial
fluid velocity $v$, electron fraction $Y_e$, nuclear composition (in
regions where nuclear statistical equilibrium (NSE) does not apply),
and pressure, $P$, as functions of radius from the progenitor data are
mapped onto the computational grid of P-HOTB. Using the pressure
guarantees that hydrostatic equilibrium in the progenitor star and the
developing core contraction at the onset of core collapse are well
represented.  The period from core collapse until core bounce is
treated using the deleptonization scheme of \citet{Lie05}, the
$Y_e(\rho)$ trajectory of Fig.~1 in \citet{Ert16a}, and the nuclear
EOS of \citet{Lat91} with an incompressibility modulus of
220\,MeV. After core bounce, neutrinos and antineutrinos of all
flavors are followed with the gray transport approximation.

About 10\,ms after bounce (specifically, at the time when the shock
encloses 1.25\,M$_\odot$), the central, high-density core of the
proto-neutron star (PNS) with a mass of 1.1\,M$_\odot$ is excluded from the
computational domain and replaced by a gravitating point mass. The
gravitational force of this point mass as well as the self-gravity of
the matter on the grid are corrected for general relativistic effects
\citep[see][]{Arc07}.  For the hydrodynamics, the inner grid boundary
corresponds to a Lagrangian mass coordinate at which the boundary
conditions of the hydrodynamic variables ensure hydrostatic
equilibrium.  As long as the evolution of the proto-neutron star
is tracked, the simulations of the present paper follow
\citet{Arc07,Ugl12,Ert16a} and \citet{Suk16} in the contraction of the
inner grid boundary and the boundary conditions for the neutrino
properties at this location.  The inner grid boundary, and with it the
whole computational grid, is moved according to an exponential
interpolation between initial radius $R_\mathrm{ib,i}$ and defined
final radius $R_\mathrm{ib,f}$: $R_\mathrm{ib} = R_\mathrm{ib,f} +
(R_\mathrm{ib,i}-R_\mathrm{ib,f})\cdot\exp{(-t/t_0)}$.  Here,
$R_\mathrm{ib,i}$ is the radius corresponding to an enclosed mass of
1.1\,M$_\odot$ at about 10\,ms after bounce. It is typically around
60\,km.  For the contraction time scale we use $t_0 = 0.4$\,s and for
the final radius $R_\mathrm{ib,f} = 20$\,km \citep[see also][]{Ugl12}.

The neutrino emission of the excised core, which defines the boundary
condition for the neutrino transport at the inner grid boundary, is
computed from an analytic one-zone model
\citep[see][]{Ugl12,Suk16}. The model equations are based on the
requirements of energy conservation and the validity of the Virial
theorem. The relative weights of neutrinos and antineutrinos are
constrained by the lepton-number loss associated with the
neutronization of the proto-neutron star core \citep{Sch06}. The values of the
parameters in this analytic model are thus constrained by (global)
physics arguments, but three of them are varied within small
ranges to obtain explosions in 1D that are in agreement with
observationally constrained energies and $^{56}$Ni masses of the well
studied SN~1987A and SN~1054-Crab. These two observational cases are
used to calibrate our ``neutrino engine'' that drives explosions in
spherical symmetry.  This choice is motivated, on the one hand, by the
fact that the Crab supernova had a low explosion energy (only around
$10^{50}$\,erg) with little nickel production (some
$10^{-3}$\,M$_\odot$), and the gaseous remnant has a high helium
abundance and relatively oxygen-poor filaments.  These characteristics
are compatible with the explosion of a low-mass star with extremely
low core compactness value. On the other hand, SN~1987A had an
explosion energy over $10^{51}$\,erg and expelled about
0.07\,M$_\odot$ of $^{56}$Ni and roughly one solar mass of oxygen,
classifying it as a more ``normal'' supernova (SN) of a massive star
greater than 15\,M$_\odot$. For a detailed discussion, see
\citet{Suk16}.

Following \citet{Suk16}, we use a 9.6\,M$_\odot$ star (Model Z9.6) for
a Crab progenitor and a set of 15, 18, and 20\,M$_\odot$ blue
supergiant stars (Models W15, W18, W20, N20) as well as a
19.8\,M$_\odot$ red supergiant star (Model S19.8) as SN~1987A
progenitors. The values of the proto-neutron star core parameters for these neutrino
engines, calibrated by comparison to Crab and SN~1987A, are listed in
Table~3 of \citet{Suk16}. In addition, we also tested several of
the binary progenitor models proposed for SN~1987A  by
\citet{Men17} and found that all of them could be exploded at least with
one of our sets of engine-parameter values from single-star
progenitors to yield reasonable agreement with the explosion energy
and nickel production of SN~1987A (see Appendix~\ref{app:comparison}).
Therefore the binary progenitors of SN~1987A do not provide any new
neutrino engines, but can be considered as variants of the already
available single-star engines.

In the present paper we will present results of all engines used by
\citet{Suk16}, but for detailed analysis of population-integrated
supernova outcomes we will mostly focus on Z9.6 in combination with
W18.  W18 provides a neutrino engine not much below the strongest ones
(N20 and S19.8), and its choice allows for detailed comparisons with
single-star results discussed by \citet{Suk16}.  In practice, we
interpolate the proto-neutron star core-model parameters between the
Z9.6 and the SN~1987A engines in terms of a parameter $M_{3000}$,
which measures the mass enclosed by a radius of 3000\,km in the core
of the pre-collapse star. This interpolation provides a smooth
transition from ``Crab-like'' to ``SN~1987A-like'' cases. For details,
see again \citet{Suk16}.

In all simulations the EOS of \citet{Lat91} with $K = 220$\,MeV is
used for $\rho\ge 10^{11}$\,g\,cm$^{-3}$. At lower densities, a 
multi-component
EOS is employed that includes radiation, electrons, positrons, and
ions.  The composition is followed using two approximations.  Above a
temperature of $7 \times 10^9$\,K, NSE is assumed and the composition
is tracked using a table that includes, as a function of $\rho$, $T$,
$Y_e$, the 13 $\alpha$-nuclei, neutrons, protons, and a ``tracer
nucleus'' (Tr) that represents neutron-rich nuclei and replaces $^{56}$Ni
in regions where $Y_e < 0.49$ and neutron-rich species are expected to
become abundant. Neutrino capture and electron and
positron capture on neutrons and protons are followed so that $Y_e$
evolves with time \citep[for the neutrino treatment, see][]{Sch06}. 
For temperatures between $7\times 10^9$\,K and $10^8$\,K, the composition
is followed using a reaction network that links the 13 $\alpha$-nuclei by the
triple-$\alpha$ reaction for helium, subsequent $(\alpha,\gamma)$-reactions for 
nuclei with atomic mass numbers, $A < 28$, the faster $(\alpha,p)(p,\gamma)$
reaction channel compared to the latter for $A \ge {28}$, and heavy-ion reactions,
$^{12}\mathrm{C}(^{12}\mathrm{C},\alpha)^{20}\mathrm{Ne}$,
$^{12}\mathrm{C}(^{16}\mathrm{O},\alpha)^{24}\mathrm{Mg}$,
$^{16}\mathrm{O}(^{16}\mathrm{O},\alpha){}^{28}\mathrm{Si}$
(original version by \citealt{Mul86}, upgraded by \citealt{Kif03}). 
In regions where lepton capture on nucleons has reduced $Y_e$ below 0.49,
the tracer nucleus Tr replaces $^{56}$Ni in the network and is produced
instead of it by the reaction $^{52}\mathrm{Fe}(\alpha,p)(p,\gamma)^{56}\mathrm{Ni}$.
The small network approximately tracks the explosive burning of carbon,
oxygen and silicon as the shock moves out. Material that has been
above $7 \times 10^9$\,K and is ejected has its composition followed in
the small network as it expands and cools down, so that, e.g., the
helium has opportunity to recombine. In regions where silicon has been
depleted, i.e., regions that have attained NSE, the final composition
consists of a mixture of helium from photodisintegration, $^{56}$Ni
from regions where $Y_e$ has remained greater 0.49 at all times, and
Tr, a partially neutronized species whose actual identity reflects the
history of weak interactions in the matter, but is not well determined
in P-HOTB. 

As in \citet{Ert16a} and \citet{Suk16}, the network and the EOS are
fully self-consistently coupled, thus taking into account the energy
release by nuclear reactions for the hydrodynamical evolution of the
explosion.

A number of improvements have been implemented in the P-HOTB code since the
work of \citet{Suk16}.\\
(1) In solving the neutrino transport equation by
an analytic method \citep[see][]{Sch06}, the energy source term associated
with neutrino-nucleon scattering (whose average energy transfer per
scattering reaction is taken into account) is not described in a closed
form as in \citet{Sch06}, but it is split into an absorption part and 
emission part of energy. This considerably improves the numerical
stability of the transport integrator.\\
(2) For the long-time simulations of supernova explosions over several weeks,
the radioactive decays of freshly produced $^{56}$Ni to $^{56}$Co
(half-life 6.077\,d) and of $^{56}$Co to stable $^{56}$Fe (half-life 77.27\,d)
are included. This has some influence on the expansion velocities of the
innermost ejecta shells \citep[see][]{Jer17b} and a minor effect 
on the late-time fallback.\\
(3) An adaptive grid-refinement algorithm is added to ensure that 
the density differences between neighboring grid cells in the proto-neutron star surface
layers obey the constraint $\Delta\rho/\rho\lesssim 6$\%. This
improves the possibility of maintaining hydrostatic equilibrium 
for the increasingly steep density gradient at the proto-NS surface.\\
(4) For the long-time simulations during phases of homologous or 
nearly homologous expansion, a steerage algorithm moves the grid with
the expanding matter. This reduces the numerical diffusion of, e.g.,
composition components, and it also reduces computational costs by 
allowing for larger time steps.

The first two points have some smaller consequences for the explosion
results published by \citet{Ert16a} and \citet{Suk16}, and in particular
point~(1) leads to slightly more optimistic conditions for explosions.
Therefore with the improved P-HOTB code we obtain a slightly larger
number of exploding cases, in particular at the edges of ZAMS-mass 
intervals where failures or fallback supernovae were obtained before. In
Appendix~\ref{app:comparison} we provide a detailed comparison of explosion
results obtained by \citet{Suk16} with the previous version of the P-HOTB
code and results obtained for all neutrino engines with the present, 
upgraded version of the code.

The stellar collapse and explosion simulations are computed with a radial 
grid of up to 2000 zones. After core bounce the grid is dynamically
adjusted and its number of zones varies between about 1300 and 2000.
After usually 10--15 seconds the neutrino-cooling of the compact
remnant is no longer followed and the inner grid boundary is moved
outward. Also the outer grid boundary is relocated to larger radii
as the supernova shock propagates through the progenitor star. We have confirmed
by tests that the radial position of the inner grid boundary has no 
relevant effect on the computation of the fallback, which is determined 
by mass escaping though the inner grid boundary with negative 
velocities \citep[see][]{Ert16b}. 
 
The supernova explosions are typically followed for about five weeks. 
In order to perform these simulations beyond
the breakout of the supernova shock from the stellar surface, the progenitor
star is embedded in a dilute circumstellar medium. For this purpose
we assume a constant mass-loss rate $\dot M_\mathrm{w}$ and 
prescribe a wind velocity of $v_\mathrm{w}(r) = v_\infty\cdot(1-R_\ast/r)$,
where $R_\ast$ is the radius of the progenitor star and 
$v_\infty$ is the asymptotic velocity at large distances. This yields
a wind-density profile of $\rho_\mathrm{w}(r) = \dot M_\mathrm{w}\cdot
[4\pi r^2 v_\mathrm{w}(r)]^{-1}$. The temperature in the wind is
computed by the assumption of hydrostatic equilibrium.
For the supernova simulations discussed in
this work we used $\dot M_\mathrm{w} = 10^{-4}$\,M$_\odot$\,yr$^{-1}$
and $v_\infty = 10^8$\,cm\,s$^{-1}$. With these values the density in
the first computational cell exterior to the star is typically one to
two orders of magnitude lower than the density in the outermost cell
of the stellar model. The corresponding total mass of the circumstellar
medium within the computational grid with outer boundary radius of 
$10^{18}$\,cm is lower than 0.04\,M$_\odot$. This mass in addition to 
the pre-collapse mass of the progenitor is so low that it has no
dynamical influence on the expanding supernova ejecta.

\subsection{Modeling with KEPLER}
\lSect{KEPLER}

KEPLER is used to postprocess the explosions calculated by P-HOTB in
order to capture details of the nucleosynthesis and to test for
consistent results. This postprocessing is facilitated by the fact
that the presupernova models used by P-HOTB were calculated using
KEPLER, so the starting points are identical. It is complicated by the
fact that P-HOTB uses an Eulerian grid, carries the proto-neutron star, and includes
neutrino transport, while KEPLER is Lagrangian and does neither. Since
regions outside what will eventually be the final mass separation have
their dynamics modified considerably by neutrino energy deposition,
postprocessing is not so simple as just taking the history of a single
mass shell in P-HOTB and using that as a piston in
KEPLER. Particularly problematic is matter that passes through the
accretion shock in P-HOTB and dwells for a considerable time between
that shock and the neutrinosphere while experiencing an irregular
radial history.  It may take up to a second for the explosion to
develop in P-HOTB.  Attempting to model this delay with a piston in
KEPLER results in multiple bounces and the artificial accumulation of
high density matter just outside the piston.

In \citet{Suk16}, the approach taken was to define a ``special
trajectory'' in P-HOTB for use as a piston in KEPLER (see their
Figures~10 and 11). This trajectory tracked the first mass shell to
move outwards vigorously once the stalled shock was revived. The
radius of that shell contracted from a few thousand km to a minimum of
80 to 140\,km (more typically 120 to 140\,km) before rapidly moving out
and being ejected. The mass inside this shell, its minimum radius, and
the time at which the matter moved out defined the initial motion of a
piston in the KEPLER calculation. The inwards motion of that piston,
from the time the core first collapsed, i.e., reached a collapse speed
of 1000\,km\,s$^{-1}$ in any zone, to the time when it reached its
minimum radius was approximated by fitting a parabolic curve to the
initial and final radius. See equation 7 of \citet{Suk16}.  Its
subsequent outward motion was given by a ballistic trajectory in a
modified gravitational field, i.e., a constant times the local
Newtonian acceleration due to gravity. Given the assumed final radius
of the piston, 10$^9$\,cm, the strength of this field, and hence the
transit time to 10$^9$\,cm, was adjusted so that the kinetic energy of
all ejecta at late time agreed with the P-HOTB simulation. By design
then, the explosion energy in KEPLER and the {\sl initial} mass
separation were the same as in P-HOTB. Subsequent fallback was
followed independently in both codes and usually agreed well.  All
remnant masses (e.g., those used in \Sect{remnants}) were taken from
the P-HOTB simulation and not from KEPLER.

For the piston trajectory just described, the nucleosynthesis of the
iron group and $^{56}$Ni usually agreed quite well between the two
codes. Most of the $^{56}$Ni made in P-HOTB was produced outside of
the special trajectory and so its synthesis was accurately captured by
KEPLER.  Between the special trajectory and the deeper final mass
separation however, there was matter in P-HOTB with an electron mole
number, $Y_e$, that had been affected by neutrino capture both during
the delayed explosion and the neutrino-powered wind that followed.  As
discussed in \Sect{PHOTB}, this matter consisted, in P-HOTB, of helium
from photodisintegration and iron-group elements, Tr, with an
uncertain isotopic composition.  For all models in \citet{Suk16} the
mass in the form of Tr was much less than that of $^{56}$Ni, so
fitting the total iron ($^{56}$Ni plus other ``neutronized'' isotopes)
well also resulted in fitting $^{56}$Ni well.  In the end, the total
iron-group synthesis in KEPLER always fell between the values of Ni
and Ni+Tr in all the matter ejected by P-HOTB.  In some cases where
this condition was not initially satisfied, the piston location was
moved inwards in KEPLER until it was. The adjustment was small, of
order 0.01\,\Msun. Ultimately, the total ejected iron-group synthesis
in KEPLER and P-HOTB agreed to within a few thousandths of a solar mass
for models below 12\,\Msun, and about 0.02\,\Msun \ for heavier
models. The average difference in total iron ejected for the two codes
was 24\% or 0.014\,\Msun \ \citep[see Figure~12 and Tables~4 and 5 
of][]{Suk16}.

While no additional modifications were made to tune the production of
$^{56}$Ni itself, it turned out that the agreement between the two
codes was even better for that isotope. $^{56}$Ni production in
KEPLER, on average, was within 9.8\% (0.005\,\Msun) of the P-HOTB
values for just $^{56}$Ni and 28\% (0.016\,\Msun) for $^{56}$Ni+Tr. In
the worst case the $^{56}$Ni to $^{56}$Ni comparison was off by
0.021\,\Msun.

In the present paper, iron and $^{56}$Ni synthesis in KEPLER were
computed this ``old way'', but an additional set of models was
calculated that used a deeper mass cut for the piston location.
Except for a few cases with large amounts of late time fallback, this
location was equal to the final mass separation in the P-HOTB
simulation.  For those cases with massive fallback, the
nucleosynthesis of iron did not matter and the final remnant mass was
taken from P-HOTB, so the special trajectory continued to be
used. Moving the mass cut in like this increases the production of
iron and $^{56}$Ni in KEPLER. Because interactions with neutrinos may
have reduced $Y_e$ in these deepest layers, the $^{56}$Ni production
is a maximum. The timing of the bounce and the minimum radius in this
other set were still that of the ``special trajectory'' and the same
agreement in final kinetic energy was enforced, but the initial 
location of the piston was deeper and was not varied.

\subsection{The Evaluation of Iron and $^{56}$Ni Yields in P-HOTB and KEPLER}
\lSect{ni56}

Because of the central role of $^{56}$Ni in producing the light curves
of Type I supernovae, we discuss here how our
best estimates were calculated and their possible error. Ultimately six
possibilities were considered, each with its own special meaning.

Part of the variation is because KEPLER and P-HOTB both calculate
iron-group nucleosynthesis. KEPLER uses a large network that is
necessary for determining isotopic and trace element
nucleosynthesis. P-HOTB uses a 13-species network ($\alpha$-nuclei
from helium through $^{56}$Ni) for
temperatures below $7 \times 10^9$\,K, supplemented by a 15-species 
nuclear statistical equilibrium (NSE) solver for higher temperatures
\citep[\Sect{PHOTB};][]{Kif03,Sch06,Ugl12,Ert16a}.  P-HOTB tracks the
dynamical history of zones near the mass cut more accurately and also
calculates a 1D approximation to the neutrino-powered wind. To do so,
it follows the evolution of ejecta near the mass cut for at least 10\,s
after core collapse. See Figure~10 of \citet{Suk16} for an example.

As discussed in \Sect{PHOTB}, P-HOTB delineates its iron-group
synthesis into three components: Ni, Tr, and $\alpha$, and there is
some ambiguity in the interpretation of each. ``Ni'' is the
$^{56}$Ni synthesized in the 13-isotope network and the 15-species
NSE solver when $Y_e$ is greater
than 0.49.  ``Tr'' is $^{56}$Ni plus other iron-group species produced
in both the network and the NSE solver when $Y_e < 0.49$,
especially in the neutrino-powered wind. In the present studies $Y_e$ was
typically $\sim$0.46--0.49 in the wind region and thus ``Tr'' would
contain essentially no $^{56}$Ni. Other more realistic studies have
shown, however, that $Y_e$ in the wind should be close to or even
greater than 0.50 \citep{Mar12,Rob12,Mir16}. Studies by
\citet[e.g.][]{Pru06} show the fraction of heavy species
that is $^{56}$Ni
is 30--80\% for $Y_e$ near 0.54.  Consequently, we take the fraction
of ``Tr'' that is $^{56}$Ni here to be some uncertain fraction that
might be near 0.5.

The $\alpha$-particles are produced by photodisintegration in the
shock, but also in the wind at late times where their mass fraction
can approach 90\%.  The total mass of material that has achieved NSE
in P-HOTB is bounded above by Ni+Tr+$\alpha$. It is an upper bound
since the sum also includes a small amount of $^{56}$Ni produced by
incomplete explosive silicon burning for shock temperatures between 4
and $5 \times 10^9$\,K. Since we are interested not just in best
estimates, but also in upper bounds (\Sect{lite}), it is possible that
some greater fraction of the $\alpha$-particles might reassemble in a
multi-dimensional model. Regardless of dimensionality, a comparable
amount of matter needs to absorb energy from neutrinos in order to
obtain the observed energy of the supernova.  In 1D this mass expands
like a wind, in 3D it will come from overturn, accretion, and
re-ejection of matter after heating.  An open question is whether the
average entropy of the bottom-most layers ejected in a 3D explosion is
less than in 1D. This would promote the assembly of more
$\alpha$-particles into $^{56}$Ni.  Coupled with the desire to explore
the sensitivity of the light curves to the $^{56}$Ni mass, this
motivates treating the $\alpha$-particles calculated here by P-HOTB as
a potential mixture of $^{56}$Ni and $\alpha$'s.  Studies with KEPLER
with the mass separation at the P-HOTB value (\Sect{KEPLER}) show that
for matter ejected solely by shock heating, the average $^{56}$Ni mass
ejected is 75\% of the matter that achieved NSE. This suggests that
0.75 times the sum of Ni+Tr+$\alpha$ from the P-HOTB calculation is an
upper bound to the $^{56}$Ni synthesis. It is an upper bound because
KEPLER does not include neutrino capture.

\Fig{ni56yields}, shown later in the paper in \Sect{W18results},
illustrates the possibilities. For the neutrino-powered models
calculated here (see \Sect{ni56max} for other possibilities), the least
amount of $^{56}$Ni is mainly the iron group material that did not 
experience any appreciable change in $Y_e$ during the explosion as 
modeled in P-HOTB. This is the light black line at the bottom of the
figure. It also contains the $^{56}$Ni made by explosive burning in
shock-heated ejecta plus $^{56}$Ni produced by nuclear recombination
in neutrino-heated wind matter for $Y_e \ge 0.49$. This last 
contribution is usually very small because $Y_e$ in neutrino-processed
matter that ends up being
ejected is $Y_e\sim 0.46$--0.49 in the P-HOTB 
models. Close to that, the dark blue line gives the KEPLER results for
calculations using the special trajectory as the mass cut, the same
approach used by \citet{Suk16} (see \Sect{KEPLER}). The two dashed
lines are for Ni+Tr/2 and Ni+Tr as calculated by P-HOTB. As discussed
in \Sect{PHOTB} it is probable that not all of Tr is neutronized iron,
but the fraction of $^{56}$Ni is undetermined when $Y_e$ in the wind
was lower than 0.49. 
The band between the two is our best estimate for $^{56}$Ni
synthesis in the current model set. The top solid black curve shows 
the KEPLER result when the mass cut is moved into the deepest possible
value for the present models. It is very similar to the broad red
curve, which is all the $^{56}$Ni, Tr, and helium from
photodisintegration ejected by P-HOTB (i.e., essentially all the
matter that has reached NSE) multiplied by 0.75. The factor 0.75 was
chosen so that the two curves would match. The best value is thus in
the gray band and the two values explored as maxima and minima for the
light curves (\Sect{lite}) are the dark red and blue lines.  The
results in the figure are for the W18 central engine, but those for
S19.8 are similar.



\begin{figure*}
\includegraphics[width=2\columnwidth]{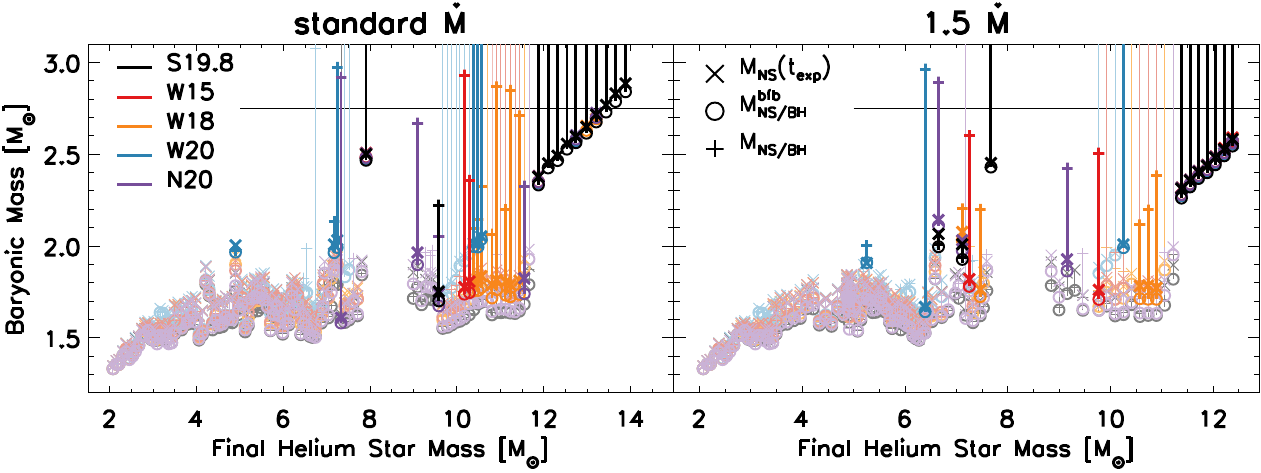}
\caption{Baryonic neutron star mass versus pre-collapse helium-star mass in
  P-HOTB simulations of helium stars with standard mass loss ({\em left}) and
  enhanced mass loss ({\em right}) for all central engines (S19.8,
  W15, W18, W20, and N20). Three different masses are indicated by
  different symbols: $M_\mathrm{NS}(t_\mathrm{exp})$ is the neutron 
  star mass at the time when the explosion sets in (crosses);
  $M_\mathrm{NS/BH}^\mathrm{bfb}$ is the mass of the compact remnant
  after mass loss by the neutrino-driven wind, but before fallback
  (open circles); and $M_\mathrm{NS/BH}$ is the final mass of the
  compact remnant including the mass accreted from fallback (plus
  signs). The dark colored symbols mark cases where at least one of
  these three masses lies between 2\,M$_\odot$ and 3\,M$_\odot$,
  whereas the light-colored symbols correspond to all other cases.
  Different colors correspond to different engines as noted by labels
  in the left panel.  Vertical bars show the fallback masses, which
  are usually too small to be visible. The thin black horizontal line
  marks the baryonic mass limit of 2.75\,M$_\odot$ assumed for cold
  neutron stars.}  \lFig{NSbaryon}
\end{figure*}

\section{Explosion Results Using P-HOTB}
\lSect{results}

Before describing specific results for the various central engines, it
is useful to define a maximum mass for cold neutron stars. This was unnecessary
in our previous study \citep{Suk16} because the progenitors there
either blew up and formed neutron stars with baryonic (gravitational) masses
below 2.15\,M$_\odot$ ($\sim$1.8\,M$_\odot$), or collapsed to compact
remnants with masses well above 3\,M$_\odot$. The latter value was
beyond any mass limit that can be stabilized using present-day nuclear
EOSs. In the new study of helium stars though, there are many cases
where the proto-neutron star has a baryonic mass between 2.5\,M$_\odot$ and
3\,M$_\odot$ at the time when the explosion sets in. This is illustrated
in \Fig{NSbaryon} which discriminates between the compact-remnant mass
at the onset of the explosion, $M_\mathrm{NS}(t_\mathrm{exp})$, the
remnant mass before fallback, $M_\mathrm{NS/BH}^\mathrm{bfb}$, and the
final remnant mass after fallback, $M_\mathrm{NS/BH}$. Revisiting our
calculations of single stars using the revised version of P-HOTB
(Appendix~\ref{app:comparison}), has also revealed
a few cases where fallback lifts the baryonic mass of the compact object to between
2.5\,M$_\odot$ and 3\,M$_\odot$ even in the single star case.  Reasons
for the variation are discussed in \Sect{fallback}.

Here we choose a baryonic mass limit for neutron stars of 2.75\,M$_\odot$,
corresponding to a gravitational mass between $\sim$2.18\,M$_\odot$
and $\sim$2.30\,M$_\odot$ for neutron star radii between 9\,km and 12\,km
according to equation~(36) of \citet{Lat01}. This limit is compatible
with estimates based on the gravitational-wave and kilonova
measurements associated with the first detection of a neutron-star
merger event in GW170817 \citep[see, e.g.][]{Mar17,Rez18}. Since the
thermal pressure of a hot neutron star can stabilize an additional mass of
several 0.1\,M$_\odot$ and thus increase the threshold mass for black hole
formation \citep{Oco11,Ste13}, this is a lower bound for proto-neutron stars. All 
cases in \Fig{NSbaryon} that exceed the cold limit of
$M_\mathrm{b,NS}(t_\mathrm{exp}) = 2.75$\,M$_\odot$ by only a small
margin (crosses above the horizontal black line), are considered
transiently stable, hot neutron stars, whose neutrino emission can trigger supernova
explosions.  The circles represent the neutron star masses after the phase of
neutrino cooling, but before fallback.  They are lower than the
corresponding masses at $t_\mathrm{exp}$ (indicated by crosses),
because of the mass that is blown away in the neutrino-driven
wind. Plus signs mark the baryonic masses of the compact objects after
fallback. They are treated as neutron stars when $M_\mathrm{b,NS/BH} \leq
2.75$\,M$_\odot$, otherwise as black holes. In the latter case we obtain
``fallback supernovae'', i.e., low-energy explosions with black hole formation due to
massive fallback, which in 1D simulations do not eject any iron-group
material.

\begin{figure}
\includegraphics[width=\columnwidth]{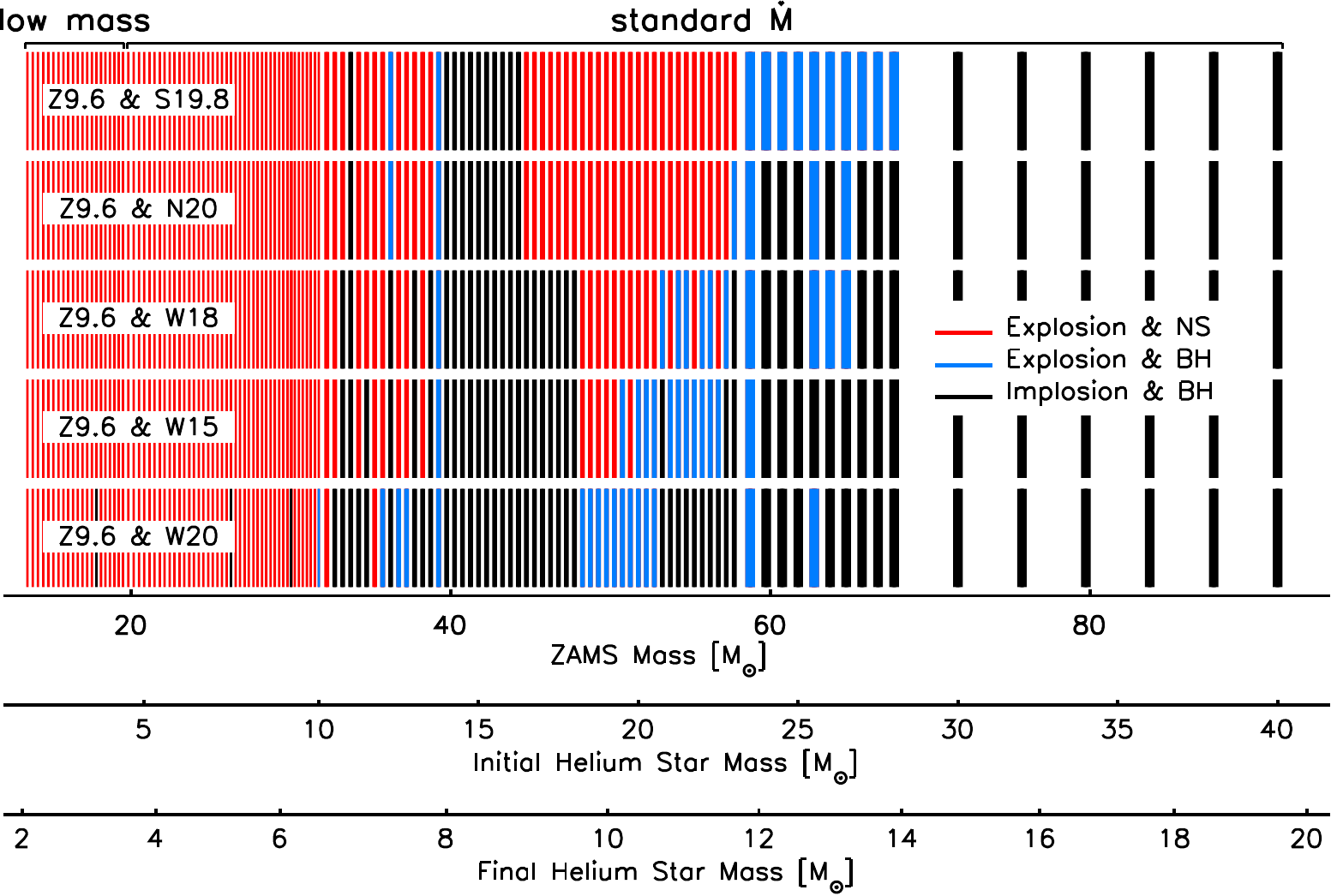}
\caption{The outcome of core collapse in calculations with P-HOTB for
  helium stars with the standard mass-loss rate and all neutrino
  engines considered. Supernova explosions with neutron star formation are indicated by
  red bars, failed explosions with black hole formation by black bars, and supernova
  explosions with black hole formation due to massive fallback accretion
  (``fallback supernovae'') by blue bars. The different engines are sorted by
  their strength. The strongest engine with the largest number of
  successful explosions is shown on top, the weakest engine at the
  bottom. The horizontal axes provide ZAMS mass (as estimated with
  equations~(4) and (5) of \citet{Woo19} as well as initial and final
  helium-star masses.  \lFig{sta-explosions}}
\end{figure}

\begin{figure}
 \includegraphics[width=\columnwidth]{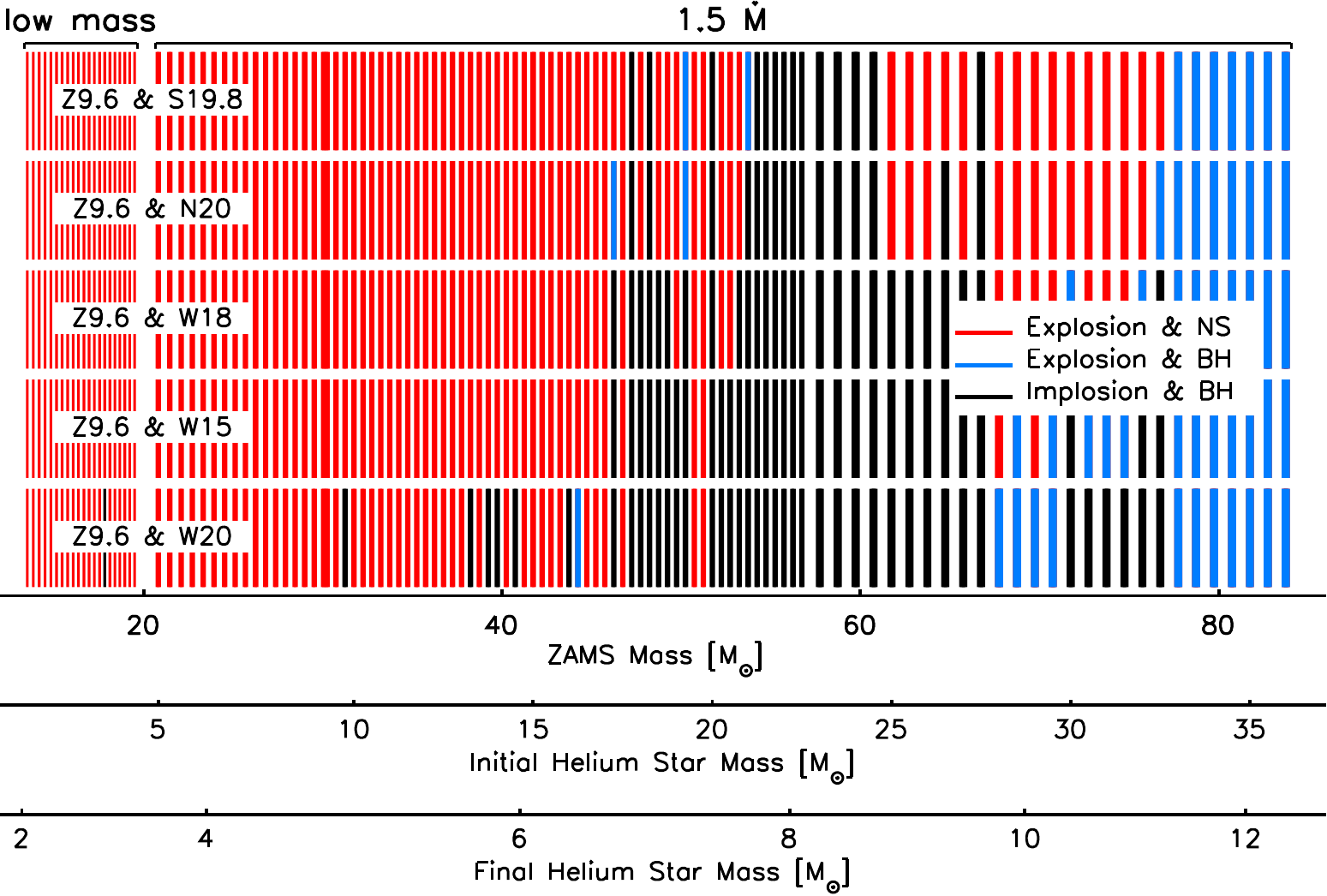}
 \caption{Same as \Fig{sta-explosions}, but for the helium-star
   models that employed an enhanced mass-loss rate during helium
   burning.  \lFig{1.5-explosions}}
\end{figure}

\subsection{Dependence on the Central Engine}
\lSect{dependence-on-engine}

Three different outcomes of stellar core collapse can be
distinguished: supernova explosions with neutron star formation, supernova explosions with black hole
formation by massive fallback (``fallback supernovae''), and failed
explosions with ``direct'' black hole formation, i.e., continuous accretion of
the transiently existing neutron star until collapse to a black hole takes place
(``Implosion \& BH'' in Figs.~\ref{fig:sta-explosions} and
\ref{fig:1.5-explosions}).

Figs.~\ref{fig:sta-explosions} and \ref{fig:1.5-explosions} show our
results from P-HOTB simulations for all the central engines employed
for the set of helium stars with standard mass loss rate and the set
with enhanced mass loss rate, respectively. The engines are named by
the combination of Crab and SN~1987A progenitors, whose sets of
parameter values are interpolated in dependence on the core
compactness (more specifically: $M_{3000}$; see \Sect{PHOTB}). The
strongest neutrino engine is shown at the top, the weakest at the
bottom. It is evident that the number of cases with supernova
explosions and NS formation (red) decreases from top to bottom,
whereas the number of cases with black hole formation, either by
failed explosions (black) or fallback supernovae (blue),
increases. While below a final helium-star mass of about 6\,M$_\odot$
essentially all progenitors blow up and give birth to neutron stars
(with very few exceptions for the weakest engine), there is a mix of
outcomes with neutron star or black hole formation for final helium-star
masses between $\sim$6\,M$_\odot$ and nearly 12\,M$_\odot$, where the
strongest engine produces mostly successful explosions and the weakest
engine mostly black holes.  Above a final helium-star mass of
$\sim$12\,M$_\odot$ for the progenitors with standard mass loss, and
of $\sim$11.3\,M$_\odot$ for enhanced mass loss, only black hole
formation by continuous or fallback accretion takes place.

A subset of the massive helium stars is especially interesting because
they produce very massive proto-neutron stars before an explosion sets in (the
baryonic neutron star masses at this time are between $\sim$2.3\,M$_\odot$ and
$\sim$2.9\,M$_\odot$). Later, fallback triggers black hole formation (see
\Fig{NSbaryon}). This is the case for final helium-star masses between
$\sim$12\,M$_\odot$ and 14\,M$_\odot$ (ZAMS mass between
$\sim$58\,M$_\odot$ and 68\,M$_\odot$) for standard mass loss, in
particular with the three strongest neutrino engines, and for final
helium-star masses between $\sim$11.3\,M$_\odot$ and 12.5\,M$_\odot$ (ZAMS
mass between $\sim$77\,M$_\odot$ and 85\,M$_\odot$) for enhanced mass
loss and all neutrino engines. Below these mass windows there is only
a single exception of a lower-mass helium-star with a similar behavior,
namely a progenitor with final helium-star mass of nearly 8\,M$_\odot$
(ZAMS mass $\sim$39\,M$_\odot$) for the standard mass loss and one
with final helium-star mass close to 7.7\,M$_\odot$ (ZAMS mass
$\sim$54\,M$_\odot$) for the higher mass loss. In all of these cases
$M_4$, i.e. the mass where the dimensionless entropy per baryon
reaches a value of $s = 4$, is large ($>1.85$\,M$_\odot$). But
the explosions set in even later, typically later than 1.5\,s after
bounce, at which time the neutron star masses are considerably bigger than
$M_4$. Consequently, the final explosion energies are low, less than
$\sim$0.5$\times 10^{51}$\,erg.

In these cases $M_4$ underestimates the neutron star mass at the time the
explosion sets in, because the star blows up only when matter whose
entropy per baryon equals about 6 in the presupernova star collapses
into the stalled shock. This corresponds roughly to the outer boundary
of the convective oxygen burning shell in the presupernova star and is
often close to strong carbon and neon burning shells in cases where
these shells have merged. At this location the mass accretion rate
makes another drop by about a factor of 2. Successful explosions in
these cases occur because $M_6$ (which is the mass enclosed by the
radius where $s = 6$) is still smaller than in other non-exploding
progenitors, namely between about 2.3\,M$_\odot$ and 2.9\,M$_\odot$.
Using $M_4$ instead of $M_6$ as a proxy of the neutron star mass and $\mu_4$ as
a parameter scaling with the mass-accretion rate, places these cases
above the two-parameter curve separating neutron star formation from black hole
formation in the $M_4\mu_4$-$\mu_4$-plane (see
Appendix~\ref{app:updatetwoparameter}, \Fig{transnewcriterion}, and
\citealt{Ert16a} for the theoretical background), i.e., on the side of
black hole formation. This is appropriate, because it reflects the fact that
these stars do not blow up when the infalling $M_4$-shell arrives at
the shock, but they explode marginally only later with low energies
and several solar masses of stellar material falling back onto the
compact remnant.

In the following discussion of explosion results and their
astrophysical consequences, we will focus on the two engines based on
the W18 and S19.8 progenitors of SN~1987A. The reason is that the
former allows for a most detailed comparison with the single-star
results discussed by \citet{Suk16}.  The latter is our strongest
neutrino engine, marginally beating N20, and thus suites to
demonstrate the range of possibilities on the optimistic side for
explosions. When this aspect is of relevance, we will therefore also
refer to our results with the S19.8 engine.

\subsection{Conversion of Baryonic to Gravitational Masses}
\label{sec:conversion-bary-to-gravmass}

In our P-HOTB simulations neutrino-energy loss from the newly formed
compact remnant is taken into account by the binding energy of the
analytic inner-core model plus the neutrino emission from the matter
that gravitationally settles in the outer proto-neutron star layers
followed directly by our hydrodynamics and neutrino-transport
simulations.  The description is very approximate. Nevertheless,
detailed comparisons (Kresse et al., in preparation) reveal
that the total release of neutrino energy, $E_{\nu,\mathrm{tot}}$,
overestimates the binding-energy computed from equation~(36) of
\citet{Lat01} \citep[see also equations~(9) and (10) of][]{Suk16} only
by a modest 10--20\% when neutron star radii of 11--12\,km are used in
these formulas. This difference does not only stem from the rough
approximations in our neutron star model, but also results from the
fact that the proto-neutron star is transiently more massive than the
neutron star after its mass loss by the neutrino-driven wind and
before fallback, i.e., $M_\mathrm{b,NS}(t_\mathrm{exp}) >
M_\mathrm{b,NS}^\mathrm{bfb}$ (see \Fig{NSbaryon}). It therefore
radiates transiently more neutrinos than expected by the final neutron
star mass.  Such time-dependent effects of the dynamical supernova
evolution in spherical symmetry are not captured by the formulas of
\citet{Lat01}.

The difference implies a small underestimation of the gravitational neutron star
mass by typically 0.02--0.04\,M$_\odot$ in the P-HOTB simulations
compared to the estimates derived on grounds of \citet{Lat01}. For
example, for neutron stars with baryonic masses of 1.4, 1.6, 1.8, 2.0,
2.2\,M$_\odot$ our P-HOTB simulations yield about 1.24, 1.40, 1.56,
1,71, 1.90\,M$_\odot$ for the gravitational masses, whereas
equation~(36) of \citet{Lat01} gives 1.27, 1.43, 1.60, 1.75,
1.90\,M$_\odot$ with $R_\mathrm{NS} = 12$\,km and 1.26, 1.42, 1.58,
1.73, 1.88\,M$_\odot$ with $R_\mathrm{NS} = 11$\,km.  In the worst
case and for a very massive neutron star, the gravitational mass of the cold
remnant estimated with our neutrino loss in the P-HOTB simulation is
lower by $\sim$0.1 (0.07)\,M$_\odot$ for a neutron star radius of 11 (10)\,km,
corresponding to an underestimation of the gravitational neutron star mass by 4--5\%
compared to the values derived from equation~(36) of \citet{Lat01}.

However, the gravitational-wave and kilonova observations of GW170817
have set new constraints on neutron star radii, which reduce the uncertainties
of the nuclear EOS in cold neutron stars. The numerical factors in
equations~(35) and (36) of \citet{Lat01} are averages over a wide
range of possibilities, some of which are not compatible with the new
radius constraints.  \citet{Abb18} concluded that the radii of the two
merger components (whose masses are most likely between
$\sim$1.1\,M$_\odot$ and $\sim$1.7\,M$_\odot$) are
$11.9_{-1.4}^{+1.4}$\,km for EOSs that permit neutron stars with masses larger
than 1.97\,M$_\odot$ as required by observations
\citep{Ant13}. \citet{Bau17} argued that the radius of a non-rotating
cold neutron star with a gravitational mass of 1.6\,M$_\odot$ is larger than
$10.68_{-0.04}^{+0.15}$\,km, and the radius of the maximum-mass
configuration must be larger than $9.60_{-0.03}^{+0.14}$\,km.  Such
small radii tend to favor higher binding energies than obtained with
the numerical factors used in equations~(35) and (36) of
\citet{Lat01}. EOSs still compatible with the new constraints, for
example WFF2, AP3, AP4, ENG, and on the very compact side WFF1, are
above the average line in Figure~8 of \citet{Lat01}. Therefore they
yield binding energies that are roughly 5--25\% higher than the mean
value when neutron stars have gravitational masses exceeding
$\sim$1.4\,M$_\odot$.

For all these reasons we will use our P-HOTB results for the
neutrino-energy loss of the new-born neutron stars in the whole paper.  They are
displayed in Figs.~\ref{fig:summaryw18-sta-exp} and
\ref{fig:summaryw18-1.5-exp} and they are employed to convert baryonic
masses to gravitational masses of the compact remnants as given in
Figs.~\ref{fig:summaryw18-sta-rem} and \ref{fig:summaryw18-1.5-rem}.
For reference, we will also provide the numbers derived with
equation~(36) of \citet{Lat01} when we discuss IMF-averaged masses.

Different from our approach in the previous works on single stars
(\citealt{Suk16}, \citealt{Ert16a}, see also
Appendix~\ref{app:comparison}), the greater number of present cases
with massive fallback motivates us now to take into account also the
release of gravitational binding energy associated with the accretion
of fallback matter. Because the neutrino-energy release calculated
with P-HOTB and the values computed from equation~(36) of
\citet{Lat01} exhibit close agreement, we use the latter formula for
estimating the energy loss from accretion. Several cases need to be
distinguished.
 
{\bf (1)} In the case of direct black hole formation by continuous accretion
without supernova explosion, the neutron star is assumed to collapse to a black hole when it
reaches a baryonic mass of $M_\mathrm{b,NS}^\mathrm{max} =
2.75$\,M$_\odot$.

{\bf (1a)} If this limiting mass is reached during the P-HOTB
simulation, we use the neutrino-energy output $E_{\nu,\mathrm{tot}}$
from the simulation to estimate the gravitational mass of the black hole by
\begin{equation}
M_\mathrm{g,BH} = M_\mathrm{b,BH} - 
\frac{1}{c^2}\,E_{\nu,\mathrm{tot}}(M_\mathrm{b,NS}^\mathrm{max})\,.
\label{eq:BHmass1}
\end{equation}
Here $M_\mathrm{b,BH}$ is the total mass of the stellar matter that
collapses into the black hole, and we assume that the loss of energy from
matter accreted after black hole formation is negligible. This assumption
holds well for radial accretion, in which case neutrinos (and, of
course, photons as well) do not have enough time to efficiently escape
from the fast inward flow. Moreover, we make the approximation that
thermal stabilization of the hot neutron star has a minor effect, i.e., we take
our assumed mass limit for cold neutron stars to also set the limiting mass of
the accreting remnant. This is valid only when the accretion proceeds
slowly and the neutron star survives for a long period of time to cool
efficiently. However, also if the neutron star collapses on a short time scale
compared to the cooling time scale, our approximation is acceptable as
well, because the neutrino-energy loss in this case is fairly small
and therefore the gravitational to baryonic mass difference accounts
only for an insignificant correction to the black hole mass.

{\bf (1b)} If the accreting neutron star does not reach the limiting mass of
$M_\mathrm{b,NS}^\mathrm{max} = 2.75$\,M$_\odot$ until our P-HOTB
simulations are terminated at $t_\mathrm{end} = 10$\,s after bounce,
we extrapolate its further energy loss by using the \citet{Lat01}
estimate. The gravitational black hole mass is therefore calculated as
\begin{equation}
M_\mathrm{g,BH} = M_\mathrm{b,BH} - \frac{1}{c^2}\,\mathrm{max}
\left\{ E_{\nu,\mathrm{tot}}(t_\mathrm{end}),\,  
E_\mathrm{b,10\,km}^\mathrm{LP01}(M_\mathrm{b,NS}^\mathrm{max})\right\}\,,
\label{eq:BHmass2}
\end{equation}
where
$E_\mathrm{b,10\,km}^\mathrm{LP01}(M_\mathrm{b,NS}^\mathrm{max})$ is
the binding energy of a cold, maximum-mass neutron star of 10\,km radius
according to equation~(36) of \citet{Lat01}. In practice, its value is
higher than the neutrino-mass decrement obtained in the P-HOTB
simulation at $t_\mathrm{end}$.

\begin{figure*}
 \includegraphics{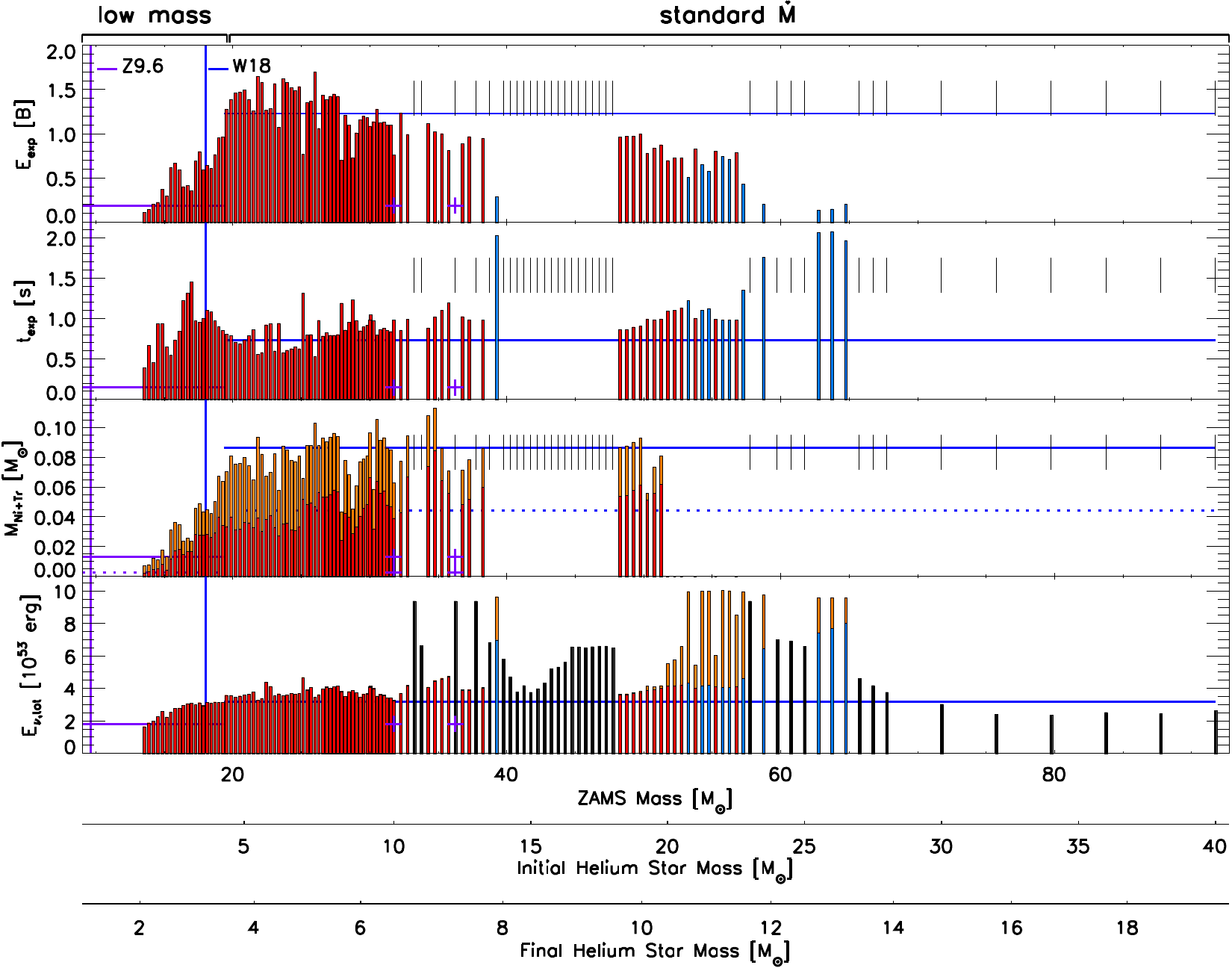}
 \caption{Overview of the explosion properties for our full set of
   helium star models with standard mass-loss rate, computed with the
   P-HOTB code and the Z9.6 \& W18 neutrino engine. From top to bottom
   the panels show the final explosion energy ($1\,\mathrm{B} =
   10^{51}\,$erg); the time when the explosion sets in (i.e., when the
   supernova shock expands beyond 500\,km); the summed mass of finally
   ejected $^{56}$Ni (red) plus tracer (orange); and the total energy
   radiated in neutrinos, which takes into account the additional
   neutrino loss from fallback accretion (orange sections of the
   histogram bars). Red histogram bars denote the cases of neutron star
   formation, black bars those of ``direct'' black hole formation by
   continuous accretion, and the blue bars correspond to cases where
   an explosion takes place but the final baryonic mass of the compact
   remnant exceeds our assumed black hole formation limit of 2.75\,M$_\odot$.
   Non-exploding cases are marked by thin, short vertical black dashes
   in the upper part of each panel. The vertical
   purple and blue lines mark the masses of the engine models,
   Z9.6 and
   W18, respectively, and the corresponding results of these engine
   models are indicated by solid and dashed horizontal purple and blue
   lines.  The mass-range spanned by the horizontal purple line of
   Model Z9.6 indicates the region of Crab-like behavior, where the
   Crab and SN~1987A engines are interpolated.}
 \lFig{summaryw18-sta-exp}
\end{figure*}

\begin{figure*}
 \includegraphics{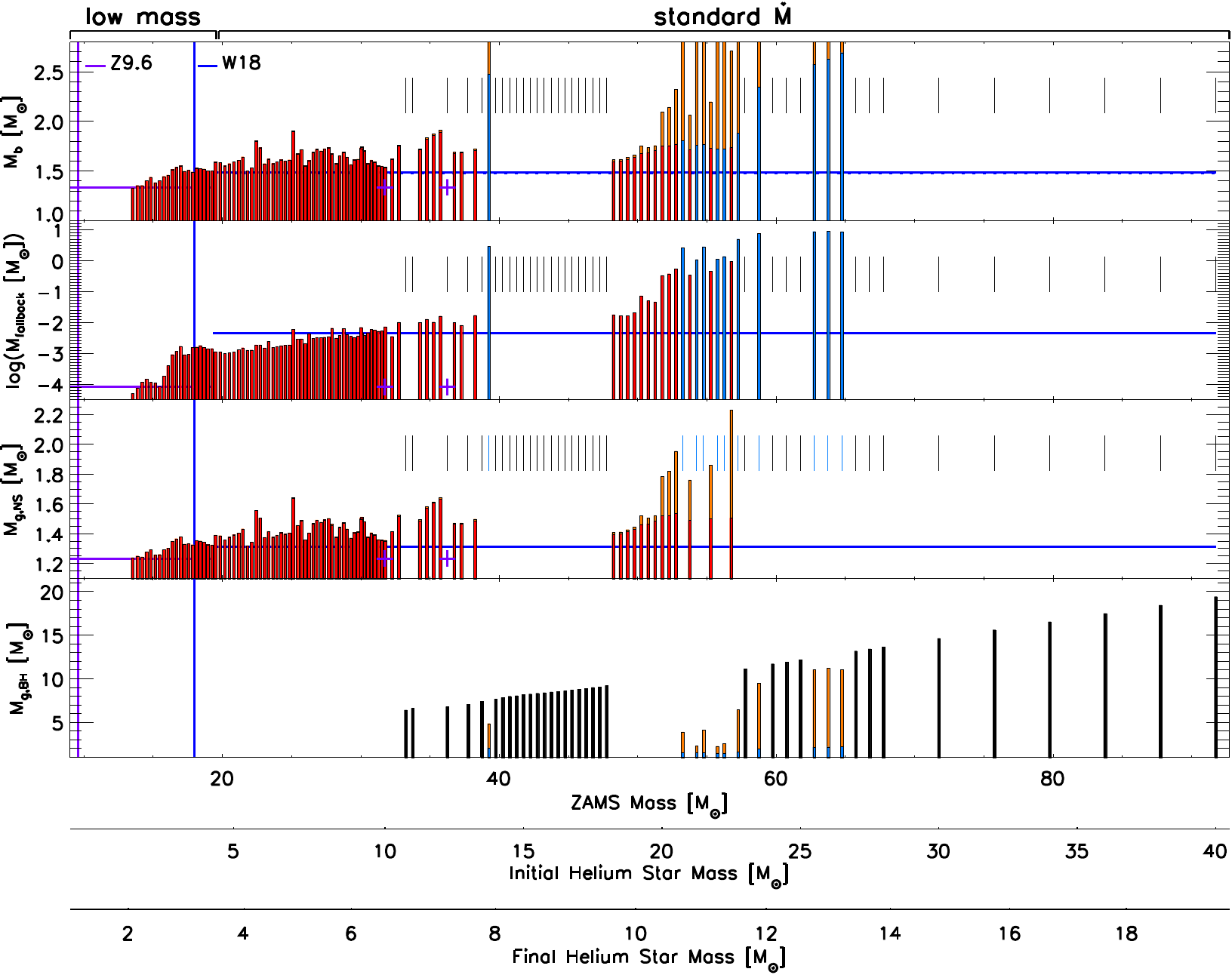}
 \caption{Overview of the properties of the compact remnants for our
   full set of helium-star models with standard mass-loss rate, computed
   with the P-HOTB code and the Z9.6 \& W18 neutrino engine. Red
   histogram bars denote the cases of neutron star formation, black bars those
   of ``direct'' black hole formation by continuous accretion, and the blue
   bars correspond to cases where an explosion takes place but the
   final baryonic mass of the compact remnant exceeds our assumed
   black hole formation limit of 2.75\,M$_\odot$.  From top to bottom the
   panels show the final baryonic mass of the remnant with the
   fallback mass indicated by the orange part of the histogram bar;
   the logarithm of the fallback mass; the gravitational mass of the
   neutron star with the fallback contribution again indicated by the orange
   section of the bar; and the gravitational mass of the black hole, also
   including the fallback contribution marked by the orange
   sections. In all the blue cases where black holes form by fallback
   accretion, the fallback mass is close to 1\,M$_\odot$ or
   higher. Gravitational masses for neutron stars and black holes are computed from
   baryonic masses as detailed in \Sect{conversion-bary-to-gravmass},
   including neutrino losses from fallback matter. Non-exploding cases
   are marked by thin, short vertical black dashes in the upper part
   of each panel.The vertical purple and blue lines
   mark the masses of the engine Models Z9.6 and W18, respectively,
   and the corresponding results of these engine models are indicated
   by solid and dashed horizontal purple and blue lines.  The
   mass-range spanned by the horizontal purple line of Model Z9.6
   indicates the region of Crab-like behavior, where the Crab and
   SN~1987A engines are interpolated.}  \lFig{summaryw18-sta-rem}
\end{figure*}

\begin{figure*}
 \includegraphics{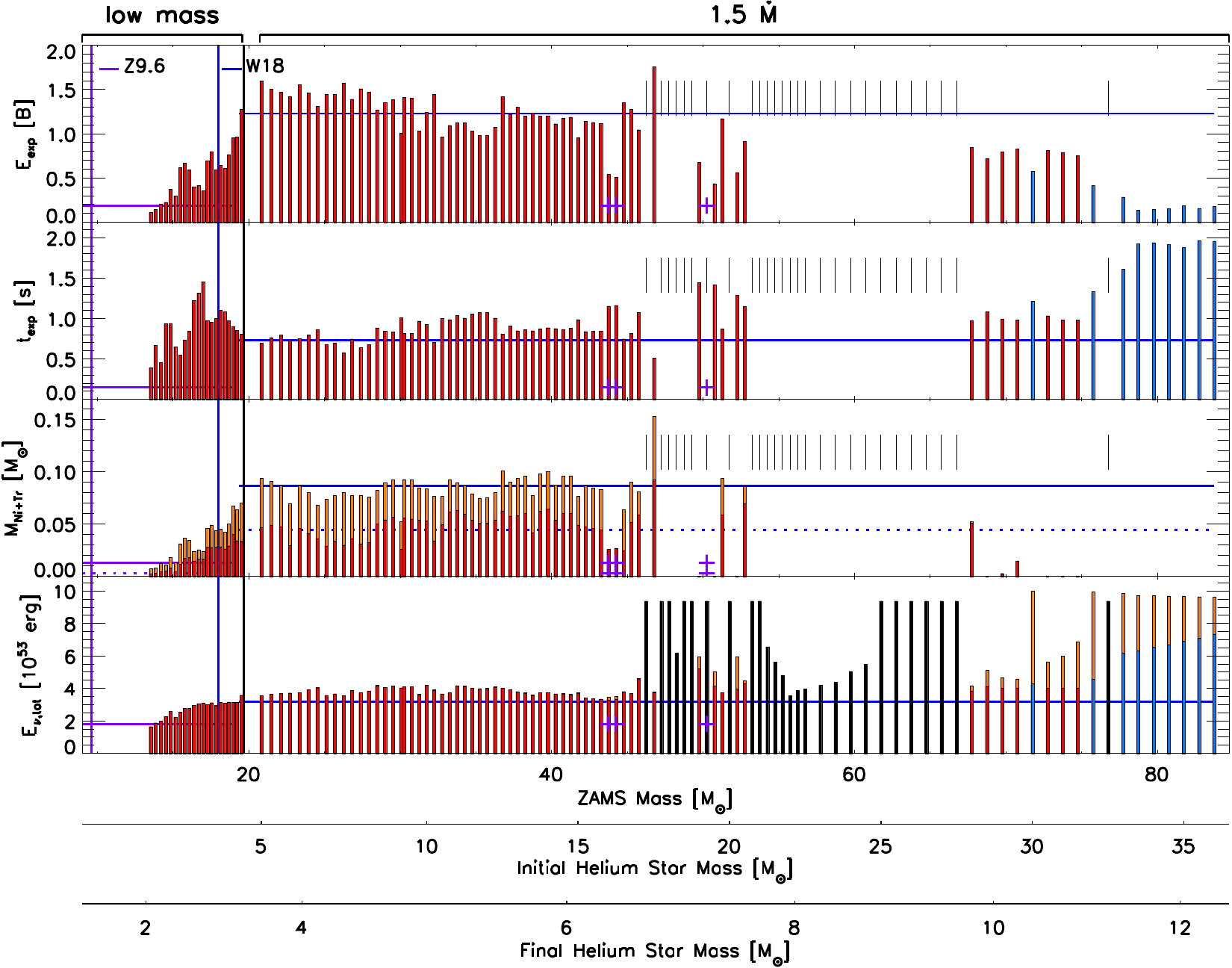}
 \caption{Same as \Fig{summaryw18-sta-exp} but for the full set of
   helium-star models with increased mass-loss rate (right of the black vertical line).}
 \lFig{summaryw18-1.5-exp}
\end{figure*}

\begin{figure*}
 \includegraphics{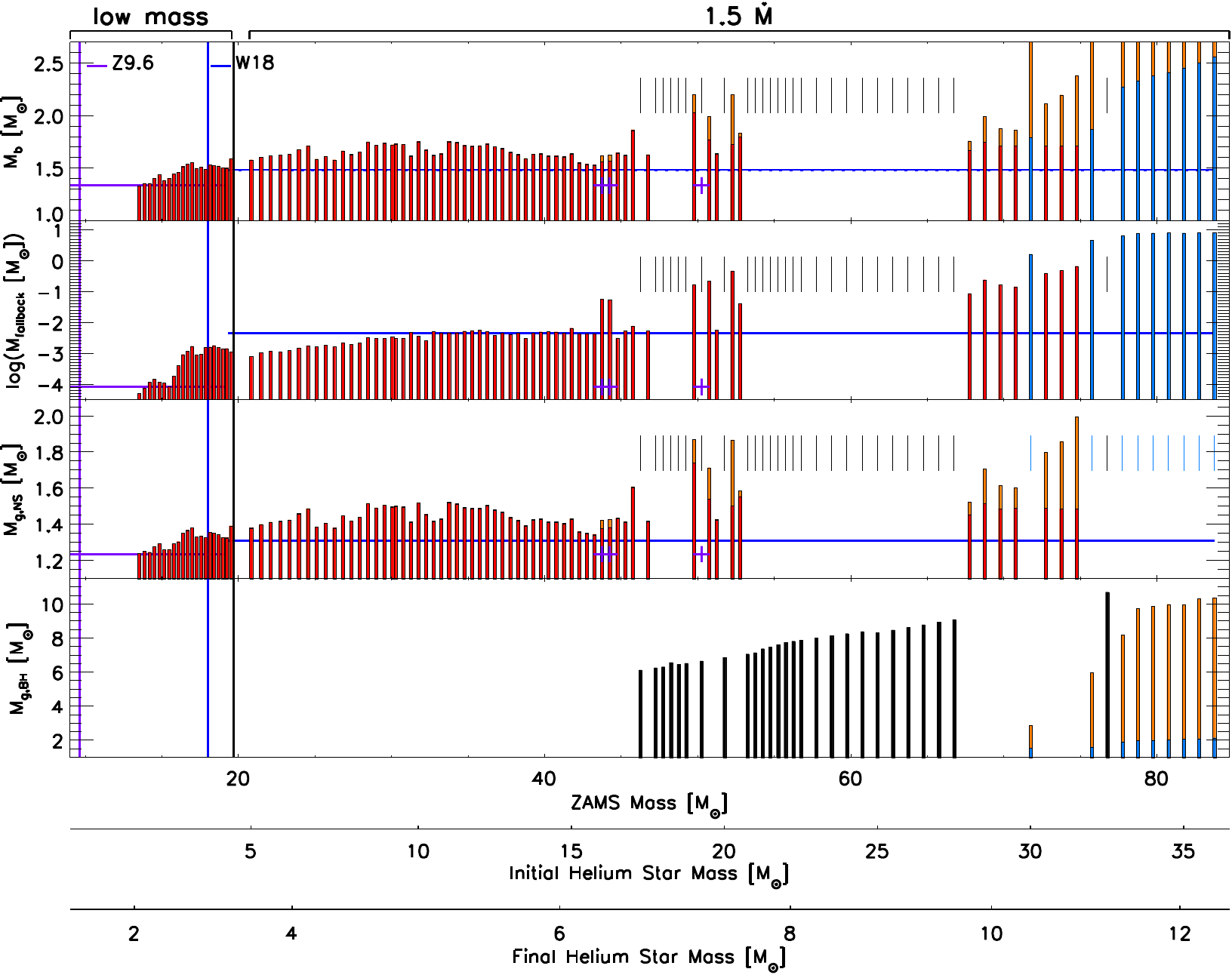}
 \caption{Same as \Fig{summaryw18-sta-rem} but for the full set of
   helium-star models with increased mass-loss rate (right of the black vertical line).}
 \lFig{summaryw18-1.5-rem}
\end{figure*}

{\bf (2)}
In the case of black hole formation by fallback accretion we compute the 
gravitational black hole mass according to
\begin{equation}
M_\mathrm{g,BH} = M_\mathrm{b,NS}^\mathrm{bfb} + M_\mathrm{fb} - 
\frac{1}{c^2}\,E_\mathrm{b,NS}^\mathrm{max}\,.
\label{eq:BHmass3}
\end{equation}
where $M_\mathrm{fb}$ is the fallback mass and 
$M_\mathrm{b,NS}^\mathrm{bfb}$ 
the baryonic mass of the transiently stable neutron star before fallback,
whose neutrino emission has triggered the supernova explosion. 
The maximum binding energy of a neutron star is assumed to be
\begin{equation}
E_\mathrm{b,NS}^\mathrm{max} = E_{\nu,\mathrm{tot}}(t_\mathrm{end})
+ E_\mathrm{b,fb}^\mathrm{LP01}\,.
\label{eq:BHmass3a}
\end{equation}
Here we estimate the neutrino-energy release associated with the
fallback, $E_\mathrm{b,fb}^\mathrm{LP01}$,
by employing equation~(36) from \citet{Lat01} for computing 
the difference between the binding energy of the final neutron star when
it collapses to a black hole and the initial neutron star mass before fallback
accretion:
\begin{equation}
E_\mathrm{b,fb}^\mathrm{LP01} =
E_\mathrm{b,10\,km}^\mathrm{LP01}(M_\mathrm{b,NS}^\mathrm{max}) -
E_\mathrm{b,12\,km}^\mathrm{LP01}(M_\mathrm{b,NS}^\mathrm{bfb})\,.
\label{eq:BHmass3b}
\end{equation}
This assumes that the fallback accretion proceeds slowly enough,
and therefore the accreting neutron star remains stable for a sufficiently
long period of time, to allow all the binding energy being 
radiated away.

Neutrino or radiation-energy loss by further radial (in 1D)
accretion after black hole formation is again assumed to be negligible.
In the few cases of the standard mass-loss set of progenitors
where $M_\mathrm{b,NS}^\mathrm{bfb} 
> M_\mathrm{b,NS}^\mathrm{max} = 2.75$\,M$_\odot$ 
(see \Fig{NSbaryon}, left panel), we expect
that the neutron star collapses to a black hole due to neutrino emission after
10\,s of simulated post-bounce evolution. Therefore in these cases
we also consider no additional neutrino-energy loss during the 
fallback accretion by the black hole. 

{\bf (3)}
For consistency, we also include mass-decrement corrections due to
neutrino and radiation loss from fallback matter accreted by 
the stable neutron stars in our simulations, although these corrections 
are minuscule in the far majority of all cases because of 
the usually small fallback masses. In order to do that, we 
follow a procedure analogue to the one applied in point~(2).
The corrected gravitational mass of the neutron star is
\begin{equation}
M_\mathrm{g,NS} = M_\mathrm{b,NS}^\mathrm{bfb} + M_\mathrm{fb} - 
\frac{1}{c^2}\,E_\mathrm{b,NS}(M_\mathrm{b,NS})\,.
\label{eq:NSmass1}
\end{equation}
where $M_\mathrm{b,NS}^\mathrm{bfb}$ is the baryonic mass of the
NS before fallback and $M_\mathrm{b,NS}$ the final baryonic neutron star mass 
including fallback. The binding energy of the final neutron star is taken
to be
\begin{equation}
E_\mathrm{b,NS}(M_\mathrm{b,NS}) = E_{\nu,\mathrm{tot}}(t_\mathrm{end})
+ E_\mathrm{b,fb}^\mathrm{LP01}
\label{eq:NSmass1a}
\end{equation}
with
\begin{equation}
E_\mathrm{b,fb}^\mathrm{LP01} =
E_\mathrm{b,12\,km}^\mathrm{LP01}(M_\mathrm{b,NS}) -
E_\mathrm{b,12\,km}^\mathrm{LP01}(M_\mathrm{b,NS}^\mathrm{bfb})\,.
\label{eq:NSmass1b}
\end{equation}
Again, we estimate the neutrino-energy release associated with the
fallback, $E_\mathrm{b,fb}^\mathrm{LP01}$,
by employing equation~(36) from \citet{Lat01}, but now we compute
the difference between the binding energy of the final neutron star and
the binding energy of the neutron star before fallback by adopting a 
radius of 12\,km both times.

We finally repeat that only in a tiny subset of our present 
simulations the assumed baryonic neutron star mass limit of 2.75\,$M_\odot$
is exceeded at the time when the explosion sets in, i.e.,
$M_\mathrm{b,NS}(t_\mathrm{exp}) > 2.75$\,M$_\odot$ (see
\Fig{NSbaryon}, left panel). However, the overshoot
is so small that we can safely assume that thermal pressure
stabilizes the neutron stars long enough for them to power the associated 
weak fallback supernova explosions by their neutrino emission. We also
treat these neutron stars such that they survive for the full simulation
time of 10\,s to radiate neutrinos. This is a crude assumption.
But since these cases have high fallback masses and give birth 
to black holes of $\sim$10\,M$_\odot$ or more, this implies only minor
errors in our estimates of the black hole masses. 

\subsection{Results for the W18 Central Engine}
\lSect{W18results}


Figs.~\ref{fig:summaryw18-sta-exp}--\ref{fig:summaryw18-1.5-rem}
summarize the P-HOTB simulations of helium stars with standard and
enhanced mass-loss rates that use the Z9.6 \& W18 engine. Cases with neutron star
formation are shown by red bars, fallback supernovae with black hole
formation by blue bars, and black hole formation without explosion by black
bars. The direct black hole formation cases are also marked by short vertical
black dashes in the upper halves of all panels, and the fallback black hole
formation cases by short vertical blue dashes in some panels, too.

Overall, the explosion energies, explosion time scales, Ni+Tr masses,
and the remnant masses for the helium stars are quite similar to those
for single stars, see \citet{Suk16} and Appendix~\ref{app:comparison},
in particular \Fig{resultsw18transnew} there.  The minimum values of the
explosion energy, $E_\mathrm{exp}$, with the Z9.6 \& W18 engine are near
$10^{50}$\,erg and the maximum values are near $1.8\times 10^{51}$\,erg. The
corresponding Ni+Tr masses range from $\sim$0.007\,M$_\odot$ for the
weakest explosions to $\sim$0.10\,M$_\odot$ for the most energetic
ones. A single outlier in the model set with enhanced mass loss
produces slightly more than 0.15\,M$_\odot$
(\Fig{summaryw18-1.5-exp}). This star, with an initial helium-star mass of
17.5\,M$_\odot$, exploded unusually early with a high energy of
$1.75\times 10^{51}$\,erg and a correspondingly high yield of Ni+Tr.
These special explosion conditions resulted because the base of the oxygen shell was
characterized by an exceptionally high entropy jump (roughly from
$s\approx 2.5$ to a value near 6), leading to a more dramatic drop of
the mass accretion rate than for neighboring stars with similar helium
star masses and similar values of $M_4$.

The lowest fallback masses of less than $10^{-4}$\,M$_\odot$ were
obtained for the helium-stars with the smallest pre-collapse masses and
the lowest explosion energies, while the more typical cases had
fallback masses between $\sim$10$^{-3}$\,M$_\odot$ and roughly
$10^{-2}$\,M$_\odot$.  Only for progenitors with initial helium-star
masses above about 15\,M$_\odot$ did fallback significantly greater
than $10^{-2}$\,M$_\odot$ become more frequent.

The estimated gravitational masses of the neutron stars ranged from about
1.24\,M$_\odot$ up to $\sim$2.23\,M$_\odot$ for a progenitor of
roughly 57\,M$_\odot$ ZAMS mass (22.50\,M$_\odot$ initial helium-star mass) in
the standard mass-loss set. For this heavy case, fallback of nearly
1.0\,$M_\odot$ lifted the baryonic mass of the compact remnant to
$\sim$2.71\,M$_\odot$, which was just below our assumed mass limit for
cold neutron stars.  A similarly high value of $M_\mathrm{g,NS} \approx
2$\,M$_\odot$ ($M_\mathrm{b,NS} \approx 2.4$\,M$_\odot$) came from
another progenitor of $\sim$75\,M$_\odot$ ZAMS mass (31.50\,M$_\odot$
initial helium-star mass) for the model set with enhanced mass loss.

The black hole masses range from values just above the threshold mass
for black hole formation to $\sim$19.5\,M$_\odot$ for the most massive
helium-stars with standard mass loss, and $\sim$10.7\,M$_\odot$ for the
set with higher mass loss rate. At the lower end the black hole mass
was set by fallback; at the upper end the entire presupernova star
collapsed.

The most obvious difference of the helium-star explosions compared to
the single-star models of \citet{Suk16} (see also
Appendix~\ref{app:comparison}) is the greater number of cases with
significant fallback. These fill in a ``gap'' in the remnant mass
distribution that might have existed
between gravitational masses of about 2 to 6\,\Msun. Most cases where
black holes are formed by fallback are associated with progenitors
that explode very late ($t_\mathrm{exp}\sim 2$\,s) at the time when
the infalling point with entropy per baryon 6.0 reaches the
shock. Such cases form massive proto-neutron stars (see \Sect{dependence-on-engine})
that explode with relatively low energies (about 1--$4\times
10^{50}$\,erg) and eject no iron-group material (in the absence of
mixing). The fallback includes much of the progenitor mass. Typically
2--3\,M$_\odot$ are still ejected in models with standard mass loss,
but only 1--1.5\,M$_\odot$ for the most massive helium stars with
enhanced mass loss.  We estimate the neutrino-energy release in such
cases (including the energy loss from fallback accretion) to be up to
$\sim$10$^{54}$\,erg (or about 0.55\,M$_\odot$), very similar to black
hole formation by continuous, slow accretion.

Also the black holes with the lowest masses originate
from fallback accretion. In the standard mass-loss set these are
just above the gravitational-mass threshold for black hole formation,
namely 2.31\,M$_\odot$, 2.29\,M$_\odot$, and 2.56\,M$_\odot$
for final helium-star masses of 10.91\,M$_\odot$, 11.23\,M$_\odot$, 
and 11.34\,M$_\odot$, respectively. In the progenitor set with
enhanced mass loss there is a similar case with 2.89\,M$_\odot$
for the fallback black hole in a star with a final helium-star mass of
11.38\,M$_\odot$.

\begin{figure}
\centering
\includegraphics[width=0.48\textwidth]{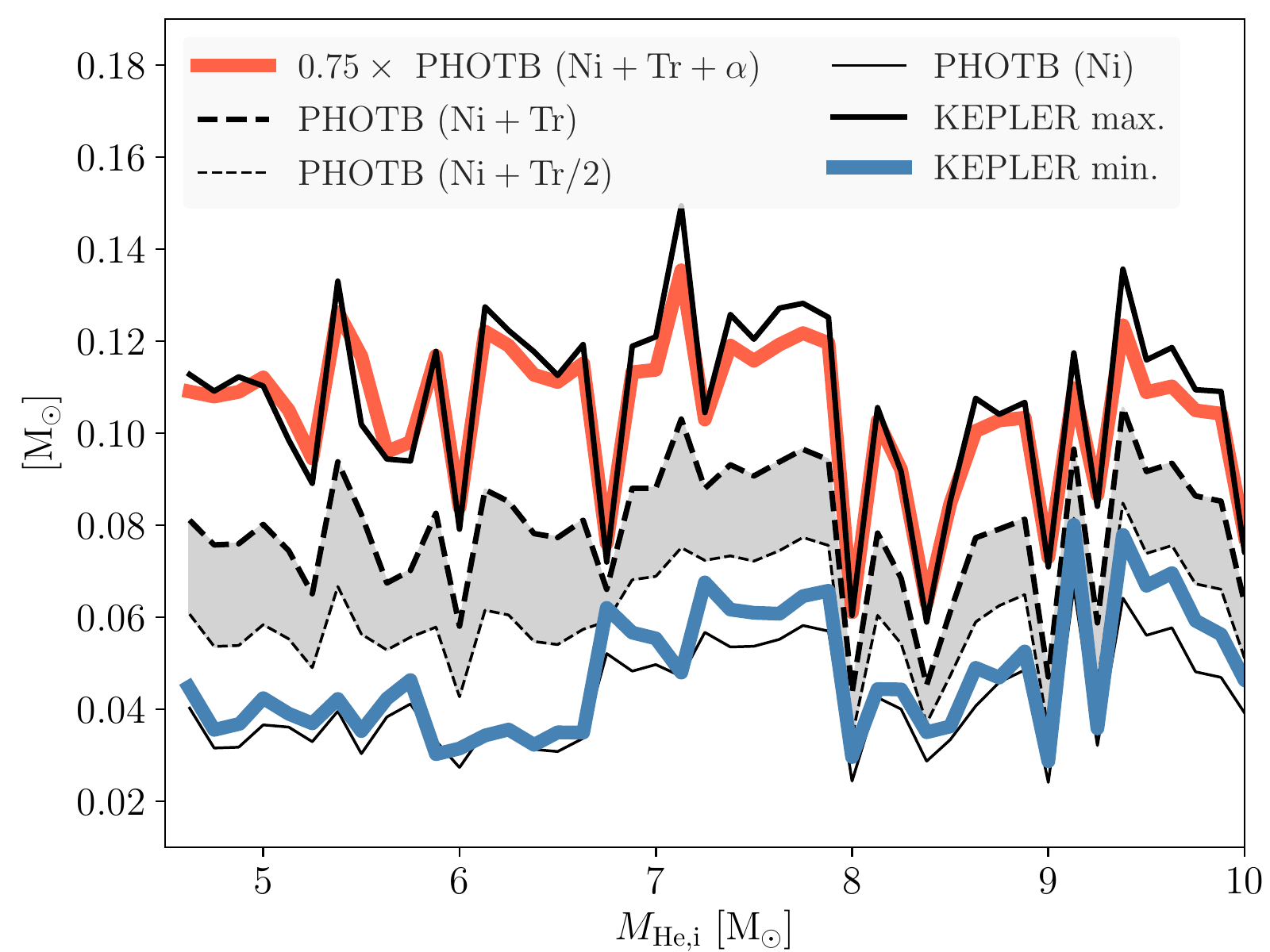}
 \caption{Production of $^{56}$Ni for various assumptions in
   P-HOTB and KEPLER versus initial helium star mass. The explosion 
   simulations of the helium-stars with standard mass-loss rate were
   performed with the Z9.6 \& W18 neutrino engine. From top to bottom:
   a) the solid dark black line is the $^{56}$Ni production in the KEPLER
   code when the piston is located at the final mass cut in the explosion.
   It is an upper bound; b) the broad red curve is 75\% of the matter that 
   has experienced NSE in P-HOTB plus a small amount made by explosive 
   silicon burning; c) the gray band bounded by dashed black lines is our
   estimate of the most likely $^{56}$Ni production in the explosion. The 
   lines are Ni+Tr ({\em top}) and Ni+Tr/2 ({\em bottom}) from the P-HOTB simulation;
   d) the broad blue line is the production in KEPLER when the special 
   trajectory \citep{Suk16} is used; and e) the thin solid black line 
   is just the $^{56}$Ni produced in 
   the P-HOTB simulation with no contribution from the wind or matter that 
   has experienced weak interactions to get ejected with $Y_e < 0.49$. See 
   the discussion in \Sect{ni56}. The broad red and blue lines were used to 
   compute maximum and minimum light curves in \Sect{lite}.}
\lFig{ni56yields}
\end{figure}

\begin{figure*}
\includegraphics[width=.495\textwidth]{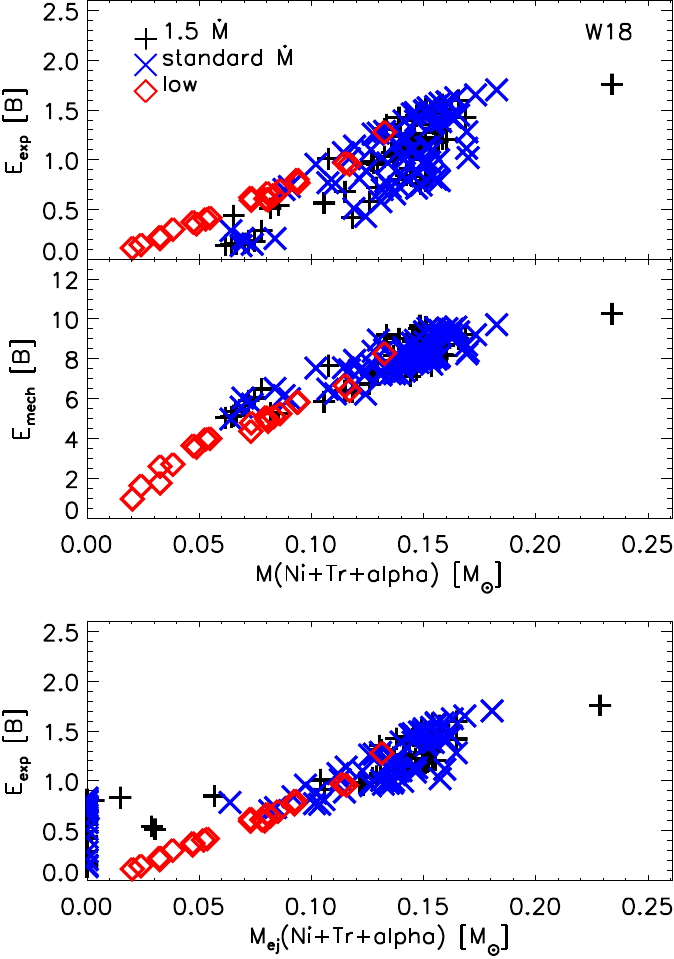} \hskip4.0pt
\includegraphics[width=.495\textwidth]{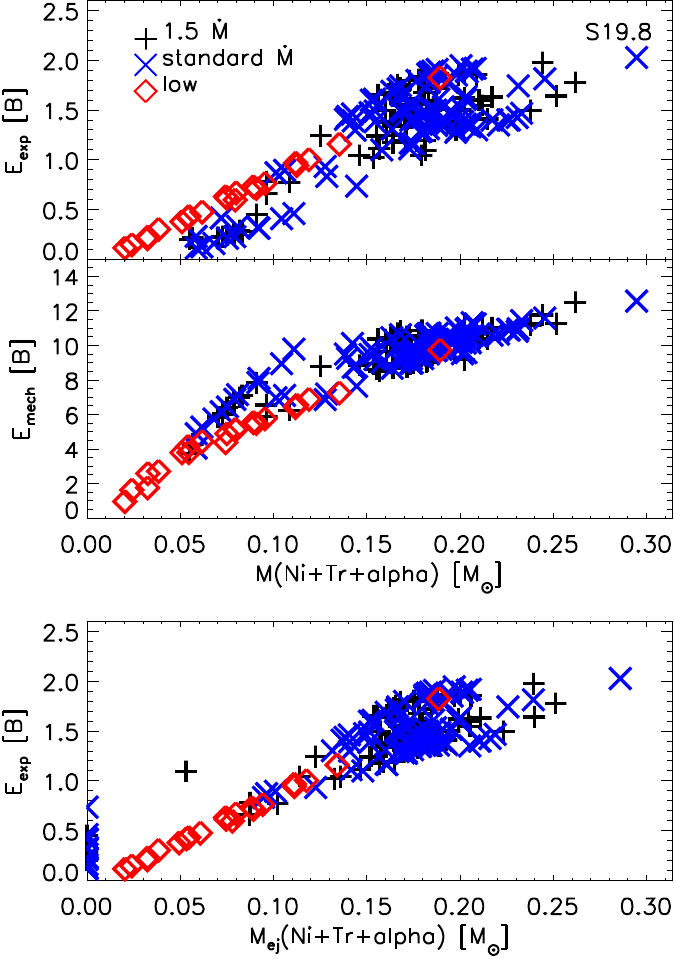}
\caption{Correlations between the mass of Ni+Tr+$\alpha$ produced by the
supernova explosion and the explosion energy, $E_\mathrm{exp}$ ({\em top and
bottom panels}), as well as the energy $E_\mathrm{mech}$ that the 
neutrino-driven mechanism has to provide to lift the ejecta out of the
gravitational potential of the neutron star and 
to unbind the whole star with an energy of $E_\mathrm{exp}$
(for the exact definition, see Equation~(\ref{eq:emech}); {\em middle panels}).
In the {\em upper four panels} $M(\mathrm{Ni}+\mathrm{Tr}+\alpha)$
on the abscissa is the mass of these species {\em before} fallback, in the
{\em bottom two panels} $M_\mathrm{ej}(\mathrm{Ni}+\mathrm{Tr}+\alpha)$ is the
corresponding mass corrected for fallback, i.e., the actually ejected mass of
the three nuclei. The
plots show P-HOTB simulations for all investigated helium-star progenitors
with the Z9.6 \& W18 ({\em left panels}) and
Z9.6 \& S19.8 ({\em right panels}) neutrino engines.} 
\lFig{NiTia-correlations}
\end{figure*}

In the context of this work the production of radioactive nickel
during the explosion is of particular interest. A detailed
discussion of associated uncertainties, which are connected to
our simplifications of the explosion modeling, was provided in 
\Sect{ni56}.
For a subset of stars with initial helium star masses between 
about 4.6\,M$_\odot$ and 10\,M$_\odot$ and standard mass loss, 
exploded with the 
Z9.6 \& W18 neutrino engine, \Fig{ni56yields} displays
lower and upper estimates of the nickel yields compared to our
best estimates. Five different quantities are shown:
$^{56}$Ni calculated using the
same approach as \citet{Suk16}, Ni from P-HOTB, Ni+Tr/2 from P-HOTB,
Ni+Tr, and 0.75 times Ni+Tr+$\alpha$ from P-HOTB.
The thin solid black line for Ni from P-HOTB represents
the results shown by the red parts of the histogram bars in the third 
panel of \Fig{summaryw18-sta-exp}, and the thick dashed black line
for Ni+Tr corresponds to the whole histogram bars in the same panel.
The analysis of \citet{Suk16} gives values which we
now believe are close to the minimum value, ``Ni only'' in the figure,
but not far below the best estimate, Ni+Tr/2. The upper bound, 0.75
times Ni+Tr+$\alpha$ is much larger than any of these, and is probably
a gross upper bound, though more extreme variations, probably not
appropriate for a purely neutrino-powered explosion, are discussed in
\Sect{ni56max}.

In neutrino-driven explosions, independent of the dimensionality of
the modeling, the explosion energy, $E_\mathrm{exp}$, is provided
by neutrino-heated matter that is expelled
from the close vicinity or from the surface of the neutron star after having 
absorbed energy from neutrinos. Neutrino heating
basically transfers the energy to lift this matter to a
gravitationally marginally bound state, i.e., to specific energy
$\epsilon_\mathrm{tot} = \epsilon_\mathrm{int}+\epsilon_\mathrm{grv}
+\epsilon_\mathrm{kin} \lesssim 0$ (with $\epsilon_\mathrm{int}$,
$\epsilon_\mathrm{grv}$, and $\epsilon_\mathrm{kin}$ being the 
specific internal, gravitational, and kinetic energies, respectively).
The excess energy that fuels the supernova explosion mainly 
stems from nuclear recombination in the neutrino-heated matter
as it expands and cools down from its initial conditions in NSE
(see discussion in Appendix~C of \citealt{Sch06}; see also
\citealt{Mar09,Mul15,Bru16,Jan17}). This leads to the simple
relation \citep[][]{Sch06}
\begin{equation}
\frac{E_\mathrm{exp}}{10^{51}\,\mathrm{erg}} \approx
\frac{M_\mathrm{rec}}{0.1\,\mathrm{M}_\odot}\cdot
\frac{\epsilon_\mathrm{rec}}{5\,\mathrm{MeV}}\,,
\label{eq:recerg}
\end{equation}
where $M_\mathrm{rec}$ is the neutrino-driven ejecta mass releasing 
the recombination energy, and $\epsilon_\mathrm{rec}$ is the mean
recombination energy per nucleon. Typical values of
$\epsilon_\mathrm{rec}$ are around 5--6\,MeV. These are below the
maximum value of 8.8\,MeV set free when nucleons recombine
to iron-group nuclei, because recombination
can be incomplete (producing $\alpha$ particles). Moreover, some 
fraction of the released energy is usually still needed to
overcome the gravitational binding energy of the ejected gas.

In our 1D simulations neutrino-heated matter has $Y_e$ values between
$\sim$0.45 and 0.49. Therefore neutrons and protons present at
NSE recombine to tracer (Tr) material
and $\alpha$ particles, depending on the entropy and expansion
time scale of the ejecta, i.e., 
$M_\mathrm{rec} = M(\mathrm{Tr}) + M(\alpha)$ (see \Sect{PHOTB}).
At conditions of $Y_e \gtrsim 0.5$ a significant mass fraction of
$^{56}$Ni would be assembled.

Because of Equation~(\ref{eq:recerg}) the explosion energy 
correlates with $M_\mathrm{rec}$, and since the mass of explosively
produced $^{56}$Ni in the shock-heated progenitor layers
with $Y_e = 0.5$ should also correlate with the explosion
energy, we expect a strong correlation between $E_\mathrm{exp}$
and the summed masses of Ni+Tr+$\alpha$. This correlation is 
visible in the top and bottom panels of the two plots in 
\Fig{NiTia-correlations} and even more tightly in
the two middle panels. In our simulations the masses of $^{56}$Ni
and Tr are roughly equal (see \Fig{ni56yields} and third 
panels of Figs.~\ref{fig:summaryw18-sta-exp} and 
\ref{fig:summaryw18-1.5-exp}).
Since Ni is formed in nuclear reactions of oxygen and silicon,
releasing only $\sim$1\,MeV per nucleon, its fusion contributes
only little to the explosion energy.

Note that in \Fig{NiTia-correlations} 
$M(\mathrm{Ni}+\mathrm{Tr}+\alpha)$ is measured {\em before fallback},
which means that it constitutes the mass of these nuclear species
that contributes to the energy provided by the explosion mechanism
through nuclear recombination and fusion. The total energy that the
mechanism would have to make available to unbind the complete star
and explode it with energy $E_\mathrm{exp}$ is denoted by
$E_\mathrm{mech}$ in \Fig{NiTia-correlations}. It is
defined as 
\begin{equation}
E_\mathrm{mech} = E_\mathrm{exp} -
E_\mathrm{bind}(m>M_\mathrm{b,NS}^\mathrm{bfb};t_\mathrm{exp})\,,
\label{eq:emech}
\end{equation}
where $E_\mathrm{bind}(m>M_\mathrm{b,NS}^\mathrm{bfb};t_\mathrm{exp})$
is the (negative) total energy (internal plus gravitational plus
kinetic), measured at the time when the explosion sets in, of all 
matter exterior to the baryonic mass of the neutron star {\em before fallback}.
Consistent with our hydrodynamic modeling, we take general relativistic 
corrections into account in calculating the gravitational energy of 
all matter $m>M_\mathrm{b,NS}^\mathrm{bfb}$.
The bounding mass shell, $m = M_\mathrm{b,NS}^\mathrm{bfb}$, is
located below the surface of the proto-neutron star at time $t_\mathrm{exp}$, and
the overlying shells are blown out by neutrino heating in the 
neutrino-driven wind to deliver the energy to the explosion as 
described above. $E_\mathrm{mech}$ is therefore
much larger than the final explosion energy $E_\mathrm{exp}$. 
On the one hand this is so because $E_\mathrm{mech}$ includes the
energy that the explosion mechanism has to spend on lifting the 
matter that contains $M(\mathrm{Ni}+\mathrm{Tr}+\alpha)$ out of
the gravitational potential trough of the neutron star, 
and also on unbinding all of the overlying stellar material 
ahead of the supernova shock. On the 
other hand, fallback removes some of the energy initially 
transferred to the transiently expanding stellar matter, which
can lead to a reduction of $E_\mathrm{exp}$. Measuring 
$E_\mathrm{bind}(m>M_\mathrm{b,NS}^\mathrm{bfb};t_\mathrm{exp})$
at $t_\mathrm{exp}$ is an approximation because the neutron star contracts
with time and its surface layers become gravitationally more 
strongly bound as time goes on. 
However, our approximation yields a useful proxy  
because the neutrino-driven outflow is strongest in the phase 
when the explosion sets in and shortly afterwards. 

\begin{figure*}
\includegraphics[width=0.49\textwidth]{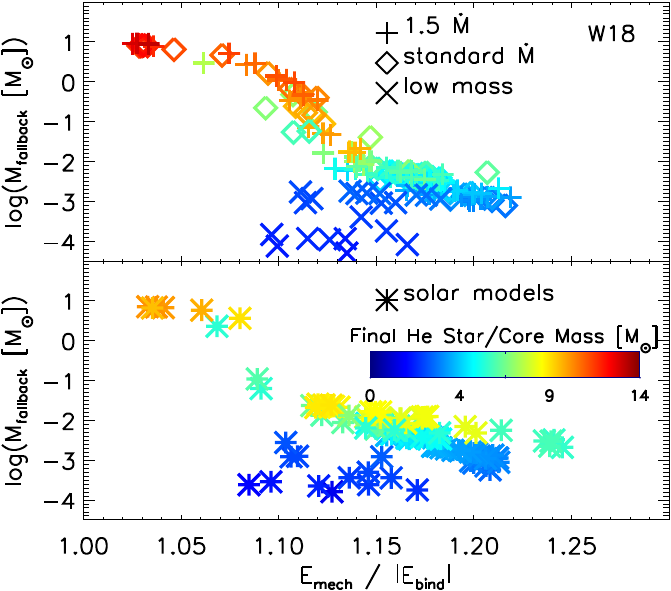}\hskip10.0pt
\includegraphics[width=0.49\textwidth]{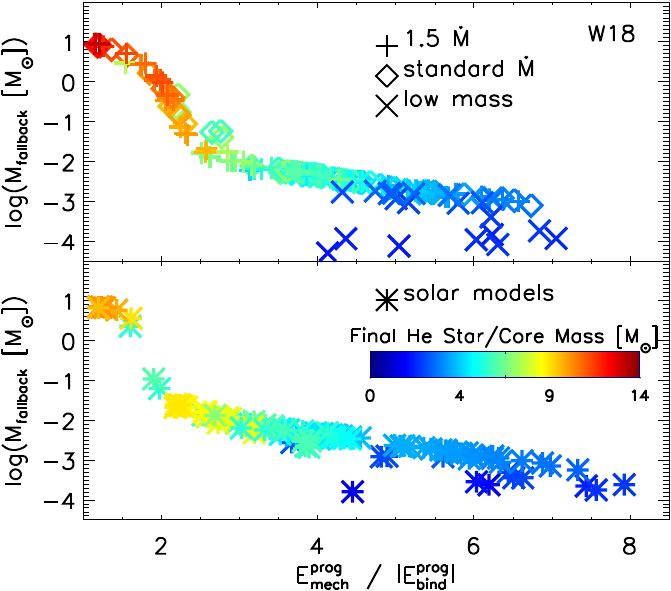}
\\$\phantom{.}$\\
\includegraphics[width=0.49\textwidth]{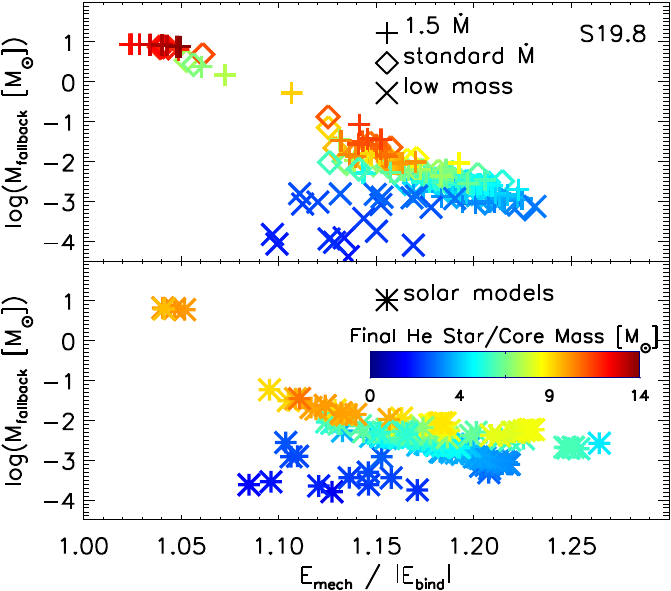}\hskip10.0pt
\includegraphics[width=0.49\textwidth]{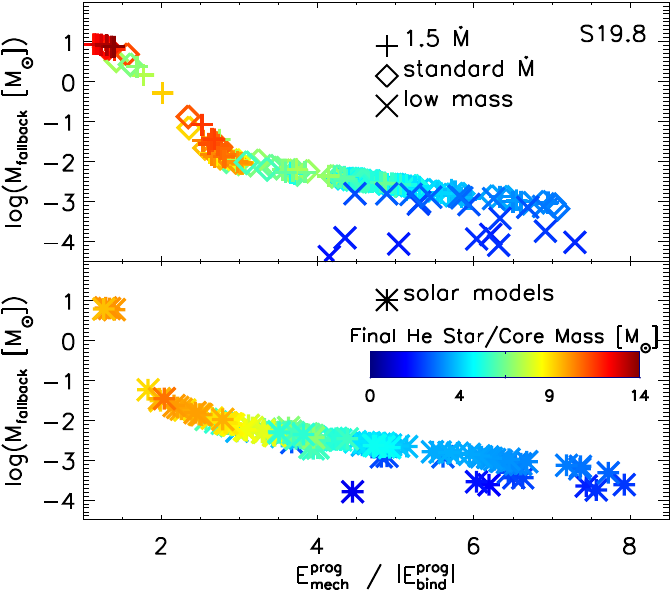}
\caption{{\em Left plots:}
Fallback masses (log-scale) versus ratio of $E_\mathrm{mech}$
(as defined in Equation~(\ref{eq:emech})) to binding energy $|E_\mathrm{bind}|$
for all helium-star progenitors ({\em upper panels} in all plots) and the
single-star progenitors from \citet{Suk16} ({\em lower panels} in all plots)
that explode in P-HOTB simulations with the Z9.6 \& W18 neutrino engine
({\em top plots}) or the Z9.6 \& S19.8 engine ({\em bottom plots}).
The ratio $E_\mathrm{mech}/|E_\mathrm{bind}|$ is computed for all mass
outside of the initial mass cut ($m > M_\mathrm{b,NS}^\mathrm{bfb}$) 
in the profile of the collapsed star at time $t_\mathrm{exp}$ 
when the explosion sets in. {\em Right plots:} Same as left plots but with
$E_\mathrm{mech}^\mathrm{prog}/|E_\mathrm{bind}^\mathrm{prog}|$ evaluated
for this mass range in the
pre-collapse profiles of the progenitors. The color coding of the symbols
in all panels corresponds to the final helium-core masses.}
\lFig{fallbackcorrelations}
\end{figure*}

The correlation between $E_\mathrm{mech}$ and 
$M(\mathrm{Ni}+\mathrm{Tr}+\alpha)$ in the middle panels of
\Fig{NiTia-correlations} is considerably tighter
than between $E_\mathrm{exp}$ and $M(\mathrm{Ni}+\mathrm{Tr}+\alpha)$
in the top two panels. In the latter panels the uppermost data points
describe the relation of Equation~(\ref{eq:recerg}) very closely.
These are the normal explosions with very little fallback (typically
much less than $\sim$10$^{-2}$\,M$_\odot$, see
Figs.~\ref{fig:summaryw18-sta-rem} and \ref{fig:summaryw18-1.5-rem}).
The cases that lie well below this imagined correlation line in the
top panels, while their correspondents in the middle panels are well 
integrated in the main data band, are supernova explosions with more
significant fallback. In these cases the binding energy of the 
collapsed stellar core is considerably larger and a bigger fraction
of $E_\mathrm{mech}$ is needed to lift the ejecta out of the 
gravitational potential trough. Therefore the resulting explosion 
energies are noticeably lower than the typical values.

The cluster of data points below the main data band 
at $M(\mathrm{Ni}+\mathrm{Tr}+\alpha)$ between $\sim$0.05\,M$_\odot$ and
$\sim$0.10\,M$_\odot$ in the top panels of \Fig{NiTia-correlations} is even more extreme.
These are our cases of fallback supernovae with black hole formation, caused by 
very large fallback masses as discussed above. They are 
characterized by collapsed stellar cores that possess particularly
high binding energies because the explosion sets in very late and
the neutron star is extremely massive and compact. These conditions permit
only marginal explosions, (a) because the binding energy in
$E_\mathrm{mech}$ is much bigger than usual, and (b) because 
$M(\mathrm{Ni}+\mathrm{Tr}+\alpha)$ is rather low when the 
explosion starts late. Since the mechanism cannot provide the
energy to unbind the whole star, massive fallback is the consequence
and $E_\mathrm{exp}$ becomes small. Therefore the data points are
located close to the bottom of the top panels in
\Fig{NiTia-correlations}.

\Fig{NiTia-correlations} demonstrates that the mass of 
Ni+Tr+$\alpha$ produced by neutrino-driven explosions is limited 
due to its correlation with the explosion energy. The latter is
not found to exceed $\sim2\times 10^{51}$\,erg even for our 
strongest neutrino engine, Z9.6 \& S19.8 (see right plot in
\Fig{NiTia-correlations}). For this reason we do not obtain
$M(\mathrm{Ni}+\mathrm{Tr}+\alpha)$ higher than $\sim$0.2\,M$_\odot$.
For the Z9.6 \& W18 engine the limit is even only around
$\sim$0.15\,M$_\odot$ (corresponding to an extreme upper bound
on $^{56}$Ni of $\sim$0.12\,M$_\odot$ indicated by the red 
line in \Fig{ni56yields}). In most cases fallback reduces
the ejected mass of Ni+Tr+$\alpha$, 
$M_\mathrm{ej}(\mathrm{Ni}+\mathrm{Tr}+\alpha)$, only slightly
compared to the nucleosynthesized mass represented by
$M(\mathrm{Ni}+\mathrm{Tr}+\alpha)$. However, the data points 
for the cases with more massive fallback are shifted considerably
in the two bottom panels of \Fig{NiTia-correlations},
where $M_\mathrm{ej}(\mathrm{Ni}+\mathrm{Tr}+\alpha)$ is plotted
on the abscissa instead of $M(\mathrm{Ni}+\mathrm{Tr}+\alpha)$.
This concerns most of the points with small explosion
energies in the lower parts of the two top panels. They are moved
closer to the correlation line describing Equation~(\ref{eq:recerg}).
This leads to a more narrow band of data points in the bottom 
panels and a correlation similarly tight as the one seen in the
two middle panels. Outliers are only
the (few) cases in which all or most of the nucleosynthesized 
Ni+Tr+$\alpha$ falls back. The corresponding data points cluster
in the left part of the bottom panels where the ejected mass
$M_\mathrm{ej}(\mathrm{Ni}+\mathrm{Tr}+\alpha)$ is very small or
zero. Since mixing is not included in our 1D simulations, it is
unclear whether multi-dimensional effects during the long-time
evolution of the supernova can change this result.

\subsection{Fallback}
\lSect{fallback}

In the new helium-star models, a considerably larger number of cases
show appreciable fallback than in our previous studies of single stars
\cite{Suk16,Ert16a}.
This fallback results in neutron stars with high
gravitational masses, up to more than 2\,M$_\odot$ and black holes with
smaller gravitational masses between 2.3\,M$_\odot$ and
6\,M$_\odot$. Such cases were absent in our previous works on
single-star explosions. Why?

\Fig{fallbackcorrelations} compares the fallback conditions of the
helium stars with the explosions of the single-star progenitors
considered by \citet{Suk16}.  All simulations, including those of
single stars, were calculated using the upgraded version of P-HOTB.
As discussed in \Sect{W18results}, large fallback often happens when a
high energy, $E_\mathrm{mech}$ (see Equation~(\ref{eq:emech})), is
required in order to gravitationally unbind the whole star and provide
an energy of $E_\mathrm{exp}$.  This quantity can be used to measure
the probability of massive fallback in the explosion. In the left
panels of \Fig{fallbackcorrelations}, the fallback mass is displayed
as a function of the ratio of $E_\mathrm{mech}$ to the absolute value
of the total energy $E_\mathrm{bind}$ as used and defined in
Equation~(\ref{eq:emech}).  In the right panels of
\Fig{fallbackcorrelations} $E_\mathrm{mech}/|E_\mathrm{bind}|$ is
replaced by the corresponding ratio of energies when the binding
energy of the star is not computed from the structure at
$t_\mathrm{exp}$, but from the pre-collapse profile of the star. This
means that $E_\mathrm{bind}^\mathrm{prog}$ is the (negative) total
energy of all progenitor shells above the initial mass cut, i.e., of
all mass shells $m > M_\mathrm{b,NS}^\mathrm{bfb}$. With that,
$E_\mathrm{mech}^\mathrm{prog} = E_\mathrm{exp} -
E_\mathrm{bind}^\mathrm{prog}$.  The upper panel of each plot shows
the results for helium stars, the lower panel the single stars for
comparison.

The figure shows, as expected, that massive fallback anti-correlates with these energy ratios. In all panels of
\Fig{fallbackcorrelations} there is a narrow main band that contains
the majority of all data points.  Outliers from this main band, where
most of the models cluster, are only low-mass helium-star and
single-star progenitors that explode with very little fallback and
which are therefore located below the main band. For helium stars,
fallback masses of more than about 0.1\,M$_\odot$ become frequent when
$E_\mathrm{mech}/|E_\mathrm{bind}| \lesssim 1.1$ or
$E_\mathrm{mech}^\mathrm{prog}/|E_\mathrm{bind}^\mathrm{prog}|
\lesssim 2.4$. This implies that in explosions with high fallback
masses the final explosion energy is less than $\sim$10\% of the
energy needed to gravitationally unbind the initial ejecta, or
$E_\mathrm{exp}$ is less than 1.4 times the gravitational binding
energy of the progenitor shells exterior to the initial mass
cut. These values are lower for single-star progenitors, where
$M_\mathrm{fallback} \gtrsim 0.1$\,M$_\odot$ for
$E_\mathrm{mech}/|E_\mathrm{bind}| \lesssim 1.07$ or
$E_\mathrm{mech}^\mathrm{prog}/|E_\mathrm{bind}^\mathrm{prog}|
\lesssim 1.7$, which implies $E_\mathrm{exp} \lesssim
0.07\,|E_\mathrm{bind}|$ and $E_\mathrm{exp} \lesssim
0.7\,|E_\mathrm{bind}^\mathrm{prog}|$.  Since $E_\mathrm{bind}$ is
similar for single stars and helium stars, this means that the former
explode less energetically when the fallback masses are
$\gtrsim$0.1\,$M_\odot$. This seems to be a consequence of the fact
that there is not only a reverse shock from the C+O/He interface but
also a second reverse shock from the interface between the helium core
and the hydrogen envelope. Both together damp the supernova blast more
strongly than just the reverse shock from the C+O/He interface.

Massive fallback occurs for helium stars with final helium star masses
between $\sim$10\,M$_\odot$ and $\sim$14\,M$_\odot$, whereas in the
set of single stars only a few cases with final helium-core masses around
10\,M$_\odot$ show this behavior.  There are two reasons for more
fallback cases in our current P-HOTB simulations of helium
stars. First, in the set of single-star progenitors used by
\citet{Suk16}, there were much fewer models with final helium-core masses
from 10.5\,M$_\odot$ to 13\,M$_\odot$, where both the helium stars and the
single stars make fallback supernovae. Unfortunately, our model set of 2016
was sparse in exactly this region \citep[see Table~2 in][]{Suk16}.
Second, the helium stars have lower compactness values for pre-collapse
helium star masses between $\sim$10\,M$_\odot$ and
$\sim$12\,M$_\odot$, namely $\xi_{2.5}\approx 0.15$--0.2
(\Fig{compact}, upper two panels), whereas the single stars possess
$\xi_{2.5}\approx 0.2$--0.3.  This favors direct black hole formation for
single stars and more fallback supernovae for helium stars.

In the panels that show single-star results in
\Fig{fallbackcorrelations} for both neutrino engines (Z9.6 \& W18 and
Z9.6 \& S19.8), one can see a corresponding gap of explosion cases
around $E_\mathrm{mech}/|E_\mathrm{bind}| \sim 1.05$ and
$E_\mathrm{mech}^\mathrm{prog}/|E_\mathrm{bind}^\mathrm{prog}| \sim
1.5$, which separates the few cases with the largest fallback from the
continuum of data with fallback masses $M_\mathrm{fallback}\lesssim
0.1$\,M$_\odot$.  For helium stars such a gap is absent (for the Z9.6 \&
W18 engine) or much less pronounced (for the Z9.6 \& S19.8 engine).

\begin{figure}[h] 
\centering 
\includegraphics[width=\columnwidth]{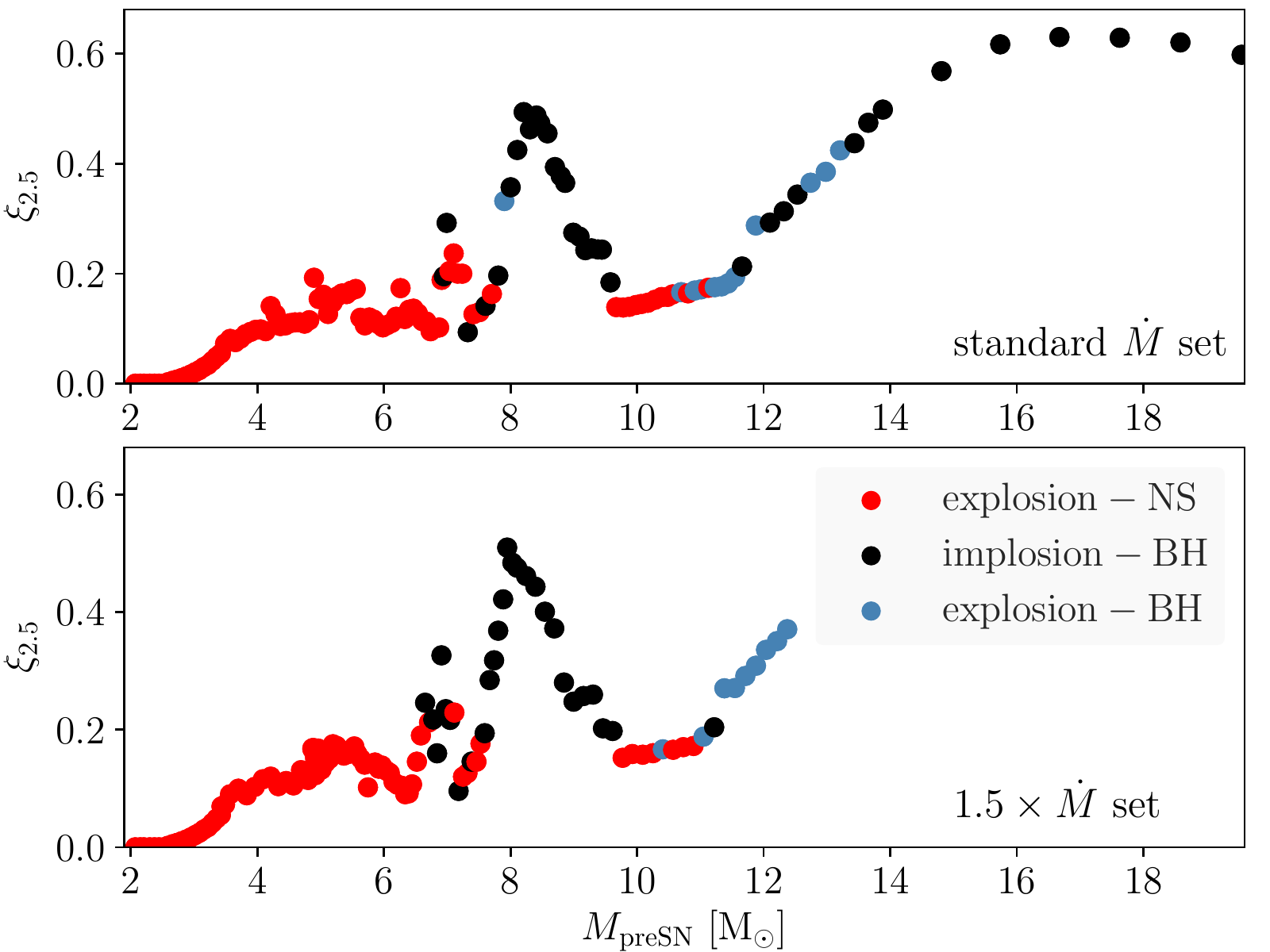}
\includegraphics[width=\columnwidth]{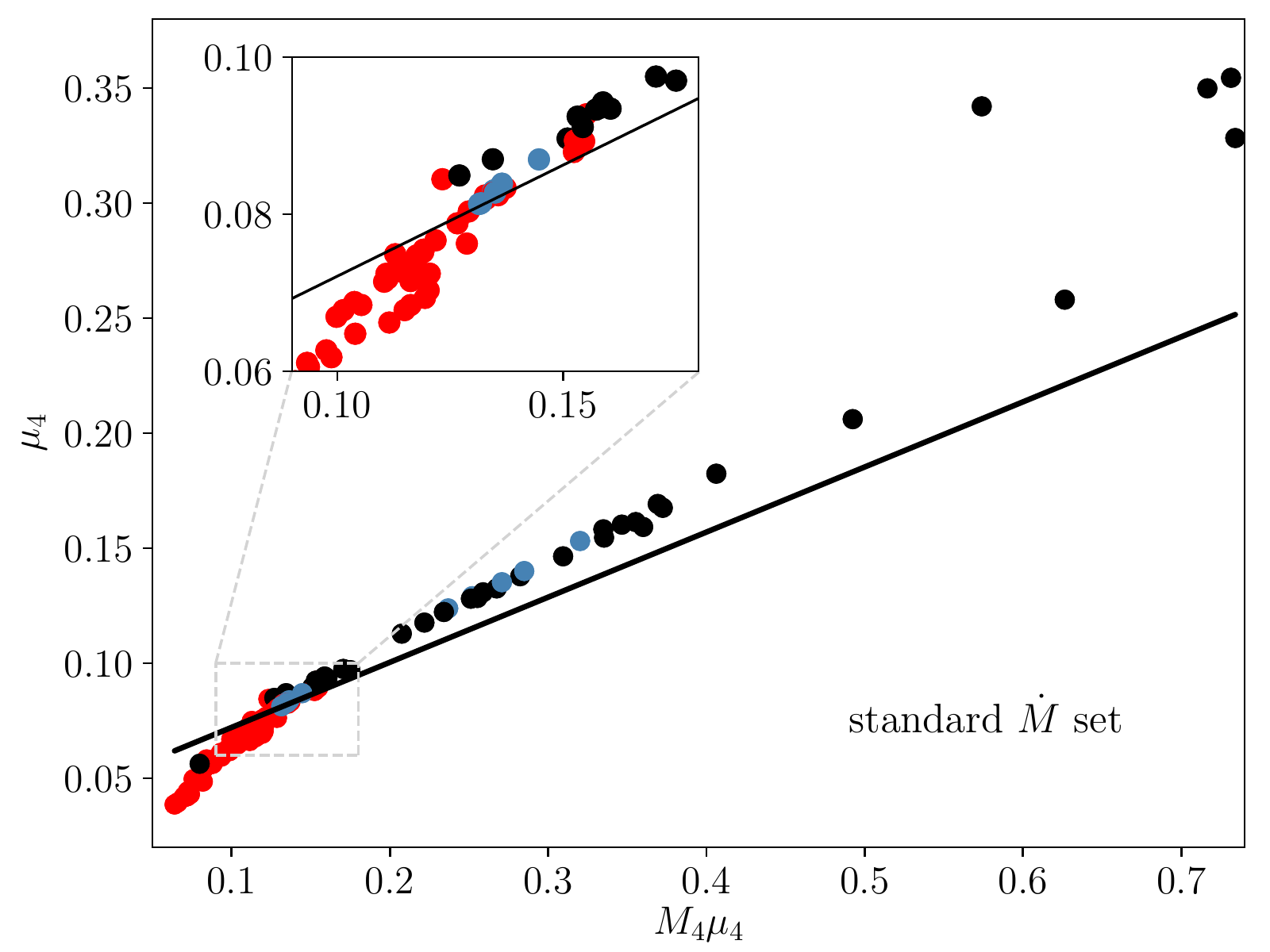}
\caption{{\em Top:} The compactness parameter \citep{Oco11}
  characterizing the inner 2.5\,\Msun\ of the presupernova star is
  shown as a function of zero-age main sequence (ZAMS) mass for all
  helium-star models.  {\em Bottom:} Ertl parametrization \citep{Ert16a}
  for the same sets of models. All panels display color-coded the
  outcomes of P-HOTB simulations with the W18 \& Z9.6 neutrino engine.
  \lFig{compact}}
\end{figure}

\begin{figure}
\includegraphics[width=\columnwidth]{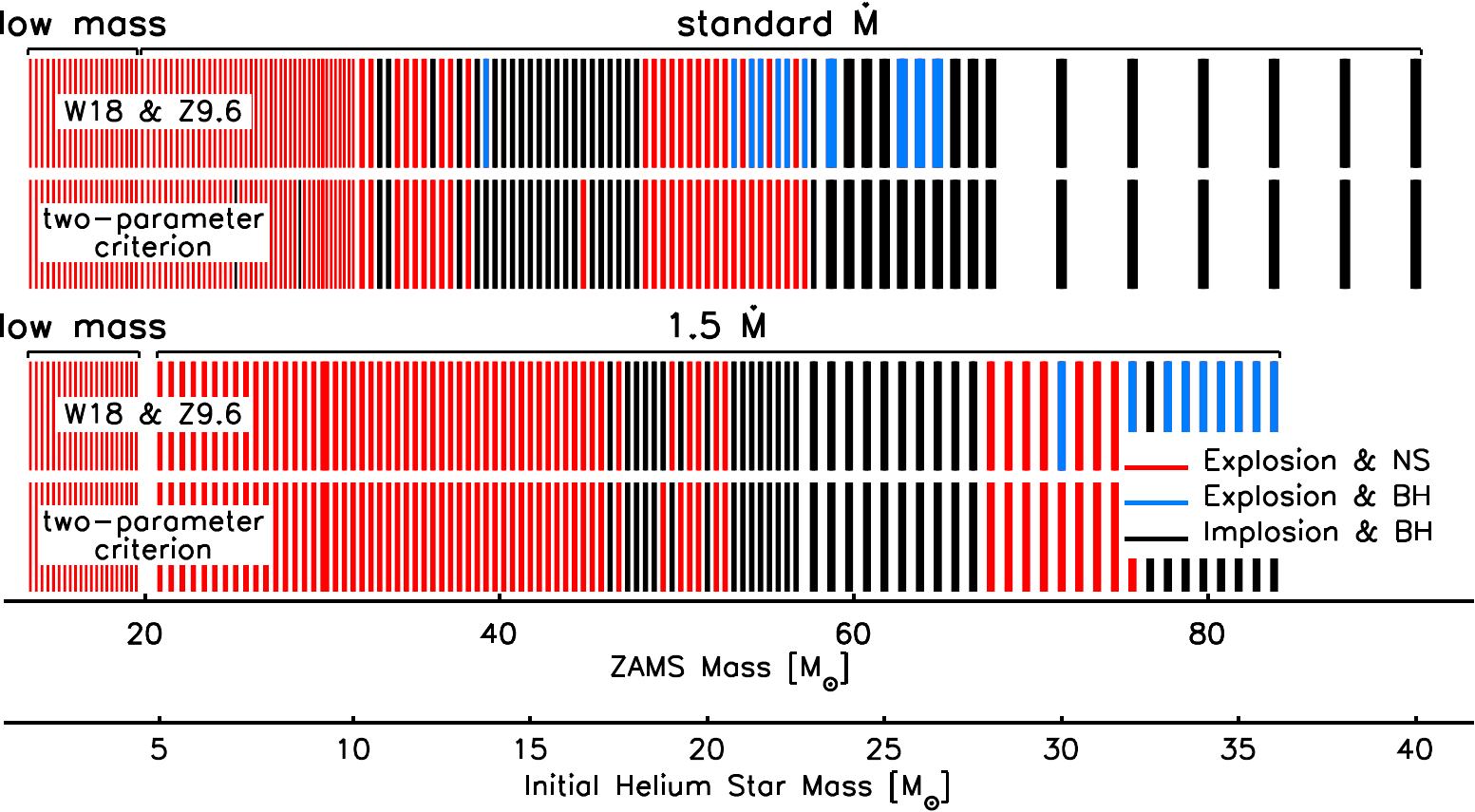}
\caption{Core-collapse outcomes color-coded for the helium stars according
  to the Ertl parametrization (\Fig{compact}, {\em bottom panel};
  \citealt{Ert16a}), compared to results directly from P-HOTB
  simulations.  \lFig{massmap}}
\end{figure}

\subsection{Comparison with Single-Star Results}
\lSect{comparison}

\Fig{compact} displays the core-collapse outcomes, i.e., 
supernova explosion and neutron-star formation (red), explosion 
and black-hole formation (blue) or collapse with black-hole
formation (black) for the Z9.6 \& W18 neutrino engine,
in dependence on the compactness parameter
$\xi_{2.5}$ as function of final helium star mass for both
model sets with standard mass loss and
enhanced mass loss \citep[top panels; a similar pattern for single stars was found by][]{Mue16,Pej15b}. In comparison, the bottom 
panel shows the separation of neutron-star formation cases
(red) from black-hole formation cases (black and blue) by
the \citet{Ert16a} two-parameter criterion for the same
neutrino engine. Compactness values $\xi_{2.5}\lesssim 0.2$
favor explosions, compactness values above this threshold 
characterize mostly core collapse with black-hole formation,
but for values between 0.1 and 0.2 both outcomes are possible.

In contrast, the \citet{Ert16a} criterion separates neutron-star
formation (red) from black-hole formation (blue and black) more reliably.
This can be seen in \Fig{massmap},
where the patterns for direct simulation results with P-HOTB
and for core-collapse outcomes predicted by the \citet{Ert16a}
criterion agree very well, both for helium stars with standard and 
enhanced mass loss. Essentially only fallback supernovae, which
are weak explosions with high fallback masses and black-hole
formation, are hard to predict with the \citet{Ert16a} criterion.

\Fig{4panel} compares the results for our current survey of
mass-losing helium stars to our previous results \citep{Suk16}. The
x-axis in each case is the presupernova mass for the helium stars or
the mass of the helium and heavy element core of the presupernova
models for single stars. The mass ranges for the two are slightly
different. Here it is assumed that helium stars below 2.50\,\Msun
\ (presupernova masses below 2.1\,\Msun) result in neon-oxygen white
dwarfs. There could be a narrow range of electron-capture supernovae
below this mass but their existence and mass range are uncertain and
they are not included here. The single stars are bounded on the lower
end by a ZAMS mass of 9.0\,\Msun \ which has a helium core of
1.57\,\Msun \ when the iron core collapses. This limit too is set by
the neglect of electron-capture supernovae. The limits are the same in
the sense that they are both the lightest models to experience iron
core collapse. They are different because radius expansion in the
helium-star models is thought to lead to a second phase of binary mass
transfer.

The models also differ at higher masses. Above about 35\,\Msun \ the
single stars lose their hydrogen envelope and experience WR mass loss,
similar to the present models. But a different, larger mass loss rate
was used in 2016, so the final masses were smaller. For example, the
presupernova mass for a 100\,\Msun \ star in \citet{Suk16} was 6.04\,\Msun.
Here it would be 22\,\Msun. Both stars collapsed to black holes
with gravitational masses approximately equal to their presupernova
masses.

\begin{figure*}
\includegraphics[width=0.49\textwidth]{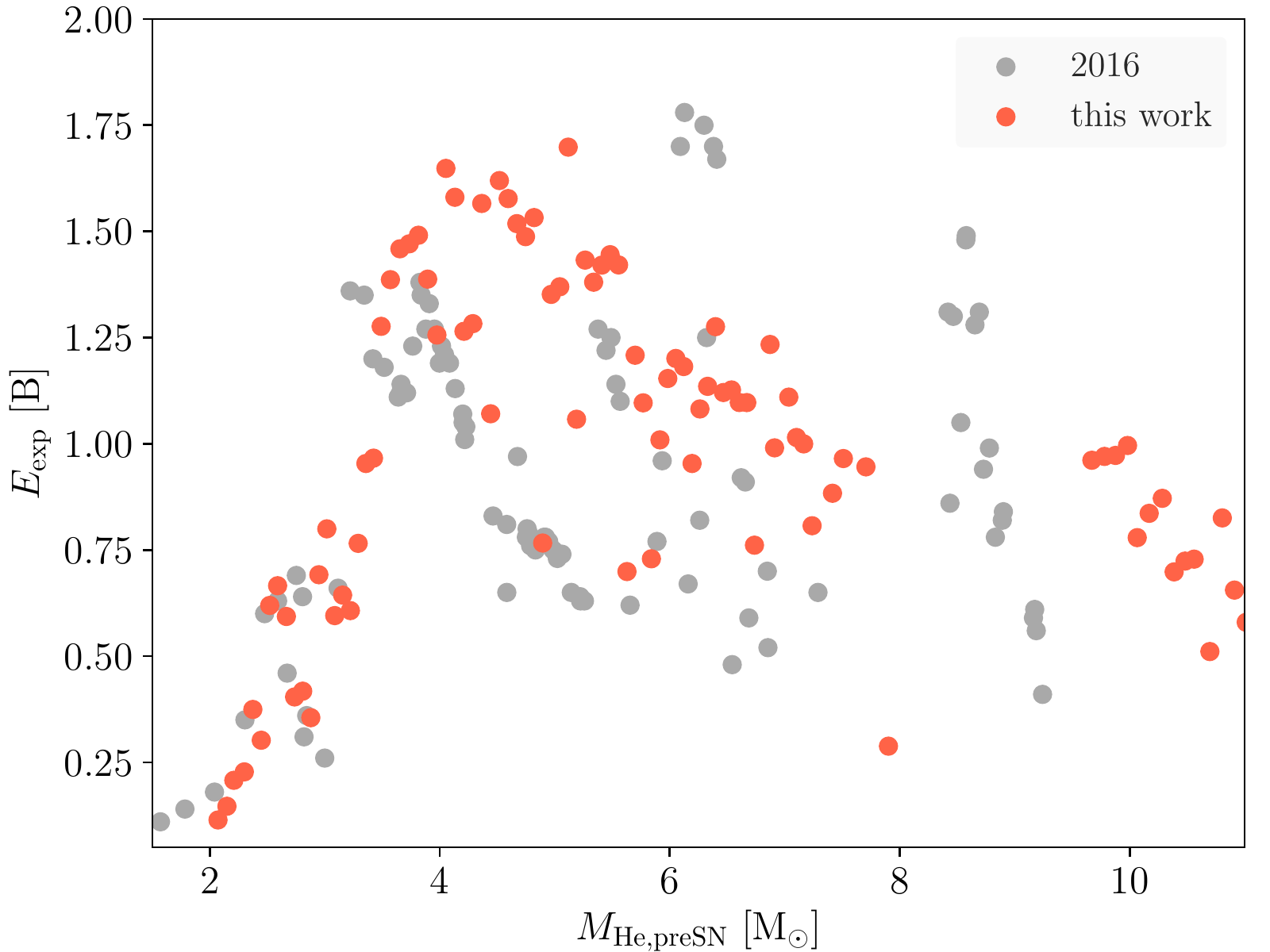}
\includegraphics[width=0.49\textwidth]{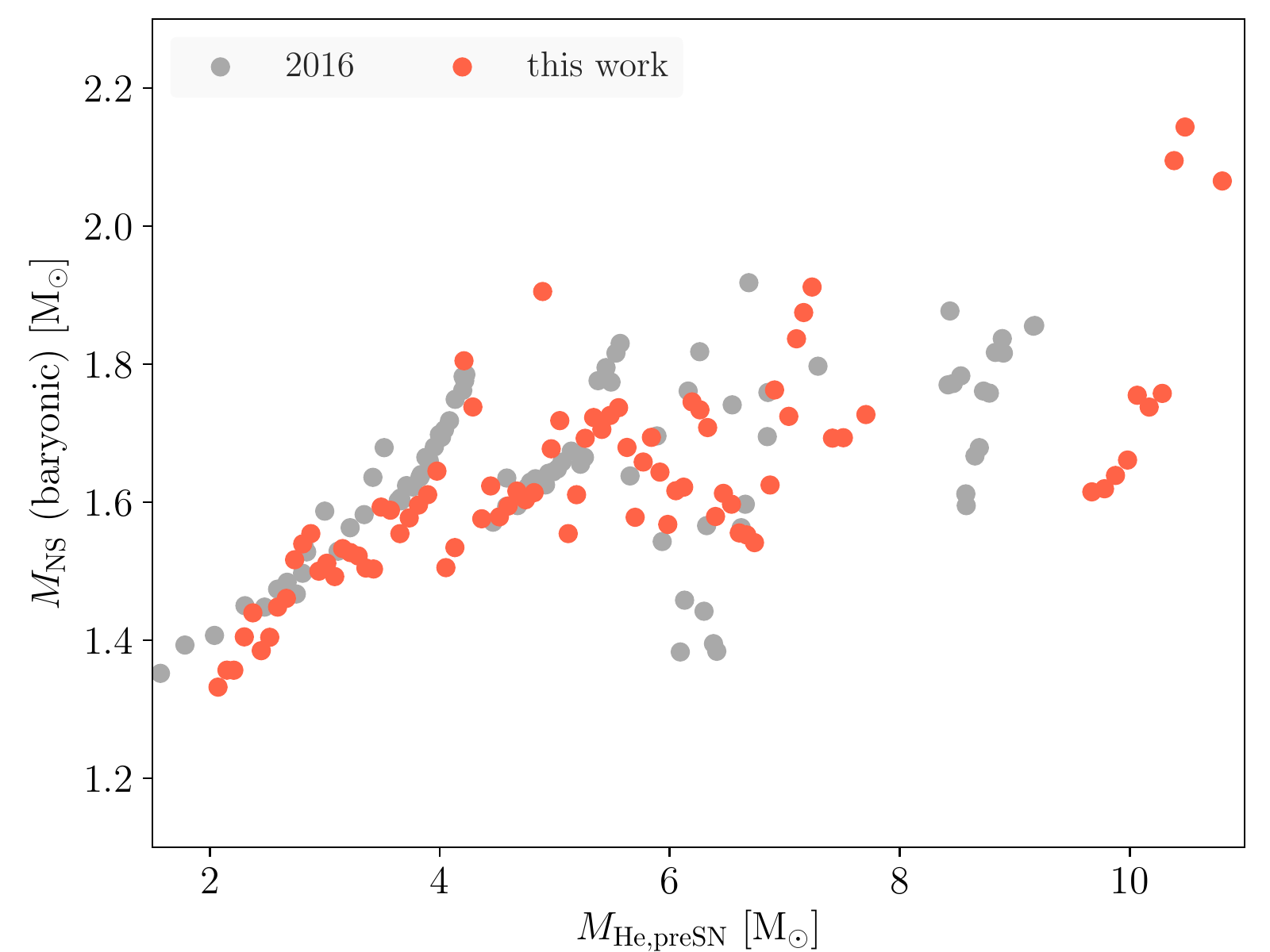}
\\
\includegraphics[width=0.49\textwidth]{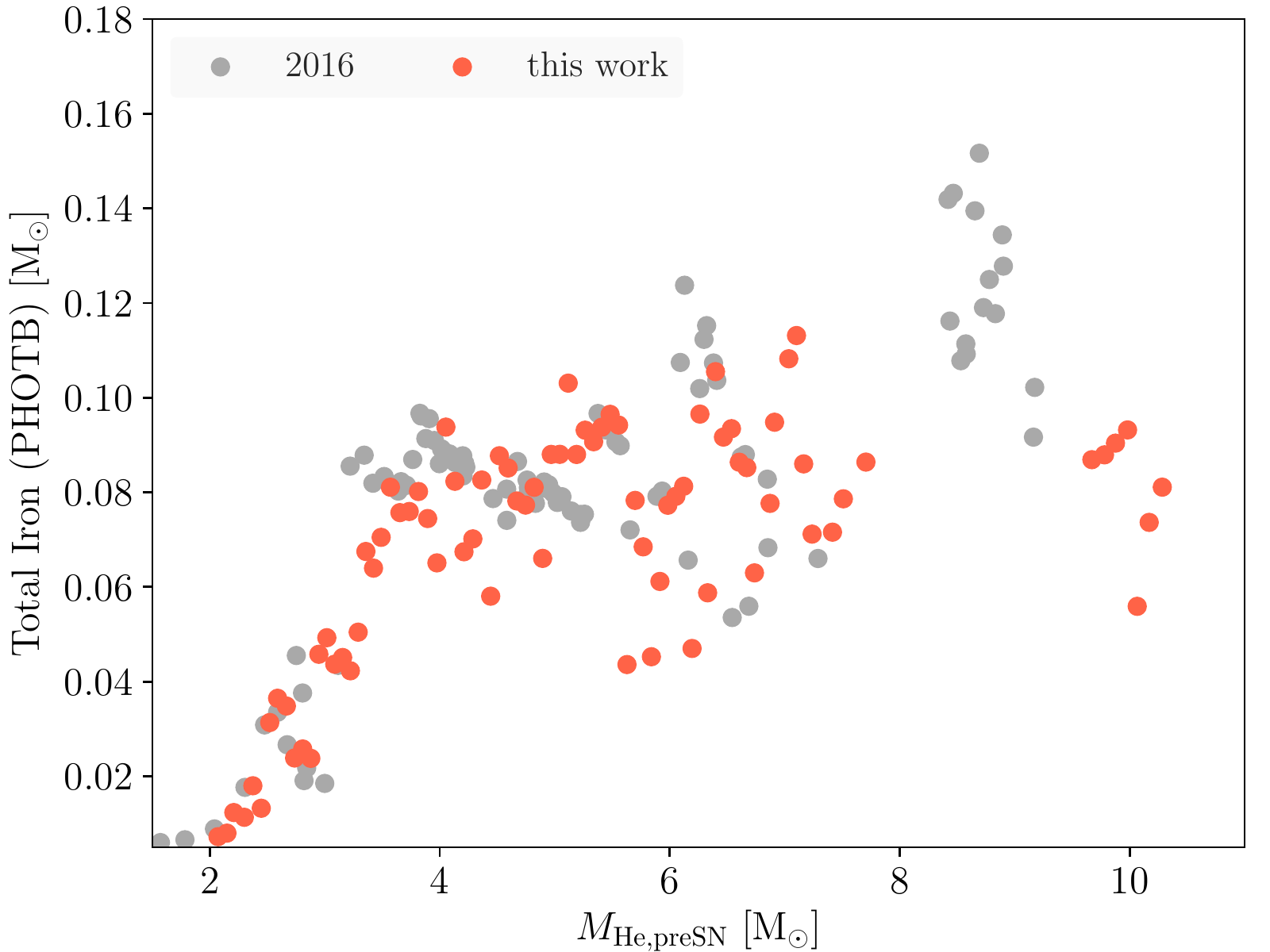}
\includegraphics[width=0.49\textwidth]{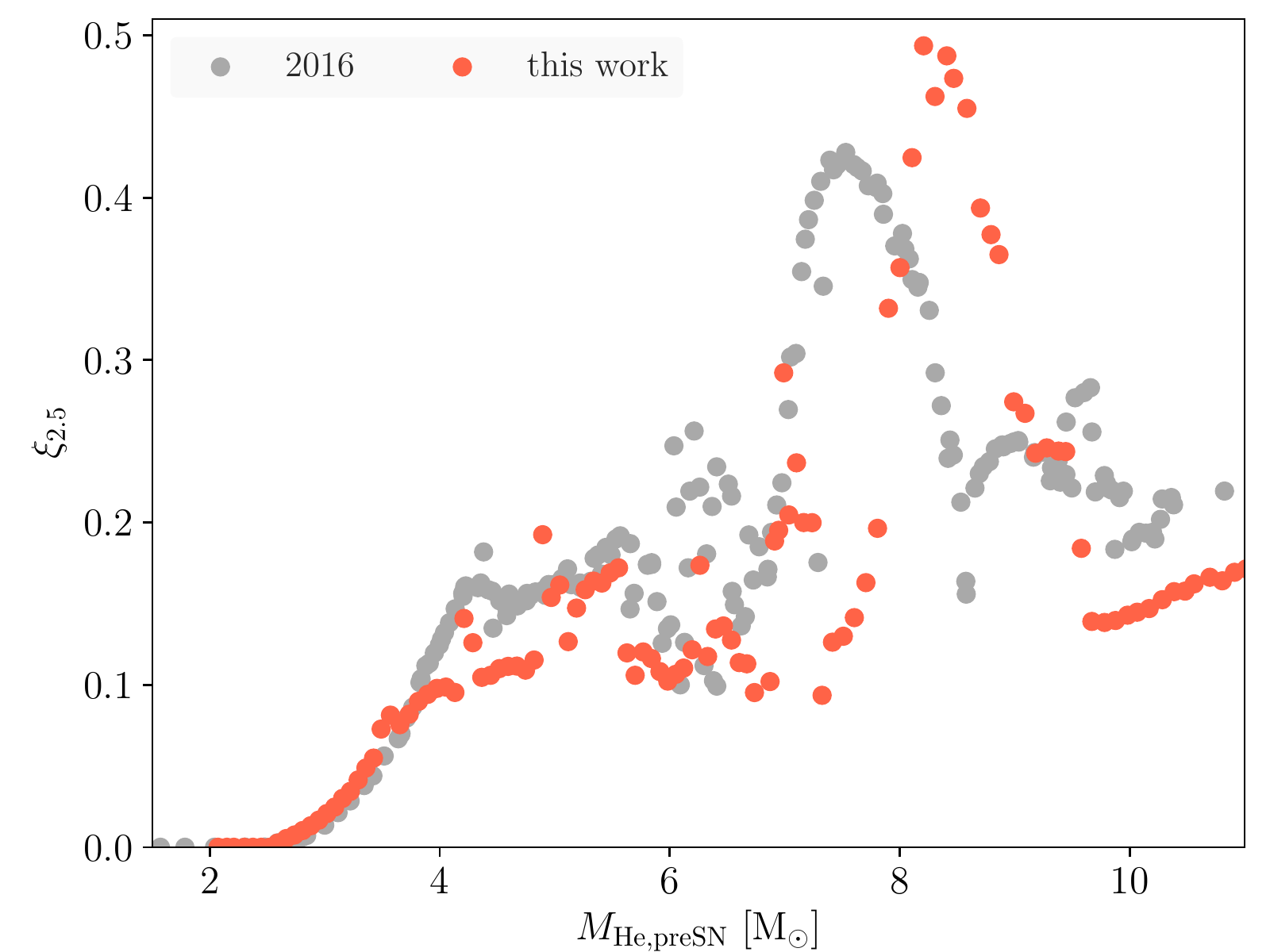}
\caption{Comparison of current results for simulated binary evolution with those of \citet{Suk16} for single star evolution.
\lFig{4panel}}
\end{figure*}

\Fig{4panel} shows a remarkable overall similarity between the earlier
single star results and the present study. To first order, it is the
presupernova mass of the helium and heavy element core (aka ``the
helium core'') that determines its explosion energy, remnant mass, and
$^{56}$Ni nucleosynthesis. This has important implications. The
minimum and maximum iron yields derived here probably characterize all
massive stars, whether single or in binaries, even though the core
masses at the time of collapse may have resulted from very different
evolutionary paths and main-sequence masses.

In detail, however, they are different. The compactness of the core,
however measured, (\Fig{compact}) is different in the two sets
(\Fig{4panel}). This reflects the different entropy and composition of
the two stars, one being derived from a helium core that lost mass,
including at the high end, most of its helium, the other from a helium
core that grew until the end. As noted by \citet{Woo19} there is an
offset in the mass giving rise to the first peak in compactness. For
single stars, the peak is at about 8\,\Msun; for the present models, it
is more like 9\,\Msun. This reflects in part the larger carbon
abundance following helium depletion in the helium stars
\citep{Woo19}, and ultimately is a consequence of central carbon burning 
is transitioning to the radiative regime at higher progenitor mass in the 
binary models \citep[see][]{Suk19}.

As a result there is a gap in neutron star production around 8\,\Msun
\ in the old models that is shifted to 9\,\Msun \ in the new ones.


\section{Nucleosynthesis}
\lSect{nucleo}

Isotopic nucleosynthesis up to atomic mass number A~=~80 was
calculated using a large reaction network in the KEPLER code for all
presupernova models and explosive nucleosynthesis was calculated for all
$\alpha$-particle nuclei up to $^{56}$Ni. These results and isotopic
nucleosynthesis for the various central engines will be discussed
elsewhere. Arguably though, the two most important nucleosynthetic
products of core-collapse supernovae are oxygen and
$^{56}$Ni. Observers of Type Ib and Ic supernovae frequently report
averages of peak luminosities and attempt to extract average $^{56}$Ni
mass fractions \citep[e.g.,][]{Lym16,Pre16}. Oxygen is primarily made
in massive stars. Observing the abundance ratio O/Fe in low
metallicity stars, presumably formed before Type Ia supernovae
contaminated the iron in the sun, can thus constrain the amount of
iron made as $^{56}$Ni in earlier generations of massive stars.
Magnesium, a representative ``$\alpha$-nucleus'' made in massive
stars, is also well studied in low metallicity stars and the Mg/Fe
ratio can give similar constraints on iron production.  It is thus
useful to compute the IMF-weighted averages for the bulk yields for
these three elements.

\Tab{nitbl} gives the average explosion energies, $^{56}$Ni yields,
oxygen and magnesium yields, and explosion frequencies for two
different central engines - W18, which is regarded as typical, and
S19.8, which is more energetic. The explosion percentages 
($f_\mathrm{SN}$) are the
fractions of all star deaths that produced neutron stars and ejected
$^{56}$Ni. One minus that fraction is the percentage that made black
holes and swallowed most of their $^{56}$Ni. The actual {\sl
  supernova} percentage would be slightly larger because some
``successful'' explosions made black holes by fallback and their
$^{56}$Ni is assumed to reimplode. All averages were computed using Salpeter-IMF with $\alpha=-2.35$ \citep{Sal55}.

The $^{56}$Ni yield in \Tab{nitbl} is calculated in five different ways
(\Sect{ni56} and \Fig{ni56yields}) that span the expected range of
minimum (Ni only) and maximum (0.75$\times$(Ni+Tr+$\alpha$)) yields.  The
most likely value is between Ni+Tr/2 and Ni + Tr from the P-HOTB
simulations. Also given for comparison are the yields calculated using
the special trajectory in KEPLER as in \citet{Suk16}. The oxygen and
magnesium yields include contributions from the winds of all stars,
even those whose cores later collapsed to black holes. $^{56}$Ni, on
the other hand, comes only from supernovae that make neutron
stars. Because of its relevance to the light curve, $^{56}$Ni
production is expressed as an average {\sl per supernova}, but because
they are made by every massive star, oxygen and magnesium yields are
expressed {\sl per massive star death} (or equivalently per massive
star birth). The $^{56}$Ni yield per star death is the yield per
supernova times the fraction of supernovae, $f_{\rm SN}$. It is important
when comparing yields that all productions be normalized to the same
event (per star death or per supernova). It is the total production of
each in a generation of massive stars that matters.

For the standard W18 engine, the average $^{56}$Ni yield is between
0.029 and 0.073\,\Msun \ with a most likely value between 0.041 and
0.053\,\Msun. These relatively small numbers are influenced by a large
number of low mass, low yield supernovae with initial helium star
masses between 2.5 and 3\,\Msun. If one examines only the mass range 5
to 8\,\Msun \ (presupernova masses 3.82--5.64\,\Msun) where most Type
Ib and Ic supernovae originate (\Sect{Ibc}), the range of $^{56}$Ni
yields becomes 0.040 to 0.11\,\Msun \ with a typical value of 0.061 to
0.081\,\Msun.  For the more energetic S19.8 engine the values are slightly
larger. The range for all masses rises to 0.036 to 0.090\,\Msun \ with
a typical value of 0.052 to 0.068\,\Msun. In summary, the favored $^{56}$Ni
production, overall, is around 0.04 to 0.06\,\Msun \ per supernova with
an upper bound about twice that. For the heaviest Type Ib and Ic
supernova, the average is probably between 0.06 and 0.11\,\Msun.

The oxygen and magnesium yields have their own uncertainties. A larger
WR mass loss rate would increase their production in stars that make
black holes. Changes in $^{12}$C($\alpha,\gamma)^{16}$O and the
treatment of convection will also change their yield. Examples using
the current physics are presented to illustrate a constraint, but are
not precise. It is also assumed that these averages characterize the
deaths of all massive stars of all metallicities, not just the solar
metallicity binaries simulated here. Oxygen and magnesium are primary
elements and their nucleosynthesis in supernovae is not so sensitive
to metallicity, though variations in the winds are expected.  The
averages in \Tab{nitbl} are very similar to those obtained by
\citet{Suk16} for single stars and the comparison is discussed further
in \Sect{comparison}.

Adopting iron and oxygen mass fractions in the sun of $1.37 \times
10^{-3}$ and $6.60 \times 10^{-3}$ respectively \citep{Lod03}, the
mass ratio of oxygen to iron in the sun is 4.8. Taking 0.840\,\Msun
\ for the mean oxygen yield per star death and 0.053\,\Msun \ per
supernova for $^{56}$Ni+Tr from W18, and multiplying the latter by the
frequency of supernovae, 0.79, gives a ratio of 20, implying that core
collapse supernovae make 4.8/20 = 24\%, by mass, of the iron in the
sun.  Using Ni+Tr/2 gives 19\%, perhaps on the low
side. Observationally the fraction of iron from core-collapse
supernovae, based on measurements of O/Fe or $\alpha$/Fe (where
$\alpha$ may be Mg, Si, S, Ca, or Ti), is usually thought to be about 1/4--1/3
\citep[e.g.,][]{Ful07,Bar18,Ama19}, though some recent
observations suggest it may be as large as 50\% \citep{Gri19}. Taking
0.073 for 0.75$\times$(Ni+Tr+a) increases the maximum average $^{56}$Ni
production for W18 and raises the fraction to 33\% which is also
acceptable. Using the more energetic explosion, S19.8, and taking the
high $^{56}$Ni yield, however, raises the fraction to 42\%. Even given
the considerable uncertainty in oxygen production in winds and
abundances in metal-poor stars, much larger average $^{56}$Ni
production than this may cause difficulty for models of galactic
chemical evolution. 

Similar restrictions come from considering magnesium. The mass
fraction of Mg in the sun is $6.6 \times 10^{-4}$ \citep{Lod03}, all
of which is believed to be made in core-collapse supernovae and
WR-winds. Taking a typical synthesis of 0.078\,\Msun \ from \Tab{nitbl}
and using the Ni+Tr value for iron, 0.053\,\Msun \ implies that 27\% of
solar iron is made in the current calculations (and presumably
core-collapse supernovae in general). Raising that value by a factor
of two would also conflict with measurements of the Mg/Fe ratio in low
metallicity stars \citep{Wei19}.

The implication is that the $^{56}$Ni abundances used in our
brightest, standard light curve calculations (\Sect{lite}; \Fig{sn1bc}
and \Fig{prentice}) are probably close to the largest values allowed
by galactic chemical evolution. Greater oxygen to iron yields in 
low-metallicity stars than calculated here for solar metallicity, or a
greater production of oxygen and magnesium in the winds of WR stars,
possibly because of a larger mass loss rate, would help ease this
restriction. Conversely, less mass loss in metal-poor stars would make
it worse. Further work is needed.

The average $^{56}$Ni yields for the most optimistic models are also
in conflict with observations of SN~1987A. The presupernova helium
core mass for SN~1987A was 5--7\,\Msun \ \citep{Utr15,Suk16},
corresponding to initial helium core masses in the range 7--11\,\Msun \
in \Tab{nitbl}. Adopting a value for the yield of $^{56}$Ni of
$0.75\times$(Ni+Tr+$\alpha$) implies that the average $^{56}$Ni
production for the W18 or S19.8 central engines would be 0.11 to
0.13\,\Msun. This is much larger than the 0.07\,\Msun \ observed for
SN~1987A \citep{Sun92}. These are average productions and there are
individual models in this mass range that make substantially less
$^{56}$Ni than the mean (\Tab{expltbl}), some even as low
as 0.07\,\Msun, but they are rare and have explosion energies considerably
less than observed in SN~1987A, about $1.5 \times 10^{51}$\,erg. This
energy is also needed to explain the maximum velocities above
3500\,km\,s$^{-1}$ observed for nickel in SN~1987A \citep{Utr15,Utr19}.

\begin{deluxetable*}{llcccccccccc}
\tablecaption{Average Nickel, Oxygen and Magnesium yields}
           \tablehead{ \colhead{}         &
            \colhead{}                     &
            \colhead{$E_{\rm exp}$}        &
            \colhead{${\rm Ni_{min}}$}     &
            \colhead{${\rm Ni+Tr/2}$}      &
            \colhead{${\rm Ni+Tr}$}        &
            \colhead{${\rm Ni_{max}}$}     &
            \colhead{${\rm Ni}\ (2016)$}   &
            \colhead{} &
            \colhead{${\rm O}$}            &
            \colhead{${\rm Mg}$}           &
            \colhead{$f_{\rm SN}$}\vspace{1mm}
            \\ 
            \colhead{}         &
            \colhead{}         &
            \colhead{[B]}      &
            \colhead{[\Msun]}  &
            \colhead{[\Msun]}  &
            \colhead{[\Msun]}  &
            \colhead{[\Msun]}  &
            \colhead{[\Msun]}  &
            \colhead{} &
            \colhead{[\Msun]}  &
            \colhead{[\Msun]}
            }\\ 
\startdata
\multicolumn{12}{c}{overall}\\
\\
\multirow{2}{*}{\rotatebox[origin=c]{90}{W18}} 
& ${\rm median}$ & 0.753 & 0.028 & 0.042 & 0.054 & 0.069 & 0.036 & & 0.382 & 0.050 & \multirow{2}{*}{0.79}\\
& ${\rm mean}$   & 0.832 & 0.029 & 0.041 & 0.053 & 0.073 & 0.035 & & 0.840 & 0.078 & \\
\\
\multirow{2}{*}{\rotatebox[origin=c]{90}{S19.8}} 
& ${\rm median}$ & 0.966 & 0.031 & 0.051 & 0.070 & 0.097 & & & 0.474 & 0.069 & \multirow{2}{*}{0.83}\\
& ${\rm mean}$   & 1.015 & 0.036 & 0.052 & 0.068 & 0.090 & & & 1.096 & 0.090 & \\
\\
\multicolumn{12}{c}{$5 > M_{\rm He,i} \geq 3$}\\
\\
\multirow{2}{*}{\rotatebox[origin=c]{90}{W18}} 
& ${\rm median}$ & 0.628 & 0.026 & 0.034 & 0.042 & 0.059 & 0.036 & & 0.236 & 0.040 & \multirow{2}{*}{1.00}\\
& ${\rm mean}$   & 0.738 & 0.023 & 0.034 & 0.044 & 0.064 & 0.032 & & 0.268 & 0.052 & \\
\\
\multirow{2}{*}{\rotatebox[origin=c]{90}{S19.8}} 
& ${\rm median}$ & 0.680 & 0.025 & 0.032 & 0.041 & 0.059 & & & 0.235 & 0.040 & \multirow{2}{*}{1.00}\\
& ${\rm mean}$   & 0.833 & 0.025 & 0.037 & 0.049 & 0.071 & & & 0.270 & 0.053 & \\
\\
\multicolumn{12}{c}{$8 > M_{\rm He,i} \geq 5$}\\
\\
\multirow{2}{*}{\rotatebox[origin=c]{90}{W18}} 
& ${\rm median}$ & 1.429 & 0.037 & 0.058 & 0.081 & 0.113 & 0.042 & & 0.972 & 0.149 & \multirow{2}{*}{1.00}\\
& ${\rm mean}$   & 1.408 & 0.040 & 0.061 & 0.081 & 0.110 & 0.045 & & 0.985 & 0.141 & \\
\\
\multirow{2}{*}{\rotatebox[origin=c]{90}{S19.8}} 
& ${\rm median}$ & 1.782 & 0.038 & 0.069 & 0.100 & 0.136 & & & 0.971 & 0.148 & \multirow{2}{*}{1.00}\\
& ${\rm mean}$   & 1.719 & 0.045 & 0.071 & 0.097 & 0.130 & & & 0.986 & 0.141 & \\
\\
\multicolumn{12}{c}{$M_{\rm He,i} \geq 8$}\\
\\
\multirow{2}{*}{\rotatebox[origin=c]{90}{W18}} 
& ${\rm median}$ & 0.969 & 0.050 & 0.064 & 0.079 & 0.100 & 0.058 & & 1.367 & 0.056 & \multirow{2}{*}{0.40}\\
& ${\rm mean}$   & 0.900 & 0.052 & 0.066 & 0.080 & 0.095 & 0.060 & & 1.566 & 0.094 & \\
\\
\multirow{2}{*}{\rotatebox[origin=c]{90}{S19.8}} 
& ${\rm median}$ & 1.363 & 0.066 & 0.087 & 0.109 & 0.130 &  & & 1.953 & 0.117 & \multirow{2}{*}{0.54}\\
& ${\rm mean}$   & 1.190 & 0.067 & 0.089 & 0.112 & 0.130 &  & & 2.282 & 0.125 & \\
\\
\cline{3-8}  \cline{10-11}
 & & \multicolumn{6}{c}{per supernova} & & \multicolumn{2}{c}{per death} & 
\enddata
\tablecomments{$\rm Ni_{min}$ is $^{56}$Ni calculated by P-HOTB ignoring
  Tr and $\alpha$. $\rm Ni_{max}$ is $0.75\times$(Ni+Tr+$\alpha$). See
  \Sect{ni56} and \Fig{ni56yields}.}  \
\lTab{nitbl}
\end{deluxetable*}

%

\section{Bound Remnants}
\lSect{remnants}

\begin{deluxetable*}{llccccccccc}
\tablecaption{Average Neutron Star Masses}
\tablehead{ \colhead{}                         &
            \colhead{}                         &
            \colhead{$M_{\rm NS,b}$}           &
            \colhead{$M_{\rm NS,g}$}           &
            \colhead{$M_{\rm NS,g,PL01}$}      &
            \colhead{} &
            \colhead{$M_{\rm NS,b}$}           &
            \colhead{$M_{\rm NS,g,PL01}$}      &
            \colhead{} &
            \colhead{$M_{\rm NS,b}$}           &
            \colhead{$M_{\rm NS,g,PL01}$}\vspace{1mm}
            \\
            \colhead{}                     &
            \colhead{}                     &
            \colhead{[\Msun]}              &
            \colhead{[\Msun]}              &
            \colhead{[\Msun]}              &
            \colhead{} &
            \colhead{[\Msun]}              &
            \colhead{[\Msun]}              &
            \colhead{} &
            \colhead{[\Msun]}              &
            \colhead{[\Msun]}
            }\\
\startdata
&
&
\multicolumn{3}{c}{$\rm P-HOTB$}   &
&
\multicolumn{2}{c}{$\rm Fe-core$} &
&
\multicolumn{2}{c}{$(2016)$} \\
\cline{3-5} \cline{7-8} \cline{10-11}\\
\multicolumn{11}{c}{overall}\\
\\
\multirow{2}{*}{\rotatebox[origin=c]{90}{W18}}
& ${\rm median}$ & 1.531 & 1.351 & 1.378 && 1.396 & 1.267 && 1.555 & 1.397 \\
& ${\rm mean}$   & 1.554 & 1.371 & 1.397 && 1.415 & 1.283 && 1.588 & 1.424 \\
\\
\multirow{2}{*}{\rotatebox[origin=c]{90}{S19.8}}
& ${\rm median}$ & 1.514 & 1.335 & 1.364 && 1.407 & 1.277 && 1.563 & 1.404 \\
& ${\rm mean}$   & 1.534 & 1.349 & 1.380 && 1.428 & 1.294 && 1.571 & 1.410 \\
\\
\multicolumn{11}{c}{$5 > M_{\rm He,i} \geq 3$}\\
\\
\multirow{2}{*}{\rotatebox[origin=c]{90}{W18}}
& ${\rm median}$ & 1.510 & 1.337 & 1.361 && 1.367 & 1.243 && 1.537 & 1.383 \\
& ${\rm mean}$   & 1.502 & 1.333 & 1.354 && 1.379 & 1.253 && 1.530 & 1.377 \\
\\
\multirow{2}{*}{\rotatebox[origin=c]{90}{S19.8}}
& ${\rm median}$ & 1.492 & 1.319 & 1.346 && 1.367 & 1.243 && 1.524 & 1.372 \\
& ${\rm mean}$   & 1.488 & 1.318 & 1.343 && 1.379 & 1.253 && 1.520 & 1.369 \\
\\
\multicolumn{11}{c}{$8 > M_{\rm He,i} \geq 5$}\\
\\
\multirow{2}{*}{\rotatebox[origin=c]{90}{W18}}
& ${\rm median}$ & 1.611 & 1.406 & 1.443 && 1.513 & 1.363 && 1.661 & 1.483 \\
& ${\rm mean}$   & 1.646 & 1.433 & 1.471 && 1.511 & 1.362 && 1.694 & 1.509 \\
\\
\multirow{2}{*}{\rotatebox[origin=c]{90}{S19.8}}
& ${\rm median}$ & 1.585 & 1.375 & 1.422 && 1.513 & 1.363 && 1.647 & 1.471 \\
& ${\rm mean}$   & 1.611 & 1.394 & 1.443 && 1.511 & 1.362 && 1.670 & 1.490 \\
\\
\multicolumn{11}{c}{$M_{\rm He,i} \geq 8$}\\
\\
\multirow{2}{*}{\rotatebox[origin=c]{90}{W18}}
& ${\rm median}$ & 1.669 & 1.457 & 1.489 && 1.431 & 1.296 && 1.706 & 1.518 \\
& ${\rm mean}$   & 1.726 & 1.486 & 1.534 && 1.466 & 1.325 && 1.760 & 1.561 \\
\\
\multirow{2}{*}{\rotatebox[origin=c]{90}{S19.8}}
& ${\rm median}$ & 1.636 & 1.414 & 1.463 && 1.469 & 1.327 && 1.688 & 1.504 \\
& ${\rm mean}$   & 1.654 & 1.427 & 1.477 && 1.497 & 1.350 && 1.698 & 1.512 \\

\enddata
\tablecomments{P-HOTB gravitational mass, $M_{\rm NS,g}$, is computed by tracking the radiated neutrino energies and a correction for late-time fallback accretion (see \Sect{conversion-bary-to-gravmass}), while $M_{\rm NS,g,LP01}$ values were computed only through the equations from \citet{Lat01}.}
\lTab{remtbl}
\end{deluxetable*}

\begin{deluxetable}{llrr}
\tablecaption{Average Black Hole Masses}
\tablehead{ \colhead{}                         &
            \colhead{}                         &
            \colhead{$M_{\rm BH,b}$}           &
            \colhead{$M_{\rm BH,g}$}\vspace{1mm}       
            \\
            \colhead{}                     &
            \colhead{}                     &
            \colhead{[\Msun]}              &
            \colhead{[\Msun]}              
            }\\
\startdata

\multicolumn{4}{c}{$90 > M_{\rm He,i} \geq 8$}\\
\\
\multirow{2}{*}{\rotatebox[origin=c]{90}{W18}}
& ${\rm median}$ & 11.42 & 10.88\\
& ${\rm mean}$   & 14.81 & 14.44\\
\\
\multirow{2}{*}{\rotatebox[origin=c]{90}{S19.8}}
& ${\rm median}$ & 10.96 & 10.49\\
& ${\rm mean}$   & 16.36 & 16.10\\
\\
\multicolumn{4}{c}{$40 > M_{\rm He,i} \geq 8$}\\
\\
\multirow{2}{*}{\rotatebox[origin=c]{90}{W18}} 
& ${\rm median}$ & 8.96  & 8.61  \\
& ${\rm mean}$   & 10.39 & 10.14 \\
\\
\multirow{2}{*}{\rotatebox[origin=c]{90}{S19.8}} 
& ${\rm median}$ & 10.07 & 9.65  \\
& ${\rm mean}$   & 10.79 & 10.42 \\
\enddata
\tablecomments{See \Sect{bh} for details.}
\lTab{bhtbl}
\end{deluxetable}

\Tab{remtbl} and \Tab{bhtbl} give the IMF averaged masses for neutron
stars and black holes produced in this study. Both sets of masses are given
before and after corrections for neutrino losses during their
formation.  Baryonic masses are determined by the final mass cut in the
P-HOTB explosion. Gravitational masses for neutron stars have been
calculated in two different ways (\Sect{conversion-bary-to-gravmass}).
In the standard case, the total neutrino losses calculated for the explosion
by P-HOTB, divided by $c^2$ are subtracted from the baryonic mass and a
further correction for late-time fallback accretion. For comparison, the
gravitational mass computed for the cold neutron star using the
analytic formula of \citet{Lat01} is also given for an assumed neutron
star radius of 12\,km. The two numbers differ slightly because of
different assumptions regarding the equation of state. The
gravitational masses of black holes are also corrected for the
neutrino emission that occurs during their formation
(\Sect{conversion-bary-to-gravmass}).

In \Tab{remtbl} and \Tab{bhtbl} entries are given for both the
median and mean masses. The median is that mass above and below which
equal numbers of remnants are expected. This is the most relevant
quantity for observations. The mean mass is the IMF weighted average
total mass (neutron star or black hole) divided by the number of
remnants. It is thus weighted towards heavier masses. Separate sets of
numbers are given for the W18 and S19.8 central engines. W18 is the
standard engine in this paper and S19.8 is an upper bound to the
explosion energy. As expected, average neutron star masses are larger
for the less energetic explosions for the self-consistent calculations
using P-HOTB. The black hole mass averages do not always follow such a
simple expectation. While a larger explosion energy reduces the amount of
fallback, it also results in stars exploding that might have become
black holes. There are fewer low mass black holes, while the heavy
ones continue to implode. Though not given in the tables, average remnant 
masses for the other central engines, N20, W15, and W20 were also calculated 
and were similar.

\subsection{Neutron Stars}
\lSect{nstar}

Of the many values in \Tab{remtbl}, we suggest the range 1.35--1.38\,\Msun
\ as the most appropriate for the median gravitational mass of
neutron stars in close binary systems. This is the range of results
for the median value using the W18 central engine and either the
P-HOTB or Lattimer and Prakash corrections to the baryonic mass. It is
is the average for systems of all initial masses. The table also gives the
partition of this overall average into intervals of initial helium
star mass: 3--5\,\Msun, 5--8\,\Msun, and $>$8\,\Msun. As expected the
average increases with progenitor mass as the core structure becomes
less centrally condensed (\Fig{nsxi}).
The weighted birth function for neutron stars is given as a function of their
gravitational mass in \Fig{nstarmass}. The mean mass (as opposed to
median mass), overall is 1.37 to 1.40\,\Msun \ with a standard deviation of 0.11\,\Msun. 

The lightest neutron star in the standard (W18) set has a baryonic
mass of 1.332\,\Msun, corresponding to a gravitational mass of 1.239\,\Msun
\ using the P-HOTB correction for neutrino losses
(\Sect{conversion-bary-to-gravmass}). If one uses instead the
\citet{Lat01} approximation, the gravitational mass is 1.214\,\Msun
\ for an assumed radius of 12\,km and 1.205\,\Msun \ for a radius of 11\,km.
Looking at all possible central engines, the overall minimum is
for the 2.50\,\Msun \ helium star using the W20 central engine.  This
combination gives a gravitational mass of 1.238\,\Msun \ using the
P-HOTB correction and 1.214\,\Msun \ (1.204\,\Msun) using the
\citet{Lat01} approximation, and an assumed radius of 12\,km (11\,km).
Slightly smaller values are in principle possible using an
artificial mass separation. For the 2.70\,\Msun \ helium star and
W20 central engine the iron core is 1.288\,\Msun \ and the equivalent
gravitational mass is 1.181\,\Msun \ (1.167 to 1.178\,\Msun \ for the
Lattimer-Prakash mass with a radius if 12 or 11\,km). This is
unrealistically small since mass separations this deep are not found
in modern models. We conclude that a gravitational mass of 1.24\,\Msun
\ is a reasonable lower bound for our models and 1.20\,\Msun\ is the
boundary of what might be reasonable with the physics employed in this
study.  Less massive neutron stars would require special conditions
for their creation.

The most massive neutron star comes from the theoretical upper limit,
taken here to be about 2.25\,\Msun (\Sect{results}).  
Thus Model He22.50 produces a neutron
star with gravitational mass 2.23\,\Msun, while Model He22.00 makes a
black hole with mass 2.29\,\Msun. This continuum of masses is due to
the action of fallback after the explosion, and low mass black holes,
while rare, should exist as should a smattering of neutron stars up to
the maximum mass. All neutron stars with gravitational mass over
1.65\,\Msun \ come from helium stars over 20.0\,\Msun.  All neutron
stars with masses below 1.40\,\Msun \ come from helium stars lighter
than 10\,\Msun, as do a lot of more massive neutron stars. All neutron
stars less massive than 1.30\,\Msun \ come from helium stars less than
3.5\,\Msun \ (\Fig{nsxi}).  Neutron stars between 1.24 and 1.7\,\Msun
\ are formed promptly in the explosion, but heavier neutron stars are
produced by fallback (e.g., the six most massive points in
\Fig{nsxi}).

\Fig{nstarmass} also suggests that the neutron star birth function has
structure with a hint of peaks around 1.35 and 1.5\,\Msun. A
larger set of models that explores variable metallicity, mass loss,
explosion characterization, and binary parameters will be needed 
before this structure is statistically significant, but an
exciting prospect is that the neutron star mass function reflects
the non-monotonic nature of the compactness parameter (\Fig{compact})
which, in turn, is sensitive to the location of critical shell burning
episodes in the presupernova star. 

The values derived here for median masses (1.35 to 1.38\,\Msun
\ favored) are in good agreement with those determined from
observations. \citet{Lat12} gives an (error-weighted) observational
mean of 1.368\,\Msun \ for X-ray and optical binaries, 1.402\,\Msun
\ for neutron star binaries, and 1.369\,\Msun \ for neutron star -
white dwarf binaries. These are not far from the average neutron star
mass found by \citet{Sch10}, $1.325 \pm 0.056$\,\Msun. 

\citet{Oze16} give the range of accurately determined neutron star
masses in 2016 as 1.17--2.0\,\Msun. Our calculated range, 1.24 to
2.25\,\Msun, is in reasonable agreement. As already remarked, a 
1.17\,\Msun \ neutron star may be difficult to produce. They also 
give 1.33\,\Msun \ with a dispersion of 0.09\,\Msun \ as the average 
for double neutron stars; 1.54\,\Msun \ with a dispersion  0.23\,\Msun 
\ for the recycled neutron stars; and 1.49\,\Msun  \ with a dispersion 
of 0.19\,\Msun \ for the slow pulsars. A recent study by \citet{Ant16}
raised the possibility of two peaks  \citep[see also][]{Val11} within 
the recycled millisecond pulsar population, with the first peak at 
1.388\,\Msun \ with a dispersion of 0.058\,\Msun \ and a second peak 
at 1.814 with dispersion 0.152\,\Msun \ \citep[see single star comparison in ][]{Rai18}.

Also given in \Tab{remtbl} are the neutron star masses resulting from
taking the mass separation to arbitrarily occur at the edge of the iron
core. This gives a lower limit neutron star mass and a maximum
production of $^{56}$Ni which is of interest for the light curve. 
These masses are consistently small compared with the observations
suggesting that such a deep mass cut is, on the average, probably not realistic.

The neutron star masses based on the same approach as in \citet{Suk16} 
are also given for comparison. Since the ``special trajectory'' (\Sect{KEPLER}) 
is more shallow, the neutron star masses are slightly higher. 

\begin{figure}
	\centering
  \includegraphics[width=0.48\textwidth]{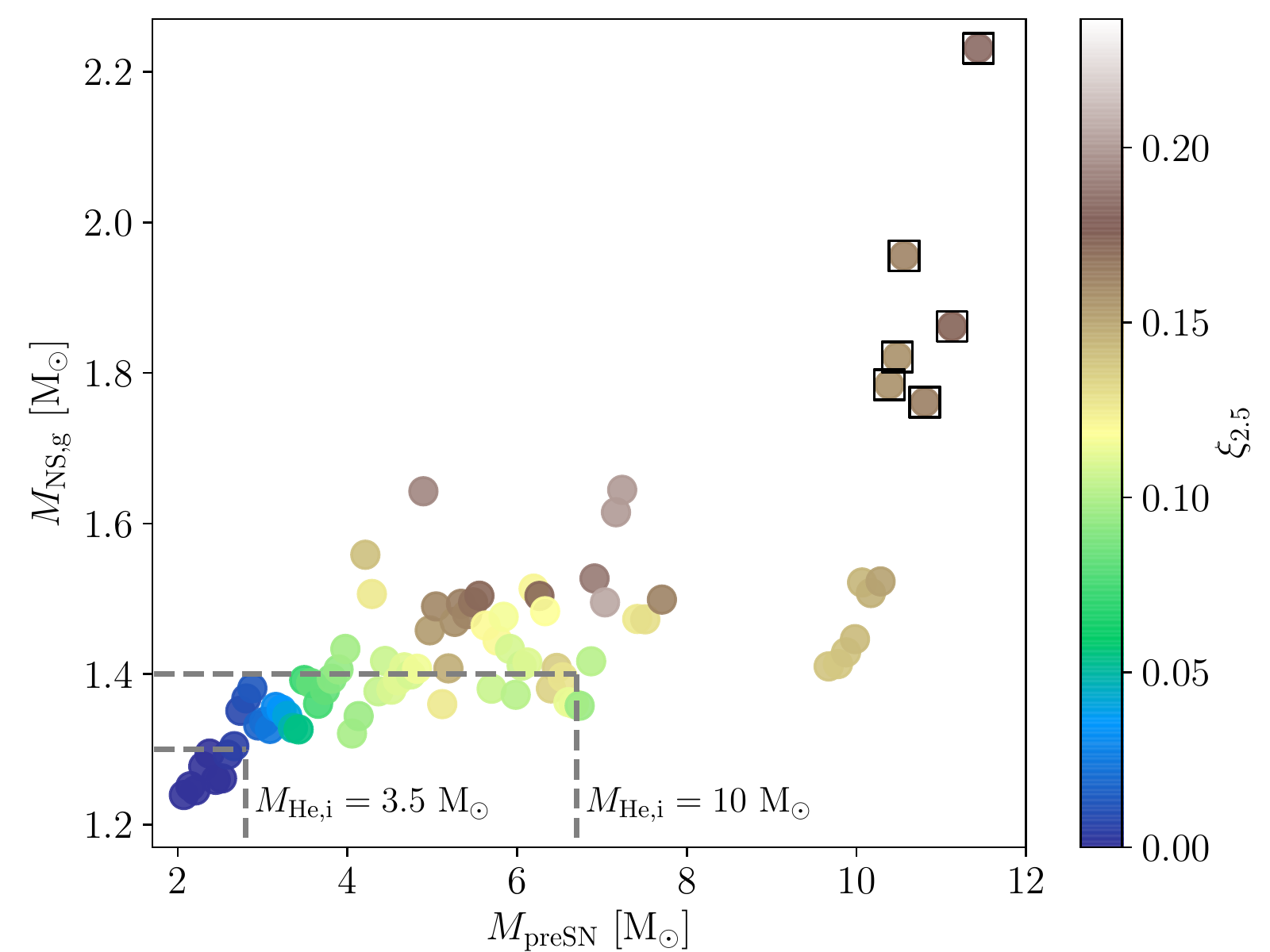}
  \includegraphics[width=0.48\textwidth]{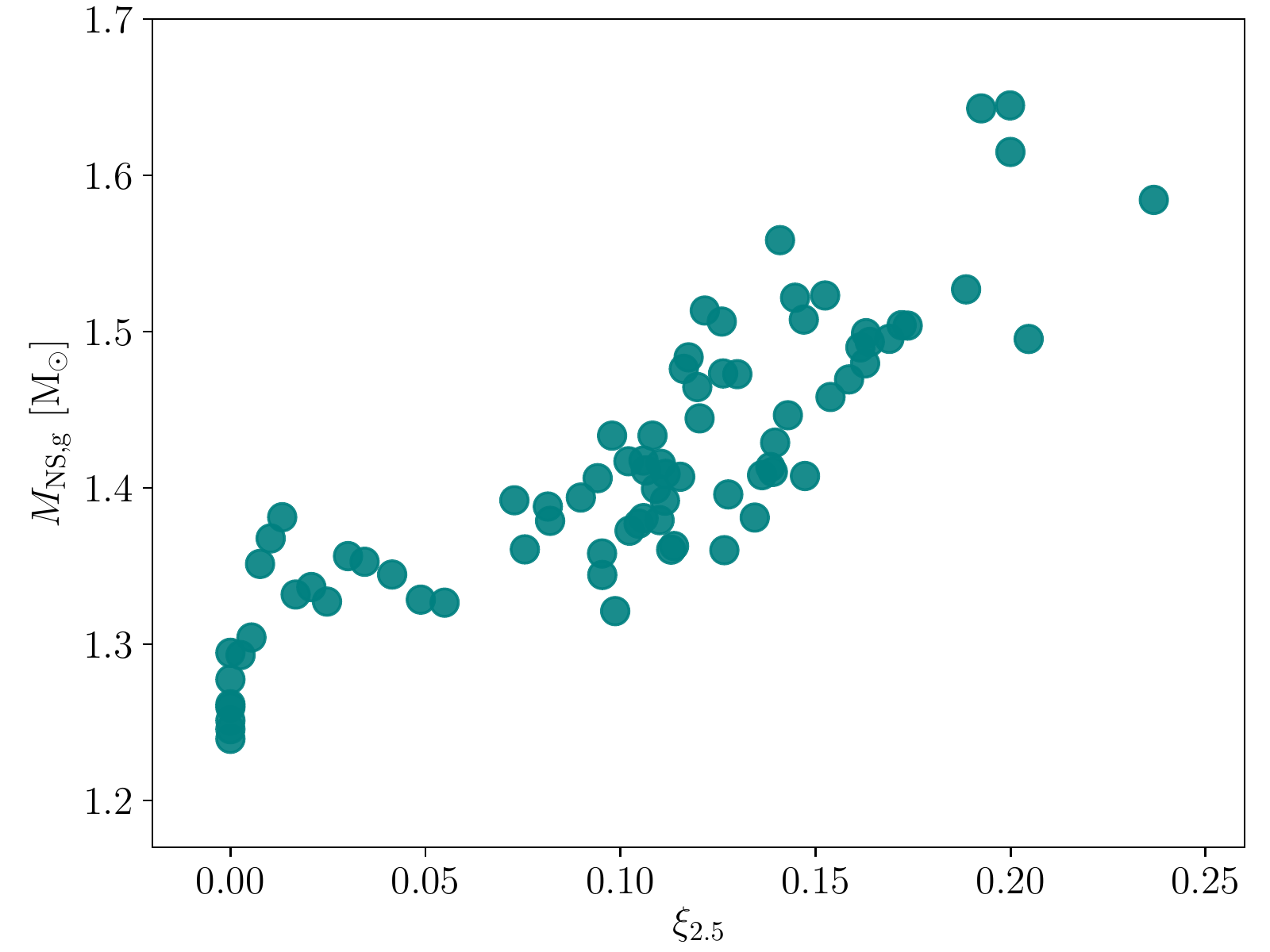}
  \caption{Neutron star mass systematics for the W18 central engine.
    {\em Top:} The gravitational mass as a function of the
    presupernova mass ($M_{\rm He,final}$). Dashed boxes show that all
    neutron stars lighter than 1.30\,\Msun \ come from helium stars
    with initial masses less than 3.5\,\Msun \ while those lighter than
    1.40\,\Msun \ come from helium stars less than 10\,\Msun.  The
    masses are color coded according to the compactness parameter,
    $\xi_{2.5}$, showing that less massive neutron stars come from
    progenitors with sharp density declines outside their iron
    cores. Note a gap in the mass distribution for $M_{\rm He,fin}$
    from 8 to 9\,\Msun \ corresponding to a local maximum in
    compactness (\Fig{compact}). The most massive neutron stars that
    experienced significant fallback ($>0.1\,\Msun$), which come from
    the most massive progenitors are highlighted with black
    boxes. {\em Bottom:} The neutron star masses (excluding those that
    experienced significant fallback) are highly correlated with the
    compactness of the presupernova star. For $\xi_{2.5}$ smaller than
    0.05 the baryonic remnant mass is essentially the cold Chandrasekhar
    mass of the presupernova core. As the entropy increases, the
    compactness parameter rises and so does the neutron star mass.
    \lFig{nsxi}}
\end{figure}

\begin{figure}
	\centering
  \includegraphics[width=0.48\textwidth]{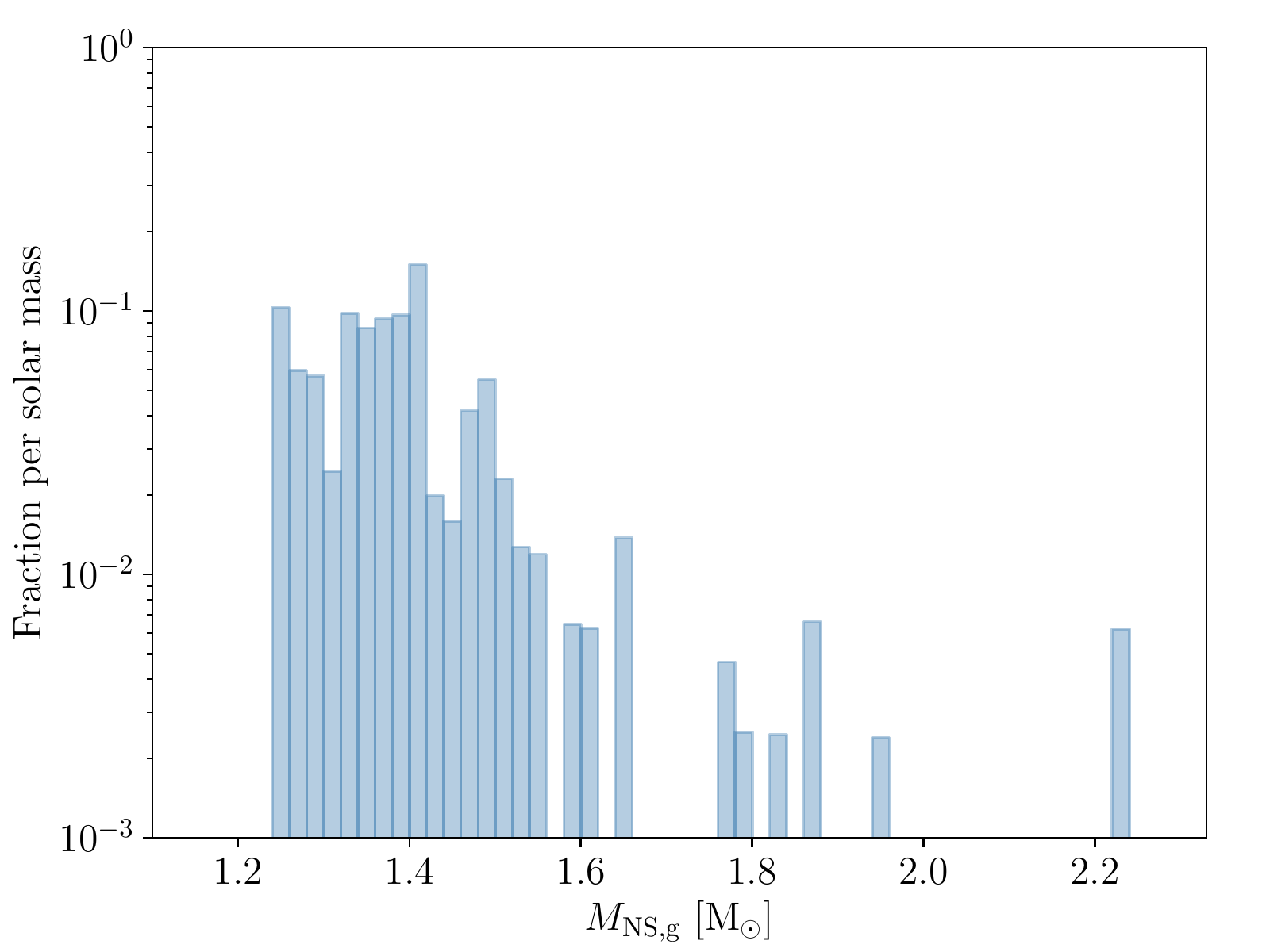}
  \caption{Neutron star gravitational mass resulting for the W18
    central engine. The average mass is 1.371\,\Msun; the median is 1.351\,\Msun. The probability has been normalized so that the integral
    under the curve is one. A few neutron stars at high mass result from 
    fallback.
    \lFig{nstarmass}}
\end{figure}

\subsection{Black Holes}
\lSect{bh}

Stellar collapses that fail to create a strong outward moving shock
before 10\,s in P-HOTB are assumed to form black
holes. While \citet{Woo19} gives results for helium cores up to 
120\,\Msun \ (presupernova masses up to 60\,\Msun), only those that had a
reasonable likelihood of leaving a neutron star were followed with
P-HOTB.  It is expected that for helium stars of up to 60\,\Msun
\ (presupernova up to 30\,\Msun) the presupernova mass will collapse to a
black hole with little mass ejection. For helium cores of 60--70\,\Msun
\ (presupernova 30--35\,\Msun) a mild pulsational instability is
encountered that does not greatly affect the mass of the star when its
iron core collapses.  To good approximation these stars too make black
holes with masses equal to their presupernova mass. For helium stars
initially above 70\,\Msun, the pair instability becomes an important
consideration. As far as remnant masses go, the chief effect of the
pair instability is to reduce the mass of the remnant. \citet{Woo19}
derives an upper limit to black hole masses coming from pulsational
pair instability supernovae of 46\,\Msun.  To good approximation the
black hole mass, for masses bigger than 20\,\Msun, for the assumed mass
loss rate is given by
\begin{equation}
  M_{\rm BH} \ = \ 0.463 M_{\rm He,i}  + 1.49   \,\Msun \,,
\end{equation}
where $M_{\rm He,i}$ is the initial mass of the helium star, and
\begin{equation}
  M_{\rm BH} \ = \ 0.232 M_{\rm ZAMS}  - 1.23   \,\Msun \,,
\end{equation}
where $M_{\rm ZAMS}$ is the main sequence mass of the star.

\Fig{bhmass} shows the resulting IMF-weighted birth function for black
holes using the P-HOTB results and an extrapolation from 19 to 46\,\Msun
\ using the above equations \citep{Woo19}. A more thorough study that explicitly
includes the PPISN results is planned.

The smaller mass black holes are made by fallback after the initial
launch of a successful shock. The lowest mass black holes made in any
models for the W18 central engine were 2.29, 2.31, and 2.56\,\Msun,
which came from stars with initial helium star masses of 22.0,
21.25, and 22.25\,\Msun, respectively.  The presupernova masses of
these systems were 11.23, 10.91, and 11.34\,\Msun \ respectively, so
most of the mass was ejected in the explosion. There is no low mass
``gap'' in the black hole mass distribution in the present
study. Fallback produces a continuous spectrum of masses from the
heaviest neutron stars to the lightest black holes. These low mass
black holes are rare, however, compared with their higher mass
counterparts.

The smallest black hole to be made directly in a failed explosion with
no outgoing shock has a baryonic mass of 6.95\,\Msun \ and a
gravitational mass if 6.42\,\Msun. The low mass peak in the birth
function (\Fig{bhmass}) below $\sim$6\,\Msun \ is due to fallback; the
remainder is mostly black holes produced by the direct implosion of
the progenitor star. From 5 to 15\,\Msun, the distribution shows
structure correlated with the compactness of the presupernova stellar
core (\Fig{compact}). A pronounced minimum at $\sim$10\,\Msun \ results
from the island of explodability near that mass.  This suppression of
black hole formation is robust for different choices of central
engine, though its location may vary depending on details of the
presupernova model, especially the rate for
$^{12}$C($\alpha,\gamma)^{16}$O and convection physics.  It should be
a target of future gravitational wave surveys.

\Tab{bhtbl} gives the average properties of the distribution. 21\% of
all models produce black holes. The median gravitational mass (after
all neutrino losses) for the W18 central engine when all black holes
up to 46\,\Msun \ are included is 10.88\,\Msun. This is influenced by
the extrapolation of the birth function from 20 to 46\,\Msun \ in
\Fig{bhmass} and is probably an overestimate. If only helium stars up
to 40\,\Msun \ (main sequence masses to 80\,\Msun) are considered, the
median is 8.96\,\Msun. Both the initial mass function for stars above
80\,\Msun \ on the main sequence and their mass loss history,
especially as luminous blue variables, are poorly known. A higher mass
loss rate would act to suppress the high mass tail in \Fig{bhmass}.
Neither the median nor mean is a well defined characteristic of the
structure seen in \Fig{bhmass}, especially if it is bimodal. There may
not even be any black holes with masses near 10\,\Msun. Using a more
energetic central engine, S19.8, actually increases the median value
because more of the lower mass models explode while the robustly
collapsing massive models remain unchanged.

Given these uncertainties, our results compare favorably with black
hole masses observed in X-ray binaries.  \citet{Oze10} gives a mean
mass of $7.8 \pm 1.2$\,\Msun.  The observed mean is consistent, but the narrow
distribution is not. \citet{Far11} give a lower limit at the 99\%
confidence level of about 4.5\,\Msun. This is inconsistent with the
presence of a substantial number of less massive black holes in our
results (\Fig{bhmass}). These black holes are all made by fallback
though, and their masses are uncertain. On the other hand, \citet{Kre12}
point out that errors in the inclination angle may have led to the overestimate
of black hole masses in the gap and that black holes may exist all the
way down to the maximum neutron star mass as our analysis suggests. 
Their analysis also suggests a peak around 7--8\,\Msun\ and a sharp
drop off above 10\,\Msun. \Fig{bhmass} shows such a peak and a fall
off, but the production remains substantial at higher masses. As
remarked earlier the abundance of such very massive black holes is
sensitive to the treatment of mass loss and the initial mass
function. All models with presupernova masses over 12\,\Msun \ collapse
directly to black holes. The issue is only how many such presupernova
stars there are.

\citet{Fry12} also explored the black hole birth function using a
different approach to modeling the explosion and found results that
were sensitive to the mass loss. For solar metallicity, they found a
maximum black hole mass around 10\,\Msun \ when using presupernova
models from \citet{Woo02}, which had a larger mass loss rate than this
present study (see their Figure~9). On the other hand, using
presupernova models from \citet{Lim06} that employed a smaller mass
loss rate gave larger black hole masses. Clearly an accurate birth
function will require careful consideration of the effects of
metallicity and mass loss.  For now, we note that the outcome 
--explosion or collapse-- and
the mass of the collapsing object is mainly determined by the
presupernova mass. It is thus feasible to estimate the black hole
birth function to good accuracy from any choice of initial mass
function, metallicity, mass loss rate, binary parameters, etc.  simply
from an estimate of the masses of the stars at carbon ignition and the
results shown in Figures~5--8.

For the models presented here (solar metallicity, W18 central engine,
standard mass loss), 21\% made black holes. This includes all stars
with final masses above about 12\,\Msun \ (initial helium star mass 25\,\Msun).
This is substantially less than the 33\% that make black holes
in single stars for the same central engine \citep[Table~6
  of][]{Suk16}. This reflects both the smaller presupernova masses for
stars of a given main sequence mass when evolved in a mass exchanging
binary, and their smaller compactness parameters \citep[Figure~10
of][]{Woo19}.

Since all presupernova stars with masses above 12\,\Msun \ collapse,
there is a maximum luminosity of any Type Ib or Ic supernova progenitor,
$10^{5.6}$\,\Lsun. The equivalent value for single stars is $10^{5.3}$\,\Lsun \
or a birth mass of about 23\,\Msun \ \citep{Suk18}, though the current
observational limit is at $10^{5.1}$\,\Lsun \ or a birth mass of about
18\,\Msun \  \citep{Sma09,Sma15}.

\begin{figure}[h]
\centering
\includegraphics[width=\columnwidth]{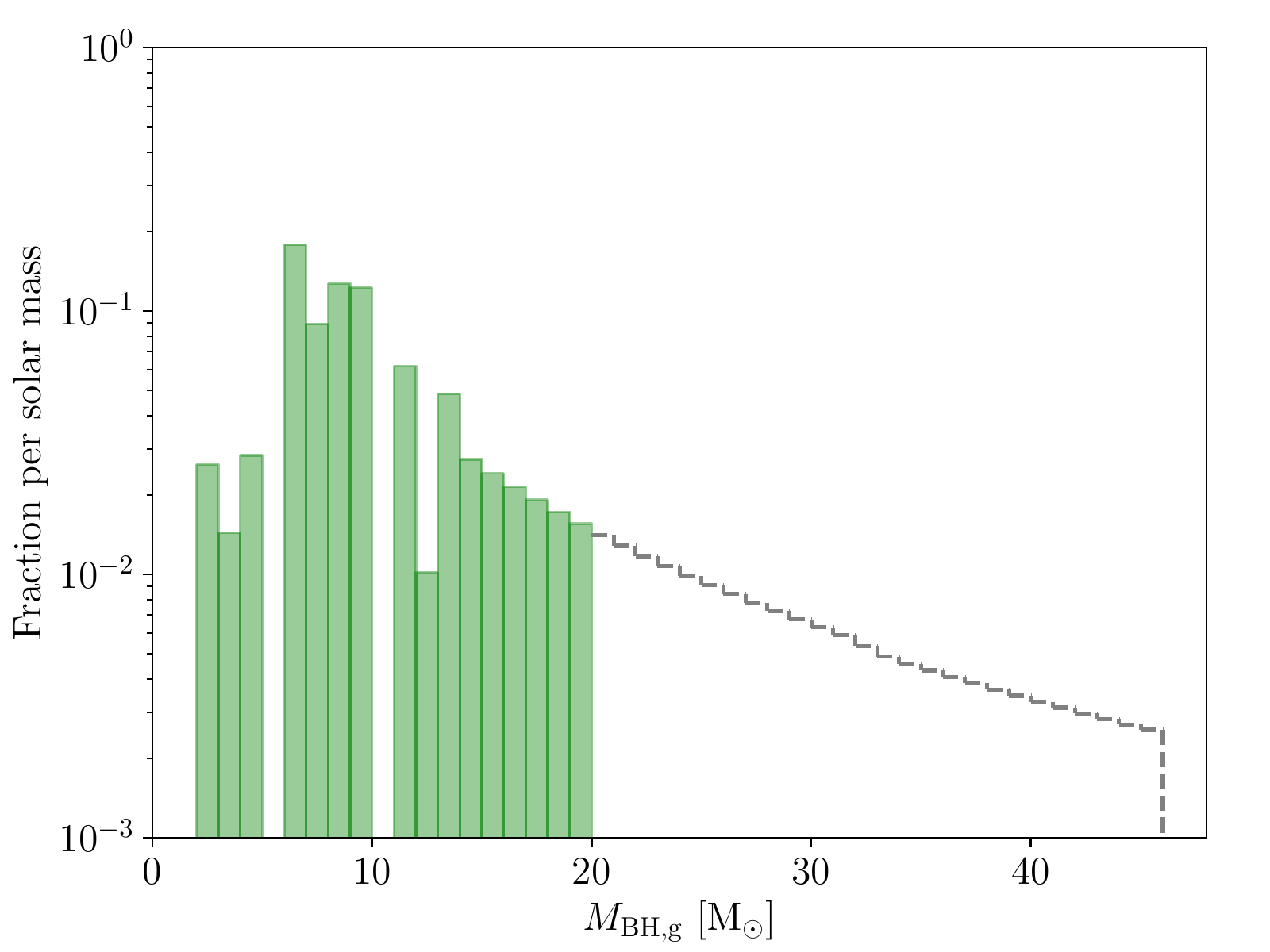}
\caption{Black hole mass distribution resulting for the W18 central
  engine.  The average black hole mass in the mass range studied (up
  to initial helium-star mass of 40\,\Msun) is 10.39\,\Msun, and the 
  median gravitational mass is 8.61\,\Msun. Note the
  existence of a dip in production around 10\,\Msun \ resulting from
  the island of explodability in \Fig{summaryw18-sta-exp}. The prominence of
  this feature depends on the central engine employed.  \lFig{bhmass}}
\end{figure}


\section{Light Curves - Code Physics}
\lSect{litephys}

The KEPLER code incorporates flux-limited radiative diffusion and thus
is capable of calculating approximate bolometric light curves for
supernovae of all types. Calculating light curves for supernovae that
have been stripped of their low density hydrogen envelopes is more
challenging than for Type IIp supernovae where hydrogen recombination
plays a dominant role. The peak brightness depends sensitively on the
amount of $^{56}$Ni created in the explosion. Line opacity is more
important; just carrying electron scattering is inadequate.  The light
curve, and especially the spectrum are more sensitive to the degree to
which that nickel has been mixed through the ejecta, and the opacity
can be highly variable with both time and location. Within its
limitations --one-dimensional, single temperature, no atomic line
physics-- KEPLER reasonably depicts the qualitative light curves of
supernovae. It is no substitute, however, for a full calculation of
radiative transfer, which eventually needs to be done. On the other
hand, the light curves are also sensitive to details of the explosion
physics, nuclear reactions, and shock hydrodynamics that are not
customarily captured in spectral synthesis codes alone.

\subsection{Opacity}
\lSect{opacity}

The opacity consists of two parts, electron scattering and other
processes that depend on details of the composition, ionization,
velocity shear, and level populations.  This second part will be
referred to loosely as the ``line opacity'' or the ``additive
opacity''. Electron scattering is calculated very well in the KEPLER
code. A Saha equation is solved in every zone at every time step using
the current temperature, density, and composition \citep{Ens88}. All
ionization stages of all the even-Z elements up to iron are calculated
assuming thermal equilibrium. The resulting electron density is used
to determine the electron scatting opacity which is corrected for
Compton scattering (including the change of a photon's energy in a
scattering event), degeneracy, and the presence of pairs.

The line opacity, on the other hand, is treated poorly. It
is represented by a single constant added to the electron scattering
opacity at all times and places. In calculations of Type IIp supernovae, this
number is given a small value, 10$^{-5}$\,cm$^2$\,g$^{-1}$, that
does not significantly affect the light curve. For Type Ia supernovae
where it is generally recognized that, owing to the large metal
content, lines play a dominant role, the constant frequently used is
$\kappa_a \sim 0.1$\,cm$^2$\,g$^{-1}$, where ``a'' stands for
``additive''.  Fig.~24 of \citet{Des15} shows the actual opacity for a
typical Type Ib supernova model calculated using the CMFGEN code
\citep{Hil12}, a state-of-the-art supernova radiation transport
code. The non-electron scattering part is by no means constant, but is
larger near the center, where the nickel is concentrated, and 
at earlier times, when the density is higher.  If one were to
attempt to assign a single value to the difference in the dashed and
solid lines in their figure (electron scattering and total opacity
respectively), it would be close to $\kappa_a \sim 0.2$ at early times
for velocities less than 6000\,km\,s$^{-1}$, but much smaller near the
peak of the light curve. There it ranges from about 0.07 near the
center ($v$ less than 3000\,km\,s$^{-1}$), to at most a few hundredths
cm$^2$\,g$^{-1}$ farther out.  Since we are more interested in the
behavior at peak than on the rise, $\kappa_a$ was chosen to be
0.03\,cm$^2$\,g$^{-1}$. This low value may slightly underestimate the
rise time compared with that of \citet{Des12}, but approximately
captures the behavior near peak.

\begin{figure}
	\centering
  \includegraphics[width=0.48\textwidth]{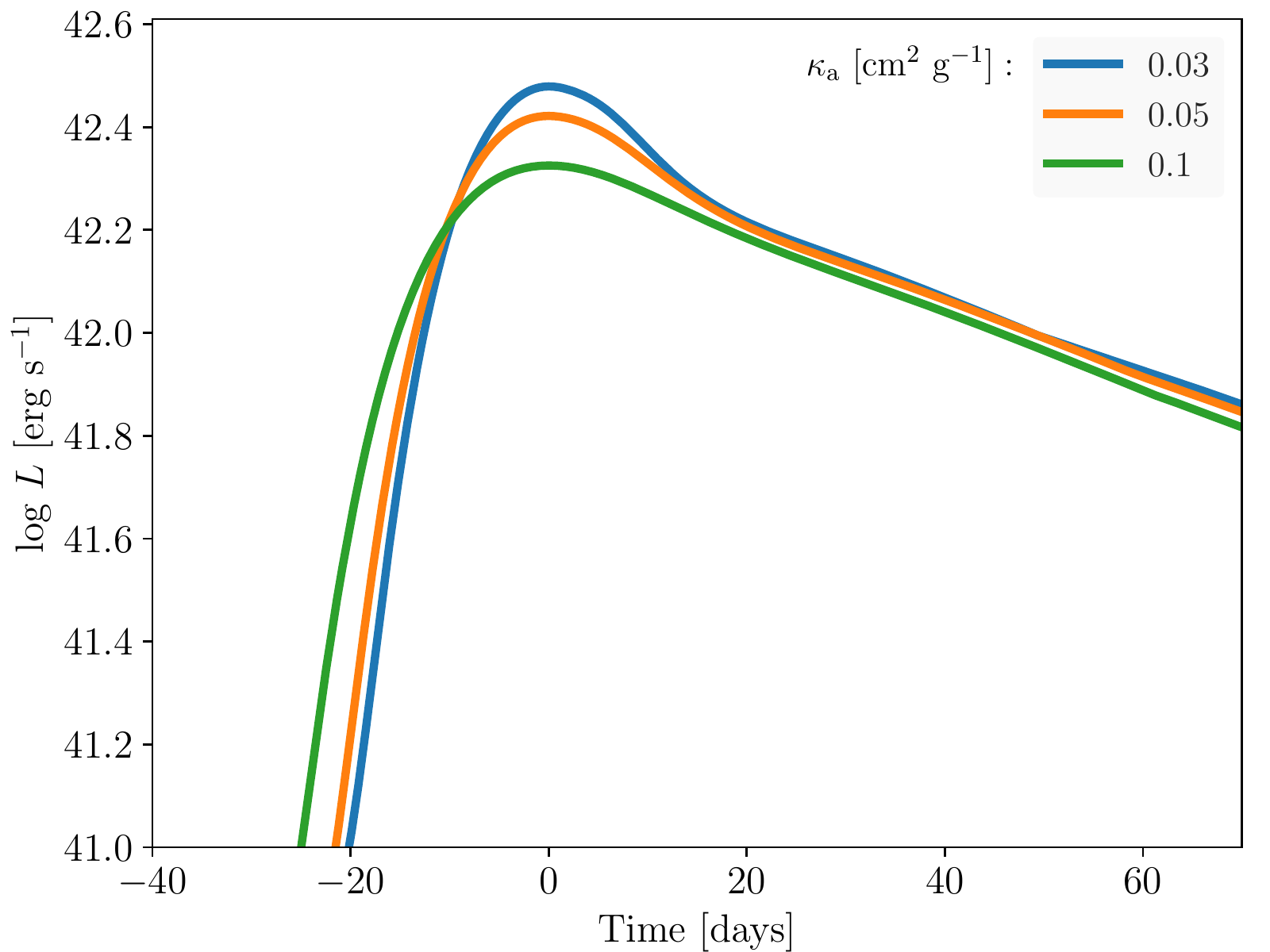}
	\caption{The light curve from the explosion of a 4.41\,\Msun
          \ model \citep{Yoo10} with a kinetic energy of 1.2 $\times
          10^{51}$\,erg, an additive opacity $\kappa_a$ = 0.03, 0.05,
          and 0.1\,cm$^2$\,g$^{-1}$ and strong mixing
          (\Sect{mix}). The smaller opacity ($\kappa_a = 0.03$\,cm$^2$\,g$^{-1}$
	  compares favorably in peak luminosity with an identical 
          model calculated by \citet{Des12} using the CMFGEN code with a
          more accurate treatment of radiation
          transport.
          \lFig{opacity}}
\end{figure}

To test this calibration, the light curve of Model 4.41 in
\citet{Des12} was recalculated using the KEPLER code and their
assumption for mixing which, in their most extreme case, is very
similar to ours (\Sect{mix}). This entailed a running boxcar average
with a box size of 0.2\,\Msun \ passed through the model of
\citet{Yoo10} four times, a procedure identical to that used by
\citet{Des12}. Several choices for $\kappa_a$ were explored
(\Fig{opacity}). In their Fig.~7, the peak for their Model 4.41x4 is
10$^{42.47}$\,erg\,s$^{-1}$, while ours is 10$^{42.48}$\,erg\,s$^{-1}$
for $\kappa_a$ = 0.03\,cm$^2$\,g$^{-1}$ and
10$^{42.42}$\,erg\,s$^{-1}$ for $\kappa_a =$ 0.05\,cm$^2$\,g$^{-1}$.
Our two light curves peak at 27.5 and 30.6 days after explosion, while
theirs peaks at 34.2 days. The discrepancy probably reflects the
larger opacity early on in their models. The decline rate at late
times is sensitive to the treatment of gamma-ray trapping which is
captured better in their full Monte Carlo simulation. All in all
though, if a single constant must be used, $\kappa_a$ =
0.03\,cm$^2$\,g$^{-1}$ seems a good value for calculating
$L_{\rm peak}$.

\subsection{Mixing}
\lSect{mix}

The principle effect of mixing is to disperse $^{56}$Ni from the
central tenth or so solar masses where it is made to masses and speeds
farther out in the star. The mixing in Type I supernovae has its origin in the
non-spherical nature of the central explosion, and not so much the
reverse shock as in Type II supernovae. Mixing leads to a light curve
that rises earlier, is a little broader, and declines
earlier. Except for filling in a gap in emission at earlier times, the
modification of the light curve near peak is not great. The effect on
the spectrum (not calculated here) though is very important
\citep{Des12,Des15}.

Mixing is inherently a three-dimensional phenomenon and does not imply
the homogenization, at the atomic scale, of any ejecta. Rather clumps
of one composition, which may carry radioactivity, are mixed out in
velocity. As noted above, the average effects of mixing are
incorporated in the KEPLER code using a ``running boxcar average''. A
specified interval of mass (the ``boxcar'') is thoroughly mixed and
the process is repeated as this boxcar is moved out, zone by zone,
through the star.  The degree of mixing depends on the size of the boxcar and
the number of times it passes through the star. The resulting
distribution is quasi-exponential in mass.

\citet{Des12} found that appreciable mixing was essential to obtain
the correct colors and spectra for Type Ib and Ic supernovae. Indeed,
the degree to which $^{56}$Ni was mixed into helium was the
determining factor in whether the supernova was of Type Ib or
Ic. Lacking the tamping influence of a massive hydrogen envelope,
mixing in an almost stripped core extends farther out in
velocity space than the 3500\,km\,s$^{-1}$ typical of models for
SN~1987A \citep{Won15,Utr19}. Three-dimensional calculations of mixing in
supernovae without hydrogen are lacking. The only exceptions are simulations of ultra-stripped supernovae by \citet{Mue18} and a Type~IIb supernova model of \citet{Won17}. The latter model, which had a hydrogen envelope
of only 0.3\,\Msun \ and therefore did not experience much  
deceleration by a reverse shock from the hydrogen-rich layer, was used to calibrate mixing in the present study.

The model they considered was the stripped down core of a 15\,\Msun
\ red supergiant with a helium core of 4.5\,\Msun. For comparison, we
used our 6.0\,\Msun \ model which had a final helium plus heavy element
core of 4.45\,\Msun.  Fig.~7 of \citet{Won17} shows the angle-averaged
$^{56}$Ni mixing from their 3D run. The \citet{Won17} model may have
experienced more mixing than most others (Annop Wongwathanarat,
private communication 2019), so it is treated as an upper bound. The
prescription finally employed in KEPLER used a boxcar width of 0.15
$M_{\rm ej}$, where $M_{\rm ej}$ is the mass ejected in the supernova
(presupernova mass minus remnant mass). This mixing region was moved
through all the ejecta three times and then a final fourth mixing was
applied using an interval half as great, i.e., 0.075 $M_{\rm ej}$. The
comparison with the model from \citet{Won17} is shown in
\Fig{mixmod}. The result of the same mixing applied to the 2.5, 6.0,
and 19.75\,\Msun \ models is shown in \Fig{mixing}.

\begin{figure}
        \centering
  \includegraphics[width=0.48\textwidth]{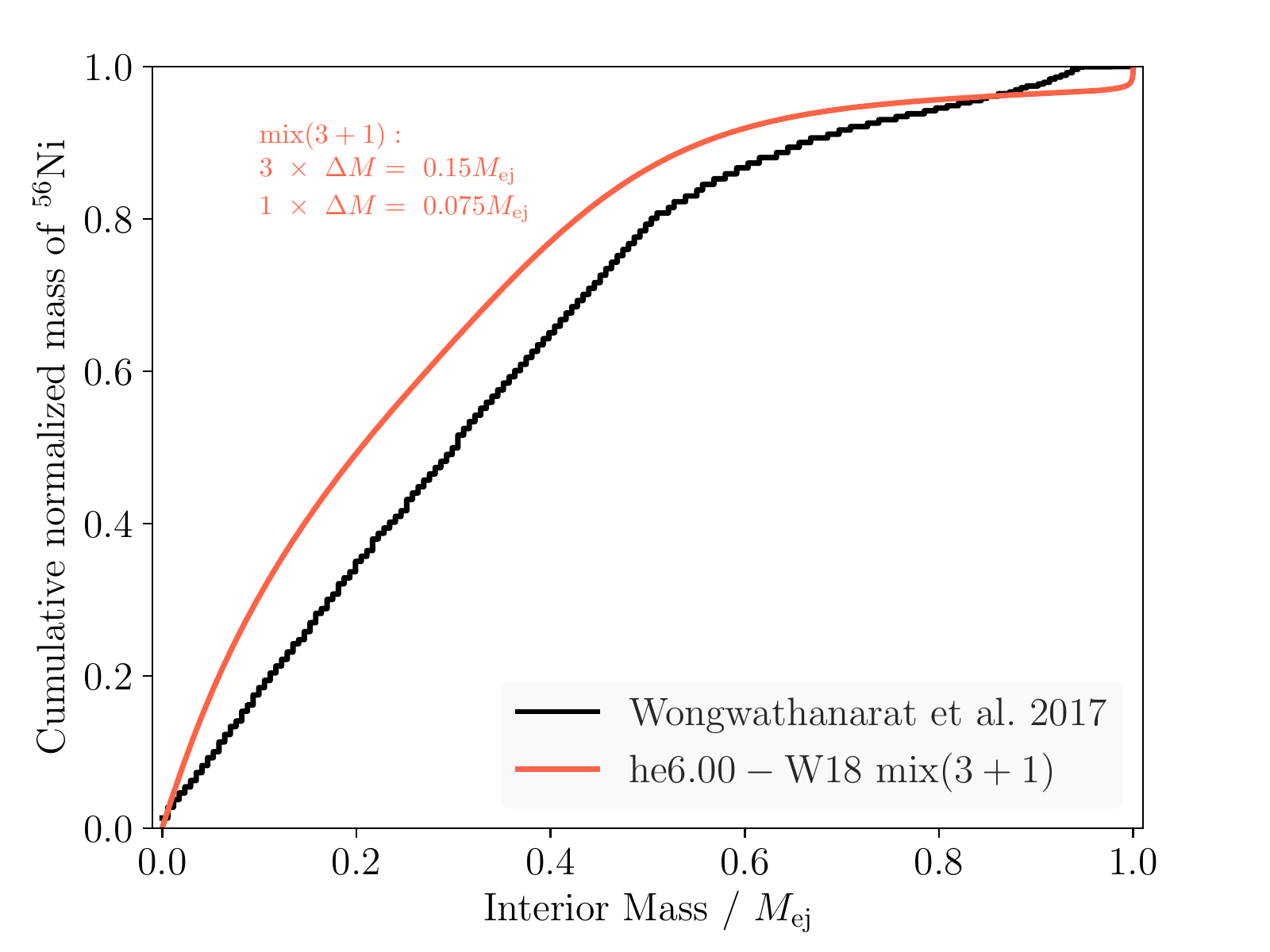}
  \caption{Mixing in our Model He6.0 (red line) which had a final mass
    of 4.45\,\Msun \ compared with mixing in the 4.5\,\Msun \ helium
    core of Model W15-2-cw-IIB of \citet[][black line]{Won17}. The
    plot shows the cumulative fraction of $^{56}$Ni contained within
    the given mass. 90\% of the $^{56}$Ni is contained within the
    inner 3.2\,\Msun \ of our Model He6.0. The velocity at that 90\%
    point is 5190\,km\,s$^{-1}$.  \lFig{mixmod}}
\end{figure}

\begin{figure}
	\centering
  \includegraphics[width=0.48\textwidth]{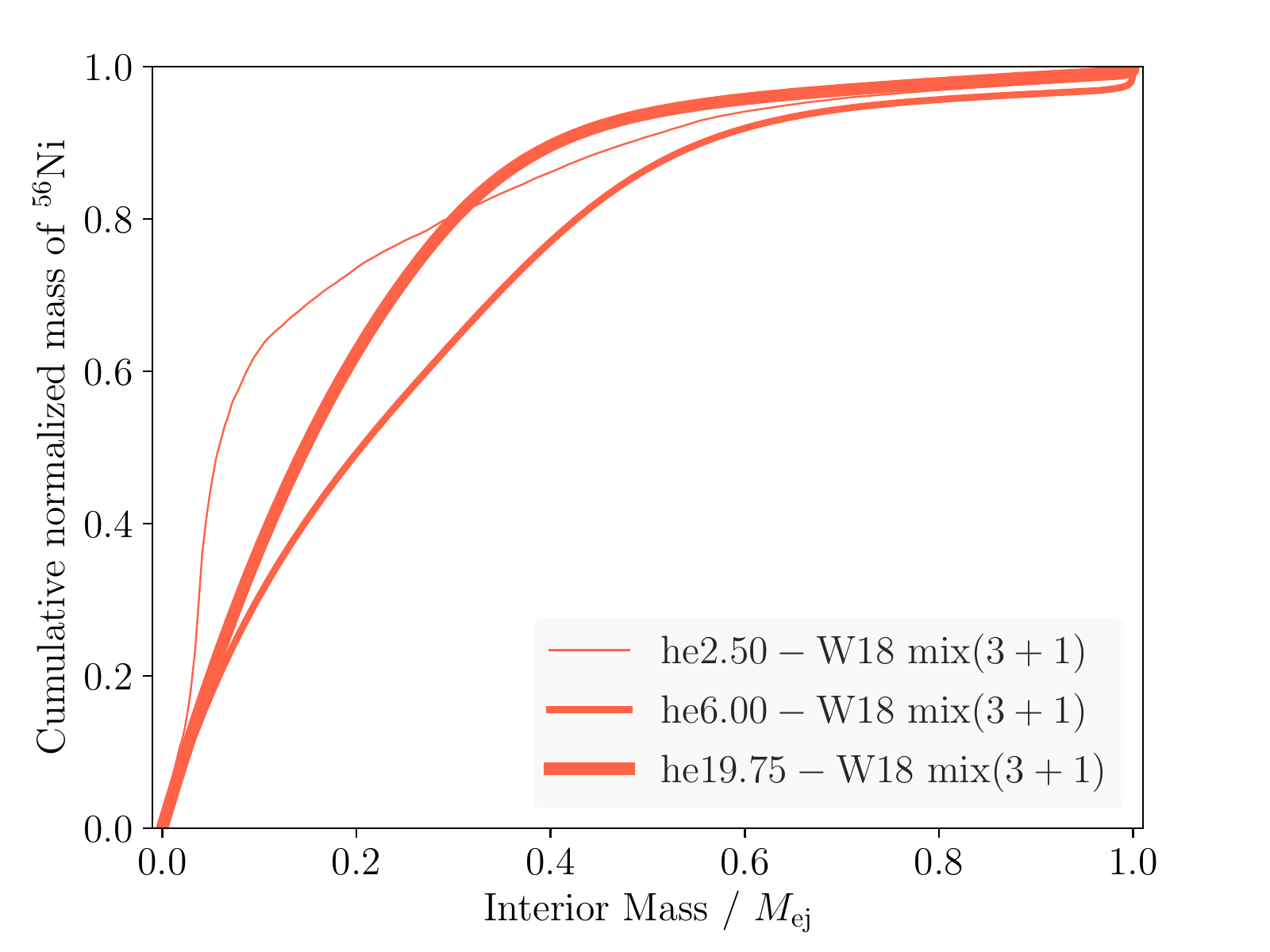}
  \includegraphics[width=0.48\textwidth]{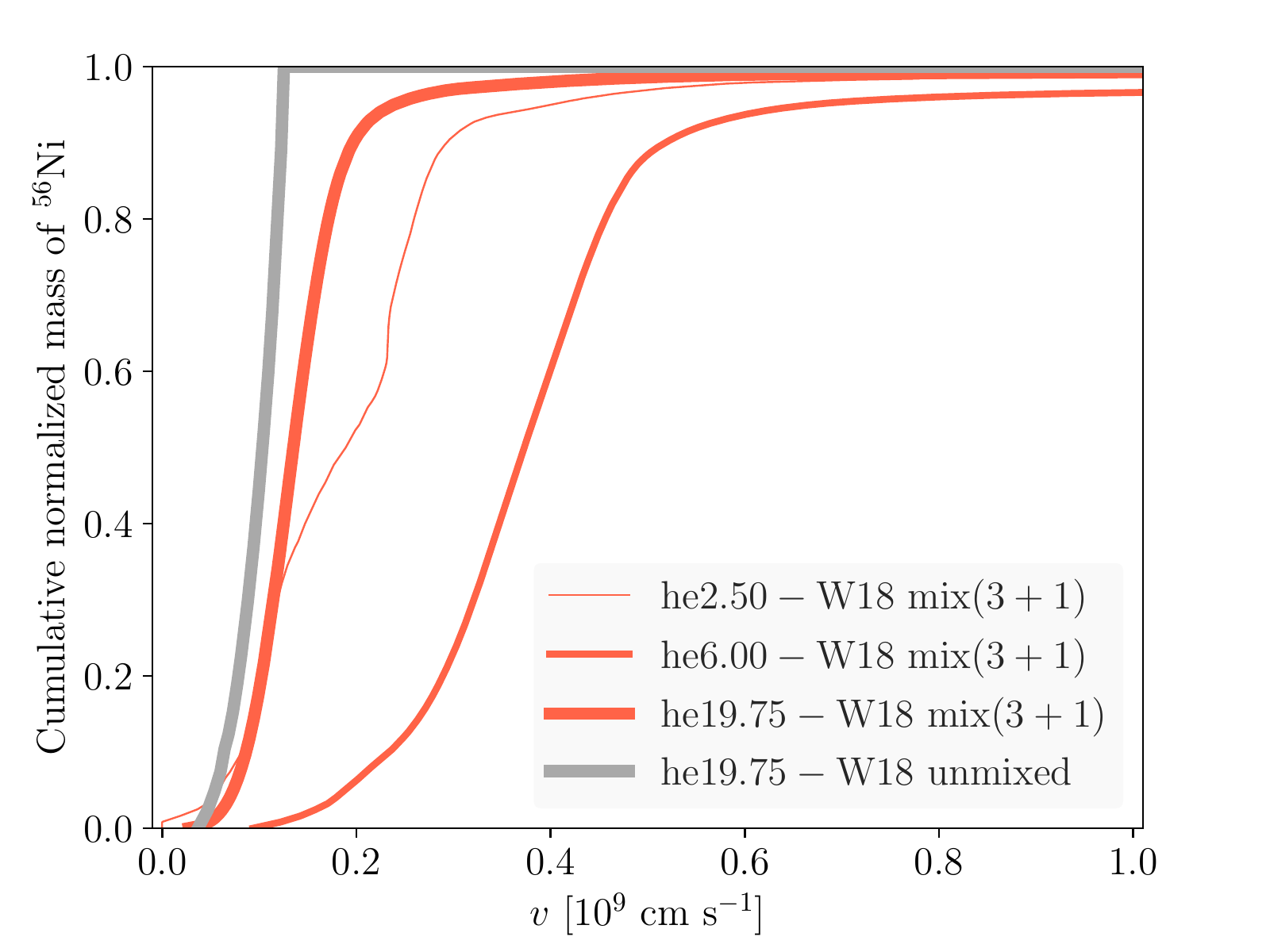}
	\caption{Mixing assumed in Models He2.5, He6.0, and He19.75 as
          a function of mass and terminal speed. The presupernova
          masses of these models were 2.07, 4.45 and 10.28\,\Msun. The
          ejected masses are 0.731, 2.780, and 8.539\,\Msun. The
          curves show the integrated fraction of $^{56}$Ni mass within
          the given mass or speed.  \lFig{mixing}}
\end{figure}


\section{Calculated Light Curves}
\lSect{lite}

Given the input physics and parametrization described in
\Sect{litephys}, light curves were calculated for all models that
exploded and left a neutron star remnant. The lowest mass models produced
less $^{56}$Ni and, in some cases, had large presupernova radii
\citep[\Tab{expltbl} and Table~3 of][]{Woo19}. A few models that
experienced a strong silicon flash had especially large radii. There
the light curves were dominated by shock interaction and helium
recombination. Higher mass models resembled more common Type Ib and Ic
supernovae. Still higher masses resulted in long faint light curves
that do not seem to be well represented in current observations. Which
of these categories characterized a given model depended mostly on its
presupernova mass. Depending on mass loss rate, this could correspond
to different initial helium core masses and main sequence masses.
Based on the standard mass loss prescription used here, our discussion
is broken into four sections: (a) Helium stars with initial masses of
2.5 to 2.9\,\Msun \ that have a large radius, but did not experience a
silicon flash, i.e., models with normal mass loss and initial helium
masses, 2.6, 2.7, 2.8, and 2.9\,\Msun; (b) other low models that {\sl
did} experience a Si flash, i.e., models with initial helium core
masses of 2.5, 3.0, 3.1, and 3.2\,\Msun; (c) ``common'' Type Ib and Ic
supernovae with initial masses 3.3 to 8.0\,\Msun; and (d) more massive
stars. Some stars below 2.5\,\Msun \ may still explode as
electron-capture supernovae, but are not included here since their
$^{56}$Ni production would be very small and the presupernova radius
large. They would resemble group (a).  The physics of the silicon
flash are sufficiently uncertain that groups (b) and (c) could perhaps
be combined. Common Type Ib and Ic supernovae would then extend down
to models with initial helium masses of 3.0\,\Msun \ with the lower
bound set by radius expansion.

\begin{deluxetable*}{ccccccccccccccc}
\tablecaption{Explosion and Light Curve Properties (subset)}
\tablehead{ \colhead{${M_{\rm He,i}}$}  &
            \colhead{${M_{\rm preSN}}$} &
            \colhead{${M_{\rm ej}}$}  &
            \colhead{${E_{\rm exp}}$}  &
            \colhead{${M_{\rm Ni}}$}  &
            \colhead{${\log\ L_{\rm peak}}$}  &
            \colhead{${t_{\rm -1/2}}$}  &
            \colhead{${t_{\rm +1/2}}$}  &
            \colhead{} &
            \colhead{${M_{\rm ej}}$}  &
            \colhead{${E_{\rm exp}}$}  &
            \colhead{${M_{\rm Ni}}$}  &
            \colhead{${\log\ L_{\rm peak}}$}  &
            \colhead{${t_{\rm -1/2}}$}  &
            \colhead{${t_{\rm +1/2}}$}\vspace{1mm}
            \\
            \colhead{[\Msun]}  &
            \colhead{[\Msun]}  &
            \colhead{[\Msun]}  &
            \colhead{[B]}  &
            \colhead{[\Msun]}  &
            \colhead{[${\rm erg\,s^{-1}}$]}  &
            \colhead{[days]}  &
            \colhead{[days]}  &
            \colhead{} &
            \colhead{[\Msun]}  &
            \colhead{[B]}  &
            \colhead{[\Msun]}  &
            \colhead{[${\rm erg\,s^{-1}}$]}  &
            \colhead{[days]}  &
            \colhead{[days]}
            }\\
\startdata
\multicolumn{15}{c}{standard $\dot{M}$}\\
\multicolumn{2}{c}{} & \multicolumn{6}{c}{W18} & & \multicolumn{6}{c}{S19.8}\\
\cline{3-8}\cline{10-15}\\
2.70  &  2.21  &  0.85  &  0.21  &  0.02  &  42.07  &  13.0  &  11.0  & &  0.85  &  0.21  &  0.02  &  42.06  &  12.9  &  11.0  \\
3.20  &  2.59  &  1.14  &  0.67  &  0.05  &  42.34  &   7.0  &  11.7  & &  1.14  &  0.67  &  0.05  &  42.33  &   6.9  &  11.8  \\
3.50  &  2.81  &  1.27  &  0.42  &  0.03  &  42.00  &   7.9  &  14.1  & &  1.28  &  0.47  &  0.04  &  42.06  &   7.8  &  13.4  \\
4.00  &  3.16  &  1.62  &  0.64  &  0.05  &  42.16  &   8.2  &  15.3  & &  1.64  &  0.76  &  0.06  &  42.23  &   8.0  &  15.1  \\
4.50  &  3.49  &  1.89  &  1.28  &  0.10  &  42.38  &   8.0  &  14.2  & &  1.96  &  1.82  &  0.15  &  42.59  &   7.5  &  13.1  \\
4.62  &  3.57  &  1.98  &  1.39  &  0.11  &  42.41  &   8.3  &  14.4  & &  2.01  &  1.65  &  0.13  &  42.48  &   7.8  &  14.3  \\
4.75  &  3.65  &  2.10  &  1.46  &  0.11  &  42.40  &   8.1  &  14.3  & &  2.11  &  1.61  &  0.12  &  42.43  &   7.9  &  13.7  \\
4.88  &  3.73  &  2.16  &  1.47  &  0.11  &  42.40  &   8.2  &  14.6  & &  2.16  &  1.60  &  0.12  &  42.42  &   8.1  &  14.2  \\
5.00  &  3.81  &  2.21  &  1.49  &  0.11  &  42.41  &   8.4  &  14.9  & &  2.24  &  1.72  &  0.13  &  42.46  &   8.1  &  14.3  \\
5.25  &  3.98  &  2.33  &  1.26  &  0.09  &  42.32  &   9.0  &  17.5  & &  2.40  &  1.77  &  0.14  &  42.49  &   8.3  &  15.0  \\
5.50  &  4.13  &  2.59  &  1.58  &  0.10  &  42.39  &   9.1  &  17.9  & &  2.61  &  1.80  &  0.10  &  42.44  &   8.7  &  16.8  \\
5.75  &  4.29  &  2.55  &  1.28  &  0.09  &  42.31  &   9.3  &  19.4  & &  2.56  &  1.41  &  0.10  &  42.34  &   9.0  &  18.9  \\
6.00  &  4.44  &  2.82  &  1.07  &  0.08  &  42.21  &  10.2  &  23.4  & &  2.85  &  1.39  &  0.10  &  42.33  &   9.7  &  20.5  \\
6.50  &  4.75  &  3.14  &  1.50  &  0.11  &  42.41  &  10.0  &  21.6  & &  3.15  &  1.70  &  0.12  &  42.44  &   9.8  &  20.3  \\
7.00  &  5.04  &  3.33  &  1.37  &  0.12  &  42.31  &  10.9  &  25.6  & &  3.38  &  1.85  &  0.16  &  42.46  &  10.3  &  22.4  \\
7.50  &  5.34  &  3.61  &  1.38  &  0.12  &  42.30  &  11.4  &  27.8  & &  3.66  &  1.79  &  0.15  &  42.42  &  10.8  &  25.0  \\
8.00  &  5.63  &  3.95  &  0.70  &  0.06  &  41.95  &  14.5  &  40.7  & &  3.96  &  0.88  &  0.07  &  42.05  &  13.4  &  36.5  \\
8.50  &  5.92  &  4.27  &  1.01  &  0.09  &  42.11  &  13.8  &  37.5  & &  4.31  &  1.30  &  0.11  &  42.21  &  12.8  &  34.0  \\
9.00  &  6.19  &  4.45  &  0.96  &  0.07  &  42.03  &  14.1  &  40.7  & &  4.51  &  1.49  &  0.11  &  42.25  &  13.1  &  33.1  \\
9.50  &  6.47  &  4.85  &  1.12  &  0.12  &  42.17  &  16.2  &  43.1  & &  4.89  &  1.41  &  0.14  &  42.26  &  14.8  &  39.0  \\
10.00  &  6.74  &  5.19  &  0.76  &  0.07  &  41.98  &  18.7  &  52.0  & &  5.22  &  0.93  &  0.09  &  42.08  &  17.9  &  46.9  \\
10.50  &  6.91  &  5.15  &  0.99  &  0.11  &  42.12  &  18.1  &  49.9  & &  5.23  &  1.47  &  0.16  &  42.29  &  15.5  &  42.6  \\
11.50  &  7.10  &  5.26  &  1.01  &  0.13  &  42.15  &  18.5  &  53.6  & &  5.42  &  1.79  &  0.21  &  42.43  &  15.4  &  41.7  \\
12.00  &  7.24  &  5.32  &  0.81  &  0.08  &  41.97  &  19.9  &  54.3  & &  5.44  &  1.50  &  0.14  &  42.26  &  16.2  &  43.5  \\
12.50  &  7.42  &  5.72  &  0.88  &  0.09  &  42.01  &  19.5  &  53.8  & &  5.83  &  1.49  &  0.15  &  42.25  &  17.2  &  44.0  \\
12.75  &  7.51  &  5.82  &  0.96  &  0.10  &  42.05  &  19.4  &  51.5  & &  5.91  &  1.54  &  0.16  &  42.26  &  17.1  &  45.0  \\
13.25  &  7.71  &  5.98  &  0.95  &  0.10  &  42.04  &  21.3  &  54.9  & &  6.08  &  1.48  &  0.16  &  42.25  &  18.0  &  48.2  \\
18.25  &  9.67  &  8.06  &  0.96  &  0.11  &  42.00  &  27.1  &  63.2  & &  8.11  &  1.29  &  0.15  &  42.13  &  23.9  &  58.6  \\
18.50  &  9.78  &  8.16  &  0.97  &  0.11  &  42.00  &  27.2  &  63.5  & &  8.22  &  1.28  &  0.14  &  42.13  &  24.2  &  61.1  \\
18.75  &  9.88  &  8.24  &  0.97  &  0.11  &  42.01  &  27.7  &  64.3  & &  8.30  &  1.31  &  0.15  &  42.14  &  24.4  &  60.2  \\
19.00  &  9.98  &  8.32  &  1.00  &  0.11  &  42.01  &  27.5  &  63.9  & &  8.39  &  1.34  &  0.15  &  42.15  &  24.8  &  60.1  \\
19.25  &  10.1  &  8.31  &  0.78  &  0.06  &  41.66  &  29.0  &  66.8  & &  8.47  &  1.35  &  0.15  &  42.15  &  25.2  &  61.1  \\
19.50  &  10.2  &  8.43  &  0.84  &  0.08  &  41.82  &  27.7  &  65.9  & &  8.56  &  1.36  &  0.16  &  42.16  &  25.7  &  59.1  \\
19.75  &  10.3  &  8.53  &  0.87  &  0.09  &  41.86  &  29.3  &  64.7  & &  8.65  &  1.40  &  0.16  &  42.17  &  25.1  &  59.9  \\
\multicolumn{15}{c}{}\\
\multicolumn{15}{c}{1.5$\times\dot{M}$}\\
\multicolumn{2}{c}{} & \multicolumn{6}{c}{W18} & & \multicolumn{6}{c}{S19.8}\\
\cline{3-8}\cline{10-15}\\
5.00  &  3.43  &  1.85  &  1.58  &  0.12  &  42.49  &   7.5  &  13.4  & &  1.88  &  1.86  &  0.14  &  42.58  &   7.2  &  12.6  \\
5.25  &  3.57  &  1.96  &  1.50  &  0.12  &  42.46  &   7.7  &  14.5  & &  2.00  &  1.81  &  0.15  &  42.52  &   7.6  &  13.4  \\
5.50  &  3.70  &  2.08  &  1.47  &  0.11  &  42.43  &   8.1  &  15.2  & &  2.12  &  1.80  &  0.14  &  42.53  &   7.8  &  14.1  \\
5.75  &  3.83  &  2.20  &  1.42  &  0.10  &  42.37  &   8.3  &  15.2  & &  2.24  &  1.76  &  0.13  &  42.47  &   7.9  &  14.7  \\
6.00  &  3.96  &  2.32  &  1.55  &  0.12  &  42.42  &   8.4  &  15.7  & &  2.35  &  1.80  &  0.14  &  42.47  &   8.0  &  15.3  \\
6.25  &  4.09  &  2.41  &  1.47  &  0.11  &  42.38  &   8.7  &  17.1  & &  2.43  &  1.68  &  0.12  &  42.44  &   8.4  &  16.7  \\
6.50  &  4.21  &  2.50  &  1.31  &  0.10  &  42.32  &   9.0  &  19.5  & &  2.55  &  1.79  &  0.14  &  42.47  &   8.5  &  16.8  \\
6.75  &  4.33  &  2.74  &  1.44  &  0.11  &  42.33  &   9.4  &  19.6  & &  2.76  &  1.66  &  0.11  &  42.39  &   8.9  &  19.0  \\
7.00  &  4.45  &  2.84  &  1.44  &  0.11  &  42.37  &   9.5  &  20.7  & &  2.86  &  1.71  &  0.14  &  42.45  &   9.3  &  19.6  \\
7.50  &  4.69  &  3.02  &  1.39  &  0.12  &  42.33  &  10.0  &  23.0  & &  3.04  &  1.65  &  0.13  &  42.39  &   9.7  &  21.5  \\
8.00  &  4.92  &  3.26  &  1.47  &  0.12  &  42.38  &  10.4  &  23.8  & &  3.30  &  1.83  &  0.14  &  42.51  &  10.0  &  21.5  \\
8.50  &  4.90  &  3.18  &  1.35  &  0.12  &  42.31  &  10.8  &  24.8  & &  3.21  &  1.61  &  0.14  &  42.37  &  10.2  &  23.0  \\
9.00  &  4.87  &  3.14  &  1.01  &  0.08  &  42.14  &  11.4  &  26.7  & &  3.16  &  1.23  &  0.09  &  42.22  &  10.7  &  25.3  \\
9.50  &  4.89  &  3.16  &  1.41  &  0.13  &  42.33  &  10.6  &  24.3  & &  3.21  &  1.87  &  0.16  &  42.46  &  10.1  &  21.8  \\
10.00  &  4.96  &  3.20  &  1.24  &  0.11  &  42.27  &  11.4  &  27.4  & &  3.26  &  1.72  &  0.15  &  42.40  &  10.5  &  23.2  \\
10.50  &  5.08  &  3.45  &  0.96  &  0.10  &  42.16  &  13.5  &  32.9  & &  3.53  &  1.48  &  0.15  &  42.35  &  11.8  &  27.2  \\
11.00  &  5.19  &  3.44  &  1.12  &  0.11  &  42.23  &  12.6  &  31.3  & &  3.49  &  1.48  &  0.14  &  42.34  &  11.8  &  27.2  \\
11.50  &  5.32  &  3.60  &  1.02  &  0.10  &  42.17  &  13.7  &  33.1  & &  3.66  &  1.47  &  0.14  &  42.33  &  12.1  &  28.2  \\
12.00  &  5.43  &  3.71  &  0.98  &  0.09  &  42.15  &  13.4  &  35.2  & &  3.78  &  1.45  &  0.13  &  42.32  &  12.4  &  29.2  \\
12.50  &  5.53  &  3.82  &  1.41  &  0.14  &  42.30  &  12.3  &  30.7  & &  3.88  &  1.86  &  0.17  &  42.44  &  11.8  &  26.6  \\
13.00  &  5.64  &  3.98  &  1.29  &  0.12  &  42.30  &  13.1  &  33.0  & &  4.00  &  1.45  &  0.14  &  42.34  &  12.8  &  31.1  \\
14.00  &  5.86  &  4.22  &  1.20  &  0.12  &  42.24  &  14.6  &  37.2  & &  4.27  &  1.52  &  0.16  &  42.32  &  13.5  &  33.4  \\
15.00  &  6.09  &  4.45  &  0.95  &  0.09  &  42.10  &  16.1  &  43.5  & &  4.53  &  1.46  &  0.14  &  42.30  &  14.3  &  34.7  \\
16.00  &  6.34  &  4.72  &  0.54  &  0.03  &  41.44  &  17.6  &  49.7  & &  4.79  &  0.66  &  0.06  &  41.91  &  19.1  &  49.2  \\
17.00  &  6.59  &  4.72  &  1.04  &  0.10  &  42.11  &  16.1  &  44.4  & &  4.76  &  1.37  &  0.12  &  42.23  &  14.8  &  39.6  \\
17.50  &  6.71  &  5.08  &  1.74  &  0.20  &  42.45  &  14.8  &  38.1  & &  5.10  &  1.95  &  0.22  &  42.49  &  14.4  &  36.6  \\
19.75  &  7.32  &  5.68  &  1.17  &  0.12  &  42.16  &  18.0  &  47.1  & &  5.73  &  1.50  &  0.15  &  42.25  &  16.5  &  43.4  \\
20.50  &  7.53  &  5.69  &  0.90  &  0.09  &  41.97  &  20.4  &  54.8  & &  5.86  &  1.59  &  0.17  &  42.30  &  17.3  &  44.3  \\
28.00  &  9.77  &  8.01  &  0.84  &  0.06  &  41.62  &  25.3  &  67.0  & &  8.10  &  1.01  &  0.11  &  42.00  &  26.9  &  62.8  \\
29.00  &  10.1  &  8.21  &  0.79  &  0.02  &  40.47  &  36.2  &  48.0  & &  8.46  &  1.37  &  0.16  &  42.16  &  25.0  &  59.6  \\
29.50  &  10.3  &  8.39  &  0.82  &  0.04  &  41.06  &  25.2  &  62.8  & &  8.61  &  1.38  &  0.16  &  42.16  &  24.9  &  62.1  \\
\enddata
\lTab{expltbl}
\end{deluxetable*}

\subsection{Low Mass Models with Radius Expansion}
\lSect{lowm}

Models with initial helium core masses less than about 3.0\,\Msun
\ (3.3\,\Msun \ if silicon flashes are strong) and standard mass loss
experience significant radius expansion during their final stages of
evolution \citep[see Table~3 and Figure~5 of][]{Woo19}. These stars
have presupernova masses less than 2.45\,\Msun \ (\Tab{bigtable}) and
eject less than 1\,\Msun. Their explosion energies and $^{56}$Ni
production are smaller than the more massive models
(\Fig{summaryw18-sta-exp}), so the radioactive portion of their light
curves is fainter and faster.  Some representative light curves are
shown in \Fig{lowmlite}.

\begin{figure}[h]
\centering
  \includegraphics[width=0.48\textwidth]{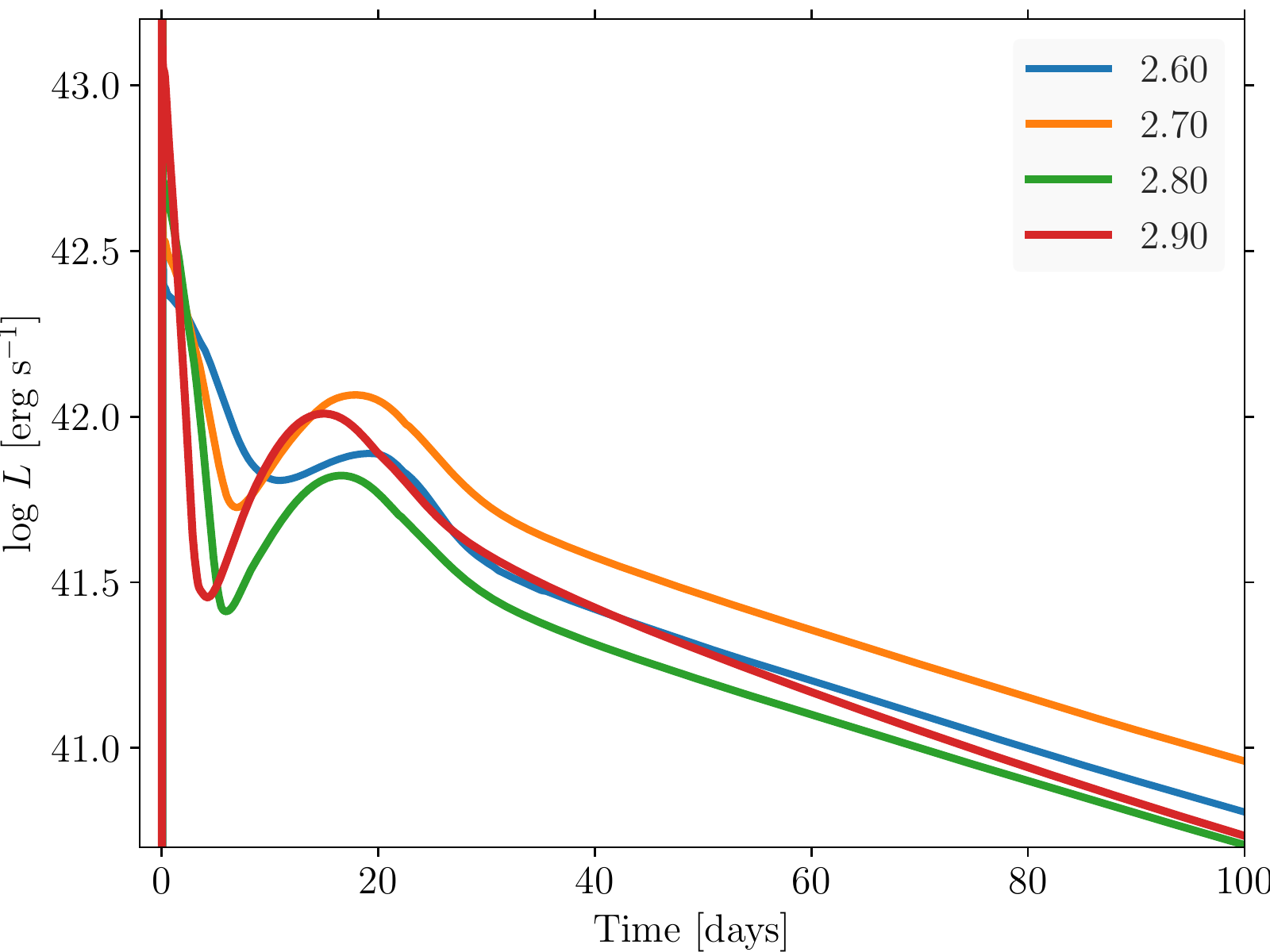}
  \includegraphics[width=0.48\textwidth]{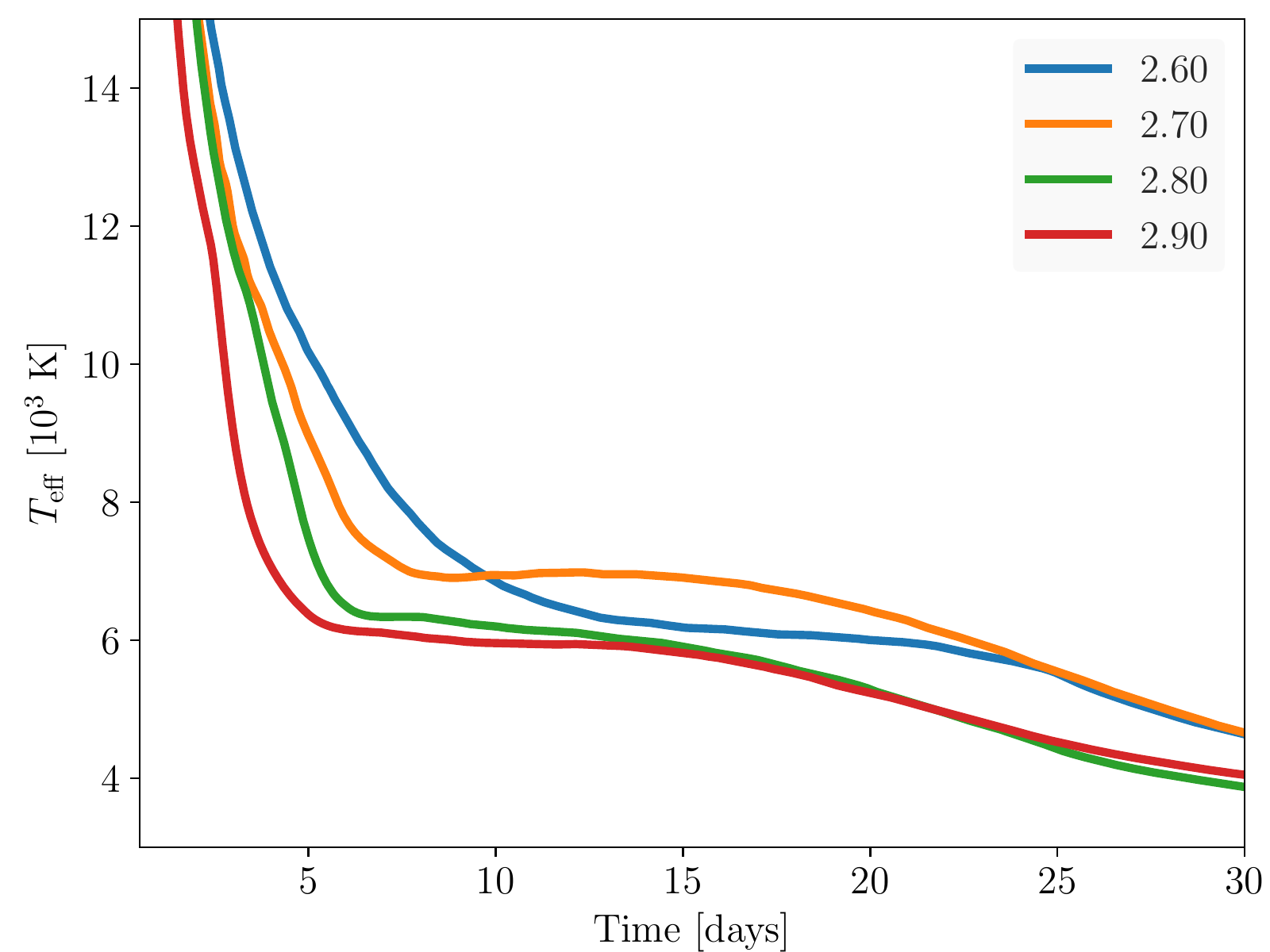}
\caption{ Light curves ({\em top}) and effective temperatures ({\em bottom})
   for low mass models ({\em colors} denote $M_{\rm He,i}$). Some with
   radius expansion show an early peak from recombination and a fainter
   radioactive peak.  None are very bright at second (radioactive) peak.
   \lFig{lowmlite}}
\end{figure}

\begin{deluxetable}{crcc}
\tablecaption{Properties of Low Mass Models}
\tablehead{ \colhead{}                     &
            \colhead{time}                 &
            \colhead{$\log\ L_{\rm bol}$}  &
            \colhead{$T_{\rm eff}$}\vspace{1mm}          
            \\
            \colhead{}                     &
            \colhead{[days]}               &
            \colhead{[erg\,s$^{-1}$]}       &
            \colhead{[$10^3$\,K]}              
            }
\startdata
\multirow{4}{*}{\rotatebox[origin=c]{90}{He2.60}}
& 1    & 42.36 & 22.8\\
& 2    & 42.32 & 16.3\\
& 5    & 42.12 & 10.2\\
& peak & 41.89 & 6.11\\
\\
\multirow{4}{*}{\rotatebox[origin=c]{90}{He2.70}}
& 1    & 42.47 & 22.0\\
& 2    & 42.36 & 15.3\\
& 5    & 41.90 & 8.81\\
& peak & 42.07 & 6.64\\
\enddata
\tablecomments{peak is the time of second light curve peak powered by
  radioactivity, which was 19.4, 18.0 days for Models He2.60 and He2.70
  respectively.}  \lTab{lowlcs}
\end{deluxetable}

These are not what we usually think of as common Type Ib or Ic
supernovae. Their early display is dominated by envelope
recombination. There are often two peaks, though the first is very
blue. Some of these might be associated with ``fast blue optical
transients'' \citep{Kle18,Woo19}. For example, Models He2.60 and
He2.70 with presupernova masses of 2.15\,\Msun \ and 2.22\,\Msun \
and radii $7.8 \times 10^{12}$ and $6.5 \times 10^{12}$\,cm,
respectively. Properties of these two supernovae are summarized in 
\Fig{lowmlite} and \Tab{lowlcs}. These large presupernova radii, small 
ejected masses, and extended epochs of high effective temperature
(\Fig{lowmlite}, lower panel) associated with shock break out are also 
similar to what was inferred for SN~2013ge by \citet{Dro16}.

\subsection{Silicon Flashes?}
\lSect{sif}

Particularly intriguing are the light curves of the low mass models
that experience a silicon deflagration and eject substantial matter
well in advance of core collapse. As discussed in \citet{Woo19}, the
amount of matter ejected and its timing  depend on
the poorly understood three-dimensional aspects of the silicon
flash and are uncertain. In cases where a lot of mass is lost months 
to years before collapse, the resulting supernova can be very bright.

\Fig{siflash} and \Tab{siflashtbl} show a representative range of
possibilities with the brightness of the light curve being mostly
determined by the timing and energy of the silicon flash.
Typically the flash occurs several weeks prior to iron core
collapse. The mean velocity of the ejecta is $v_{\rm ej,Si} \sim (2 \,
KE_{\rm ej,Si}/M_{\rm ej,Si})^{1/2}$ and models where the shell
expands to $v_{\rm ej,Si} t_{\rm ej,Si} \sim 10^{14}-10^{15}$\,cm
will be especially bright. This criterion explains the great
brightness of Models He3.0 and He3.1 which have $v_{\rm ej,Si} t_{\rm
  ej,Si}$ = 0.85 and 0.32 $\times 10^{15}$\,cm respectively. 
The amount of mass ejected and the radius reached by that ejecta are not so extreme in the other two models (only to a few times
10$^{13}$\,cm), so they are brighter than typical Type Ib and Ic supernovae only during the first few days after the explosion.
At their second bright peaks,
Models He3.0 and He3.1 had peak luminosities of 10$^{44.54}$ and
10$^{44.88}$\,erg\,s$^{-1}$ at days 4.5 and 8.3. The effective
temperatures there were not well determined since they were due to
shock interaction, but were calculated to be 21,000 and 32,000\,K. Models He2.5
and He3.2, on the other hand, had weak secondary maxima due to
radioactivity of 10$^{41.90}$ and 10$^{42.34}$\,erg\,s$^{-1}$ at days
22 and 15. Their effective temperatures then were only 5900 and 6600\,K.

The mass and energy of the ejected shell and the delay time are
determined mostly by the amount of silicon that burns explosively in
the flash and not, directly, by the mass of the star that
explodes. Brighter explosions could be generated by 2.5 and 3.2\,\Msun
\ models and fainter ones by 3.0 and 3.1\,\Msun \ models. The four
choices here are a very limited sampling meant to illustrate a broad
possibility of outcomes.  In cases where the silicon flash is weak or
absent, these supernovae would resemble the models of \Sect{lowm}. See
\citet{Woo19} for further details and a comparison to SN~2014ft \citep{De18}.

\begin{deluxetable}{ccccccc}
\tablecaption{Properties of Silicon Flash Models}
\tablehead{ \colhead{$M_{\rm He,i}$}      &
            \colhead{$M_{\rm preSN}$}     &
            \colhead{$M_{\rm ej,Si}$}     &
            \colhead{$t_{\rm ej,Si}$}     &                
            \colhead{$KE_{\rm ej,Si}$}    &
            \colhead{$E_{\rm exp}$}       &
            \colhead{$M_{\rm Ni}$}\vspace{1mm}
            \\
            \colhead{[\Msun]}          &
            \colhead{[\Msun]}          &
            \colhead{[\Msun]}          &
            \colhead{[days]}           &
            \colhead{[10$^{47}$\,erg]}   &
            \colhead{[10$^{51}$\,erg]}   &
            \colhead{[\Msun]}                   
            }
\startdata
2.5 & 2.07 &  0.25 &  16 &  0.93 &  0.115 & 0.027\\
3.0 & 2.45 &  0.74 &  62 &  190  &  0.303 & 0.035\\
3.1 & 2.52 &  0.42 &  34 &   49  &  0.621 & 0.080\\
3.2 & 2.59 &  0.02 &  15 &  0.47 &  0.668 & 0.079\\
\enddata
\tablecomments{$M_{\rm ej,Si}$ is the mass of the shell ejected by the
  silicon flash and $KE_{\rm ej,Si}$ is its kinetic energy. $t_{\rm
    ej,Si}$ is the delay between the flash and iron core
  collapse. Many other combinations are possible.}  \lTab{siflashtbl}
\end{deluxetable}

\begin{figure}[h]
  \centering
  \includegraphics[width=\columnwidth]{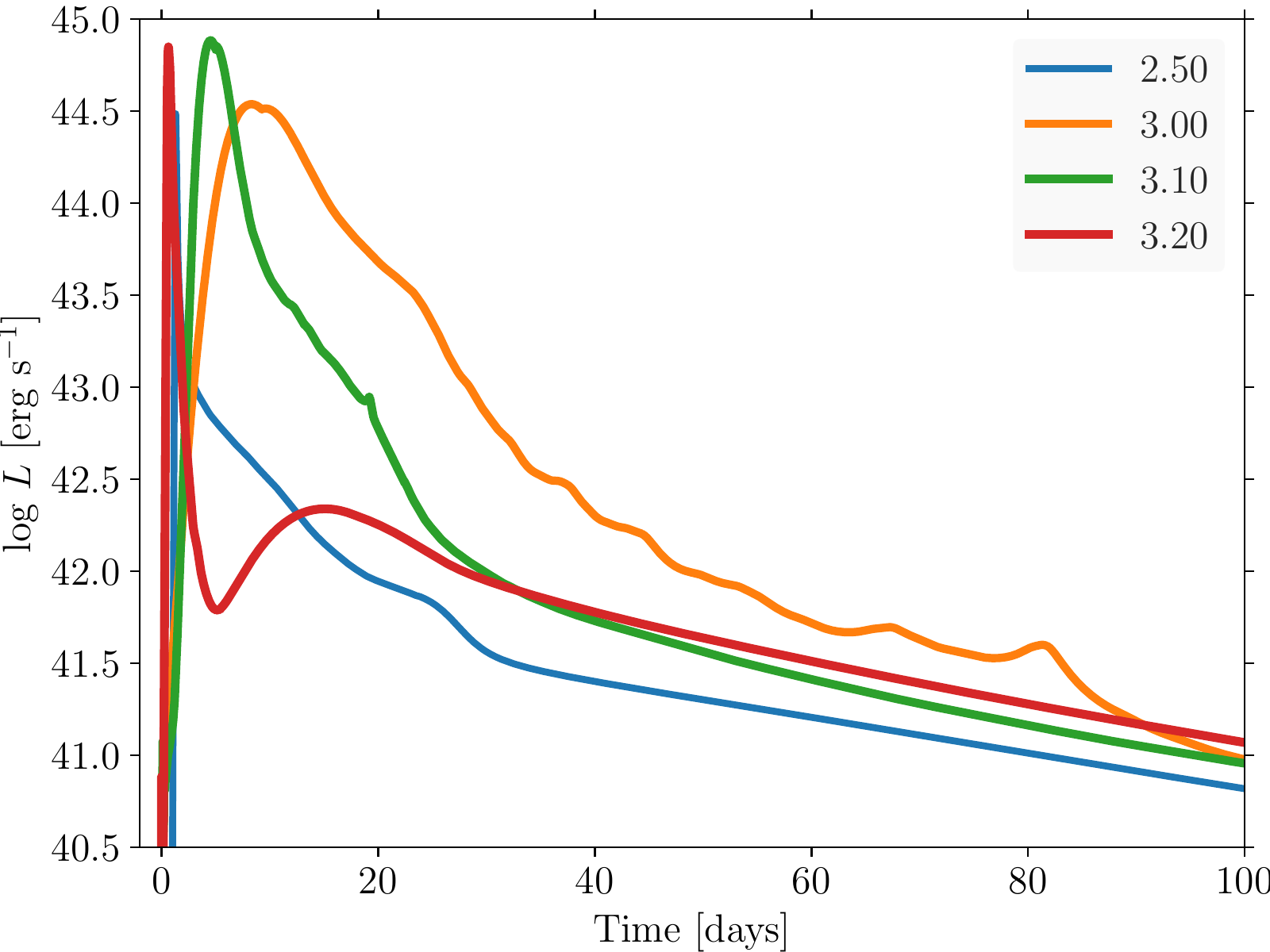}
\caption{Light curves for supernovae that experienced silicon flashes
  accompanied by mass ejection prior to core collapse. The bright
  displays are due to the interaction of matter ejected by core
  collapse with the previously ejected shell. {\em Colors} denote
  $M_{\rm He,i}$. \lFig{siflash}}
\end{figure}

\subsection{Common Type Ib and Ic Supernovae}  
\lSect{Ibc}

For the standard mass loss rate, the most likely candidates for common
Type Ib and Ic supernovae have initial helium-star masses between 3.0
and 8.0\,\Msun. These produce presupernova stars with masses of 2.45 to
5.63\,\Msun. Lower mass stars experience radius expansion, produce
little $^{56}$Ni, and do not look like common events. The sub-interval
3.0 to 3.2\,\Msun \ might make normal Ib and Ic supernovae, but is
complicated by the possibility of strong silicon flashes.  More
massive stars expand too slowly and make broad, faint events
(\Sect{massive}).  Both boundaries, 3.0 and 8.0\,\Msun \ are imprecise
and depend on the assumed mass loss rate.

\begin{deluxetable}{llcccc}
\tablecaption{Average Light Curve Properties}
\tablehead{ \colhead{}                         &
            \colhead{}                         &
            \colhead{$\log\ L_{\rm peak}$}     &
            \colhead{$t_{\rm +1/2}$}           &
            \colhead{$t_{\rm -1/2}$}           &
            \colhead{$\log\ L_{\rm peak} (2016)$}\vspace{1mm}           
            \\
            \colhead{}                     &
            \colhead{}                     &
            \colhead{[$\rm erg\,s^{-1}$]} &
            \colhead{[days]}               &
            \colhead{[days]}               &
            \colhead{[$\rm erg\,s^{-1}$]}
            }\\
\startdata
\multicolumn{6}{c}{overall}\\
\\
\multirow{2}{*}{\rotatebox[origin=c]{90}{W18}} 
& ${\rm median}$ & 42.15 & 14.6 & 8.33 & 41.87 \\
& ${\rm mean}$   & 42.17 & 21.2 & 10.4 & 41.88 \\
\\
\multirow{2}{*}{\rotatebox[origin=c]{90}{S19.8}} 
& ${\rm median}$ & 42.23 & 14.3 & 8.18 & \\
& ${\rm mean}$   & 42.23 & 22.4 & 10.9 & \\
\\
\multicolumn{6}{c}{$8 > M_{\rm He,i} \geq 3$}\\
\\
\multirow{2}{*}{\rotatebox[origin=c]{90}{W18}} 
& ${\rm median}$ & 42.29 & 15.0 & 8.29 & 41.96 \\
& ${\rm mean}$   & 42.25 & 16.9 & 8.74 & 41.94 \\
\\
\multirow{2}{*}{\rotatebox[origin=c]{90}{S19.8}}
& ${\rm median}$ & 42.33 & 14.6 & 8.07 & \\
& ${\rm mean}$   & 42.30 & 16.0 & 8.51 & \\
\\
\multicolumn{6}{c}{$5 > M_{\rm He,i} \geq 3$}\\
\\
\multirow{2}{*}{\rotatebox[origin=c]{90}{W18}} 
& ${\rm median}$ & 42.15 & 14.1 & 7.99 & 41.94 \\
& ${\rm mean}$   & 42.17 & 14.0 & 7.96 & 41.93 \\
\\
\multirow{2}{*}{\rotatebox[origin=c]{90}{S19.8}}
& ${\rm median}$ & 42.22 & 13.4 & 7.93 & \\
& ${\rm mean}$   & 42.21 & 13.8 & 7.86 & \\
\\
\multicolumn{6}{c}{$8 > M_{\rm He,i} \geq 5$}\\
\\
\multirow{2}{*}{\rotatebox[origin=c]{90}{W18}} 
& ${\rm median}$ & 42.37 & 20.2 & 9.85 & 41.98 \\
& ${\rm mean}$   & 42.36 & 21.6 & 10.0 & 41.95 \\
\\
\multirow{2}{*}{\rotatebox[origin=c]{90}{S19.8}} 
& ${\rm median}$ & 42.46 & 18.7 & 9.47 & \\
& ${\rm mean}$   & 42.46 & 19.7 & 9.56 & \\
\\
\multicolumn{6}{c}{$M_{\rm He,i} \geq 8$}\\
\\
\multirow{2}{*}{\rotatebox[origin=c]{90}{W18}} 
& ${\rm median}$ & 42.11 & 44.5 & 16.3 & 41.87 \\
& ${\rm mean}$   & 42.08 & 47.3 & 17.9 & 41.88 \\
\\
\multirow{2}{*}{\rotatebox[origin=c]{90}{S19.8}} 
& ${\rm median}$ & 42.24 & 42.2 & 16.0 & \\
& ${\rm mean}$   & 42.22 & 46.8 & 18.5 & \\
\\
\enddata
\tablecomments{The rise time from 50\% to peak luminosity, $t_{\rm -1/2}$, the decline time from peak to 50\% of peak luminosity, $t_{\rm +1/2}$, and the peak luminosity, $L_{\rm peak}$, are based on models with maximum $^{56}$Ni ($\rm Ni_{max}$ of \Tab{nitbl}).}
\lTab{lcstbl}
\end{deluxetable}

\Fig{sn1bc} shows the expected light curves for all the models in this
mass range that do not experience a strong silicon flash.  The
interval of initial helium core masses between 3.0 and 8.0\,\Msun \ has been further divided into two
subintervals, 3.0 to 5.0\,\Msun \ ($2.5<M_{\rm preSN}<3.8$) and 5.0 to 8.0\,\Msun \ ($3.8<M_{\rm preSN}<5.6$) to highlight
some systematic variations with mass. More realistic calculations of
radiation transport than possible with KEPLER will be needed to tell
if, spectroscopically, these two regions correspond to Type Ib and Ic
supernovae. From \Tab{bigtable} one expects that all the models in
this mass range would be Type Ib since $Y_s$ remains close to 1, but
it may not be necessary to remove all the helium to make a Type
Ic. Similar presupernova models result from higher mass stars with
greater mass loss; \Tab{bigtable} shows that if the mass loss rate is
multiplied by 1.5, stars with presupernova masses over 5\,\Msun \ will
lose their helium.

\Fig{meject} shows the mass ejected by supernovae in approximately
this same mass range. These calculations used the W18 engine.
Including every star that exploded, the median mass ejected was 2.21\,\Msun
\ and the average was 2.97\,\Msun, but for the limited range
shown here and relevant to Type Ib and Ic supernovae, the median and
mean masses were 2.31\,\Msun \ and 2.30\,\Msun.  This does not include
the low mass explosions with radius expansion, or the high mass ones
that produce unusual light curves (\Sect{massive}). These values are
similar to those for observed Type Ib and Ic supernovae
\citep{Dro11,Lym16,Pre19}. In particular, \citet{Pre19} report
observed median and mean ejected masses for Type Ib of 2.0\,\Msun \ and
2.20\,\Msun. For Type Ic, the median is 2.2\,\Msun \ and the mean is
3.2\,\Msun, though the error bars in all cases are large (of
order 1\,\Msun). \Fig{meject} shows that this may reflect an actual 
large spread of masses in nature.

For a typical explosion energy of 10$^{51}$\,erg, our masses imply
average expansion speeds of order 7000\,km\,s$^{-1}$, but substantial
material moves much faster than this average and dominates the
spectrum at early times. The lower panels of \Fig{meject} show the
effective temperature and the speed of the zone at the photosphere at
the time of peak light. Low mass models below 3.5\,\Msun \ have not
been included because of the lingering effects of radius
expansion. The 8.0\,\Msun \ model was also dropped because of an
anomalously low explosion energy ($7 \times 10^{50}$\,erg).
That is not to say these explosions do not exist, just that they are
atypical.  The remaining velocities at peak light lie between 6500\,km\,s$^{-1}$ and
10000\,km\,s$^{-1}$. Speeds above the photosphere are larger and may
affect the spectrum. Generally though, these predictions agree with
what is observed for Type Ib and Ic at peak \citep[e.g., Table 6
  of][]{Lym16}. Photospheric temperatures are between 5000 and 6000\,K,
and that also agrees reasonably well with observations
\citep{Pre19}, though perhaps on the cool side.

Generally speaking, the models also have light curve shapes --rise times
and durations-- similar to observations (\Fig{WLR}; \Tab{lcstbl}), but
are too faint at peak, on the average, by a factor of about 1.5,
even using our most optimistic $^{56}$Ni yields (0.75 times
Ni+Tr+$\alpha$; \Fig{ni56yields}; \Sect{ni56}). \citet{Pre16} give a
median luminosity at peak for common Type Ib and Ic supernovae of
10$^{42.50}$ and 10$^{42.51}$\,erg\,s$^{-1}$,
respectively. \citet{Lym16} give a similar value for Type Ib, but
10$^{42.6}$\,erg\,s$^{-1}$ for Type Ic (our \Fig{sn1bc}; their
Fig.~3). Generally only our brightest, most energetic supernovae
approach these medians. The discrepancy is worse if the $^{56}$Ni
yield is estimated more conservatively using the approach of
\citet{Suk16}. A more recent observational study by \citet{Pre19} gives
substantially reduced values, $10^{42.3 \pm 0.2}$\,erg\,s$^{-1}$ for the
mean of Type Ib supernovae and 10$^{42.3 \pm 0.3}$\,erg\,s$^{-1}$ for
narrow-lined Type Ic. These values are within reach of our more
optimistic (highest energy, greatest $^{56}$Ni mass) models, but are only
for the emission from 4000 to 10000\,\r{A}. A larger value is expected
when the ultraviolet and infrared emission is included (Simon Prentice, private 
communication). All in all though, it seems that neutrino-powered explosions 
as parametrized here, have trouble making enough $^{56}$Ni to explain the light
curves of many of the brighter ordinary Type Ib and Ic supernovae
reported in the literature. A similar conclusion was reached by
\citet{And19}. This flies in the face of current convention which
states that all such supernovae come from binaries, are neutrino
powered, and have radioactive-powered light curves, and warrants some
discussion.

\begin{figure}[h]
  \centering
  \includegraphics[width=\columnwidth]{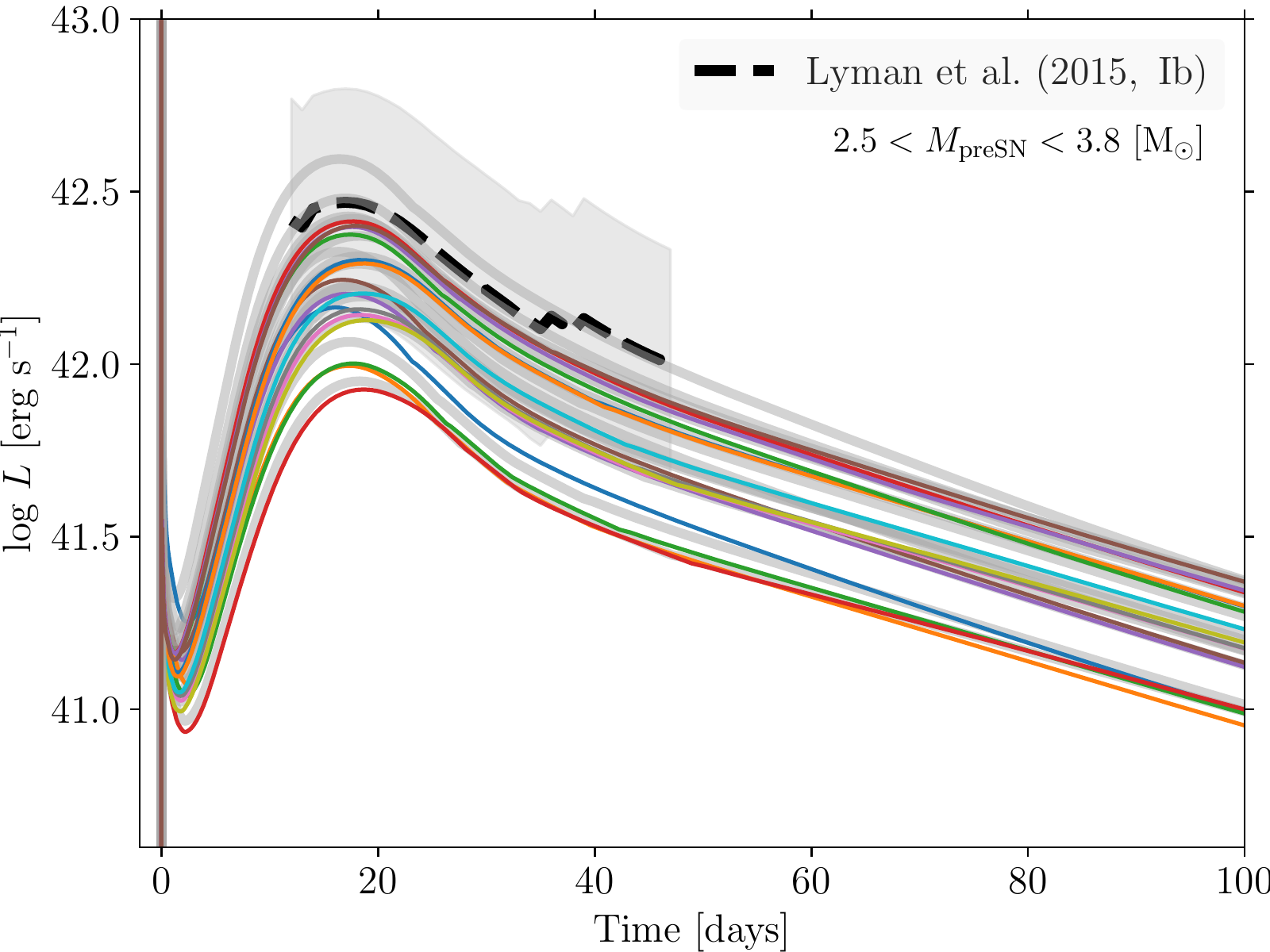}
  \includegraphics[width=\columnwidth]{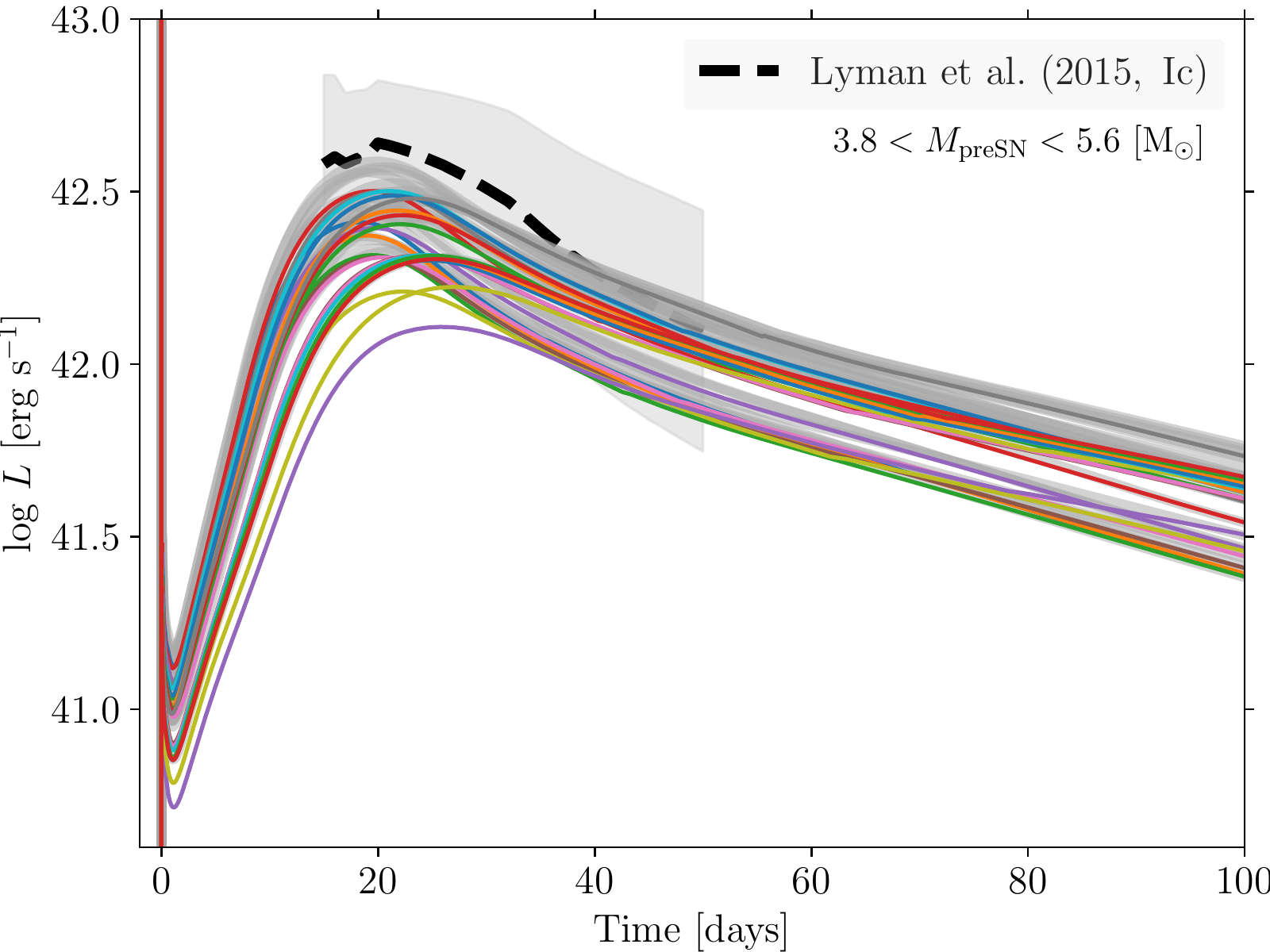}
  \caption{Light curves for Type Ib and Ic supernovae. The mass range
    is broken into two intervals of $M_{\rm He,i}$, 3 to 5\,\Msun \ ({\em upper panel}) and 5
    to 8\,\Msun \ ({\em lower panel}) to show systematics.  \lFig{sn1bc}}
\end{figure}

\begin{figure}[h]
  \centering
  \includegraphics[width=\columnwidth]{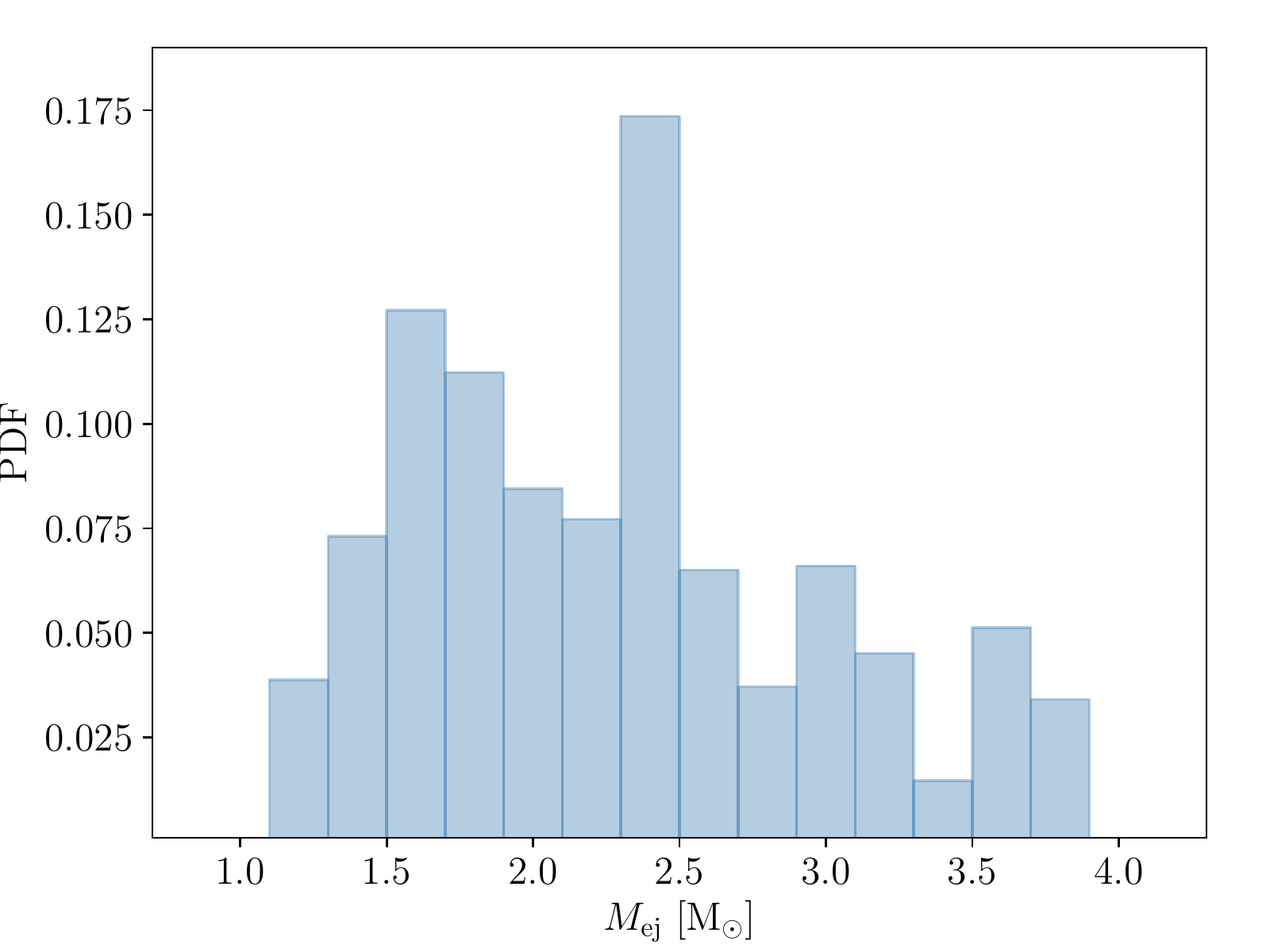}
  \includegraphics[width=\columnwidth]{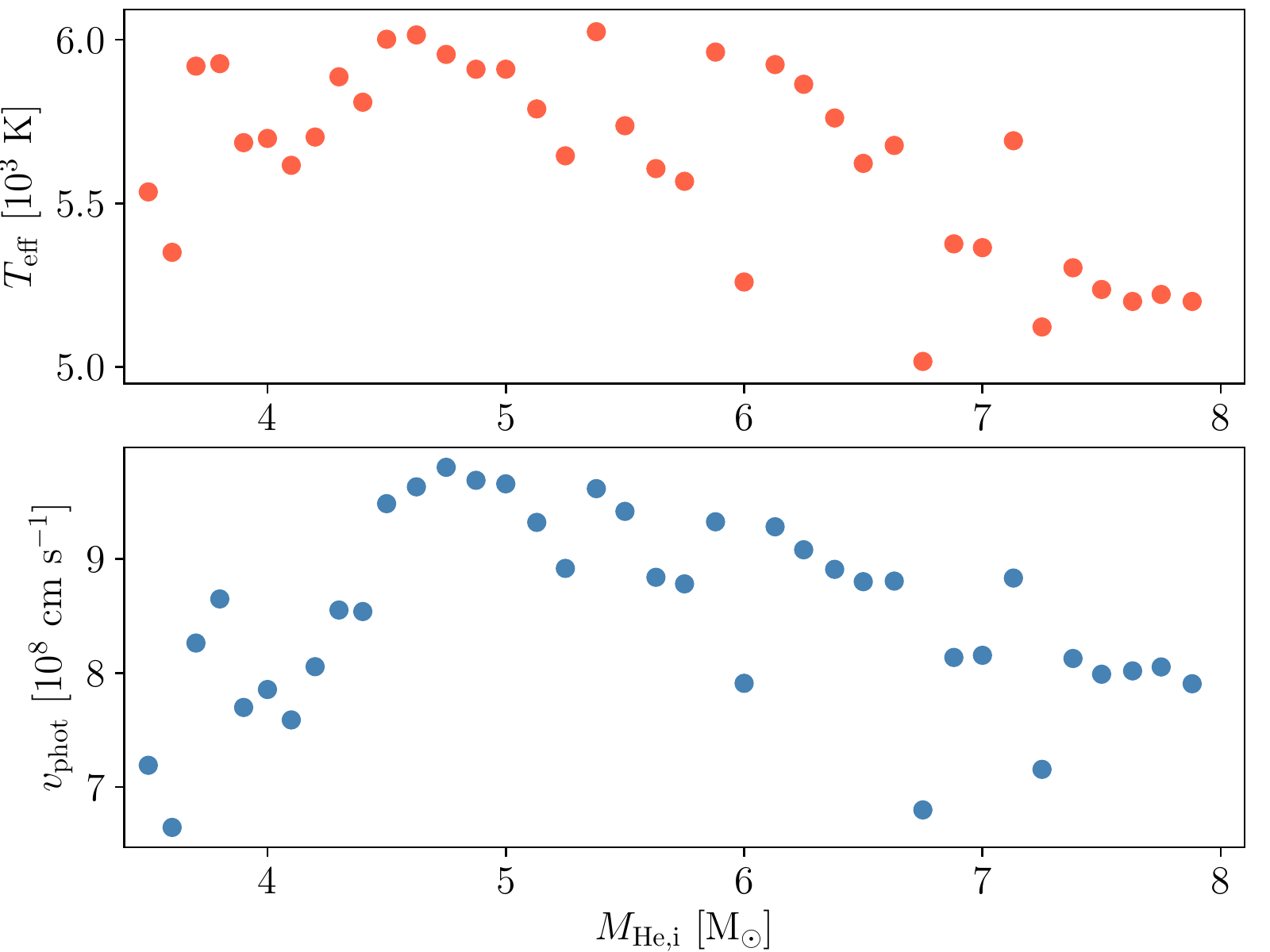}
  \caption{{\em Top:} The mass ejected by Type Ib and Ic supernovae
    weighted by the initial mass function. The initial helium star
    masses are in the range 3.05 to 7.9\,\Msun, so the presupernova
    stars have masses 2.45 to 5.64\,\Msun. The median mass ejected is
    2.31\,\Msun \ and the average is 2.30\,\Msun. Higher presupernova
    mass models that might be Type Ic supernovae eject more mass than those with
    lower presupernova mass. {\em Middle:} The photospheric velocity at
    light curve maximum for this same set of models. {\em Bottom:} their
    effective temperatures. \lFig{meject}}
\end{figure}

\begin{figure}[h]
  \centering
  \includegraphics[width=\columnwidth]{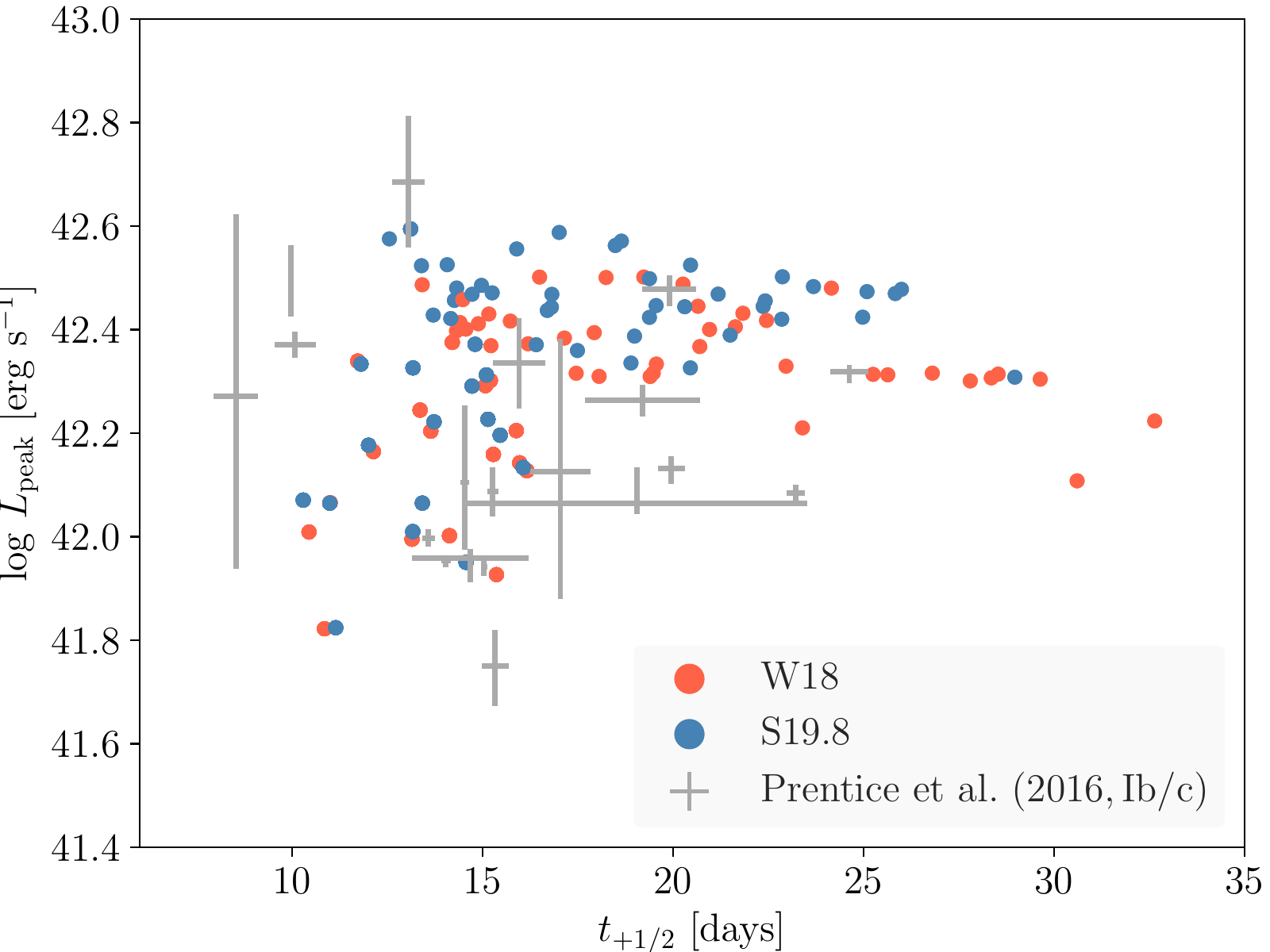}
  \includegraphics[width=\columnwidth]{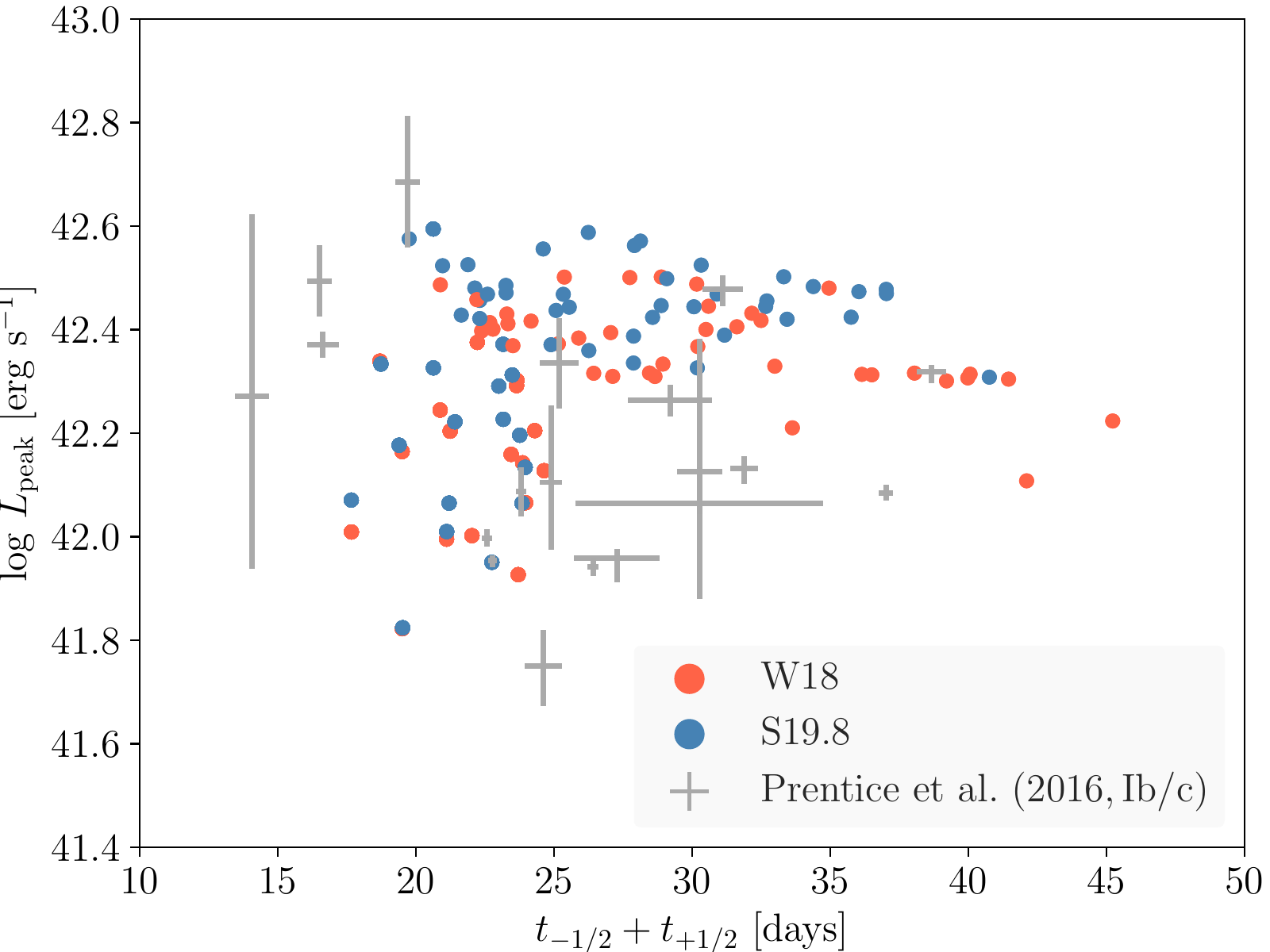}
  \caption{Width-luminosity relations for Type Ib and Ic supernovae
    with initial helium star masses between 3.0 and 8\,\Msun
    \ (presupernova masses 2.45 to 5.64\,\Msun) excluding the models
    with silicon flash before final explosion. {\em Top:} Peak luminosity
    vs time for the light curve to decline to 50\% of its peak
    value. {\em Bottom:} Peak luminosity as a function of the full width
    at half maximum. \lFig{WLR}}
\end{figure}

\begin{figure}[h]
  \centering
  \includegraphics[width=\columnwidth]{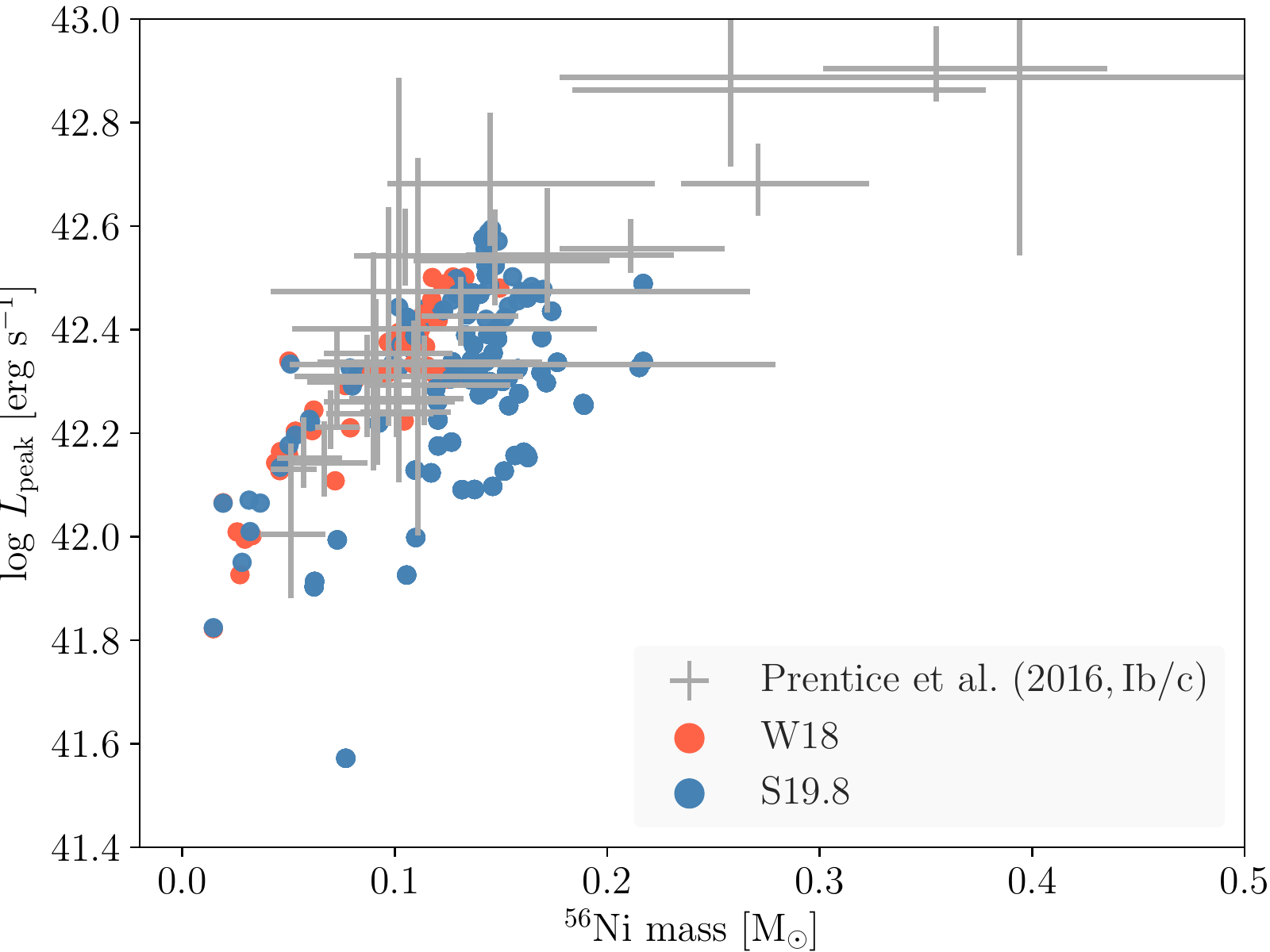}
 \includegraphics[width=\columnwidth]{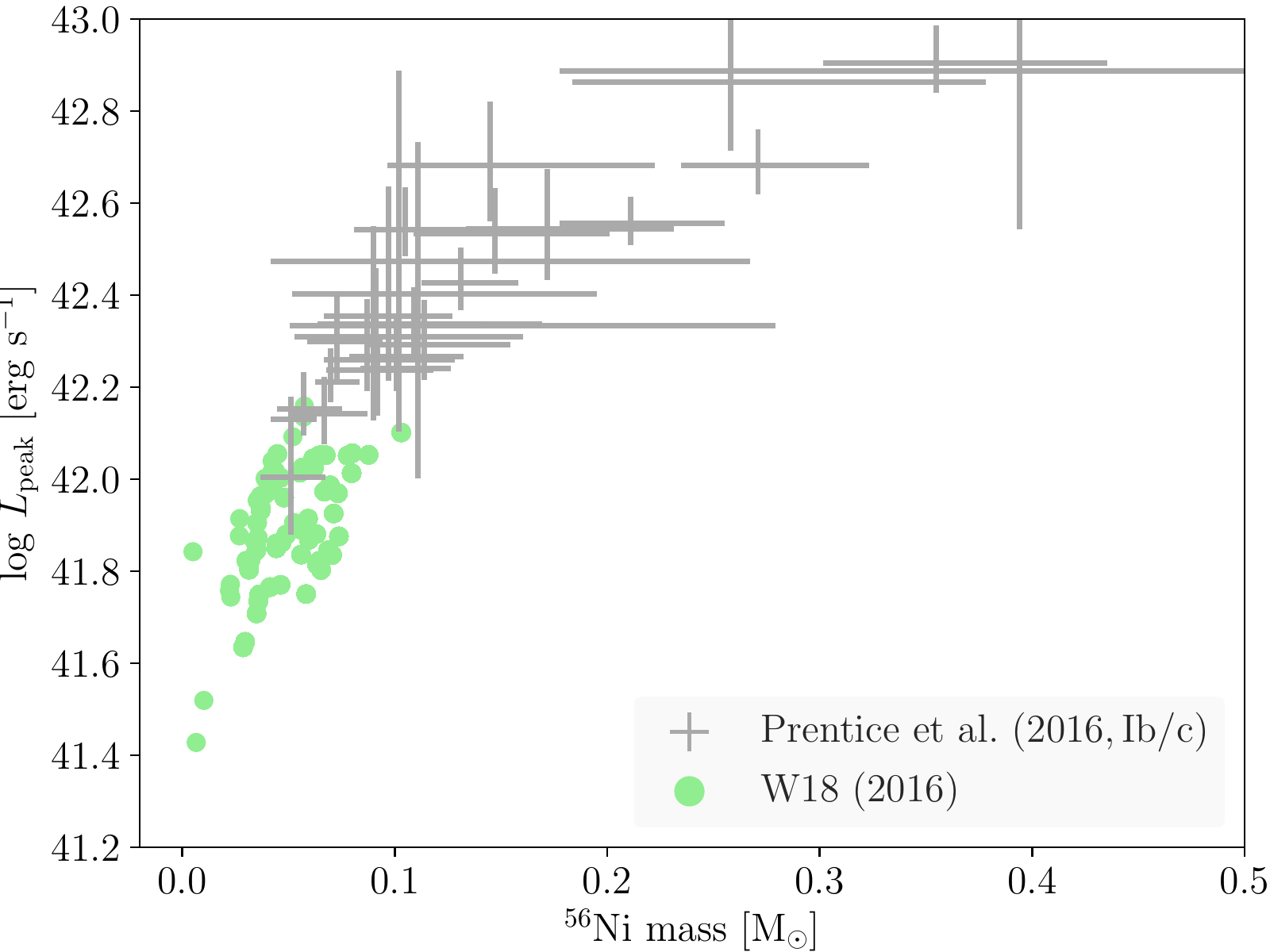}
  \caption{Peak luminosity as a function of $^{56}$Ni mass for the
    present models compared with the bolometric luminosities at peak
    as determined for a large set of normal Type Ib and Ic supernovae by
    \citet{Pre16}. {\em Top:} Using the current analysis approach which
    maximizes the $^{56}$Ni yield (see text), the observations at a given
    $^{56}$Ni mass are typically 1.5 times brighter than the
    models. Blue and red points represent two choices for the central
    engine. S19.8 is more energetic. There are also many very luminous
    supernovae that are observed to have luminosities and inferred
    $^{56}$Ni mass greater than any of our standard models. {\em Bottom:}
    The discrepancy is worse if we analyze the $^{56}$Ni yields using
    the approach of \citet{Suk16}. Here green points are for the W18
    central engine. The discrepancy in averages is now about a factor
    of three. \lFig{prentice}}
\end{figure}

\subsection{Towards Brighter Type Ib and Ic Supernovae}
\lSect{starbrite}

The simplest, traditional way to make a Type I supernova brighter is
to increase its $^{56}$Ni yield. Though numerically convenient, such
variations are actually severely limited by physics and other
observations. Whatever is done for Type Ib and Ic models must
also be done for Type IIp models which have very similar core
structure and explosion characteristics (\Sect{comparison} and
Appendix~\ref{app:comparison}).  There are limits from iron
nucleosynthesis (\Sect{nucleo}), the $^{56}$Ni produced in SN 1987A
(\Sect{nucleo}), the tails of the light curves of Type IIp supernovae 
\citep{And19},  and the masses of neutron stars (\Sect{nstar}) that 
prohibit moving the mass cut in too deep. So other explanations of 
the bright light curves of Type Ib and Ic supernovae are also 
explored here - an embedded magnetar and circumstellar interaction.

\subsection{A Deeper Cut - Some Limits on $^{56}$Ni Production}
\lSect{ni56max}

The required average $^{56}$Ni production for common Type Ib and Ic
supernovae, based upon observations and assuming the luminosity at
peak is derived solely from radioactivity and given by Arnett's Rule
\citep[a questionable assumption;][]{Des15} is
0.14$^{+0.04}_{-0.04}$\,\Msun \ for Type Ib and
$0.16^{+0.03}_{-0.10}$\,\Msun \ for Type Ic according to
\citet{Pre16}. \citet{Lym16} give slightly larger values, $0.17 \pm
0.16$\,\Msun \ for Type Ib and $0.22 \pm 0.16$\,\Msun \ for Type Ic.
\citet{And19} also gives similar values based on a literature survey,
0.163\,\Msun \ for Type Ib and 0.155\,\Msun \ for Type Ic. In mild
tension with these numbers, recent studies by \citet{Pre19} give mean
$^{56}$Ni productions of 0.09 $\pm$ 0.06\,\Msun \ for Type Ib and 0.11
$\pm$ 0.09\,\Msun \ for Type Ic, but as noted previously, these values
are based on observations in a limited wavelength range and are lower
bounds. The larger values for the average
$^{56}$Ni masses \citep[i.e., those other than][]{Pre19} are
appreciably greater than computed here for models in the initial mass
range 3.0 to 8\,\Msun, the ones in our study deemed most likely to make
these sorts of supernovae. As noted in \Sect{nucleo} and \Tab{nitbl},
the average $^{56}$Ni production for the W18 central engine for stars
likely to become Type Ib or Ic supernovae ranges from 0.034\,\Msun,
for the lower mass stars, weaker engine, and conservative approach to
evaluating $^{56}$Ni production, to 0.13\,\Msun \ for the highest
mass, most energetic models, evaluated in the most optimistic way. Our
best estimate would be between Ni and Ni+Tr/2 in \Tab{nitbl} for the W18
and S19.8 central engines which is between 0.034 to 0.097\,\Msun, with
the smaller value appropriate for the lower mass supernovae and weaker
engine.

Since a large fraction of the observations {\sl are} explained by
models within our error bar, it is natural to ask if still larger
values might be possible.  Physically, the $^{56}$Ni yield is
constrained by the amount of matter that attains a peak temperature in
excess of about $5 \times 10^9$\,K and is ejected with a small neutron
excess. This is an upper bound because other iron-group isotopes will
be produced and the fraction of helium from the alpha-rich freeze out
is not small. A much smaller contribution to $^{56}$Ni also comes from
matter experiencing explosive silicon burning between 4 and $5 \times
10^9$\,K. A popular ansatz for the explosion temperature,
$T_\mathrm{exp}$, is given by \citep{Woo02}
\begin{equation}
  \frac{4}{3} \pi r^3 f a T_\mathrm{exp}^4 \approx E_\mathrm{exp}
  \lEq{texplosion}
\end{equation}
where $E_\mathrm{exp}$ is the energy of the explosion, usually taken to be
the final kinetic energy, $r$ is the radius, at the time the supernova
shock begins to move outwards, of the zone that eventually achieves
$T_{\exp}$, $a$ is the radiation constant, and $f$ is a factor that accounts for the presence of
electron-positron pairs. Often the radius of the given mass shell is
evaluated using a presupernova model, assuming that the adjustment
before the shock arrives is negligible and $f$ is taken to be 1.

In the present situation, even an approximate result requires some
improvements.  First, the abundance of pairs is not negligible. For
the temperatures of relevance, $4 \times 10^9$\,K and higher, $f
\approx 11/4$, similar to its value in the early Big Bang
\citep{Pee93}. This has been verified for our models using the KEPLER
equation of state. Second, the settling of mass shells located only a
few thousand km out as the explosion develops cannot be neglected. The
gravitational acceleration at 4000\,km is about 10$^9$\,cm\,s$^{-2}$ and
the time from the presupernova stage to the launch of the shock is
typically 0.5 to 1 s.  Even though the layers are not far from
hydrostatic equilibrium, a contraction of 35\% is typical. Finally,
the full energy of the explosion may not have been developed when the
shock temperature declines to $5 \times 10^9$\,K. Since the explosion
energy only enters as $E^{1/3}$ and most of the explosion energy
usually {\sl has} been deposited, this last factor is not critical.

Still, appropriately corrected, \eq{texplosion} works very well.  For
an explosion energy of $1.2 \times 10^{51}$\,erg and $f = 2.75$, the
equation says that a temperature greater than $5 \times 10^9$\,K is
achieved inside a radius of 2800\,km.  For $f = 1$, the boundary radius
is increased to 3900\,km. Examining the runs themselves, 90\% of
$^{56}$Ni is typically synthesized inside a radius that, if the
structure is evaluated when the shock is launched, has a value of 2970
$\pm$ 9\%\,km.  The same 90\% criterion gives a radius of 4470 $\pm$5\%\,km
using radii in the presupernova star. Of course it is not realistic
to neglect pairs, but it turns out that the correction to the radius
for pairs, $(11/4)^{1/3}$ $\approx$ 1.40 very nearly cancels the
contraction factor, 4470/2970 $\approx$ 1.50. Thus \eq{texplosion} can
be used approximately with either $f = 1$ and $r$ the presupernova
radius or $f = 2.75$ and $r$ the radius in the star when the shock is
launched. This Lagrangian mass shell in the presupernova star where
the shock temperature declines below $5 \times 10^9$\,K, the value
required to achieve NSE on a hydrodynamic time scale, will be referred
to as $M_{\rm T9=5}$.

It is important to note that $M_{\rm T9=5}$ is mostly a function of
presupernova structure. It depends only weakly on explosion energy and
not on the mass cut. The mass of the ejecta that is heated above $5
\times 10^9$\,K is then just $M_{\rm T9=5} - M_{\rm rem}$ where $M_{\rm
  rem}$ is the baryonic mass of the gravitationally bound
remnant. \Fig{niyield} shows the good agreement of this estimate with
the mass of ejected material that actually achieved NSE in those
models likely to make Type Ib or Ic supernovae. The agreement suggests
that the $^{56}$Ni yield is not sensitive to details of how the
explosion is calculated, but mostly to the final mass cut and,
secondarily, the explosion time.

This semi-analytic model allows a qualitative exploration of how
$^{56}$Ni synthesis might change as the explosion energy and mass cut
are varied for a given presupernova star. \Fig{maxyield} shows the
results if the explosion energy is varied between 1 and $5 \times
10^{51}$\,erg, or if the mass cut is moved in to the edge of the iron
core. These are extreme excursions. None of our models had explosion
energy exceeding $2.0 \times 10^{51}$\,erg, even with the energetic
S19.8 engine and most had less (\Fig{summaryw18-sta-exp}). No modern
calculation of a purely neutrino-powered model that we know of has
greatly exceeded that \citep[e.g.][]{Fry99,Fry01,Sch06,Bru16,Mue17}. Similarly the mass cut in
neutrino-powered models rarely, if ever, occurs as deep as the edge of
the iron core. The reduction in remnant mass might also cause an
unacceptable decrease in the mean neutron star mass (\Sect{nstar}) and
could make the contribution of Type Ia supernovae to iron synthesis a
minor component. Nevertheless, neglecting these constraints, the synthesis
of iron group plus alpha's might be pushed to $\sim 0.20$\,\Msun. For
reasonable explosion energies, the placement of the mass cut has more
leverage on the yield than the explosion energy. The maximum $^{56}$Ni
mass, even in models that have unreasonable mass cuts, is
0.33\,\Msun. Some past studies, e.g. \citet{Des12} and \citet{Des15}
arbitrarily placed the mass cut at the edge of the iron core and thus
overestimated the $^{56}$Ni yield for purely neutrino-powered
explosions.

In summary, larger $^{56}$Ni yields capable of resolving the
discrepancy between the models and the observed brightness of Type Ib
and Ic supernovae might conceivably be possible, and certainly can be prepared in
parametrized explosions, but they are not achievable in today's
neutrino-powered models. The yields may violate limits on iron
nucleosynthesis, and the neutron stars may be too small compared to
observations. The same approximations applied to SN~1987A would overproduce
$^{56}$Ni there. An additional source of energy seems to be required to explain
supernovae brighter than about 10$^{42.5}$\,erg\,s$^{-1}$.

\begin{figure}
	\centering
  \includegraphics[width=0.48\textwidth]{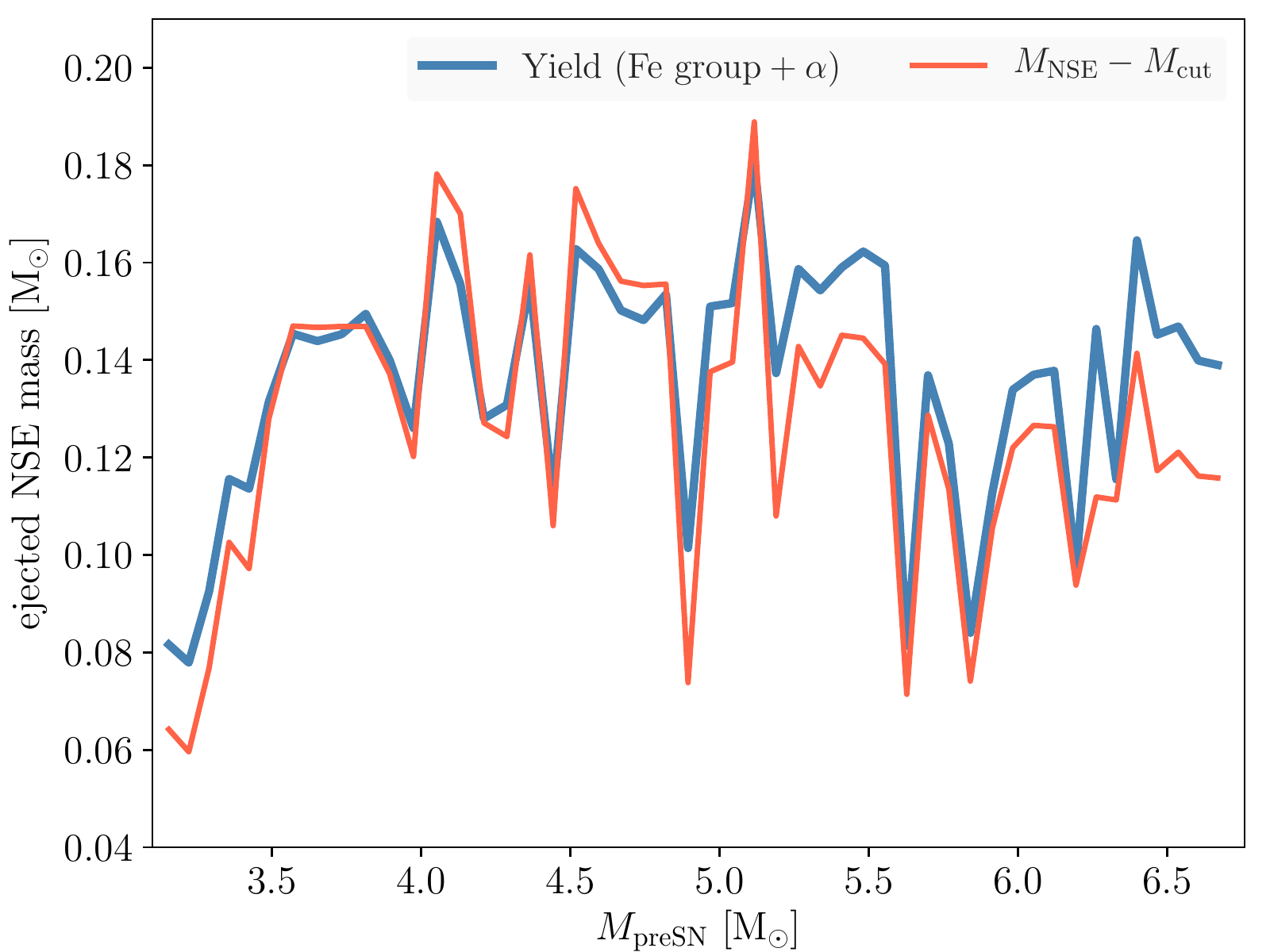}
	\caption{The estimated mass of material that attains NSE and
          is ejected is compared to the actual ejected mass of
          iron-group nuclei plus helium from photodisintegration in
          the models based on the S19.8 central engine.  The estimated
          mass is the difference between the mass, $M_{\rm T9=5}$, where the
          shock temperature falls below $5 \times 10^9$\,K, and the
          baryonic remnant mass when the explosion is over.  Here the
          radius used for estimating the shock temperature was
          calculated using \eq{texplosion} with $f = 2.75$ and the radii
          in the model when the shock was launched. A very similar
          plot would result if one used $f = 1$ in \eq{texplosion} and
          the presupernova zonal radii. Final helium presupernova
          masses corresponding to these initial masses are given in
          Table~4 of \citet{Woo19} and \Sect{physandmods}.
          \lFig{niyield}}
\end{figure}

\begin{figure}
	\centering
\includegraphics[width=0.48\textwidth]{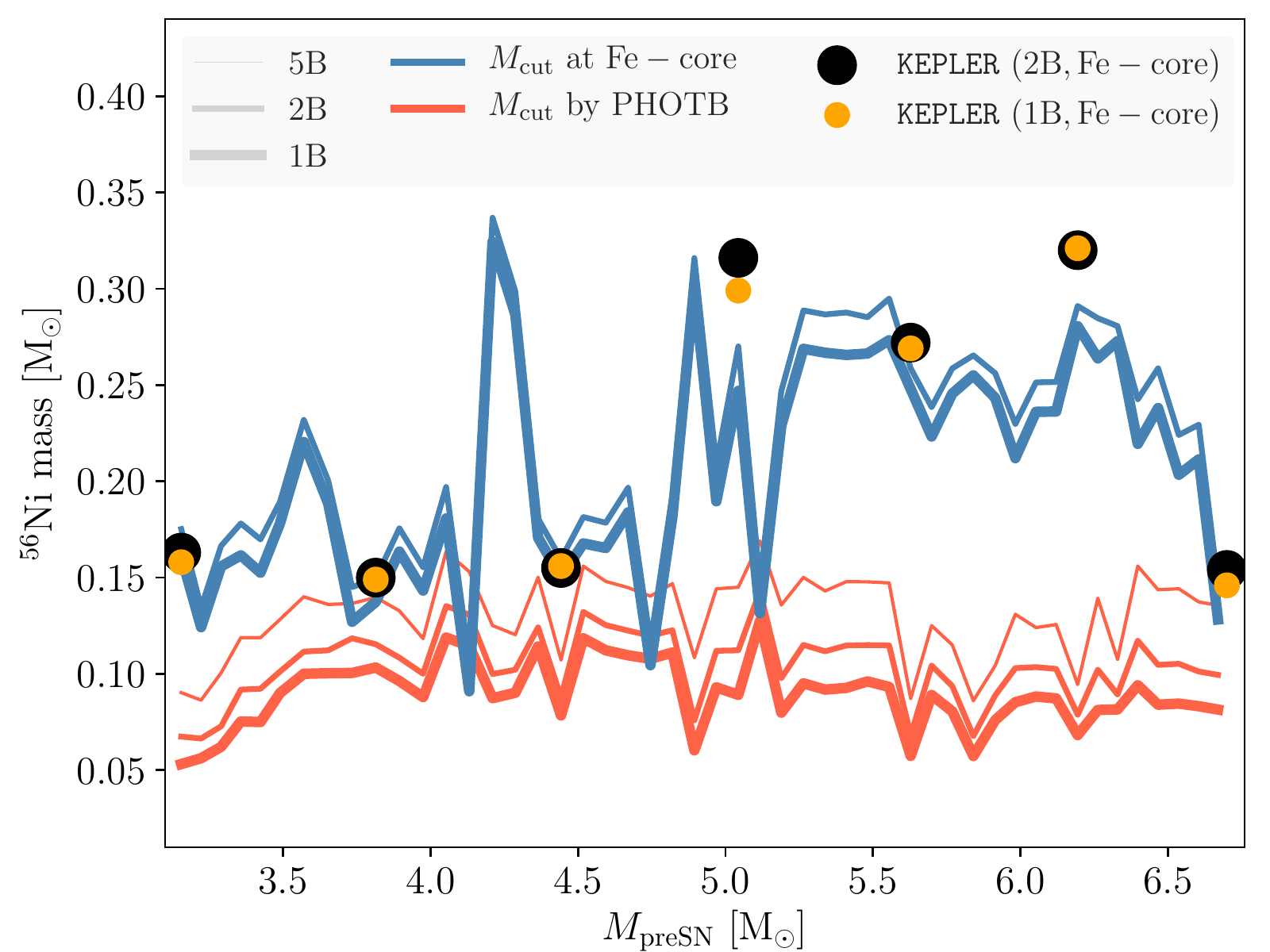}
\caption{Maximum yields for $^{56}$Ni. The lower (red) lines explore
  the effect of varying the explosion energy from 1 to $5 \times
  10^{51}$\,erg. Values larger than $2 \times 10^{51}$\,erg are
  incompatible with neutrino-powered models using known physics. As in
  \Fig{niyield} the ejected mass is estimated by subtracting $M_{\rm
    rem}$ from $M_{\rm T9=5}$. The latter is calculated using the
  stars density structure at bounce and \eq{texplosion} with $f =
  2.75$. The upper (blue) lines show that greater variation is
  obtained by artificially reducing the remnant mass, here taken to be
  the iron core. Black and orange circles represent actual
  KEPLER models that used the iron core mass for the location of the
  piston (which became the mass cut). In both cases the mass that
  achieves NSE and is ejected has been multiplied by an efficiency
  factor, 0.75, to give $^{56}$Ni synthesis.  Potentially large
  $^{56}$Ni yields, up to 0.33\,\Msun, are numerically possible, but
  are incompatible with neutrino-powered explosions and may give
  results that are inconsistent with constraints from stellar
  nucleosynthesis and neutron star masses.  \lFig{maxyield}}
\end{figure}

\subsubsection{Circumstellar Interaction}
\lSect{csm}

The brightness of the supernova could be amplified by colliding with a
circumstellar gas, presumably the outcome of mass loss during the
final year of the star's life \citep[e.g.][]{Shi14,Woo17}. On the
positive side, ample energy exists in the outer edge of the ejecta and
the required circumstellar mass (CSM) is not large. In a typical Ib
progenitor, e.g. Model He6.0, the outer 0.001\,\Msun \ moves at speeds in
excess of 0.1\,c implying a kinetic energy in this small mass of
$\sim10^{49}$\,erg, more than enough to explain the full light curve of
a typical Type Ib supernova. This would only need to encounter
$\sim$0.001\,\Msun \ of CSM matter inside a radius of a few times
10$^{15}$\,cm.

On the other hand getting the time structure of the light curve is
problematic and requires the introduction of many uncertain parameters
\citep{Cha13,Vre17}. The spectrum from such high velocity material
might not resemble common Type Ib and Ic supernovae.  In the
simplest case of a steady wind in which $\rho \propto r^{-2}$, the
luminosity
\begin{equation}
L_{\rm csm} \approx 0.5 \dot M  v_{\rm shock}^3/v_{\rm wind}
\end{equation}
where $\dot M$ is the mass loss rate, $v_{\rm wind}$, the wind speed
and $v_{\rm shock}$, the speed of the fast moving supernova
ejecta. For $\dot M$ = $2 \times 10^{-4}$\,\Msun \ yr$^{-1}$, $v_{\rm
  wind}$ = 500\,km\,s$^{-1}$, and $v_{\rm shock}$ = 30,000\,km\,s$^{-1}$,
the luminosity is $3 \times 10^{42}$\,erg\,s$^{-1}$. This luminosity
would persist and slowly decline until the fast moving ejecta
encountered about half its mass, or $\sim$20 days. But the light
curve would commence immediately and decline slowly, quite unlike what
is observed.

Assuming a CSM shell with constant density only improves things
marginally. Now the luminosity
  \begin{align}
    L_{\rm csm} & \approx 2 \pi r^2 \rho v_{\rm shock}^3 \\
    & \approx 2 \pi \rho v_{\rm shock}^5 t^2 \\
    & \approx 4 \times 10^{42} \left( \frac{v_{\rm shock}}{3 \times
      10^9 }\right)^5 \left(\frac{\rho}{10^{-17}
      }\right) \ {\rm erg\,s^{-1}}
    \end{align}
increases quadratically with time and declines once the edge of the
shell is reached. A density structure, $\rho(r)$, could be found to
match any given light curve \citep{Cha13}, but its construction seems
artificial. Given the lack of narrow CSM lines in the spectrum of
common Type Ib supernovae, CSM interaction seems an unlikely estimate
for their luminosity excess.

\subsubsection{A Magnetar}
\lSect{magnetar}

A significant fraction, substantially more than 10\%, of all
neutron stars are born with magnetic fields 0.3 to $1 \times 10^{15}$\,G
\citep{Ben19}. This is roughly consistent with the fraction of
core-collapse supernovae that are of Type Ib and Ic. It is thus
possible that a substantial fraction of this class of supernovae make
magnetars. In one of the few studies of Type Ib progenitors that
included rotation, \citet{Yoo10} found periods of $\sim$10--20\,ms,
rapid enough for rotational energy to affect the light curve of a
supernova, though not its explosion energy. Loss of the hydrogen
envelope diminishes the rotational braking the helium core would have
experienced in a red giant, though angular momentum is still lost to
winds. The explosion would still be neutrino powered as calculated
here, but the light curve dominated by the magnetar.

As an example, consider Model He6.0. The unmodified version of this
supernova, using the S19.8 central engine, exploded with a kinetic
energy of $1.2 \times 10^{51}$\,erg and produced 0.083\,\Msun \ of
$^{56}$Ni. \Fig{magnetar} shows the effect on the light curve of
embedding a magnetar with field strength 5, 7 and 10 $\times
10^{14}$\,G and initial rotational energy 2.5, 3.0, and 3.5 $\times
10^{49}$\,erg. The prescription for energy deposition here was the
same as \citet{Woo10}, which is very similar to \citet{Kas10}, but uses
slightly different field strengths to achieve the same luminosity.
Energy from radioactive decay was included in the three magnetar
models.

The corresponding initial rotational periods are 28, 26, and 24\,ms.
Other parameters could be chosen to give different peak
luminosities and slopes on the tail but these three peaked near
10$^{42.7}$\,erg\,s$^{-1}$, similar to a brighter than average Type Ib
or Type Ic supernova \citep{Pre16}. This compares with
10$^{42.21}$\,erg\,s$^{-1}$ for the same model with no magnetar. The
rise time from 50\% to peak luminosity ($t_{-1/2}$) for the three models varied from 10.4 to 12.4
days with the models with greater rotational energy but weaker
magnetic field rising more slowly. The decline time from peak to 50\% of peak luminosity ($t_{+1/2}$) were
19.0 to 28.5 days for the same models. For the model without the
magnetar, the rise and decline timescales were 10.4 and 23.4 days. The
ejected mass here was 2.816\,\Msun, the neutron star remnant was 1.417\,\Msun
\ (gravitational). The radioactive contribution came from the
decay of 0.084\,\Msun \ of $^{56}$Ni.

\begin{figure}
	\centering
\includegraphics[width=0.48\textwidth]{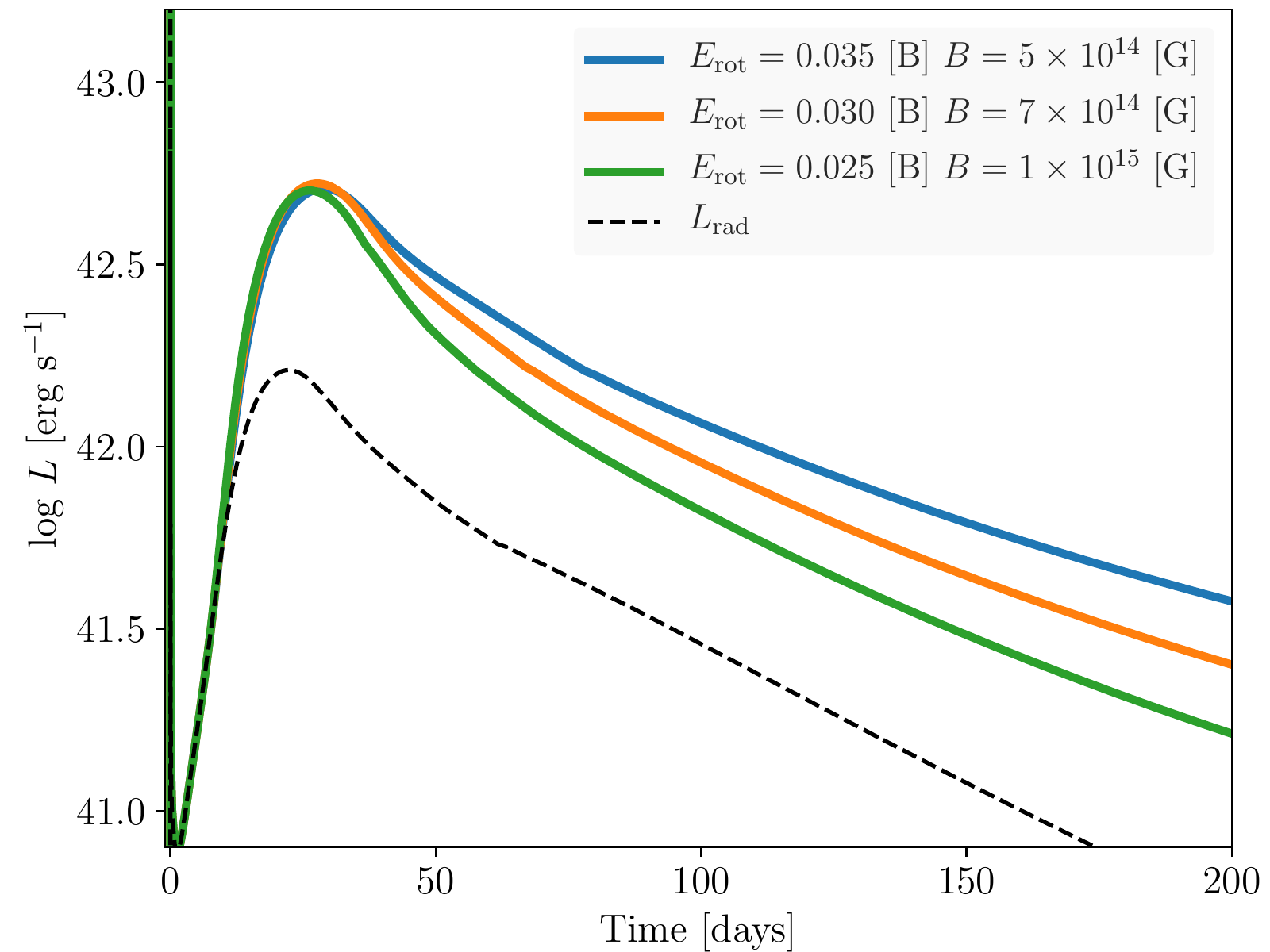}
	\caption{Light curve of Model He6.0 with and without a
          contribution from an embedded magnetar. The dashed black
          line shows the light curve of the standard model powered
          only by $^{56}$Ni decay. The orange, blue and green curves show
          the modification caused by a magnetar with initial
          rotational energy and magnetic field strength of $2.5 \times
          10^{49}$ erg and $1 \times 10^{15}$\,G (blue); $3.0 \times
          10^{49}$ erg and $7 \times 10^{14}$\,G (green); and $3.5 \times
          10^{49}$ erg and $5 \times 10^{14}$\,G (orange)
          respectively. \lFig{magnetar}}
\end{figure}

The idea that many ``normal'' Type Ib and Ic supernovae might not be
powered at peak by radioactive decay is novel and would have important
implications for the role of magnetars in many diverse kinds of
supernovae. Since they are much more common than superluminous
supernovae and a more regular class of events, they might be a good
laboratory for the study of magnetar birth. After 100 years the
magnetars in \Fig{magnetar} would have periods of 1.26, 0.88, and 0.63\,s
respectively and luminosities of $4 \times 10^{36}$, $8 \times
10^{36}$, and $1.6 \times 10^{37}$\,erg\,s$^{-1}$, assuming the field
strength remains constant.

The tails of the light curves of Type Ib and Ic supernovae might be
studied carefully to ascertain $^{56}$Ni abundances, but this could be
complicated by the fact that gamma-rays are already escaping shortly
after peak in the radioactive model. For some field strengths, the
tail of magnetar models also declines at a similar rate as the
radioactive model (\Fig{magnetar}).  On the other hand, light curves
declining slower than the 77.3 day half life of $^{56}$Co, i.e., full
gamma-ray trapping, might be indicative of a magnetar energy source.
If the iron abundance could be accurately inferred from late time
spectra, the magnetar model predicts a much smaller abundance than
needed to explain the brightness of the supernova. 

\subsection{Type Ic Supernovae from More Massive Stars}  
\lSect{massive}

Helium stars with presupernova masses greater than 6\,\Msun \ require
extraordinary explosion energies or gamma-ray escape to avoid
producing light curves that are too faint and too broad to be common
Type Ib and Ic supernovae \citep{Ens88}, yet our present formalism
predicts many such explosions (\Tab{expltbl},
\Fig{sta-explosions}). This is particularly true when the more
energetic S19.8 central engine is employed. \Fig{sn1c} shows a
selection of light curves.  Though some are as bright as their lower mass
counterparts in \Fig{sn1bc}, the curves all peak appreciably later and
decline much more slowly. The decline time to 50\% peak luminosity
($t_{+1/2}$) takes more than 40 days for the lightest models in this
group and longer for the more massive ones (\Tab{expltbl}). The
three broadest, faintest light curves in \Fig{sn1c} came from helium
stars with initial masses 19.25, 19.50 and 19.75\,\Msun \ which had
presupernova masses from 10.07 to 10.29\,\Msun. These supernovae took two
months to decline to half their peak luminosity. Their effective
temperatures at peak were near 3500\,K, so they would be quite
red. Since they lost most of their helium prior to dying, these
supernovae would probably be Type Ic.

We are not aware of any observed supernovae with these
characteristics.  Perhaps they have been missed since they are faint,
slow, and red, or these massive progenitors, essentially all the
stripped stars with presupernova masses greater than about 6\,\Msun,
collapsed to black holes. This is an issue that gravitational
radiation studies might some day address.  Are there indications in
the black hole birth function of cores with masses greater than 6\,\Msun
\ exploding and not leaving black hole remnants?

Or our models may be deficient. It is hard to see how anything other
than a major alteration of the neutrino-powered,
radioactive-illuminated paradigm could greatly increase the
luminosities or decrease the light curves' widths for such massive
progenitors.  The alterations again include circumstellar interaction
or extra energy from a rotationally powered central engine.

\begin{figure}[h]
  \centering
  \includegraphics[width=\columnwidth]{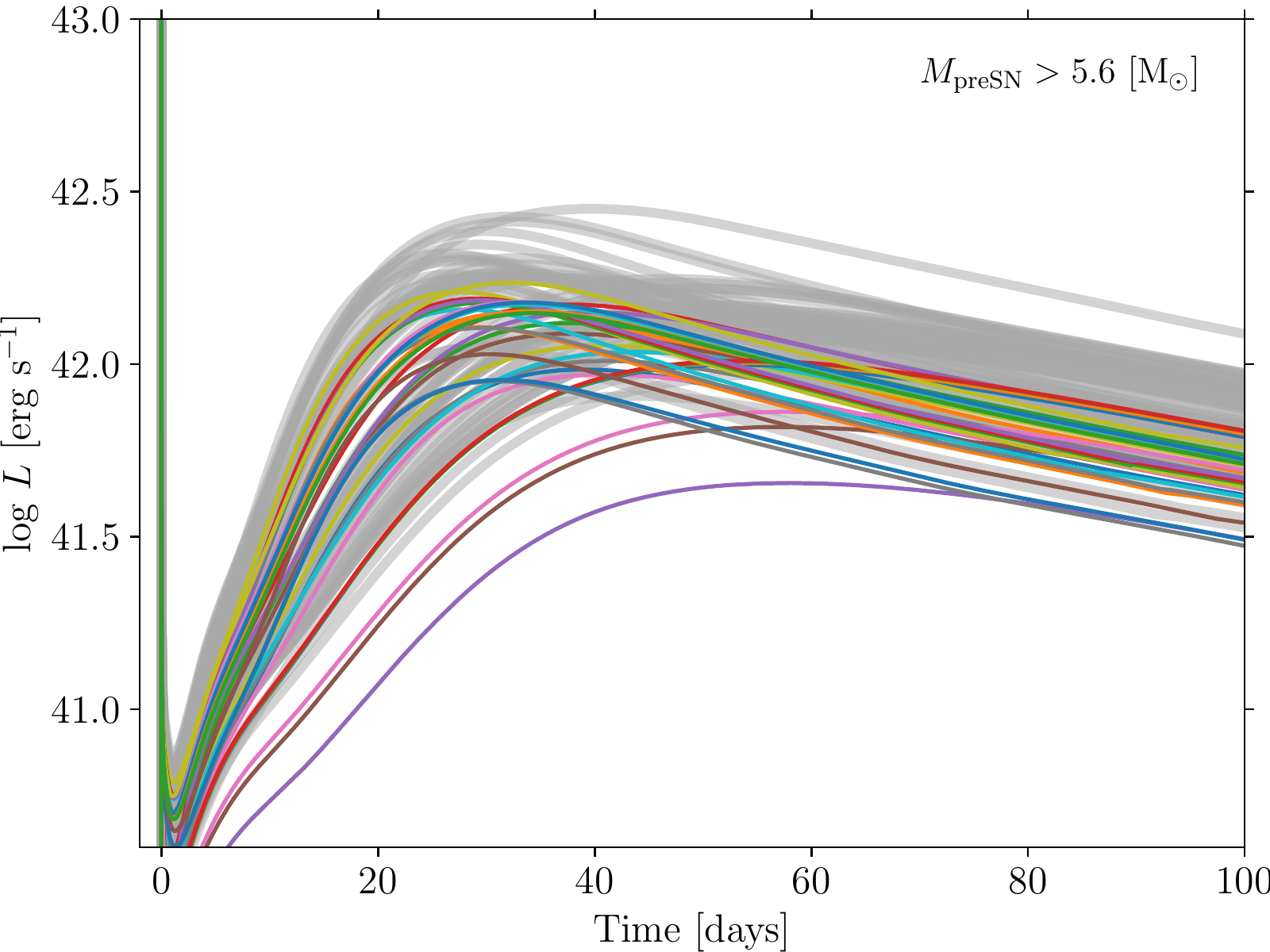}
  \caption{Light curves for the more massive models.  The masses
    displayed have initial helium-star masses greater than 8.0\,\Msun
    \ and final presupernova masses greater 5.64\,\Msun. They
    correspond to stars with ZAMS masses greater than 30\,\Msun. These
    would be Type Ic, but, if powered only by neutrinos and illuminated
    only by radioactive decay, have longer, fainter light curves than
    common Type Ic supernovae. Colored lines indicate models powered
    by the W18 central engine, whereas gray lines come from the more
    energetic S19.8 set which makes more $^{56}$Ni. \lFig{sn1c}}
\end{figure}

In favor of a magnetar explanation is the fact that, faint or not,
many of these massive models do explode (\Tab{expltbl}). A successful
neutrino-powered explosion during the first second after the iron core
collapses facilitates the subsequent formation and operation of a
magnetar by evacuating a region around the proto-neutron star and
allowing it time to cool, contract, and spin up. Since the initial
explosion makes only $\lesssim0.15$\,\Msun \ of $^{56}$Ni and the
magnetar would make none, the luminosity both at peak and, usually, on
the tail would not be due to radioactivity \citep[see e.g.][]{Woo10},
but the iron lines might be strong in the remnant. While the
calculation of rotating presupernova models is deferred, we note the
tendency of more massive stars, especially those that lose their
envelopes early on, to produce rapidly rotating neutron stars
\citep{Heg05}. Most of the angular momentum loss in the core occurs,
in single stars, during helium burning when the star's hydrogen
envelope expands to become a giant. The essentially stationary
envelope saps angular momentum from the rapidly rotating core, so
stars stripped of their envelopes, as these are, end up rotating more
rapidly at death.

There are two sorts of magnetar models -- those where the magnetar's
initial rotational energy is small compared with the kinetic energy of
the neutrino-powered explosion, and those where it is not. In the
former case the magnetar just illuminates a coasting configuration
whose dynamics were determined by the neutrinos.  These are the
magnetars that were studied in \Sect{magnetar}.  In the latter case,
energy input by the magnetar appreciably modifies the velocity profile
of the supernova. These two cases are illustrated with Model
He10.00 (\Fig{IcBL}) and He19.75 (\Fig{bigmagnetar}). Both
used the W18 explosion model (\Tab{expltbl}) for their initial
explosions. Model He10.00, which had a presupernova mass of
6.74\,\Msun, is a bit heavier than usually deemed responsible for
common Type Ibc supernovae. Model He19.75, with a presupernova mass of
10.3\,\Msun, was in the island of high mass explodability in
\Fig{massmap}, and was one of the heaviest models to explode in the
current study.

\begin{figure}[h]
  \centering
  \includegraphics[width=\columnwidth]{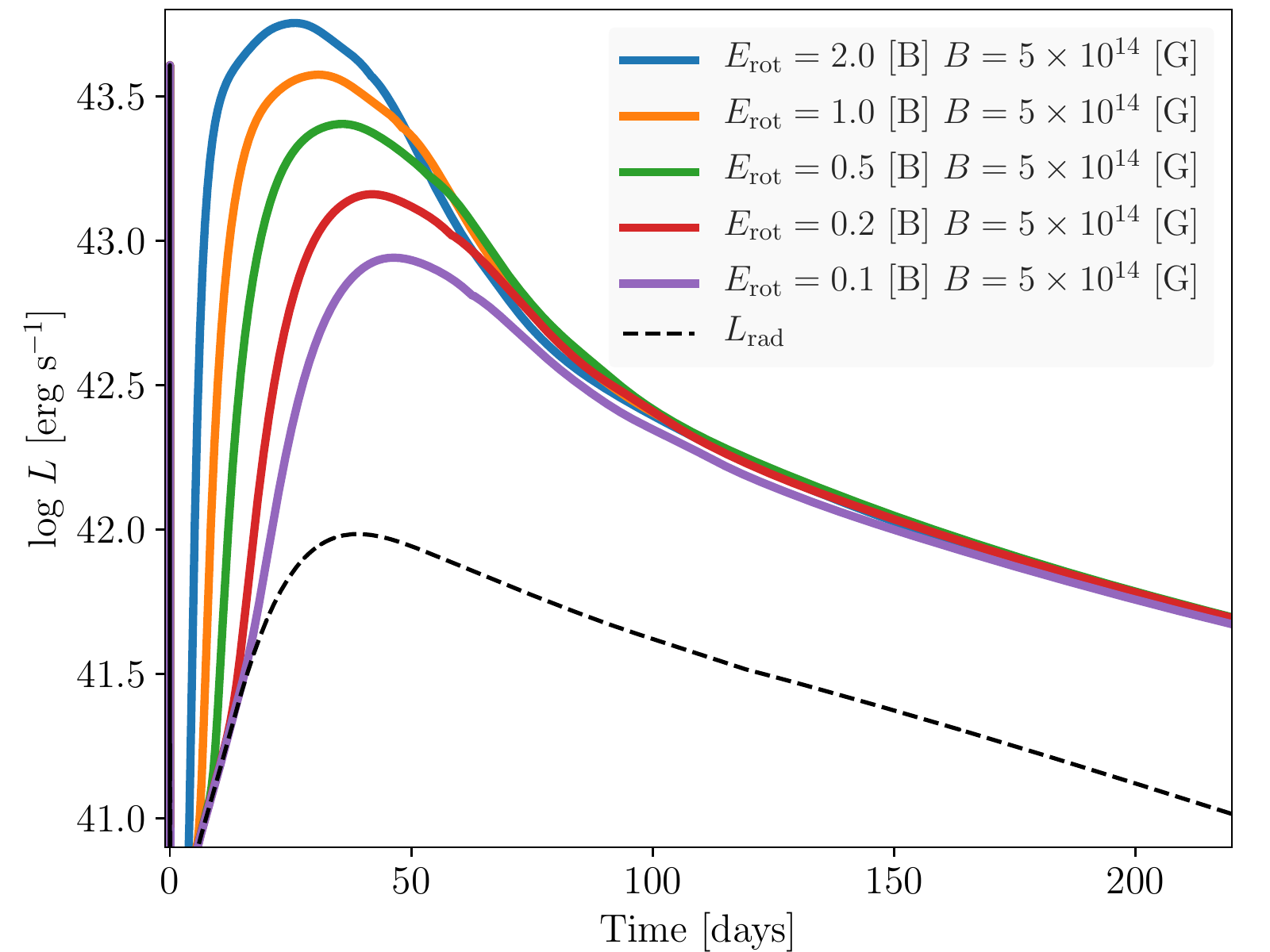}
  \includegraphics[width=\columnwidth]{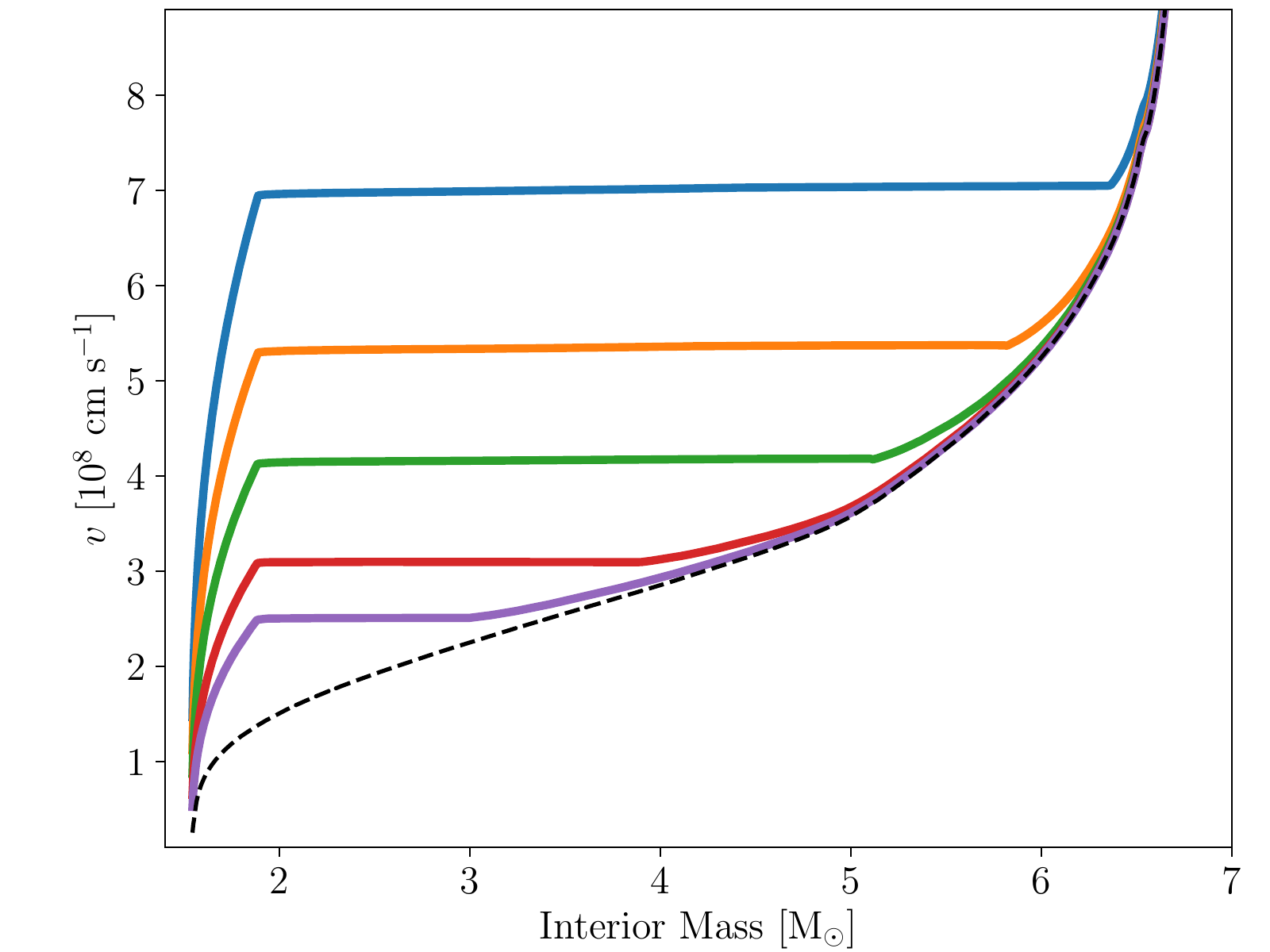}
  \caption{{\em Top:} Light curves for Model He10.00 with
    and without an embedded magnetar. Because all calculations used
    the same magnetic field, the light curves converge to similar
    values at late times. {\em Bottom:} Terminal velocities for the
    models in the upper panel. Regions of constant velocity reflect
    the presence of a dense thin shell containing all that mass. In a
    multi-dimensional simulation, these shells would be spread over a
    larger radius and some of the magnetar energized matter would
    penetrate into the overlying, faster moving material. \lFig{IcBL}}
\end{figure}

\begin{figure}[h]
  \centering
  \includegraphics[width=\columnwidth]{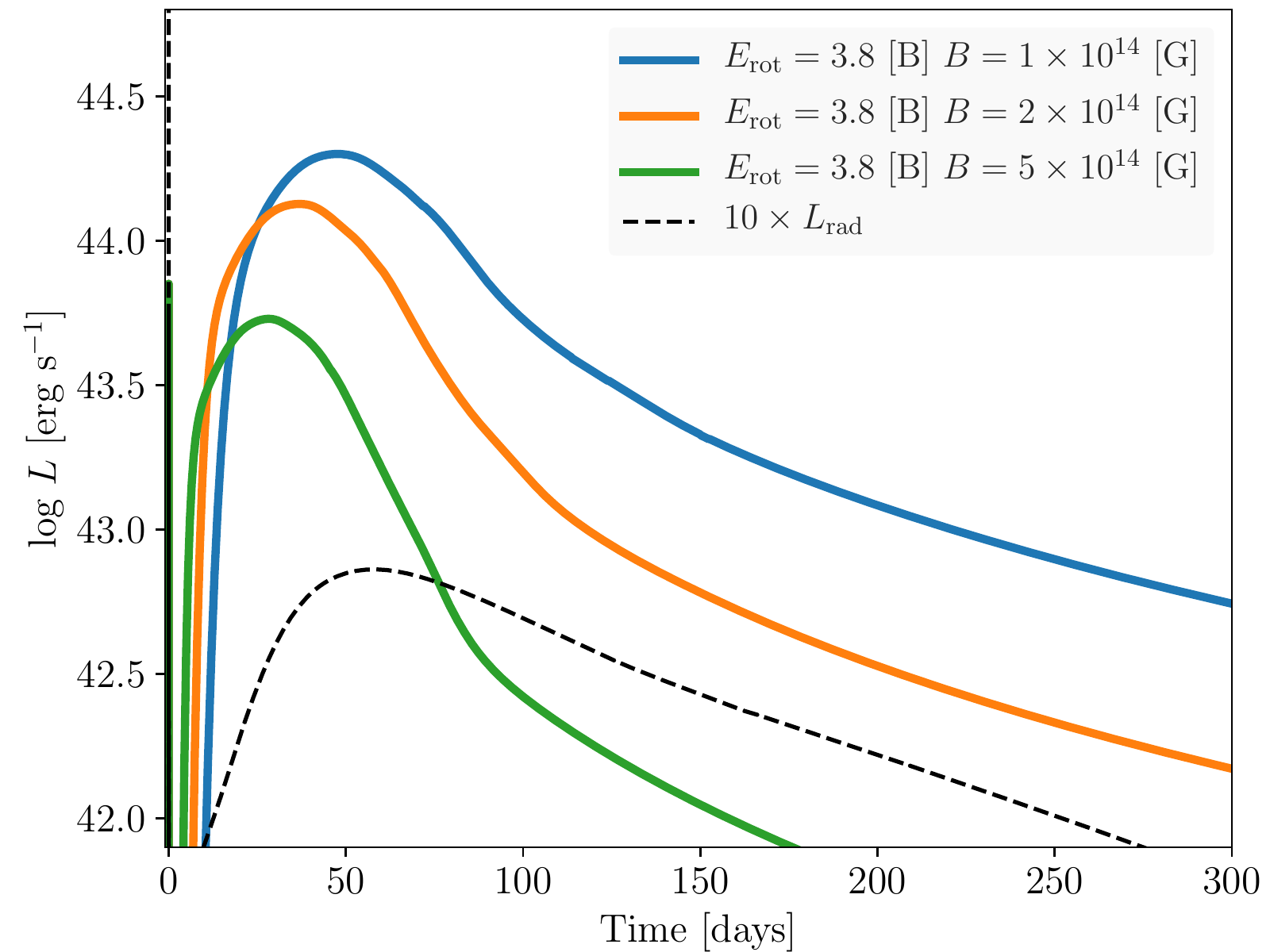}
  \caption{Light curves for Model He19.75 with and without an embedded
    magnetar.  \lFig{bigmagnetar}}
\end{figure}

The five magnetars introduced in Model He10.00 all had magnetic
fields of $5 \times 10^{14}$\,G and rotational energies of 0.1, 0.2,
0.5, 1.0, and $2 \times 10^{51}$\,erg. These rotational energies were
chosen to span a range where the rotational energy was either a small
or major adjustment to the energy of the neutrino-powered supernova,
$7.6 \times 10^{50}$\,erg. Given the single value of magnetic field
considered, the light curves are all very similar on their tails
(\Fig{IcBL}), but vary greatly in peak luminosity and rise time. All
are substantially brighter than the purely radioactive case. For the
smaller rotational energies, the light curves are similar to those for
the lower mass models (\Fig{magnetar}), but broader due to the larger
mass. They could potentially account for some of the brighter, broader
Type Ib and Ic supernovae in \Fig{prentice}. For larger magnetic
fields, the light curve would decline more rapidly. The light curves
in \Fig{IcBL} assume full trapping of the magnetar-deposited
energy. Depending on the spectrum of the magnetar and the pulsar wind
interaction inside the supernova, a substantial fraction of the energy
might escape at late times and not contribute to the optical light
curve. The column depth at 200\,d for the five magnetar energies shown
are 28, 22, 14, 9 and 6\,g\,cm$^{-2}$. If the opacity were like
gamma-rays from radioactivity, $\sim$0.05\,cm$^2$\,g$^{-1}$, then the
light curve at 200\,d for the most energetic case would be about three
times fainter. The optical depth at earlier times would scale
approximately as $t^{-2}$.

The light curves for the more energetic cases bear some similarity to
Type Ic-BL supernovae.  A potential problem with such a scenario
though is that, except for a small mass near the surface (\Fig{IcBL})
the velocities are quite low, hence no ``broad'' lines.  Typical Ic-BL
are estimated to have velocities characteristic of 4\,\Msun \ exploding
with $7 \times 10^{51}$\,erg with $^{56}$Ni masses near 0.3\,\Msun
\ \citep{Tad19}, or over 10,000\,km\,s$^{-1}$. \citet{Dro11} adopt a
velocity of 20,000\,km\,s$^{-1}$ at peak luminosity. Both this extreme
energy and mass of $^{56}$Ni are well out of reach of our standard
models.  There are three possible solutions to this difficulty: 1) The
masses of Type Ic-BL supernovae may be lighter or their explosion
energies greater than considered here.  A different explosion
mechanism would need to be invoked to raise the energy. 2) The light
at peak could come from circumstellar interaction. The outer
0.015\,\Msun \ of Model He10.00 moves faster than 12,000\,km\,s$^{-1}$
and has kinetic energy $3 \times 10^{49}$\,erg, more than enough to
provide the entire light curve of most Type Ic-BL supernovae, if the
matter interacted with a comparable mass of circumstellar matter at a
few $\times 10^{15}$\,cm. The problems with this scenario are
explaining why the necessary circumstellar matter is there with the
right density distribution to give the observed light curve shape
(\Sect{csm}), and why narrow lines from the circumstellar matter are
not usually seen in Type Ic-BL supernovae. This may be a more
promising model for Type Ibn supernovae \citep{Pas16} (though typical
``broad line'' components are slower for this class). 3) The light
curve and most of the explosion energy could come from a magnetar or
accretion into a black hole. The explosion would then probably be 
asymmetric.

Also shown in \Fig{IcBL} are the velocity profiles resulting from the
various magnetars. If plotted vs radius instead of mass, the regions
of near constant coasting speed would show up as thin, dense shells.
For large magnetar energy and 1D simulations, most of the mass of the
explosion is piled up in this thin shell. That the shell is thin and
dense is an artifact. In a multidimensional simulation, the hot pulsar
wind for the more energetic magnetars would break through the slower
moving denser material outside \citep{Che16} and surround the supernova
with an envelope of faster moving material. The explosion would be
asymmetric. The problem here is that a definitive calculation of the
spectrum is difficult. At late times the typical velocities would be
slower, more like the broad flat spots in \Fig{IcBL}. Velocities of
4000--5000\,km\,s$^{-1}$ are typical of some long duration Type Ic SLSN
\citep{Jer17} and may be similar to what is seen in some Type Ic-BL
supernovae. \citet{Mau10} find that the velocities for Type Ic-BL
supernovae are highly variable and might depend on observing angle.
They derive an average of about 6000\,km\,s$^{-1}$.

As a second example, consider Model 19.75. Without a magnetar, this
model exploded, using the W18 central engine, with a terminal kinetic
energy $8.4 \times 10^{50}$\,erg and ejected 0.088\,\Msun \ of
$^{56}$Ni. The presupernova mass was 10.28\,\Msun \ of which 8.5\,\Msun
\ was ejected. Only 0.25\,\Msun \ of that was helium. This would have
been a Type Ic supernova. The original ZAMS mass of the star before it
lost its envelope in the binary would have been 51\,\Msun. This
explosion was recomputed with an embedded magnetar with an initial
rotational energy of $3.8 \times 10^{51}$\,erg (rotational period 2.3\,ms)
and magnetic field strengths of 1, 2, and $5\times 10^{14}$\,G. Energy
deposition was calculated according to \citet{Woo10} and full
absorption was assumed. The case with the smallest field strength is similar 
to magnetar parameters used by \citet{Vre17} to fit the Type I SLSN iPTF12dam 
and are rather typical of the large spread observed in SLSN
\citep{Nic17,Vil18}. \Fig{bigmagnetar} shows the light curves both
with and without the magnetar. With the magnetar, the model
essentially replicates the results of \citet{Vre17} and observations
of iPTF12dam. As noted by those authors, the magnetar model, even with
some approximation for escape is too bright at very late times
compared with the observations, but magnetar energy transport and
break out are poorly understood. The improvement here over
\citet{Vre17} is that the explosion is derived from an actual
stellar evolution and explosion calculation.

The light curve shown in \Fig{bigmagnetar} might also rise more
rapidly than would actually be observed in the optical because the
magnetar pushes the ejected matter into a thin shell that has an
unphysically short diffusion time. The bolometric light curve also
peaks earlier than the optical light curve because of its high
effective temperature \citep{Nic17}.

\section{Conclusions} 
\lSect{conclude}

Using an approach to the explosion physics similar to that previously
used to study single massive stars \citep{Suk16},
we have modeled the explosion of mass-losing helium stars.
Presupernova models are taken from \citet{Woo19}. It is assumed that
the initial helium star is revealed in a close, mass exchanging binary
system when the star loses its entire envelope just prior to or
shortly after central helium ignition. While simplistic, the models
capture an essential aspect of stripped star evolution. The
helium core and its central convection zone shrink during helium
burning due to mass loss, while in single stars the helium core grows
due to hydrogen shell burning.  Because the helium core shrinks rather
than growing, the main sequence masses producing supernovae in binary
systems are typically greater and their lives are shorter than for
single stars.

As in \citet{Suk16}, an explosion is not artificially forced, but a
consequence of presupernova structure and the choice of a ``central
engine''. A realistic, albeit 1D and parametrized, calculation of
neutrino transport is done for each explosion. Several central engines
of various strengths were explored with parameters calibrated to SN~1987A
or the Crab Supernova.  The standard choice was a combination of
the W18 engine for helium stars with final presupernova masses above
about 3.5\,\Msun, and an engine interpolating between W18 and Z9.6 for
stars with smaller final helium-core mass. See \citet{Suk16} for a
discussion of these engines. W18 was calibrated against SN~1987A and
Z9.6, the Crab. The effects of varying the mass loss rate for the
presupernova stars were also explored. A more energetic central engine, S19.8,
was also extensively explored to test the limits of $^{56}$Ni production.

Despite being derived from main sequence stars with larger masses, the
compact remnant masses, $^{56}$Ni yields, and explosion energies for
simulated binaries are very similar to single stars with a helium core
mass equal to the presupernova mass of the stripped star (compare
\Fig{summaryw18-sta-exp} and \Fig{summaryw18-1.5-exp} with
\Fig{resultsw18transnew} and \Fig{resultsn20transnew} in Appendix A).
This did not have to be so. The compositions, entropies, and
compactness parameters are different in the two sets. The stripped
stars in the present study, explode a bit more frequently with
a bit more energy, but for those that explode, the observable
outcomes are robust. This implies that the iron core masses and
gradients just outside those cores are to first order, set by the
final mass of the helium and heavy element core, and only to second order
by how that core came to exist.

Except for a few cases with the weakest central engine, stars with
presupernova masses from 2.1 to 6.5\,\Msun \ explode robustly,
(\Fig{summaryw18-sta-exp} and \Fig{1.5-explosions}).  For final masses
between 6.5 and 12\,\Msun, the results are more nuanced, with a mixture
of black holes and neutron stars sensitive to the explosion
model. Presupernova masses up to 8\,\Msun \ mostly explode, and there
is also an island of explosions centered around 10.5\,\Msun \ that grows
with the relative strength of the central engine. Black holes are
invariably produced for presupernova masses around 8 to 9\,\Msun \ 
and above 12\,\Msun. These outcomes correlate strongly with both the 
compactness parameter and Ertl parametrization of the presupernova 
stars (\Fig{compact}).

Similar systematics were seen for single stars \citep{Suk16,Suk18},
but with a shift upwards in the presupernova mass scale for the more
massive helium-star 
models of about 2\,\Msun \ \citep[\Fig{compact} here and Figure
 10 of][]{Woo19}.  This shift is partly due to a higher mass fraction
of carbon following helium burning in the mass losing helium
stars. Results will presumably also be sensitive to the rate for
$^{12}$C($\alpha,\gamma)^{16}$O, though we have not yet undertaken a
sensitivity study. They are also a consequence of the different
placement of interfering convective shells and a larger carbon-oxygen
core for a given helium core in the mass losing stars \citep[e.g.,][]{Suk19}.

No final kinetic energy greater than $2.0 \times 10^{51}$\,erg was
found in any model. Combined with our previous results in
\citet{Suk16}, we think this represents a fundamental upper
limit to what can be obtained with neutrino-energy input alone, at
least for the physics in our 1D models and assuming that SN~1987A was
a neutrino-driven explosion. Other similar studies support this
conclusion \citep{Mue16,Ebi19}. \citet{Pej15b} used a parametrized
prescription of a neutrino-driven wind and obtained explosions up to
$\sim 6\times 10^{51}$\,erg. Though their estimated $^{56}$Ni masses were
similar to our results, our smaller limit may be more realistic
because of a more physical treatment of neutrino
transport. Multi-dimensional simulations with neutrino transport and
parametric luminosities of the neutron star \citep{Sch06} as well as
self-consistent neutrino transport \citep[e.g.][]{Fry01,Bru16,Mue17}
have never produced neutrino-driven explosions with converged energies
above $\sim 2\times 10^{51}$\,erg.

Remnant masses were determined for neutron stars produced in the
successful explosions and black hole masses for the failures. These
are particularly appropriate for binaries where most masses are
measured and for gravitational radiation signals. For the standard W18
engine, black holes formed in 21\% of collapses and the median neutron
star gravitational mass was 1.351\,\Msun \ when the mass decrement due
to neutrino losses was calculated self-consistently using P-HOTB
(\Tab{remtbl}; \Fig{nstarmass}). The mean value was 1.371\,\Msun.
Slightly larger values were obtained using the \citet{Lat01}
prescription for the gravitational mass correction. The values derived
using the Lattimer-Prakash correction are very similar to the what was
obtained for equivalently modeled single stars by \citet{Suk16}, 1.40\,\Msun.
Other average values are given in \Tab{remtbl} and the
predicted distribution functions are given in \Fig{nstarmass} and
\Fig{bhmass}.

Since there is some disagreement in the literature
\citep{Bel12,Kre12}, we want to emphasize that we have no
problem producing low mass black holes right down to the maximum
neutron star mass (\Sect{bh}). These low mass black holes are made by
fallback in a partially successful explosion. There is a deficiency of
black holes below 6\,\Msun \ (\Fig{bhmass}) because that is the lowest
mass black hole produced by prompt implosion, but no empty ``gap''. On
the other hand, our calculations do show the possibility of a gap at
an unexpected location, around 10--11\,\Msun. The width and significance of
this void is model dependent (\Fig{sta-explosions}). It is broader for
central engines with greater power. Its presence would
reveal some dependency of the explosion energy on the compactness of
the presupernova star and hence its shell burning history. For
presupernova masses above 12\,\Msun, all models produced black holes
with masses nearly equal to their presupernova mass. This resulting
distribution function is thus sensitive to the treatment of mass loss,
and hence to metallicity and will be explored further in a future
paper.

Because of its importance to the light curves of Type~I supernovae,
particular attention was paid to $^{56}$Ni synthesis.  Using the same
approach as in \citet{Suk16} similar average yields were found. The
IMF averaged $^{56}$Ni production (Ni+Tr) per supernova for our
standard W18 engine, was 0.054\,\Msun. In 2016 it was 0.053\,\Msun.  For
the most energetic S19.8 engine, the value rises to 0.068\,\Msun. These
values are all skewed towards low values by the large contribution of
low mass supernovae to the number statistics. The possibility of
producing larger values was explored (\Sect{ni56} and
\Sect{ni56max}). Taking liberal estimates for $^{56}$Ni production by
$\alpha$-recombination in the neutrino-powered wind and assuming most
of the iron group isotopes, even in deep layers that experience
neutrino interactions, are $^{56}$Ni, these yields can be increased to
0.073 and 0.090\,\Msun \ per supernova for the W18 and S19.8
engines. This liberal estimate corresponds to assuming that 75\% of
all material that achieves NSE in the neutrino-transport calculation
using P-HOTB is ejected as $^{56}$Ni. This is our upper bound. Still
greater values exceed the reasonable error bar for neutrino-powered
explosions, but are possible if greater energies and deeper mass cuts
are forced in the explosion. The greatest $^{56}$Ni observed in any
calculation even with unrealistic, artificial variation of these
parameters, especially the mass cut, was 0.33\,\Msun
\ (\Sect{ni56max}).

Placing the mass cut that close to the edge of the iron core seems to
be ruled out though, since it would imply a median gravitational mass
for the neutron star of only 1.28\,\Msun \ (\Tab{remtbl}). What is
added to $^{56}$Ni gets subtracted from the remnant mass. It is thus
very difficult to reconcile $^{56}$Ni yields of 0.2\,\Msun
\ \citep{Dro11,Tad19} and more with observations, even for models with
additional (non-neutrino) energy sources. Doubling the iron yield of
massive stars would also mean they account for most of the iron seen
in the sun. This would be inconsistent with observations showing that
the majority of iron has been created later in the history of the
Galaxy by Type Ia supernovae (\Sect{nucleo}). It would also grossly 
overproduce the $^{56}$Ni observed in SN~1987A.

Ordinary Type Ib and Ic supernovae are attributed here to models with
final masses between 2.45 and 5.63\,\Msun, which, depending on
uncertain mass loss rates, correspond to initial helium-star masses of
3.0 to 8.0\,\Msun. This is for the (possibly conservative) estimates of
mass loss rate employed. These in turn correspond to main sequence
masses 15 to 29\,\Msun.  These supernovae come from a population that,
on the average, is more massive than Type IIp supernovae. They may
thus be found more tightly correlated with star forming regions
\citep{Mau18}.  The range of ages for 15 to 29\,\Msun \ stars is
10$^{7.12}$\,yr to 10$^{6.81}$\,yr \citep{Suk18} . Maund gives typical
ages of 10$^{7.05}$\,yr and $10^{6.57}$\,yr for Type Ib and Ic
respectively. More massive, shorter lived stars will also produce Type
Ic supernovae if the mass loss rates were greater, but the
presupernova mass should not exceed 12\,\Msun \ (\Sect{bh}). For the
standard mass loss rates, the ZAMS mass corresponding to a 12\,\Msun
\ progenitor is 57\,\Msun. Such a star would have a lifetime of
10$^{6.62}$\,yr, though at very high mass the lifetime becomes
insensitive to the mass. There should be a tendency, however, for
highest mass models to produce faint, broad light curves \citep{Ens88}
unless the explosion energy is much greater than calculated here. The
maximum luminosity of a Type Ib or Ic progenitor star is 10$^{5.6}$\,\Lsun.
Brighter progenitors are over 12\,\Msun \ and collapse to black
holes (\Sect{bh}) or explode using energy sources in addition to
neutrinos from core collapse.

Using our most optimistic, but plausible $^{56}$Ni abundances,
0.75$\times$(Ni+Tr+$\alpha$) (\Sect{ni56}), reproducing the light
curves of the brighter common Type Ib and Ic supernovae is 
problematic.  Using the standard W18 central engine,
it is possible to reproduce the light curve peak luminosity and rise
and decline rates for events with peak luminosities below the {\sl
  averages} quoted by the observers (\Fig{sn1bc}), but not for any
with peak luminosities in excess of 10$^{42.5}$\,erg\,s$^{-1}$
(\Fig{prentice}). Using our most energetic central engine, S19.8 only
raises this limit to 10$^{42.6}$\,erg\,s$^{-1}$ and these upper limits
are only achieved in a few cases (\Tab{expltbl}). For $^{56}$Ni yields
evaluated more conservatively, like in \citet{Suk16}, the upper bound
for the W18 central engine falls to 10$^{42.1}$\,erg\,s$^{-1}$ and
{\sl most} observed Type Ib and Ic supernova luminosities cannot be
achieved. Our medians are of course lower than these peak estimates
with a value of 10$^{42.0}$ to 10$^{42.3}$\,erg\,s$^{-1}$ typical for
the full range of stars that make Type Ib and Ic supernovae (3--8 \Msun; 
\Tab{lcstbl}).  The upper limit comes from using the energetic S19.8 central
engine and evaluating the $^{56}$Ni yield optimistically. The lower
value is for the W18 central engine with $^{56}$Ni yields evaluated
conservatively. 

Many supernovae classified by the
observers as ordinary Type Ib and Ic are simply too bright to be made
by our unassisted neutrino-driven models. There are several ways out
of this dilemma, but we have argued that simply increasing, without
bound, the mass of $^{56}$Ni ejected in the explosion violates several
basic constraints (\Sect{nucleo}; \Sect{nstar}; \Sect{ni56max}). 
Maybe our simple calculation of light curves using
single-temperature flux-limited diffusion in KEPLER is inadequate to
capture the peak of the light curve. We expect that studies in the
near future will address this issue. Maybe the observed brightness
distribution for Type Ib and Ic supernovae
\citep[e.g.,][]{Lym16,Pre16,And19} will drift downwards in the
future \citep[e.g.,][]{Pre19}. A more careful treatment e.g., of bolometric corrections or Malmquist bias, might lead to a significant decrease in published mean values. Or a significant fraction of 
what the observers have called Type Ib and Ic supernovae, the 
brightest ones, have a non-radioactive power source at their peak. 
We explored the possibilities.

Circumstellar interaction is one way to boost the luminosity and must
play a role in some rare forms of Type I supernovae, e.g., Type Ibn
\citep[e.g.,][]{Pas08}, but it requires fine tuning of the
circumstellar mass distribution to mimic the light curve shape of an
ordinary Type Ic supernova. If the light were produced by a small mass
of high velocity material with slow moving presupernova mass loss,
there would presumably be spectroscopic signatures that are not
reported. There might be a superposition of high and low velocity
lines.

If radioactivity and circumstellar interaction are ruled out, the
remaining possibility is magnetars. Given the great diversity of Type
I supernovae in nature, including SLSN, Type Ic-BL supernovae, and
GRBs, it is clear that some other source besides $^{56}$Ni does, at
least occasionally, provide the light of Type I supernovae, and
magnetars are frequently invoked.  Might they also contribute, at a
reduced level, to ordinary events?  The energy and field requirements
are modest, less than 10$^{50}$\,erg (corresponding to neutron-star
birth rotation periods longer than 14\,ms) and greater than $5 \times
10^{14}$\,G. The explosion is still neutrino powered, but the light
both at peak and on the tail, comes, at least partly, from the
magnetar.  Distinguishing characteristics besides the presence of a
magnetar in the remnants of known supernovae of Type Ib and Ic, would
be light curves that declined slower on the tail than $^{56}$Co decay,
or evidence for a smaller abundance of $^{56}$Fe and $^{56}$Co in the
late time spectrum than necessary to explain the peak of the light
curve. Energetic magnetars might be easier to make in binaries since
removing the envelope removes a large sink for the angular momentum of
the rapidly rotating helium core. Most angular momentum in the
evolution of a single star is lost between hydrogen depletion and
helium ignition \citep{Heg05} as the star becomes a red
supergiant. Removing the envelope would diminish this braking. If they
were present in Type IIp supernovae, magnetars of the sort proposed
here would have little influence on the light curve except at very
late times. The energy that they deposit would be adiabatically
degraded during the expansion while on the plateau.

Outside of the mass range that we attribute to normal Type Ib and Ic
supernovae, unusual events are predicted.  Lower mass models with
presupernova masses less than 2.59\,\Msun \ coming from initial helium
stars with masses 2.5 to 3.2\,\Msun \ have distinctive properties. For
helium-star masses below 3.0\,\Msun \ the presupernova star is a helium
blue supergiant, not a WR star. Following explosion, the expansion and
recombination of this envelope produces a broad bright blue initial
peak before declining and, depending on the $^{56}$Ni yield, rising
again later to a second (radioactive) peak (\Fig{lowmlite}). Given their
small ejecta masses, the models have high velocities even though their
kinetic energies are low. Some subset might be fast blue optical
transients \citep{Kle18,Woo19}, SN~2014ft-like objects \citep{De18},
Type Ibn supernovae, or even Type Ic-BL supernovae. Supernovae in this
mass range may also be complicated by the effects of a degenerate
silicon flash weeks to months before iron core collapse. If the flash
is strong and substantial material is ejected, depending on the time
of the ejection, the supernova could be very bright (\Fig{siflash}),
like some Type Ibn \citep{Pas08}. If the silicon flash is weak
though, they would look more like ordinary faint Type Ib supernovae
(3.0, 3.1, 3.2\,\Msun; \Fig{sn1bc}) or supernovae with radius expansion
(2.5\,\Msun; \Fig{lowmlite}). The specific masses and strengths of the
flashes here are sensitive to uncertain flame physics during the
silicon flash.

Presupernova models more massive than about 6\,\Msun \ produce light
curves that are too faint and too broad to be common Type Ib and Ic
supernovae, if neutrinos are the cause of their explosion and
radioactivity is their only illuminating power. Perhaps most of these
collapse to black holes. Others may simply have escaped detection
since their light curves are both red and faint. Ultimately gravitational wave surveys will offer some constraints.

Some of them might harbor more energetic magnetars though with similar
field strengths to that needed for common events, but faster initial
rotation rates. \Fig{IcBL} and \Fig{bigmagnetar} show some
possibilities. It is important that neutrinos alone can explain the
initial explosion of such stars, if not their light curves. A
successful explosion allows the necessary time for the proto-neutron
star to contract, speed up, and develop a strong magnetic field. The
magnetar energy input can either be a small fraction of the final
kinetic energy of the supernova or the dominant contributor. In the
latter case, pulsar wind break out might lead to asymmetric ejecta and
poking holes in the slower moving ejecta, possibly contributing to the
high velocity material present in the spectra of Type
Ic-BL supernovae. Alternatively, supernovae of Type Ic-BL could
be lower mass stars in which circumstellar interaction is important.
If a magnetar is powering both the explosion and the light curve, one
might expect there to be a correlation between peak luminosity and
velocity (\Fig{IcBL}), though mass is clearly an important second parameter.

\section{Acknowledgements}

We thank the anonymous referee for useful suggestions. We also 
thank Simon Prentice for providing light curve data, and 
Bernhard M\"uller for feedback. TE and HTJ thank G.~Stockinger 
for the improved treatment of energy transfer by neutrino-nucleon 
scattering in the neutrino-transport solver, A.~Menon and 
A.~Heger for providing the data of their SN~1987A binary 
progenitors, and R.~Hix and F.-K.~Thielemann for the NSE solver 
introduced into P-HOTB by K.~Kifonidis. At UCSC, this work was 
supported by NASA (NNX14AH34G).  At Garching, funding by the 
European Research Council through grant ERC-AdG No.~341157-COCO2CASA
and by the Deutsche Forschungsgemeinschaft through grants
SFB-1258 ``Neutrinos and Dark Matter in Astro- and Particle Physics
(NDM)'' and EXC~2094 ``ORIGINS: From the Origin of the Universe 
to the First Building Blocks of Life'' is acknowledged. TS was 
supported by NASA through a NASA Hubble Fellowship grant \#60065868 
awarded by the Space Telescope Science Institute, which is operated 
by the Association of Universities for Research in Astronomy, Inc., 
for NASA, under contract NAS5-26555.

\clearpage

\appendix

\setcounter{table}{0}
\setcounter{figure}{0}

\section{Improved Single-star Results}
\label{app:comparison}

\begin{deluxetable*}{ccccccccccc}[h]
  \tablecaption{Results from \citet{Suk16} (``old'') for all neutrino engines applied
                to SN~1987A and Crab calibration models, compared to simulations with 
                the upgraded P-HOTB code (``new'')}
  \tablehead{\colhead{Calibration}                                 &
             \colhead{$t_\mathrm{exp}$$^a$}                        &
             \colhead{$E_\mathrm{exp}$$^b$}                        &
             \colhead{$E_\mathrm{exp}/M_\mathrm{ej}$$^c$}          &
             \colhead{$M_{^{56}\mathrm{Ni}}$$^d$}                  &
             \colhead{$M_\mathrm{Tr}$$^e$}                     &
             \colhead{$M_\mathrm{NS}$$^f$}                         &
             \colhead{$M_\mathrm{wind}$$^g$}                       &
             \colhead{$M_\mathrm{fb}$$^h$}                         &
             \colhead{$t_{\nu,90}$$^i$}                            &
             \colhead{$E_{\nu,\mathrm{tot}}$$^j$}                  \\
             \colhead{Model}                                       &
             \colhead{[ms] }                                       &
             \colhead{[B] }                                        &
             \colhead{[B/M$_\odot$] }                              &
             \colhead{[M$_\odot$] }                                &
             \colhead{[M$_\odot$] }                                &
             \colhead{[M$_\odot$] }                                &
             \colhead{[M$_\odot$] }                                &
             \colhead{[$10^{-2}$\,M$_\odot$] }                     &
             \colhead{[s]}                                         &
             \colhead{[100\,B]}   }
  \startdata
     S19.8${}^{k}$ old & 750  & 1.30 & 0.100     & 0.072   & 0.034 & 1.551 & 0.096 & 0.00298  & 3.78  & 3.58    \\
     S19.8${}^{k}$ new & 790  & 1.34 & 0.103     & 0.052   & 0.046 & 1.560 & 0.089 & 0.00409  & 3.61  & 3.64    \\
     \\
     W15 old           & 580  & 1.41 & 0.103     & 0.045   & 0.046 & 1.317 & 0.088 & 0.00018  & 3.53  & 2.67    \\
     W15 new           & 550  & 1.44 & 0.105     & 0.040   & 0.051 & 1.315 & 0.084 & 0.00060  & 3.19  & 2.60    \\
     \\
     W18 old           & 730  & 1.25 & 0.081     & 0.056   & 0.036 & 1.484 & 0.081 & 0.00310  & 3.57  & 3.22    \\
     W18 new           & 730  & 1.23 & 0.080     & 0.044   & 0.043 & 1.489 & 0.083 & 0.00444  & 3.54  & 3.21    \\
     \\
     W20 old           & 620  & 1.24 & 0.070     & 0.063   & 0.027 & 1.562 & 0.089 & 0.00168  & 4.15  & 3.50    \\
     W20 new           & 580  & 1.43 & 0.080     & 0.049   & 0.046 & 1.543 & 0.103 & 0.00259  & 3.94  & 3.49    \\
     \\
     N20 old           & 560  & 1.49 & 0.100     & 0.036   & 0.052 & 1.549 & 0.117 & 0.00243  & 3.47  & 3.38    \\
     N20 new           & 560  & 1.59 & 0.107     & 0.032   & 0.057 & 1.543 & 0.123 & 0.00290  & 3.27  & 3.34    \\
     \\
     Z9.6  old         & 155  & 0.165 & 0.020     & 0.006   & 0.006 & 1.338 & 0.019 & 0.00016  & 7.03  & 1.82    \\
     Z9.6  new         & 145  & 0.190 & 0.023     & 0.003   & 0.010 & 1.335 & 0.020 & 0.00008  & 7.03  & 1.83    \\
     \\
     \multicolumn{11}{l}{SN~1987A-like explosions with upgraded P-HOTB and different engines
                         for binary progenitors of \cite{Men17}: } \\
     \\
     M16+7b (W20)      & 490 & 1.49 & 0.073      & 0.051   & 0.051 & 1.474 & 0.110 & 0.00250  & 3.88  & 3.23    \\
     M16+4a (W18)      & 923 & 1.22 & 0.070      & 0.068   & 0.030 & 1.751 & 0.067 & 0.00517  & 3.60  & 4.19    \\
     M16+4a (W15)      & 893 & 1.22 & 0.070      & 0.071   & 0.031 & 1.745 & 0.078 & 0.00523  & 3.53  & 4.04    \\
     M15+8b (W20)      & 481 & 1.38 & 0.065      & 0.038   & 0.049 & 1.378 & 0.087 & 0.00164  & 3.83  & 2.86    \\
     M15+7b (S19.8)    & 994 & 1.00 & 0.052      & 0.039   & 0.030 & 1.695 & 0.085 & 0.00212  & 3.71  & 4.09    \\
  \enddata

  \tablecomments{ ${}^{a}$~Post-bounce time when explosion sets in
    (i.e., when the shock expands beyond 500\,km).  ${}^{b}$~Final
    explosion energy (with binding energy of progenitor star taken
    into account; 1\,B\,=\,1\,bethe\,=\,$10^{51}$\,erg).
    ${}^{c}$~Explosion energy divided by the final ejecta mass (with
    fallback taken into account).  ${}^{d}$~Ejected $^{56}$Ni mass
    (with fallback taken into account).  ${}^{e}$~Mass of neutron-rich
    tracer nucleus ejected in neutrino-heated matter with neutron
    excess (with fallback taken into account).  ${}^{f}$~Final
    baryonic neutron star mass (with late-time fallback included).
    ${}^{g}$~Neutrino-driven wind mass measured by mass between gain
    radius at $t_\mathrm{exp}$ and preliminary mass cut {\em before}
    fallback.  ${}^{h}$~Fallback mass.  ${}^{i}$~Emission time of 90\%
    of the total radiated neutrino energy.  ${}^{j}$~Total radiated
    neutrino energy (at 10\,s after bounce, when typically $\sim$99\%
    of the neutrino energy has been radiated).  ${}^{k}$~Red
    supergiant progenitor from \cite{Woo02}.}
  \lTab{explodability_calibration}
\end{deluxetable*}

\begin{figure*}
  \includegraphics{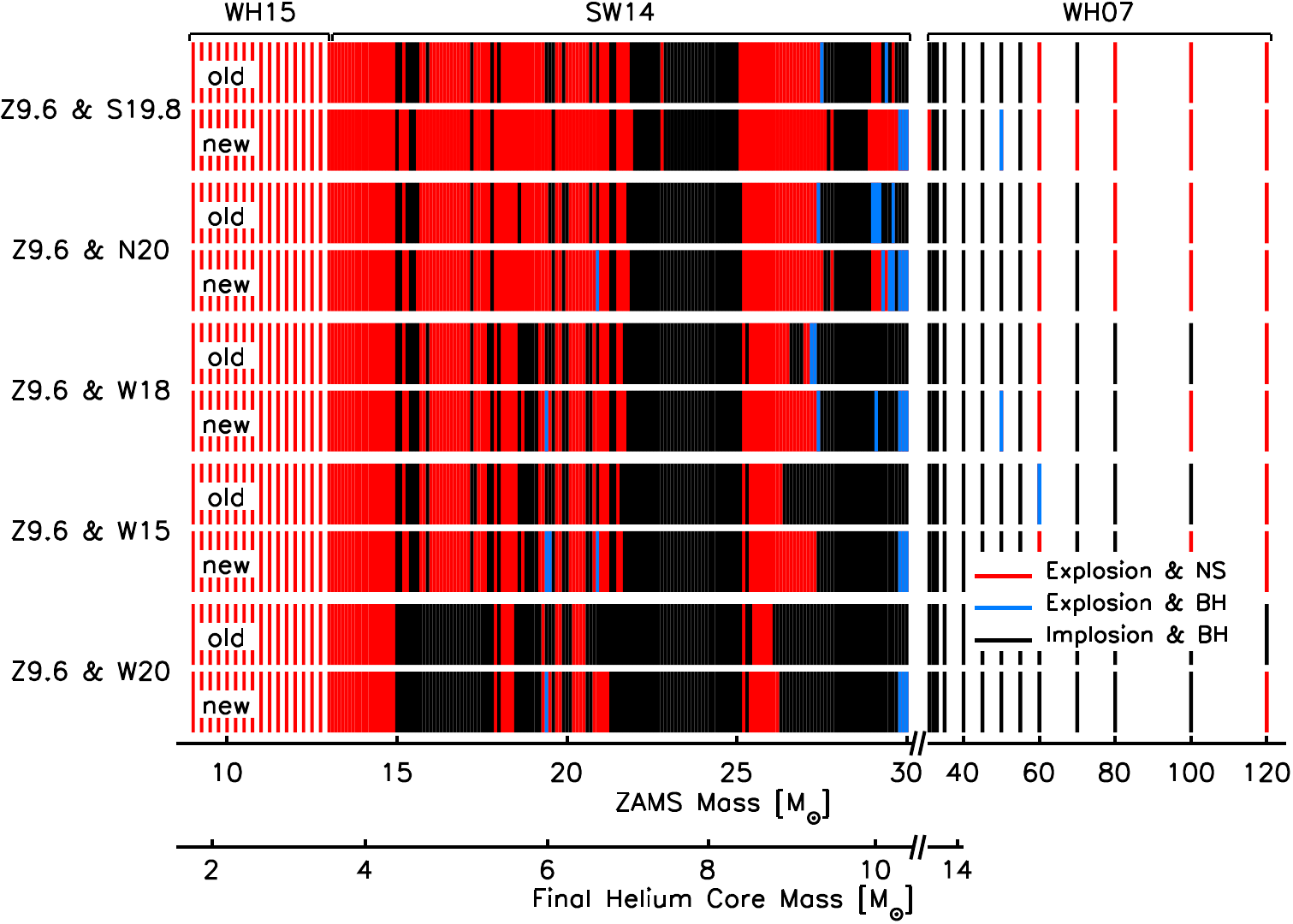}
  \caption{Outcomes of stellar core collapse for five different
    neutrino engines applied in 1D simulations of the whole set of
    single-star progenitors investigated by \citet{Suk16}. The engines
    are sorted by strength (measured by the number of successful
    explosions) from top to bottom. The results with the P-HOTB code
    version used by \citet{Suk16} are labeled by ``old'', and the
    models recomputed with the upgraded P-HOTB version used in the
    present work are labeled by ``new''. The different engines are
    named by the combination of Crab-like and SN~1987A-like
    progenitors used for calibrating the neutrino-engine
    parameters. Cases of supernova explosion and neutron star
    formation are marked by red, cases of black hole formation without
    explosion by black, and cases of supernova explosion and black
    hole formation by massive fallback (``fallback supernovae'') by
    blue.  We set the limit of black hole formation at a baryonic mass
    of 2.75\,M$_\odot$, corresponding to a gravitational mass of about
    2.18--2.30\,M$_\odot$ \citep{Lat01}. Compact remnants whose mass
    is pushed by fallback above this limit are considered as black
    holes.}  \lFig{comptransnewold}
\end{figure*}

A number of improvements have been made to the P-HOTB code since its
use in \citet{Suk16}. They chiefly concern the treatment of
neutrino-nucleon scattering in the transport solver, the inclusion of
weak decays for $^{56}$Ni and $^{56}$Co, and some some refinements to
the treatment of the radial grid (\Sect{PHOTB}). In this Appendix we
briefly explore the results of core-collapse and explosion simulations
when the improved code is applied to the same single-star progenitors
investigated by \citet{Suk16}. Generally, the differences are small.

\Tab{explodability_calibration} gives some basic results for the
supernova, neutrino, and remnant properties of the SN~1987A and Crab
progenitors employed as calibration models in both papers.  The
results from the previous version of the P-HOTB code (``old'') are
taken from \citet{Suk16}, the simulations with the upgraded code
version (``new'') were performed with exactly the same values of the
engine parameters as used in \citet{Suk16} (see Table~3 there). The
differences between old and new simulations are minor.  The time when
the explosion sets in can be slightly different, both earlier or later
by some 10\,ms. The explosion energy with the new code is $\sim$3\% up
to $\sim$15\% (for W20 and Z9.6) greater, but can also be marginally
smaller (W18). Interestingly, the ejected mass of $^{56}$Ni is
reduced, but the ejecta mass of the tracer nucleus ($M_{\rm Tr}$), is
increased by roughly the same amount so the sum remains nearly
unchanged. Similar small differences are found in the masses of the
neutrino wind and the final neutron star mass. The latter is slightly
lower when the explosion with the new code starts earlier and slightly
higher when the explosion sets in later. The small differences in
these results are a consequence of opposite trends in different
neutrino quantities. With the improved neutrino treatment the radiated
$\nu_e$ luminosity decreases by $\sim$5--10\%, but the mean energy of
the emitted $\bar\nu_e$ increases by roughly 1\,MeV. The first effect
decreases the postshock heating by electron neutrinos, whereas the
second effect increases the heating by electron
anti-neutrinos. Depending on which effect dominates, the explosion can
commence slightly later or earlier.

\begin{figure*}
  \includegraphics{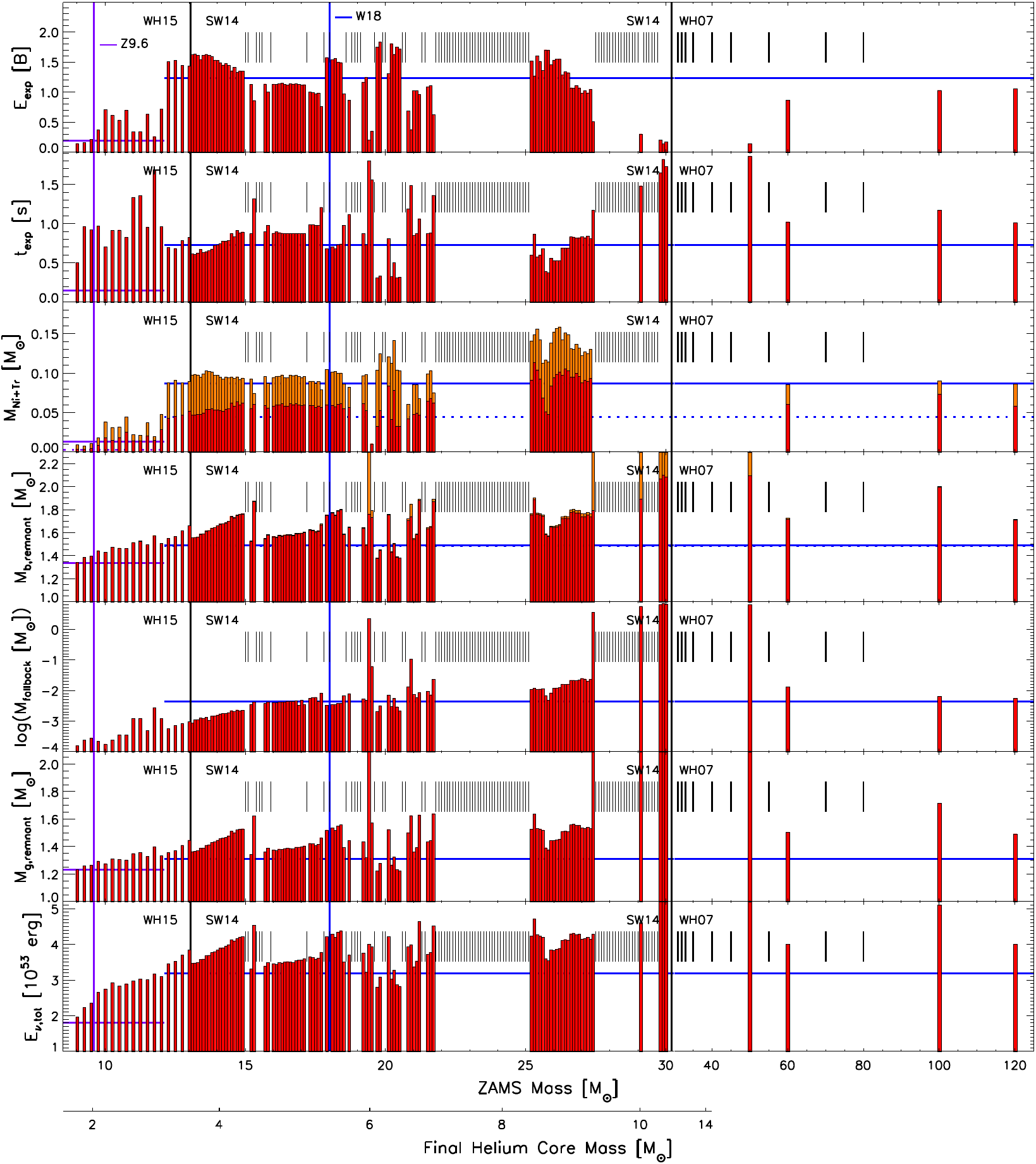}
  \caption{Explosion properties for all single-star, solar-metallicity
    progenitors of \citet{Suk16}, computed with the upgraded P-HOTB
    code and the combined Z9.6 and W18 neutrino engine. The plot can
    be directly compared with Figure~8 of \citet{Suk16}. From top to
    bottom the panels show the final explosion energy ($1\,\mathrm{B}
    = 10^{51}\,$erg); the time of onset of the explosion (when the supernova
    shock expands beyond 500\,km); the summed mass of finally ejected
    $^{56}$Ni (red) plus tracer (orange); the baryonic mass of all neutron stars
    that exist at least transiently, with the fallback mass indicated
    by orange extensions on the histogram bars; the logarithm of the
    fallback mass; the gravitational mass of the compact remnants,
    most of which remain neutron stars except the cases with more than
    1\,$M_\odot$ of fallback, which are expected to become black holes; and
    the total energy radiated in neutrinos, which we determine only
    from the proto-neutron star cooling phase without taking into account possible
    neutrino emission during fallback accretion \citep[for better
      comparison with][]{Suk16}. In contrast to Figure~8 of
    \citet{Suk16}, where the neutrino radiation loss was shown for a
    post-bounce evolution of 15\,s, the current models have been
    simulated only for 10\,s after bounce. The differences in the
    neutrino-energy loss would not be visible on the scale of the
    plot. Non-exploding cases are marked by thin, short vertical black
    dashes in the upper part of each panel. The thick vertical black
    lines separate the different sets of single-star progenitor models
    investigated by \citet{Suk16}, the vertical purple and blue lines
    mark the masses of the engine Models Z9.6 and W18, respectively,
    and the corresponding results of these engine models are indicated
    by solid and dashed horizontal purple and blue lines. The
    mass-range spanned by the horizontal purple line of Model Z9.6
    indicates the region of Crab-like behavior, where the Crab and
    SN~1987A engines are interpolated.}  \lFig{resultsw18transnew}
\end{figure*}

\begin{figure*}
  \includegraphics{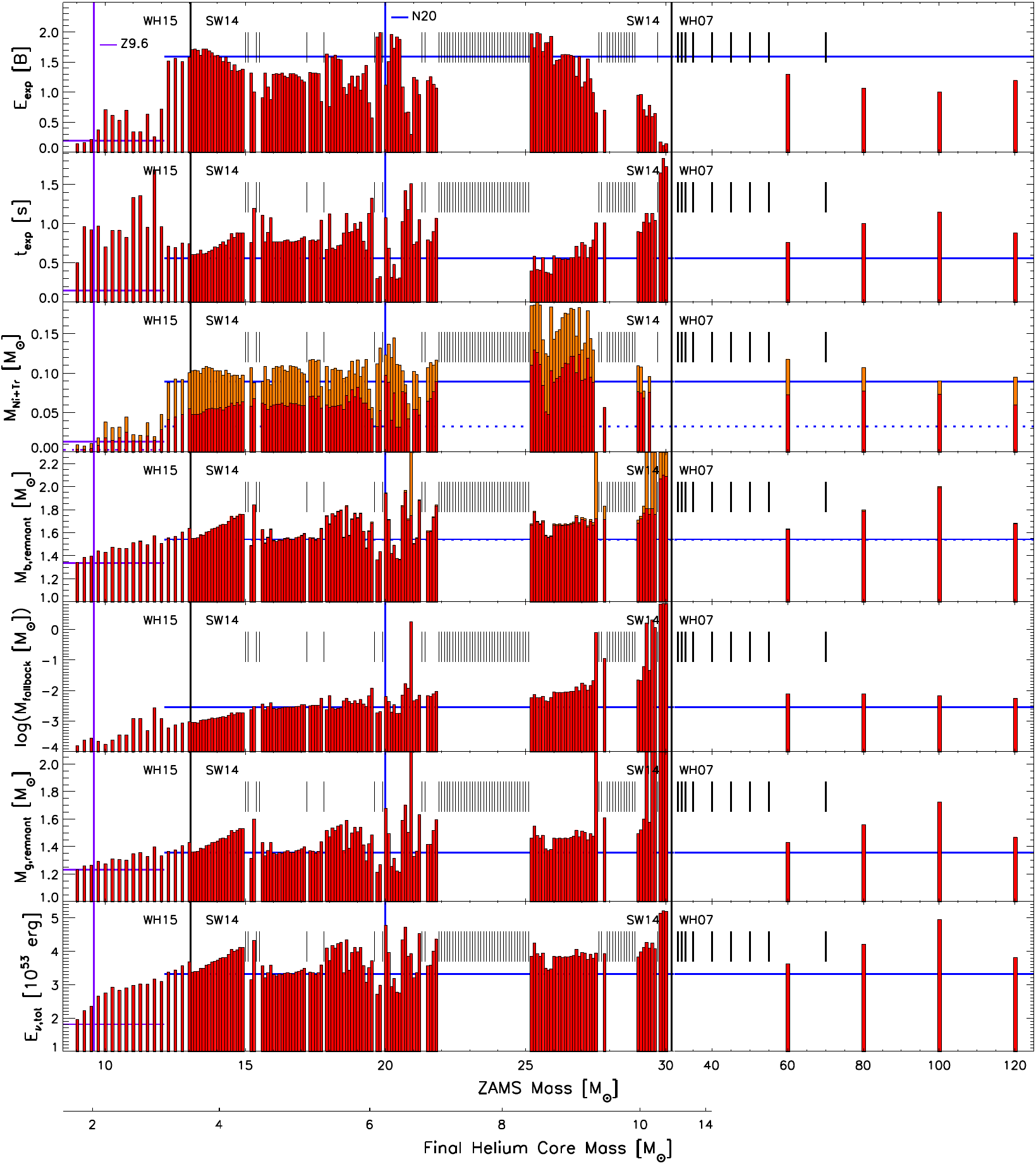}
  \caption{Same as \Fig{resultsw18transnew}, but for simulations with
    the upgraded P-HOTB code and the combined Z9.6 and N20 neutrino
    engine.  The plot can be directly compared with Figure~9 of
    \citet{Suk16}.  The exploding case at a ZAMS mass of
    27.5\,M$_\odot$ has a fallback mass of nearly 0.8\,M$_\odot$ and
    the compact remnant is likely to survive as neutron star, whereas
    in all other cases with massive fallback the proto-neutron star
    accretes more than 1\,M$_\odot$ of fallback matter and must be
    expected to collapse to a black hole.}  \lFig{resultsn20transnew}
\end{figure*}

The total energy radiated in neutrinos within 10\,s of post-bounce
evolution is nearly identical in the old and new runs, as is the time
$t_{\nu,90}$ over which 90\% of this energy is radiated.  The reason
for the near-equality of old and new results of these quantities is
the fact that the luminosity of heavy-lepton neutrinos and
antineutrinos increases by essentially the same amount as the $\nu_e$
luminosity drops.  The only quantity in
\Tab{explodability_calibration} that shows significant difference is
the fallback mass, $M_\mathrm{fb}$, which is greater in the new
simulation runs by several tens of percent up to a factor of 3 (for
Model W15), except for Model Z9.6, where it is smaller by a factor of
2.  However, these differences are only affecting small
quantities. The fallback masses are only between about
$10^{-4}\,{\mathrm M}_\odot$ and some $10^{-3}\,{\mathrm
  M}_\odot$. The larger fallback masses seem to be connected to our
tracking of the neutrino emission and of the neutrino-driven wind of
the proto-neutron star only for 10\,s after core bounce in the new
simulations instead of 15\,s in the old models. This reduces the push
of the neutrino wind on the inner ejecta and leads to some increase of
the fallback. However, both the neutrino emission and the wind were
artificially enhanced in the old simulations at late times. The
improved transport treatment and the new adaptive mesh refinement in
the proto-neutron star surface layers allow for a higher accuracy with
the upgraded P-HOTB code also at late times after bounce. The new
models show that at times later than 10\,s the neutrino luminosities
are very low and the neutrino-driven wind is correspondingly weak, so
that a termination of the neutrino modeling is justified. We therefore
consider the new results as more reliable.

In the bottom part of \Tab{explodability_calibration} we also present
results obtained for some of the recent binary progenitor models of
SN~1987A from \citet{Men17}. For all listed cases we could find an
engine calibration (named by the single-star progenitor of SN~1987A in
parentheses behind the name of the binary model) for which the
explosion energy and nucleosynthesized mass of Ni+Tr/2 are roughly
compatible with those of SN~1987A. Application of other engines to the
binary models, in particular those of the single-star progenitors N20
and S19.8, leads to overestimated explosion energies and nickel masses
compared to SN~1987A. Only binary Model M15+7b is harder to explode,
and even our strongest single-star engine S19.8 is able to produce
explosion results that are only marginally consistent with
SN~1987A. Therefore, because the binary progenitors explode with
SN~1987A-like properties for our single-star engines, we can consider
them as variants of the already large range of SN~1987A engines applied
in the present paper and in \citet{Suk16} and \citet{Ert16a}.

\Fig{comptransnewold} gives an overview of the outcome of our
simulations with the upgraded P-HOTB code compared to the previous
results of \citet{Suk16} for the whole set of single-star progenitors
considered there. Results for all five central engines are shown.  Red
means supernova explosions with neutron star formation, black marks
black hole formation with no explosion, and blue indicates fallback
supernovae where the outer part of the star is ejected while massive
fallback of inner stellar shells lifts the new-born neutron star
beyond the mass limit for black hole formation. The corresponding
supernova is most of the times fairly weak and, at least in 1D, no
ejection of radioactive nickel is expected.

In the present paper we assume that neutron stars collapse to black
holes when their baryonic mass exceeds 2.75\,M$_\odot$, which
corresponds to a gravitational mass of $\sim$2.18--2.30\,M$_\odot$
according to equation~(36) of \citet{Lat01} for neutron star radii
between 9\,km and 12\,km. This limiting mass for cold neutron stars is
roughly compatible with estimates derived from the gravitational-wave
and kilonova observations of the first detected binary neutron star
merger, GW170817 \citep[see, e.g.,][]{Mar17,Rez18}. For the old
simulations of \citet{Suk16} we did not have to assume any such mass
limit for neutron stars, because the progenitors either exploded and
formed neutron stars with baryonic (gravitational) masses below
2.15\,M$_\odot$ ($\sim$1.8\,M$_\odot$), or either continuous or
fallback accretion increased the baryonic mass of the compact remnant
well beyond 3\,M$_\odot$, clearly classifying it as a black hole.  In
the simulations with the upgraded P-HOTB code we obtain, for all
single-star simulations with all employed central neutrino engines, a
few cases where fallback brings the baryonic remnant mass between
2.5\,M$_\odot$ and 3\,M$_\odot$, namely one near 2.5\,M$_\odot$ (for
the N20 engine), one just above 2.75\,M$_\odot$ (for engine W15), and
two close to 3.0\,M$_\odot$ (one for W15 and one for N20). The first
two cases may be disputable as being either a neutron star or a black
hole, the last two cases are most likely black holes.
 
Comparing the ``landscapes'' of the old and new simulations in
\Fig{comptransnewold}, it is obvious that the patterns with mass
intervals of explosions and neutron star formation alternating with
mass intervals of failed explosions and black hole formation remain
essentially unchanged when applying the upgraded version of
P-HOTB. However, we see that some black hole forming cases in the old
simulations explode now with the new code, forming either neutron
stars or fallback black holes. Similarly, some fallback supernovae in
the old runs convert to explosions with neutron star formation in the
new models. These changes occur mostly near the boundaries between
mass intervals of exploding and non-exploding cases, or for
progenitors that formed islands of successful explosions or failures
as single, isolated cases.  This suggests that progenitors that
marginally evolved in the one or the other direction could flip their
behavior when applying the new version of P-HOTB. The slightly
increased number of explosions with neutron star or fallback black
hole formation in the new runs implies that the new code has the
tendency to produce explosions a bit more readily. Such a trend can be
guessed also from \Tab{explodability_calibration}, where all
calibration models except W18 blow up with a slightly higher explosion
energies in the new runs.

Figs.~\ref{fig:resultsw18transnew} and \ref{fig:resultsn20transnew}
display results of a number of characteristic quantities for our
simulations of the solar-metallicity single-star progenitors of
\citet{Suk16} with the upgraded version of P-HOTB. The two plots can
be directly compared with Figures~8 and 9 in \citet{Suk16}.  The
previous results and the new results look extremely similar, even the
hills and valleys of the landscape pattern display great similarity
except for the slightly larger number of exploding cases in the new
simulations already discussed in connection with
\Fig{comptransnewold}. The main differences of the new runs compared
to the previous ones are, first, the larger number of cases with
massive fallback instead of failed explosions, narrowing some of the
windows of direct black hole formation, and, second, for a dominant
fraction of cases slightly higher explosion energies (by about
0.1\,B), marginally earlier explosions, and a somewhat lower $^{56}$Ni
mass and at the same time an increased tracer mass by roughly the same
amount.

Our conclusion from this comparison is that we can safely compare the
results of the helium-star simulations in the present paper with the
single-star models published by \citet{Suk16}. Although the new version
of P-HOTB improves the numerical performance and efficiency of the code,
the physics results for the exactly same choice of values of the 
neutrino-engine parameters have changed only in inessential details.

\begin{figure*}
\includegraphics[width=.495\textwidth]{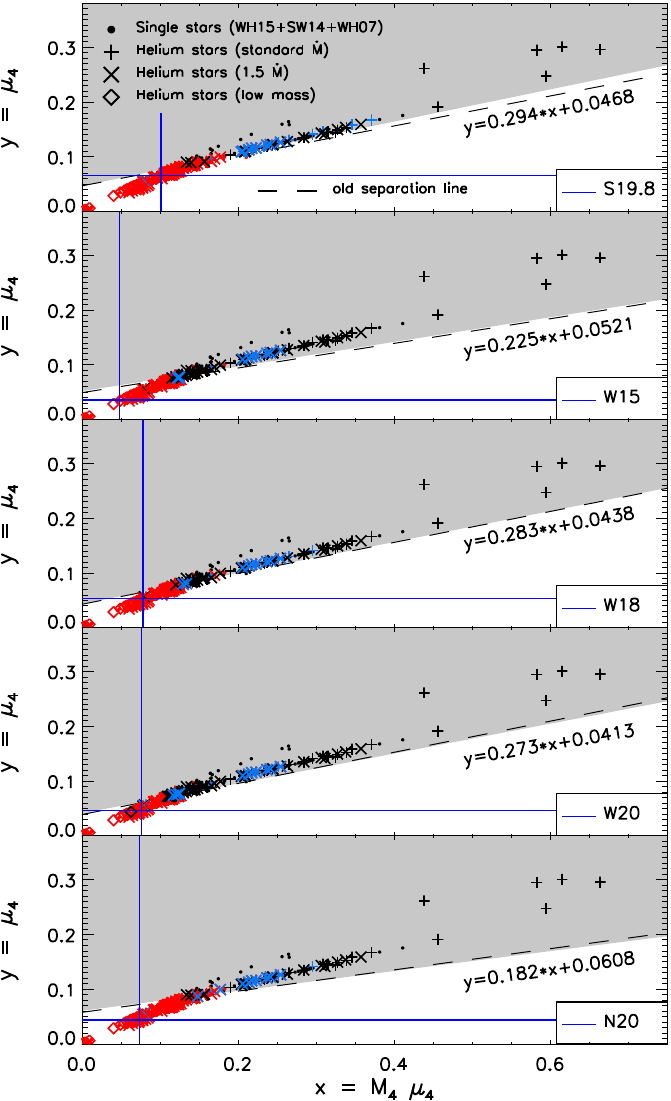} \hskip4.0pt
\includegraphics[width=.495\textwidth]{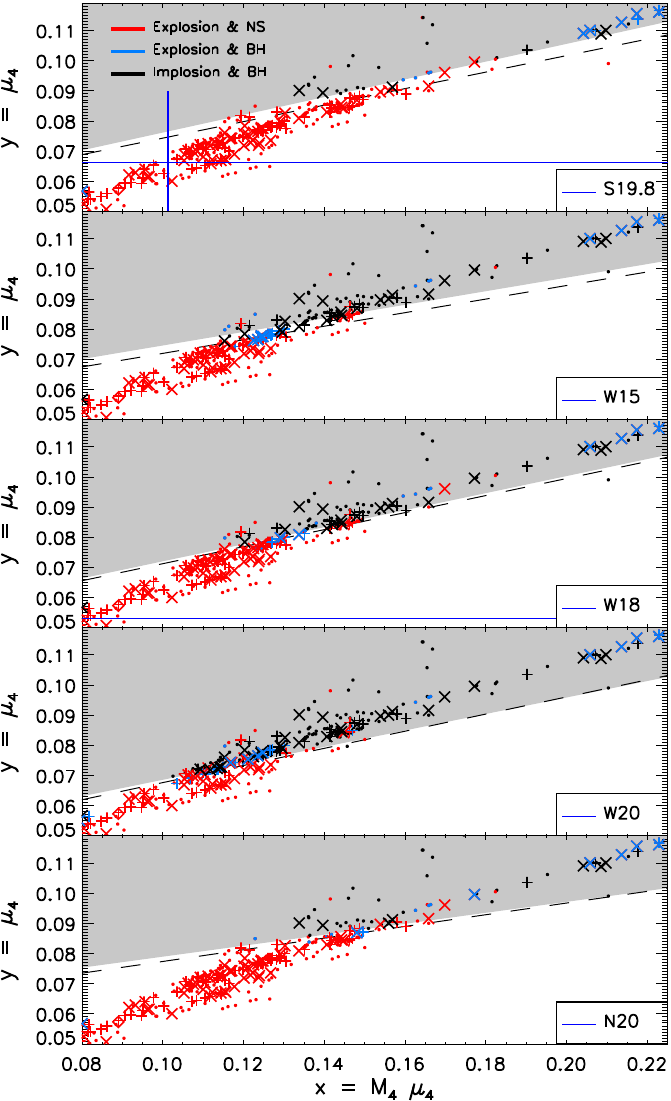}
\caption{Separation curves between supernova explosions with neutron
  star formation (white region, red symbols) and black hole formation
  without supernova explosion (gray region, black symbols) or black
  hole formation in fallback supernovae (gray region, blue symbols) in
  the plane of parameters $x=M_4\mu_4$ and $y=\mu_4$. The different
  panels correspond to the different neutrino engines used for the 1D
  core-collapse modeling. The panels in the right column display
  zoom-ins to the region where the band of high model density crosses
  the separation line. Each simulated progenitor star corresponds to
  one symbol located at the $(x,y)$ position computed from the stellar
  density profile. The different symbols indicate the different sets
  of progenitor stars with dots marking the solar-metallicity
  single-star models investigated by \citet{Suk16} and all other
  symbols belonging to the helium stars studied in the present
  paper. All simulations were performed with the upgraded version of
  the P-HOTB code. The separation lines deduced by \citet{Ert16a} from
  simulations of a large set of single-star models with the old
  version of the P-HOTB code are displayed by dashed black lines. The
  crossing points of the blue solid horizontal and vertical lines
  correspond to the locations of the SN~1987A calibration models in
  the $x$-$y$-plane.}  \lFig{transnewcriterion}
\end{figure*}

\begin{deluxetable}{cccccc}
\tablecolumns{3}
\tabletypesize{\scriptsize}
\tablecaption{
Coefficients of BH-NS separation curves for all neutrino engines
from \citet{Ert16a} (``old'') and from simulations with the upgraded
P-HOTB code (``new'')
}
\tablehead{
  \colhead{Engine Model}                                           &
  \colhead{$k_1{}^{a}$     }                                      &
  \colhead{$k_2{}^{a}$     }                                      &
  \colhead{\msf${}^{b}$    }                                       &
  \colhead{\scc${}^{b}$    }                                       &
  \colhead{\msf\scc${}^{b}$}
  }
\startdata
Z9.6 \& S19.8${}^{c}$ old & 0.274  &  0.0470 & 1.529  & 0.0662  & 0.101 \\
Z9.6 \& S19.8${}^{c}$ new & 0.294  &  0.0468 & 1.529  & 0.0662  & 0.101 \\
\\
Z9.6 \& W15${}^{d}$ old   & 0.225  &  0.0495 & 1.318 & 0.0176  & 0.023 \\
Z9.6 \& W15${}^{d}$ new   & 0.225  &  0.0521 & 1.318 & 0.0176  & 0.023 \\
\\
Z9.6 \& W18 old           & 0.283  &  0.0430 & 1.472  & 0.0530  & 0.078 \\
Z9.6 \& W18 new           & 0.283  &  0.0438 & 1.472  & 0.0530  & 0.078 \\
\\
Z9.6 \& W20 old           & 0.284  &  0.0393 & 1.616  & 0.0469  & 0.076 \\
Z9.6 \& W20 new           & 0.273  &  0.0413 & 1.616  & 0.0469  & 0.076 \\
\\
Z9.6 \& N20 old           & 0.194  &  0.0580 & 1.679  & 0.0441  & 0.074 \\
Z9.6 \& N20 new           & 0.182  &  0.0608 & 1.679  & 0.0441  & 0.074 \\
\enddata
  \tablecomments{
${}^{a}$~Best-fit values for the coefficients of the separation line (Equation~(\ref{eq:sepcurve})) when
$x$ and $y$ are measured at a time when the central stellar density is $5\times 10^{10}$\,g\,cm$^{-3}$.
${}^{b}$~Measured for SN~1987A calibration models when the central stellar density is $5\times 10^{10}$\,g\,cm$^{-3}$.
${}^{c}$~Red supergiant progenitor from the model series of \cite{Woo02}.
${}^{d}$~\msf \ and \scc \ measured roughly at core bounce, because pre-collapse data are not available.
}
\lTab{separation}
\end{deluxetable}

\section{Update of the two-parameter criterion}
\label{app:updatetwoparameter}

Based on a large set of 1D explosion simulations for
solar-metallicity, ultra-metal poor, and zero-metallicity single-star
progenitors, \citet{Ert16a} introduced a two-parameter criterion that
separates cases of supernova explosions with neutron star formation
from cases of black hole formation in a 2D plane that is spanned by
the parameters
\begin{equation}
x = M_4\mu_4 
\label{eq:x}
\end{equation}
and
\begin{equation}
y = \mu_4  \,.
\label{eq:y}
\end{equation}
Here 
\begin{equation}
M_4\equiv m(s=4)/M_\odot
\label{eq:ms4}
\end{equation}
is the mass enclosed by the radius where the dimensionless entropy per 
nucleon has a value of $s=4$, and 
\begin{equation}
\mu_4\equiv\left.\frac{\mathrm{d}m/M_\odot}{\mathrm{d}r/1000\,\mathrm{km}}\right|_{s=4}\,,
\label{eq:dmdr}
\end{equation}
is the derivative of the enclosed mass at this location. 
Both parameters are determined
from the density profiles of the presupernova star and thus measure
properties of the supernova progenitor, which means that each occupies one point
in the $x$-$y$-plane. \citet{Ert16a} found that the explosion
behavior of the vast majority (over 97\%) of their simulated models could
be correctly predicted by the following criterion. For low values of $y$, i.e.\
below an inclined line 
\begin{equation}
y_\mathrm{sep}(x)=k_1\cdot x+k_2\,,
\label{eq:sepcurve}
\end{equation}
in the $x$-$y$-plane, almost all stars explode and form neutron stars,
whereas above this line and therefore for high values of $y$ basically
all stars collapse to black holes. \citet{Ert16a} explained this
finding by demonstrating that the separation line can be interpreted
as a correspondent of a generalized form of the critical luminosity
condition for explosions introduced by \citet{Bur93}. \citet{Ert16a}
argued that $x = M_4\mu_4$ determines the neutrino luminosity, which
has a strong accretion contribution, at the time when the explosion
sets in, and $y = \mu_4$ scales with the mass accretion rate $\dot M$
at this time. They supported this argument through very tight
correlations between these quantities seen in their numerical
models. Therefore the separation line $y_\mathrm{sep}(x)$ can be
associated with the critical luminosity, $L_{\nu,\mathrm{crit}}(\dot
M)$, of \citet{Bur93}.  The line is inclined in the $x$-$y$-plane with
a positive slope $k_1$, because higher values of the mass accretion
rate and thus $y$ also lead to higher values of the neutrino
luminosity and thus $x$, which leads to stronger neutrino heating and
therefore an enlarged range of $y$-values where explosions are
possible.

In \Fig{transnewcriterion} we display the separation curves
$y_\mathrm{sep}(x)$ for the five neutrino engines used in our study,
applied to all solar-metallicity single-star models investigated by
\citet{Suk16} and to the two sets of helium-star models considered in
the present paper.  All 1D simulations were conducted with the
upgraded new version of P-HOTB.  \Fig{transnewcriterion} can be
directly compared with Figure~8 of \citet{Ert16a}. The white lower
part of each panel is the region of supernova explosions with neutron
star formation, the upper, gray-shaded part of each panel highlights
the region where black hole formation occurs.  The ``old'' separation
curves, indicated by a dashed black line in each panel of
\Fig{transnewcriterion}, correspond to the gray-white boundaries in
the panels of Figure~8 of the \citet{Ert16a} paper.  In
\Tab{separation} the values for the slope coefficient $k_1$ and the
shift coefficient $k_2$ are listed for all of the employed neutrino
engines, comparing the optimal values derived from the present
simulations with the values provided by \citet{Ert16a} in their Table~2.

The best-fit values of $k_1$ and $k_2$ are nearly unchanged, and the
old and new separation lines differ only very little.  In the cases of
the S19.8 and W15 engines the zoom-ins of the right panels in
\Fig{transnewcriterion} reveal the biggest differences, caused by a
slight increase of $k_1$ for S19.8 and a slight increase of $k_2$ for
W15. In the case of W18 the old and new values for both coefficients
are effectively identical, and for W20 and N20 a minor change of one
coefficient combined with the change of the other coefficient yields a
hardly visible shift of the separation line in the region of interest.

The accuracy of predictions of explosion or non-explosion of stellar
progenitors with the revised separation lines is similarly good as
with the old ones. For the S19.8 engine the fraction of
misclassifications is as low as 1.8\%, for W15 it is 5\% (but these
cases lie de facto on the separation line), for W18 we find 3.3\%, for
W20 4.8\%, and for N20 2.6\%. Therefore in total again about 97\% of
all cases are correctly predicted in their explosion behavior by our
$y_\mathrm{sep}(x)$ criterion.

It is worth to note that both the black hole formation cases without
supernova explosions and the black hole formation cases due to massive
fallback (defined as in the caption of \Fig{comptransnewold}), which
occur in fallback supernovae and are marked by blue symbols in
\Fig{transnewcriterion}, are chosen to lie above the separation
curves, i.e. in the black hole region. These cases barely explode, and
their behavior is closer to failed explosions than to successful
supernovae with neutron star formation.

\clearpage



\clearpage


\newpage 
\begin{thebibliography}{99}
\expandafter\ifx\csname natexlab\endcsname\relax\def\natexlab#1{#1}\fi

\bibitem[Abbott et al.(2018)]{Abb18}
 Abbott, B.~P., Abbott, R., Abbott, T.~D., et al.\ 2018, \prl, 121, 161101
 
\bibitem[Amarsi et al.(2019)]{Ama19} 
Amarsi, A.~M., Nissen, P.~E., \& Sk{\'u}lad{\'o}ttir, {\'A}.\ 2019, 
arXiv e-prints, arXiv:1908.10319

\bibitem[Anderson(2019)]{And19}
Anderson, J.~P.\ 2019, \aap, 628, A7

\bibitem[Antoniadis et al.(2013)]{Ant13}
 Antoniadis, J., Freire, P.~C.~C., Wex, N., et al.\ 2013, Science, 340, 448

\bibitem[Antoniadis et al.(2016)]{Ant16} 
Antoniadis, J., Tauris, T.~M., Ozel, F., et al.\ 2016, 
arXiv e-prints, arXiv:1605.01665

\bibitem[Arcones et al.(2007)]{Arc07}
Arcones, A., Janka, H.-T., \& Scheck, L.\ 2007, \aap, 467, 1227

\bibitem[Barbuy et al.(2018)]{Bar18} 
Barbuy, B., Chiappini, C., \& Gerhard, O.\ 2018, \araa, 56, 223

\bibitem[Bauswein et al.(2017)]{Bau17}
  Bauswein, A., Just, O., Janka, H.-T., et al.\ 2017, \apjl, 850, L34

\bibitem[Belczynski et al.(2012)]{Bel12}
Belczynski, K., Wiktorowicz, G., Fryer, C.~L., et al.\ 2012,
\apj, 757, 91

\bibitem[Beniamini et al.(2019)]{Ben19}
Beniamini, P., Hotokezaka, K., van der Horst, A., et al.\ 2019,
\mnras, 1336

\bibitem[Branch, \& Wheeler(2017)]{Bra17}
Branch, D., \& Wheeler, J.~C.\ 2017, Supernova Explosions: Astronomy
and Astrophysics Library

\bibitem[Bruenn et al.(2016)]{Bru16}
Bruenn, S.~W., Lentz, E.~J., Hix, W.~R., et al.\ 2016, \apj, 818, 123

\bibitem[{Burrows \& Goshy(1993)}]{Bur93}
Burrows, A., \& Goshy, J. 1993, ApJL, 416, L75

\bibitem[Chatzopoulos et al.(2013)]{Cha13}
Chatzopoulos, E., Wheeler, J.~C., Vinko, J., et al.\ 2013, \apj, 773,
76

\bibitem[Chen et al.(2016)]{Che16}
Chen, K.-J., Woosley, S.~E., \& Sukhbold, T.\ 2016, \apj, 832, 73

\bibitem[De et al.(2018)]{De18} 
De, K., Kasliwal, M.~M., Ofek, E.~O., et al.\ 2018, Science, 362, 201 

\bibitem[De Marco \& Izzard(2017)]{Dem17} 
De Marco, O., \& Izzard, R.~G.\ 2017, Pub. Astron. Soc. Australia, 34, 
e001 

\bibitem[Dessart et al.(2012)]{Des12} 
Dessart, L., Hillier, D.~J., Li, C., \& Woosley, S.\ 2012, \mnras,
424, 2139

\bibitem[Dessart et al.(2015)]{Des15}
 Dessart, L., Hillier, D.~J., Woosley, S., et al.\ 2015, \mnras, 453, 2189 

\bibitem[Drout et al.(2011)]{Dro11}
Drout, M.~R., Soderberg, A.~M., Gal-Yam, A., et al.\ 2011,
\apj, 741, 97

\bibitem[Drout et al.(2016)]{Dro16} 
Drout, M.~R., Milisavljevic, D., Parrent, J., et al.\ 2016, \apj, 821, 57

\bibitem[Ebinger et al.(2019)]{Ebi19}
Ebinger, K., Curtis, S., Fr{\"o}hlich, C., et al.\ 2019, \apj, 870, 1

\bibitem[Ensman \& Woosley(1988)]{Ens88} 
Ensman, L.~M., \& Woosley, S.~E.\ 1988, \apj, 333, 754

\bibitem[Ertl et al.(2016a)]{Ert16a}
Ertl, T., Janka, H.-T., Woosley, S.~E., et al.\ 2016a, \apj, 818, 124

\bibitem[Ertl et al.(2016b)]{Ert16b}
Ertl, T., Ugliano, M., Janka, H.-T., et al.\ 2016b, \apj, 821, 69

\bibitem[Farr et al.(2011)]{Far11}
Farr, W.~M., Sravan, N., Cantrell, A., et al.\ 2011, \apj, 741, 103

\bibitem[Fryer(1999)]{Fry99}
Fryer, C.~L.\ 1999, \apj, 522, 413

\bibitem[Fryer \& Kalogera(2001)]{Fry01}
Fryer, C.~L. \& Kalogera, V.\ 2001, \apj, 554, 548

\bibitem[Fryer et al.(2012)]{Fry12} 
Fryer, C.~L., Belczynski, K., Wiktorowicz, G., et al.\ 2012, \apj, 749, 91 

\bibitem[Fryxell et al.(1989)]{Fry89}
Fryxell, B., M{\"u}ller, E., \& Arnett, D. 1989, in Proceedings of the 5$^\mathrm{th}$ Workshop on Nuclear Astrophysics, ed. W. Hillebrandt \&  E. M{\"u}ller, 100

\bibitem[Fryxell et al.(1991)]{Fry91}
Fryxell, B., M{\"u}ller, E., \& Arnett, D.\ 1991, \apj, 367, 619 

\bibitem[Fulbright et al.(2007)]{Ful07} 
Fulbright, J.~P., McWilliam, A., \& Rich, R.~M.\ 2007, \apj, 661, 1152

\bibitem[Gessner \& Janka(2018)]{Ges18}
Gessner, A., \& Janka, H.-T.\ 2018, \apj, 865, 61

\bibitem[Griffith et al.(2019)]{Gri19}
Griffith, E., Johnson, J.~A., \& Weinberg, D.~H.\ 2019,
arXiv e-prints, arXiv:1908.06113

\bibitem[Hammer et al.(2010)]{Ham10}
Hammer, N.~J., Janka, H.-T., \& M{\"u}ller, E.\ 2010, \apj, 714, 1371

\bibitem[Heger et al.(2005)]{Heg05}
Heger, A., Woosley, S.~E., \& Spruit, H.~C.\ 2005, \apj, 626, 350

\bibitem[Hillier \& Dessart(2012)]{Hil12}
Hillier, D.~J., \& Dessart, L.\ 2012, \mnras, 424, 252 

\bibitem[Janka(2017)]{Jan17}
Janka, H.-Th.\ 2017, in: Handbook of Supernovae,
eds.\ A.~Alsabti and P.~Murdin (Springer International
Publishing AG), p.~1095

\bibitem[Janka \& M\"uller(1996)]{Jan96}
Janka, H.-T. \& M\"uller, E.\ 1996, \aap, 306, 167

\bibitem[Jerkstrand et al.(2018)]{Jer17b}
Jerkstrand, A., Ertl, T., Janka, H.-T., et al.\ 2018, \mnras, 475, 277

\bibitem[Jerkstrand et al.(2017)]{Jer17}
Jerkstrand, A., Smartt, S.~J., Inserra, C., et al.\ 2017, \apj, 835,
13

\bibitem[Kasen \& Bildsten(2010)]{Kas10}
Kasen, D. \& Bildsten, L.\ 2010, \apj, 717, 245

\bibitem[Kifonidis et al.(2003)]{Kif03}
Kifonidis, K, Plewa, T., Janka, H.-T., \& M\"uller, E.\ 2003,
\aap, 408, 621

\bibitem[Kleiser et al.(2018)]{Kle18} 
Kleiser, I.~K.~W., Kasen, D., \& Duffell, P.~C.\ 2018, \mnras, 475, 3152 

\bibitem[Kreidberg et al.(2012)]{Kre12} 
Kreidberg, L., Bailyn, C.~D., Farr, W.~M., \& Kalogera, V.\ 2012, \apj, 757, 36 

\bibitem[Langer(2012)]{Lan12}
Langer N., 2012, ARA\&A, 50, 107

\bibitem[Lattimer \& Prakash (2001)]{Lat01}
Lattimer, J. M., \& Prakash, M. 2001 \apj, 550, 426

\bibitem[{Lattimer \& Swesty(1991)}]{Lat91}
Lattimer, J.~M., \& Swesty, F.~D. 1991, NuPhA, 535, 331

\bibitem[Lattimer(2012)]{Lat12} Lattimer, J.~M.\ 2012, Annual 
Review of Nuclear and Particle Science, 62, 485 

\bibitem[Limongi \& Chieffi(2006)]{Lim06}
Limongi, M., \& Chieffi, A. 2006, ApJ, 647, L483

\bibitem[Liebend\"orfer(2005)]{Lie05}
Liebend\"orfer, M.\ 2005, \apj, 633, 1042

\bibitem[Lodders(2003)]{Lod03}
 Lodders, K.\ 2003, \apj, 591, 1220

\bibitem[Lyman et al.(2016)]{Lym16}
Lyman, J.~D., Bersier, D., James, P.~A., et al.\ 2016, \mnras,
457, 328
 
\bibitem[Marek \& Janka(2009)]{Mar09}
Marek, A., \& Janka, H.-T.\ 2009, \apj, 694, 664

\bibitem[Margalit \& Metzger(2017)]{Mar17}
Margalit, B., \& Metzger, B.~D.\ 2017, \apjl, 850, L19

\bibitem[Mart{\'\i}nez-Pinedo et al.(2012)]{Mar12} 
Mart{\'\i}nez-Pinedo, G., Fischer, T., Lohs, A., et al.\ 2012, 
\prl, 109, 251104

\bibitem[Maund(2018)]{Mau18}
Maund, J.~R.\ 2018, \mnras, 476, 2629

\bibitem[Maurer et al.(2010)]{Mau10}
Maurer, J.~I., Mazzali, P.~A., Deng, J., et al.\ 2010, \mnras,
402, 161

\bibitem[Menon \& Heger(2017)]{Men17}
Menon, A., \& Heger, A.\ 2017, \mnras, 469, 4649 

\bibitem[Mirizzi et al.(2016)]{Mir16} 
Mirizzi, A., Tamborra, I., Janka, H.-T., et al.\ 2016, Nuovo Cimento 
Rivista Serie, 39, 1

\bibitem[M\"uller(2015)]{Mul15}
M\"uller, B.\ 2015, \mnras, 453, 287

\bibitem[M{\"u}ller et al.(2018)]{Mue18}
M{\"u}ller, B., Gay, D.~W., Heger, A., et al.\ 2018, \mnras, 479, 3675

\bibitem[M{\"u}ller et al.(2016)]{Mue16}
M{\"u}ller, B., Heger, A., Liptai, D., et al.\ 2016, \mnras, 460, 742


\bibitem[M{\"u}ller et al.(2017)]{Mue17}
M{\"u}ller, B., Melson, T., Heger, A., et al.\ 2017, \mnras, 472, 491

\bibitem[M\"uller(1986)]{Mul86}
M\"uller, E. 1986, \aap, 162, 103

\bibitem[M\"uller et al.(1991)]{Mue91}
M\"uller, E., Fryxell, B., \& Arnett, D.\ 1991, \aap, 251, 505 

\bibitem[Nicholl et al.(2017)]{Nic17}
Nicholl, M., Guillochon, J., \& Berger, E.\ 2017, \apj, 850, 55

\bibitem[O'Connor \& Ott(2011)]{Oco11}
 O'Connor, E., \& Ott, C.~D.\ 2011, \apj, 730, 70

\bibitem[{\"O}zel et al.(2010)]{Oze10}
{\"O}zel, F., Psaltis, D., Narayan, R., \& McClintock, J. E. 2010, \apj, 725, 1918

\bibitem[{\"O}zel \& Freire (2016)]{Oze16} 
{\"O}zel, F., \& Freire, P.\ 2016, ARAA, 54, 401

\bibitem[Pastorello et al.(2008)]{Pas08}
Pastorello, A., Mattila, S., Zampieri, L., et al.\ 2008, \mnras, 389,
113

\bibitem[Pastorello et al.(2016)]{Pas16}
Pastorello, A., Wang, X.-F., Ciabattari, F., et al.\ 2016,
\mnras, 456, 853

\bibitem[Peebles(1993)]{Pee93}
Peebles, P.J. E. 1993, {\sl Principles of Physical Cosmology},
Princeton University Press, p. 131

\bibitem[Pejcha \& Thompson(2015)]{Pej15b}
 Pejcha, O., \& Thompson, T.~A.\ 2015, \apj, 801, 90 

\bibitem[Podsiadlowski(1992)]{Pod92}
Podsiadlowski, P.\ 1992, \pasp, 104, 717

\bibitem[Prentice et al.(2016)]{Pre16}
Prentice, S.~J., Mazzali, P.~A., Pian, E., et al.\ 2016, \mnras,
458, 2973

\bibitem[Prentice et al.(2019)]{Pre19}
Prentice, S.~J., Ashall, C., James, P.~A., et al.\ 2019, \mnras, 485,
1559

\bibitem[Pruet et al.(2006)]{Pru06}
Pruet, J., Hoffman, R.~D., Woosley, S.~E., et al.\ 2006, \apj, 644,
1028

\bibitem[Raithel et al.(2018)]{Rai18}
Raithel, C.~A., Sukhbold, T., \& {\"O}zel, F.\ 2018, \apj, 856, 35

\bibitem[Roberts(2012)]{Rob12} 
Roberts, L.~F.\ 2012, \apj, 755, 126

\bibitem[Rezzolla et al.(2018)]{Rez18}
Rezzolla, L., Most, E.~R., \& Weih, L.~R.\ 2018, \apjl, 852, L25

\bibitem[{{Salpeter}(1955)}]{Sal55}
{Salpeter}, E.~E. 1955, \apj, 121, 161

\bibitem[Sana \& Evans(2011)]{San11} 
Sana, H., \& Evans, C.~J.\ 2011, Active OB Stars: Structure,
Evolution, Mass Loss, and Critical Limits, 272, 474

\bibitem[Sana et al.(2012)]{San12} 
Sana, H., de Mink, S.~E., de Koter, A., et al.\ 2012, Science, 337,
444

\bibitem[Scheck et al.(2006)]{Sch06}
Scheck, L., Kifonidis, K., Janka, H.-T., M\"uller, E.\ 2006, \aap,
457, 963

\bibitem[Schneider et al.(2019)]{Sch19}
Schneider, A.~S., Roberts, L.~F., Ott, C.~D., et al.\ 2019, arXiv e-prints, arXiv:1906.02009

\bibitem[Schwab et al.(2010)]{Sch10}
 Schwab, J., Podsiadlowski, P., \& Rappaport, S.\ 2010, \apj, 719, 722

\bibitem[Shiode \& Quataert(2014)]{Shi14} 
Shiode, J.~H., \& Quataert, E.\ 2014, \apj, 780, 96

\bibitem[{{Smartt}(2009)}]{Sma09}
{Smartt}, S.~J. 2009, \araa, 47, 63

\bibitem[Smartt(2015)]{Sma15} 
Smartt, S.~J.\ 2015, Pub. Astron. Soc. Australia, 32, e016

\bibitem[Suntzeff et al.(1992)]{Sun92}
Suntzeff, N. B., Phillips, M. M., Elias, J. H., Walker, A. R., \&
Depoy, D. L.\ 1992, \apj, 384, L33

\bibitem[Steiner et al.(2013)]{Ste13} 
Steiner, A.~W., Hempel, M., \& Fischer, T.\ 2013, \apj, 774, 17

\bibitem[Sukhbold \& Woosley(2014)]{Suk14}
 Sukhbold, T., \& Woosley, S.~E.\ 2014, \apj, 783, 10

\bibitem[Sukhbold et al.(2016)]{Suk16}
Sukhbold, T., Ertl, T., Woosley, S.~E., et al.\ 2016, \apj, 821, 38

\bibitem[Sukhbold et al.(2018)]{Suk18}
Sukhbold, T., Woosley, S.~E., \& Heger, A.\ 2018, \apj, 860, 93

\bibitem[Sukhbold \& Adams(2019)]{Suk19}
Sukhbold, T., \& Adams, S.\ 2019, arXiv e-prints, arXiv:1905.00474

\bibitem[Taddia et al.(2019)]{Tad19}
Taddia, F., Sollerman, J., Fremling, C., et al.\ 2019, \aap, 621, A71

\bibitem[Ugliano et al.(2012)]{Ugl12}
 Ugliano, M., Janka, H.-T., Marek, A., \& Arcones, A.\ 2012, \apj, 757, 69

\bibitem[Utrobin et al.(2015)]{Utr15}
Utrobin, V.~P., Wongwathanarat, A., Janka, H.-T., \& M{\"u}ller, E.\ 2015, \aap, 581, A40 

\bibitem[Utrobin et al.(2019)]{Utr19}
Utrobin, V.~P., Wongwathanarat, A., Janka, H.-T., et al.\ 2019, \aap, 624, A116 

\bibitem[Valentim et al.(2011)]{Val11} 
Valentim, R., Rangel, E., \& Horvath, J.~E.\ 2011, \mnras, 414, 1427

\bibitem[Villar et al.(2018)]{Vil18}
Villar, V.~A., Nicholl, M., \& Berger, E.\ 2018, \apj, 869, 166

\bibitem[Vreeswijk et al.(2017)]{Vre17}
Vreeswijk, P.~M., Leloudas, G., Gal-Yam, A., et al.\ 2017, \apj, 835,
58

\bibitem[Weaver et al.(1978)]{Wea78}
Weaver, T.~A., Zimmerman, G.~B., \& Woosley, S.~E.\ 1978, \apj, 225,1021

\bibitem[Weinberg et al.(2019)]{Wei19}
Weinberg, D.~H., Holtzman, J.~A., Hasselquist, S.,
et al.\ 2019, \apj, 874, 102

\bibitem[Wellstein \& Langer(1999)]{Wel99}
Wellstein, S., \& Langer, N.\ 1999, \aap, 350, 148

\bibitem[Wongwathanarat et al.(2010a)]{Won10a}
Wongwathanarat, A., Hammer, N.~J., \& M{\"u}ller, E.\ 2010, \aap, 514, A48

\bibitem[Wongwathanarat et al.(2010b)]{Won10b}
Wongwathanarat, A., Janka, H.-T., \& M{\"u}ller, E.\ 2010, \apjl, 725, L106 

\bibitem[Wongwathanarat et al.(2013)]{Won13}
Wongwathanarat, A., Janka, H.-T., \& M{\"u}ller, E.\ 2013, \aap, 552, A126

\bibitem[Wongwathanarat et al.(2015)]{Won15}
Wongwathanarat, A., M{\"u}ller, E., \& Janka, H.-T.\ 2015, \aap, 577, A48 

\bibitem[Wongwathanarat et al.(2017)]{Won17}
Wongwathanarat, A., Janka, H.-T., M{\"u}ller, E., Pllumbi, E., \&
Wanajo, S.\ 2017, \apj, 842, 13

\bibitem[Woosley(2010)]{Woo10}
Woosley, S.~E.\ 2010, \apj, 719, L204

\bibitem[Woosley(2017)]{Woo17}
Woosley, S.~E.\ 2017, \apj, 836, 244
  
\bibitem[Woosley(2019)]{Woo19} 
Woosley, S.~E.\ 2019, \apj, 878, 49

\bibitem[Woosley et al.(1988)]{Woo88}
Woosley, S.~E., Pinto, P.~A., \& Ensman, L.\ 1988, \apj, 324, 466

\bibitem[Woosley et al.(2002)]{Woo02}
Woosley, S.~E., Heger, A., \& Weaver, T.~A.\ 2002, Reviews of Modern
Physics, 74, 1015

\bibitem[Woosley \& Heger(2007)]{Woo07}
Woosley, S.~E., \& Heger, A.\ 2007, \physrep, 442, 269

\bibitem[Woosley \& Heger(2015)]{Woo15}
Woosley, S.~E., \& Heger, A.\ 2015, \apj, 810, 34 

\bibitem[Yoon et al.(2010)]{Yoo10}
Yoon, S.-C., Woosley, S.~E., \& Langer, N.\ 2010, \apj, 725, 940

\bibitem[Yoon(2017)]{Yoo17} 
Yoon, S.-C.\ 2017, \mnras, 470, 3970

\end{thebibliography}
\end{document}